\PassOptionsToPackage{table}{xcolor}

\documentclass[acmtog,nonacm]{acmart}

\usepackage{booktabs} 
\usepackage[utf8]{inputenc}
\usepackage{float}
\usepackage{afterpage}
\usepackage{amsmath}
\usepackage{color}
\usepackage{algorithm}
\usepackage{multicol}
\usepackage{makecell}
\usepackage{multirow}
\usepackage{array}
\usepackage{subfig}
\usepackage{stfloats}
\usepackage{graphicx}
\usepackage{tikz}
\usepackage{enumitem}
\usepackage{fontawesome5}
\usepackage{manfnt}
\usepackage[export]{adjustbox}
\usepackage[font=small,skip=5pt]{caption}
\usepackage[noend]{algpseudocode}
\usepackage{soul}
\usepackage{xcolor}   

\newcommand{\bC}{\mathbf{C}}

\newcommand{\bI}{\mathbf{I}}

\newcommand{\cG}{\mathcal{G}}

\newcommand{\cL}{\mathcal{L}}

\newcommand{\cR}{\mathcal{R}}

\usepackage{amsfonts}
\usepackage{xspace}     

\makeatletter
\DeclareRobustCommand\onedot{\futurelet\@let@token\@onedot}
\def\@onedot{\ifx\@let@token.\else.\null\fi\xspace}

\makeatother

\usepackage{xcolor}
\definecolor{darkred}{rgb}{0.7,0.2,0.1}
\definecolor{darkgreen}{rgb}{0.2,0.7,0}
\definecolor{orange}{RGB}{255,127,0}
\definecolor{ourpurple}{RGB}{127,127,204}
\definecolor{ourteal}{RGB}{60,160,160}
\definecolor{palgreen}{RGB}{51,179,179}
\definecolor{magenta}{RGB}{199,21,133}
\definecolor{olive}{RGB}{100,150,85}

\newcommand{\ourmodel}{GLOSS}

\AtBeginDocument{%
  }

\setcitestyle{square}

\newif\ifcomments
\commentsfalse

\begin{document}


\title{{\ourmodel}: Geometric Local Self-Similarity Learning for Faithful Reference-Guided Texture Fill}

\author{Chenyue Cai}
\affiliation{%
  \institution{Princeton University}
  \city{Princeton}
  \country{USA}}
\authornote{Work done during an internship at NVIDIA.}

\author{Anita Hu}
\affiliation{%
  \institution{NVIDIA}
  \city{Toronto}
  \country{Canada}}

\author{James Lucas}
\affiliation{%
  \institution{NVIDIA}
  \city{London}
  \country{UK}
}

\author{Szymon Rusinkiewicz}
\affiliation{%
 \institution{Princeton University}
 \city{Princeton}
 \country{USA}}

\author{Maria Shugrina}
\affiliation{%
  \institution{NVIDIA}
  \city{Toronto}
  \country{Canada}}
\affiliation{%
  \institution{University of Toronto}
  \city{Toronto}
  \country{Canada}}
   \authornote{In support of more inclusive conference locations, this author removed herself from the SIGGRAPH Asia 2026 publication.}
\authorsaddresses{}

\renewcommand{\shortauthors}{Cai, Hu, Lucas, Rusinkiewicz, and Shugrina}


\begin{abstract}

Using conditional image generators, texture artists can explore many single-view looks for an existing 3D shape. Despite impressive progress, state-of-the-art generative methods still struggle to generate a full object texture while closely adhering to fine scale geometric detail and single view references, leaving little room for artist guidance. Furthermore, current automatic models lack the flexibility for artists to explore multiple textures from varied sources in an interactive and controllable manner. Unlike methods trained on large 3D datasets that generate full object textures from global guidance, our work explores a local and less data-hungry approach to texture with explicit artist control. 
We leverage the geometric self-similarity and geometry-texture correlation existing in many natural and man-made shapes, and train a shape-specific local texture generation and completion model. This model learns from existing image model priors and a single 3D shape, and is guided by attending to a set of geometry-aware reference patches. 
The trained shape-specific
network can transfer any novel reference to the full target object texture through patchwise inpainting.
We show improved or comparable quality to strong image-conditioned texture generation baselines,
suggesting local texturing as a promising research direction.
Our model also enables local geometry-conditioned texture inpainting, guided by artist-selected references, and generalizes to PBR materials and unseen meshes for texture transfer. We piloted our novel texture fill capability as a Blender add-on with several 3D texturing professionals who reported positive feedback on the model’s controllability, practical usefulness, and creative affordances. We will release the code, the model checkpoint and Blender add-on on project page: \url{https://chenyuecai.github.io/gloss-page/}.

\end{abstract}



\begin{teaserfigure}
\centering
  \includegraphics[width=0.85\textwidth]{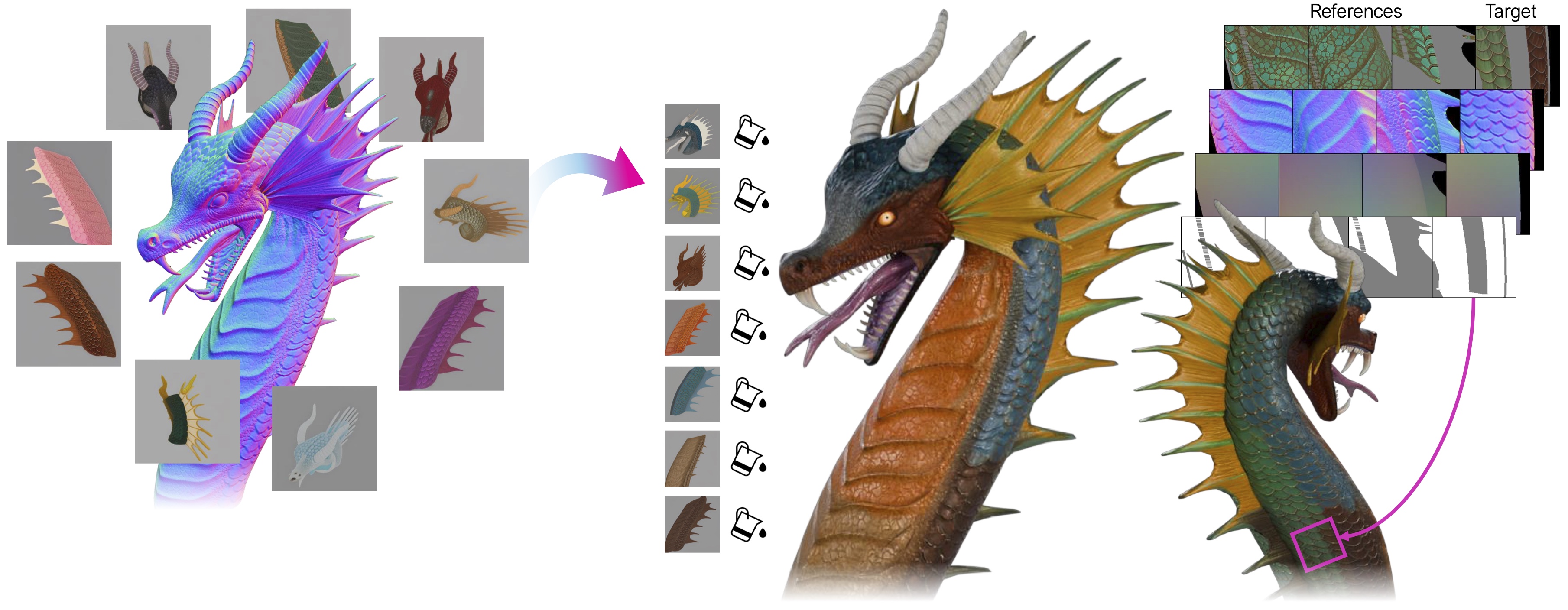}
  \caption{We train a {\ourmodel} network on many single-view generated images of one untextured 3D shape (left), allowing it to learn how to texture local shape regions by attending to local geometry (such as scales) and
  faithfully following conditional texture patches. Once trained, artists can use {\ourmodel} to mix and match any number of references for interactive texture completion of the target shape (right), enabled through reference-guided local texture fill. We also show fully automated texturing, PBR channel completion, and transfer results across shapes. }
  \label{fig:teaser}
\end{teaserfigure}

\maketitle

\section{INTRODUCTION}


Recently, new workflows for texturing 3D shapes are emerging through the application of generative models, but the control and flexibility of existing approaches remain limited.
Generative image models conditioned on normals and other channels (e.g.\ \cite{zhang2023controlnet,ye2023ipadapter}) and configurable pipelines like ComfyUI \cite{comfyui} allow artists to explore diverse single-view renderings of a given 3D shape, respecting its geometric details (e.g., generated views in Fig.~\ref{fig:teaser} follow the scales on the dragon skin). However, building consistent and faithful textures for whole meshes from such single-view references remains challenging.
While recent texture generative models \cite{deng2024flashtex, zhang2024dreammat, yu2024texgen, hunyuan3d2025hunyuan3d, xiang2025native} achieve compelling results and often support single image conditioning, these methods are not designed to preserve local details
of the reference. The global nature of these techniques precludes local artist guidance or editing and limits overall texture resolution regardless of geometry scale. In contrast, techniques tackling local and artist-guided texture generation \cite{hu2024diffusion,decatur20243d} ignore the geometric details of the target geometry, making them less suitable for realistic use cases. Our work addresses reference-guided workflows in texturing, enabling greater local control and geometry-texture consistency, while circumventing the need for large-scale 3D datasets. We focus on showing promising results of artistic control and the underexplored small-data training, both highly relevant to production workflows, where custom models trained in-house on small task-specific data are common in startups and studios (e.g., InterPositive AI acquired by Netflix, Cuebric, Corridor Digital, Wonder Dynamics). 

We introduce {\ourmodel}, a technique that learns local texturing priors from a single shape's self-similarity and off-the-shelf image generative models. Once trained, the shape-specific texture generation
network can automatically texture the target shape using any novel reference view by progressively synthesizing texture patches across the surface, attending to local geometric and appearance references. {\ourmodel} enables novel interactive texturing workflows, wherein artists in-fill textures for any selected regions on the target shape, using any combination of references. For example, an artist might use one part of a reference view to texture the fins of a fish, but another for the body. Trained to inpaint, the local texturing model generates seamless and geometry-aware transitions between artist-specified regions. Thus, artists can edit or iteratively create textures. Moreover, unlike global texture generation techniques, patch-based texture generation allows our model to produce textures of any resolution. Preliminary results show {\ourmodel} outperforms or matches strong generative baselines trained on large 3D data sets, both quantitatively and qualitatively. 
Indeed, 3D professionals using a prototype Blender add-on in a pilot study reported that {\ourmodel} complements their existing workflows and could be used regularly.

Our work is inspired by classical patch-based synthesis, but with improved blending and natural variations shown by patch-based generative models \cite{hu2024diffusion}. 
Given an input shape $M$, we leverage image generative models to automatically create shape-specific training data. 
Despite their diversity, such individual views (Fig.~\ref{fig:teaser}) exhibit strong local self-similarity for many classes of shapes, including natural (plants, amphibians and reptiles, terrain) and man-made (pottery, objects with weathering effects, architecture). We leverage this insight to train a local texture inpainting model that learns to generate texture patches in one area, while attending to references from another area of the same single-view rendering using \emph{batch multi-attention}. At run-time, this allows the model to closely follow multiple reference patches when generating and inpainting new textures. Unlike existing patch-based generative models, {\ourmodel} generates textures conditioned on local geometry like bumps and wrinkles, and attends to such correlations in the provided references, ensuring faithful and controllable local texture generation. This enables our method to support previously impossible reference-guided texture-fill operations on the mesh. Fine-tuning for a similar class of objects can quickly adapt a model to new meshes.

To summarize, the contributions of our work include: 
\begin{itemize}
  \item a novel geometry-conditioned texture inpainting model with batch multi-attention that learns from geometry-texture self-similarity within a single 3D object.
  \item the design of a data-generation pipeline for training a local texturing model without any textured 3D models.
  \item support for novel interactive generative texture fill with \emph{close adherence to reference patches} and \emph{strong geometry-texture consistency}, with preliminary zero-shot transfer on unseen meshes and PBR materials.
  \item a patch-based mesh texture generation pipeline showcasing global consistency and competitive results against generative texturing models trained on large 3D datasets, demonstrating the promise of patch-based generative texturing.
\end{itemize}
Following related work (\S\ref{sec:related}), we detail our local texturing model design and data generation in \S\ref{sec:model}. In \S\ref{sec:applications}, we show applications for automatic and interactive texturing, and evaluate results in \S\ref{sec:results}. 


\section{RELATED WORK}\label{sec:related}

\paragraph{Synthesis via self-similarity}
Classical texture synthesis methods on 3D surfaces blend patches in a coarse-to-fine manner while traversing the mesh geometry \cite{turk2001texture, praun2000lapped, knoppel2015stripe}. Similarly, symmetries have been exploited in classical shape analysis \cite{pauly2008discovering, harary2014context}.
Recent approaches extend patch-based synthesis ideas into learning-based frameworks for 3D shape understanding, generation and reconstruction \cite{berkiten2017learning, liang2022meshmae, erler2020points2surf, tretschk2020patchnets, fogarty2025self}.
Point2Mesh \cite{hanocka2020point2mesh}, for example, employs mesh convolutions to capture local geometric self-similarity across the surface of a single shape. Similar ideas are explored in the image domain \cite{shocher2018zero, kulikov2023sinddm, shaham2019singan} and recent work \cite{mitchel2024single} applied this to 3D shapes for texture transfer and editing. Inspired by these ideas, our work applies self-similarity to the joint distribution of local geometry and texture, learning an explicit self-similarity prior in the joint space between geometry and texture.

\paragraph{Texture generation via Pretrained Diffusion Models}
Recent advances in 3D generation have used 2D diffusion models in place of large-scale 3D datasets. Score distillation sampling (SDS) \cite{poole2023dreamfusion, wang2023score} uses a frozen, pretrained diffusion model to score rendered geometry and update the texture representation.
Subsequent work has improved visual fidelity \cite{lin2023magic3d, yi2024gaussiandreamer, metzer2023latent, youwang2024paint}, material fusing \cite{han2024vfusion3d}, and multi-view consistency \cite{Kwak_2024_CVPR, Ren_2025_CVPR, voleti2024sv3d}. Alternatively, geometry, lighting and material conditioning have been provided explicitly \cite{chen2023fantasia3d, zhang2024dreammat, deng2024flashtex}.

Other methods formulate texture synthesis as a view-conditioned inpainting task over the mesh, progressively generating textures across multiple views \cite{richardson2023texture, chen2023text2tex, zeng2024paint3d}. These methods offer faster inference than SDS-based pipelines, but often struggle with multi-view consistency and limited 3D awareness. Past work improved view sampling to enforce consistency \cite{cao2023texfusion}, or used view-consistent noise sources and multi-view attention mechanisms \cite{liu2024text, he2025materialmvp}.

\paragraph{Generative Texture Models Learned from 3D Datasets}
Several approaches train texture generation models directly on textured meshes or multi-view renderings. Early approaches used texture fields to predict per-point color values directly from spatial coordinates \cite{OechsleICCV2019}. Later methods adopted StyleGAN-inspired latent texture codes~\cite{karras2019style} to generate textures using either differentiable rendering \cite{chen2023shaddr} or operating directly on surfaces \cite{siddiqui2022texturify, bokhovkin2023mesh2tex}. More recent methods focus on explicitly generating UV textures \cite{cheng2023tuvf, Yu_2023_ICCV}. All of these techniques generate category-specific textures, but their generalization to arbitrary shapes remains limited.
Recent works further leverage large curated 3D datasets for multi-view geometry-conditioned diffusion training, scaling output UV map resolution and improving textural diversity \cite{bensadoun2024meta, zhang2024clay, zhao2025hunyuan3d, Zeng_2024_CVPR, yu2024texgen, yuan2025seqtex, yan2025flexpainter, huang2025mv, xiang2025native}. 
While these methods produce high-quality texture on a variety of shapes, they rely on carefully curated and often proprietary 3D datasets for training. This data dependence limits their scalability and generalization to out-of-distribution shapes or rare object categories. In contrast, our method does not require large-scale 3D data and instead learns from the self-similarity within a single object, enabling high-fidelity texture synthesis without external supervision.

\section{LOCAL TEXTURE INPAINTING MODEL} \label{sec:model}

\begin{figure*}[ht!]
	\centering
		\includegraphics[width=0.95\textwidth]{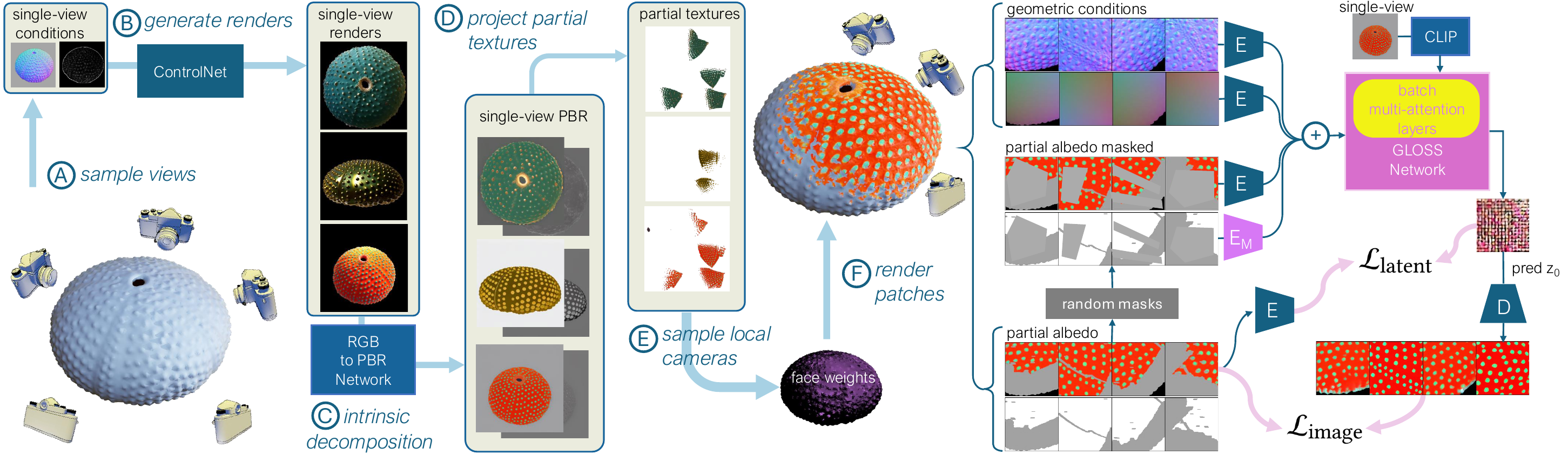}
	\caption{\textbf{Data and Training}: we use off-the-shelf models to generate training data (\S\ref{ssec:data}) for our local texture inpainting model (\S\ref{ssec:model}, \S\ref{ssec:training}). Steps A to F show the patch data generation pipeline. Then we show how the generated data is used for model input as well as the training for the model.}
	\label{fig:data}
\end{figure*}


Our core idea is to train an image diffusion model $\cG_M$ to complete textures in the local render space of a specific mesh $M$, 
conditioned on local geometry and paired geometry-texture references. We hypothesize that $\cG_M$ can learn
locally consistent texturing from only generated single-view renders of $M$, e.g.\ \emph{given that these dragon scales have this texture, similar scales in another region should have a similar texture}. We describe our automatic data generation pipeline (\S\ref{ssec:data}), model (\S\ref{ssec:model}) 
and training details (\S\ref{ssec:training}). See \S\ref{sec:applications} for inference.

\subsection{Preliminaries}
Our input is an untextured 3D mesh $M=\{F, V, U, T_{N}\}$ with faces $F$, vertices $V$, UV-map $U$ mapping face vertices to texture coordinates, 
and an optional normals texture image $T_{N}$. For high-resolution shapes, geometric details may reside in $F, V$, or be baked into
$T_{N}$, and we accept either. Our goal is to develop a model that can fill the missing albedo texture image $T_{\rho}$, and define
$T_{\alpha}$ as the alpha (or known) part of $T_{\rho}$. While we focus on albedo, we also
define $T_{mat}$ as the optional multi-channel image of other material properties
like roughness in the standard PBR model \cite{burley2012physically}.

To transition between texture and render spaces, we define several functions for forward rendering and backprojection:
{
\setlength{\abovedisplayskip}{3pt} 
    \setlength{\belowdisplayskip}{3pt}
    \setlength{\jot}{2pt} 
\begin{subequations}
    \begin{align}
         I_{N}, I_{\mathrm{geo}} & \gets \cR_g(M, c) \label{eq:rR_geo} \\
         I_{\rho}, I_{\alpha}, I_{mat} & \gets \cR_{mat}(M, c, T_{\rho}, T_{\alpha}, T_{mat}, W) \label{eq:R_mat} \\
         T'_{\rho}, T'_{\alpha}, T'_{mat} & \gets \cR_{mat}^{-1}(M, c, I_{\rho}, I_{mat}, W') \label{eq:R_inv}
    \end{align}
\end{subequations}
}
where the \emph{geometric} rendering $\cR_g$ rasterizes $M$ from viewpoint $c$ at resolution $W$, rendering world-space positions $I_{\mathrm{geo}}$ 
and texture-mapping with $T_{N}$ to produce a normals image $I_N$ (e.g.\ Fig.~\ref{fig:teaser}, left). If no $T_N$ is
provided, only geometric normals are used, and in either case $I_N$ is converted to \emph{camera-relative coordinate frame}, important for 
network generalization later. The material rendering $\cR_{mat}$ rasterizes and texture-maps $M$, producing albedo image
$I_{\rho}$, alpha map $I_{\alpha}$ (texture-mapping of $T_{\alpha}$) and an optional multi-channel image of other material
properties $T_{mat}$.
We also define an inverse function denoted $\cR_{mat}^{-1}$ that back-projects renderings $I_{\rho}$, $I_{mat}$ from
camera $c$, back into partial textures of resolution $W'$, producing the back-projected albedo texture $T'_{\rho}$, $T'_{\alpha}$ denoting areas that received a projection and optional materials $T'_{mat}$. 
In practice, all functions share an implementation and are only split here for clarity.
We use
a similar approach to \cite{hu2024diffusion}; please refer to it for details.

\subsection{Training Data Generation}\label{ssec:data}

Our training data consists of paired local geometry patches and local texture patches. Such data is generated automatically using an ensemble of off-the-shelf image and language models to obtain diverse textures conditioned on mesh surface geometry. 
We begin by sampling $n$ \emph{global} cameras $\bC^{(M)} = \{c_1...c_n\}$ around $M$ (Fig.~\ref{fig:data}\textcircled{\footnotesize A}) and render geometric inputs $I^i_{N}, I^i_{\mathrm{geo}} \gets \cR_g(M, c_i)$ (Eq.~\ref{eq:rR_geo}). To obtain diverse and visually rich textures, we use an LLM to generate descriptive prompts for variations in color, material, and texture details. Using an off-the-shelf diffusion model with multiple ControlNets \cite{zhang2023controlnet}, we generate single-view images $\tilde{I}^i...\tilde{I}^n$ (Fig.~\ref{fig:data}\textcircled{\footnotesize B}) following the geometry and prompts. We then apply an off-the-shelf de-lighting network to convert each $\tilde{I}^i$ to albedo $\tilde{I}^i_{\rho}$ and materials $\tilde{I}^i_{mat}$ images (Fig.~\ref{fig:data}\textcircled{\footnotesize C}). We leave more prompt and single view generation details in Appendix~\ref{app:datagen}.

As shown in Fig.~\ref{fig:data}, even one albedo view $\tilde{I}^i$ can contain many local self-similarities. To learn from these without suffering from distortions in the rendered view, we opt to re-render $M$ using partial textures from many close-up views. 
We backproject each $\tilde{I}^i_{\rho}$ using $\cR_{mat}^{-1}$ (Eq.~\ref{eq:R_inv}) to obtain a partial texture map $\tilde{T}^i_{\rho}$ and alpha $\tilde{T}^i_{\alpha}$ for the visible area (Fig.~\ref{fig:data}\textcircled{\footnotesize D}). We use a heuristic to compute sampling face weights $\omega_i$, prioritizing faces by both area and number of visible pixels in $\tilde{I}^i_{\rho}$. During training, each batch is confined to one view $c_i$, and its corresponding texture $\tilde{I}^i_{\rho}$ and $\omega_i$ are used
to sample local cameras $\hat{c}_1...\hat{c}_b$ that fall within the known texture region of the single view $c_i$ (Fig.~\ref{fig:data}\textcircled{\footnotesize E}). For each $\hat{c}_j$, we importance sample a face $f \in F$, get camera \emph{at} vector by sampling a point $p$ within $f$, set camera \emph{up} vector as $v - p$ for a random vertex $v$ of $f$, sample distance $d$ along the normal $\vec{n}$ direction at $p$ and get camera position as $p + d \vec{n}$, and finally sample a field of view within a range. By scaling $M$ to a unit cube, we can share camera settings across all experiments. Local renderings from these views are used to train our model. 

\subsection{Model Architecture}\label{ssec:model}

Our image diffusion model $\cG_M$ completes texture in the
\emph{local render space} of $M$. 
As shown in Fig.~\ref{fig:data}, the input is a batch of albedo patches $\bI_{\rho}$ with masks $\bI_{\alpha}$ rendered from local cameras, 
corresponding normal $\bI_{N}$ and position $\bI_{\mathrm{geo}}$ renderings (we use normalized world positions relative to the center pixel to denote orientation relative to the object, but not exact location). 
The objective of $\cG_M$ is to output $\hat{\bI}_{\rho}$, inpainting missing albedo, while attending to $\bI_{N}, \bI_{\mathrm{geo}}$ and aligned geometry-texture references within the batch. 

As our backbone, we adopt Stable Diffusion v2.1 unCLIP \cite{openai2022unclip}, a latent diffusion variant accepting a CLIP image embedding in addition to the text prompt. For this, we use a full reference albedo view ($\tilde{I}^i_{\rho}$) to provide loose global context in the cross-attention layers, and pass an empty string for the text prompt.
To accommodate additional input conditions, we expand the UNet \cite{ronneberger2015unet} with extra input channels, which are zero-initialized. The 3-channel $\bI_{N}$, $\bI_{\mathrm{geo}}$, $\bI_{\rho}$ are all encoded with the default pretrained image encoder $E$ for our backbone, and the inpainting mask is downsampled to the latent space size. All conditioning channels are concatenated with the noisy latent input along the channel dimension and passed into the UNet. 

\paragraph{Batch Multi-Attention} To capture the long-range self-similarity across the mesh $M$, we extend the standard self-attention mechanism to operate across all patches within a batch.  This allows the model to leverage similarities beyond the local neighborhood of any patch, effectively enabling texture inpainting given reference patches from any part of the mesh or other sources at run-time.  

We implement batch multi-attention across the spatial positions of all image patches in the batch. Let \( X \in \mathbb{R}^{B \times C \times H \times W} \) denote a batch of feature maps with height \( H \), width \( W \), and \( C \) channels. We first flatten the spatial dimensions and stack the batch to obtain:
\[
X' = \text{reshape}(X) \in \mathbb{R}^{(B \cdot H \cdot W) \times C}
\]
We then compute full self-attention across all spatial positions in the batch:
\[
\text{Attn}(X') = \text{softmax} \left( \frac{QK^\top}{\sqrt{C}} \right)V
\]
\noindent where
\[
Q = X'W_Q, \quad K = X'W_K, \quad V = X'W_V, \quad W_Q, W_K, W_V \in \mathbb{R}^{C \times C}
\]
This formulation enables each pixel in a patch to attend to all other pixels across the batch, allowing the model to leverage long-range structural repetition and appearance cues across the object surface.

During training, we sample a set of local patches from the same textured mesh and treat them as a single attention context as illustrated in Fig.~\ref{fig:data}. The choice of batch size now imparts a quadratic computational complexity on model training. While a larger batch size increases the likelihood that the model can attend to geometrically similar regions across patches, this benefit comes with significantly higher memory usage during attention computation and increased training time cost per batch. This can be alleviated by implementing batch multi-attention only across subgroups of the patch-level batch. In practice, we train the model with a batch size of 8 and allow attention across the entire batch. Once trained, the attention mechanism generalizes beyond the fixed training batch size and can operate on arbitrary batch sizes during inference, providing flexibility for downstream applications.



\subsection{Model Training}\label{ssec:training}

\paragraph{Training Batches} For each batch we sample a partial texture $\tilde{I}^i_{\rho}$ with alpha $\tilde{I}^i_{\alpha}$ and local cameras $\hat{c}_1...\hat{c}_b$ sampled for that view. We next render both geometric and material images (Fig.~\ref{fig:data}\textcircled{\footnotesize F}):
{
\setlength{\abovedisplayskip}{2pt} 
    \setlength{\belowdisplayskip}{3pt}
    \setlength{\jot}{2pt} 
\begin{subequations}
    \begin{align}
    \bI_{N}, \bI_{\mathrm{geo}} & \gets \cR_g(M, \hat{c}_1...\hat{c}_b) \label{eq:rR_geo_b} \\
    \bI_{\rho}, \bI_{\alpha} & \gets \cR_{mat}(M, \hat{c}_1...\hat{c}_b, \tilde{I}^i_{\rho}, \tilde{I}^i_{\alpha}, w)
    \end{align}
\end{subequations}
} 
where $\bI_{\rho}$ is the ground truth with only unmasked known areas. To train, we apply additional masks
to $\bI_{\rho}$ and $\bI_{\alpha}$ using a combination of full masks and irregular masks from LaMa \cite{suvorov2021lama}. 

\paragraph{Diffusion Training} During training, we follow the standard denoising diffusion of Stable Diffusion v2.1~\cite{rombach2022high}. Given a ground truth albedo latent \( z_0 \), we randomly sample a timestep \( t \sim \mathcal{U}(1, T) \) and compute noisy $z_t$ using the pre-defined DDPM scheduler. We train with v-prediction~\cite{salimans2022progressive}, restricting the loss to the masked regions in the latent space:
\begin{equation}
\mathcal{L}_{\text{latent}} = \left\| M_{D} \cdot (\hat{v}_t - v_t) \right\|_2^2,
\end{equation}
\noindent where \( M_{D} \) is the downsampled mask $\bI_{\alpha}$, $\hat{v}_t$ is the model output on $z_t$, and $v_t$ is the target velocity. See Appendices~\ref{app:model} and~\ref{app:training} for more model and training details.

\paragraph{Masking Details} The downsampled mask does not fully prevent unknown pixels in the albedo channel from influencing the loss computation through convolutional kernels. This unintended information leakage, where masked (unknown) pixels affect neighboring activations, leads to a ``fuzzy'' supervision signal that often fails to preserve fine-grained texture details. To mitigate this, we apply two strategies. First, we replace the unknown pixels in the input albedo with the average color of the known texture regions, reducing the impact of unknown texture values during encoding. Second, we introduce an auxiliary loss in image space. For near-zero diffusion timesteps (e.g., \( t < 10 \)), we decode the predicted latent \( \hat{z}_0 \) back to the image space using the pretrained decoder $D$, and compute the LPIPS \cite{zhang2018unreasonable} loss between the masked prediction and the masked ground truth albedo:
\begin{equation}
\mathcal{L}_{\text{image}} = LPIPS (\bI_{\alpha} \cdot \hat{\bI}_{\rho}, \bI_{\alpha}  \cdot \bI_{\rho} ),
\end{equation}
\noindent where \( \hat{\bI}_{\rho} = D(\hat{z}_0) \). This perceptual loss complements the latent-space supervision, encouraging both structural coherence and high-frequency detail in the synthesized textures.

\begin{figure*}[ht!]
	\centering
	\subfloat[Brush construction and inference applications \label{fig:inference}]{
		\includegraphics[width=0.62\linewidth]{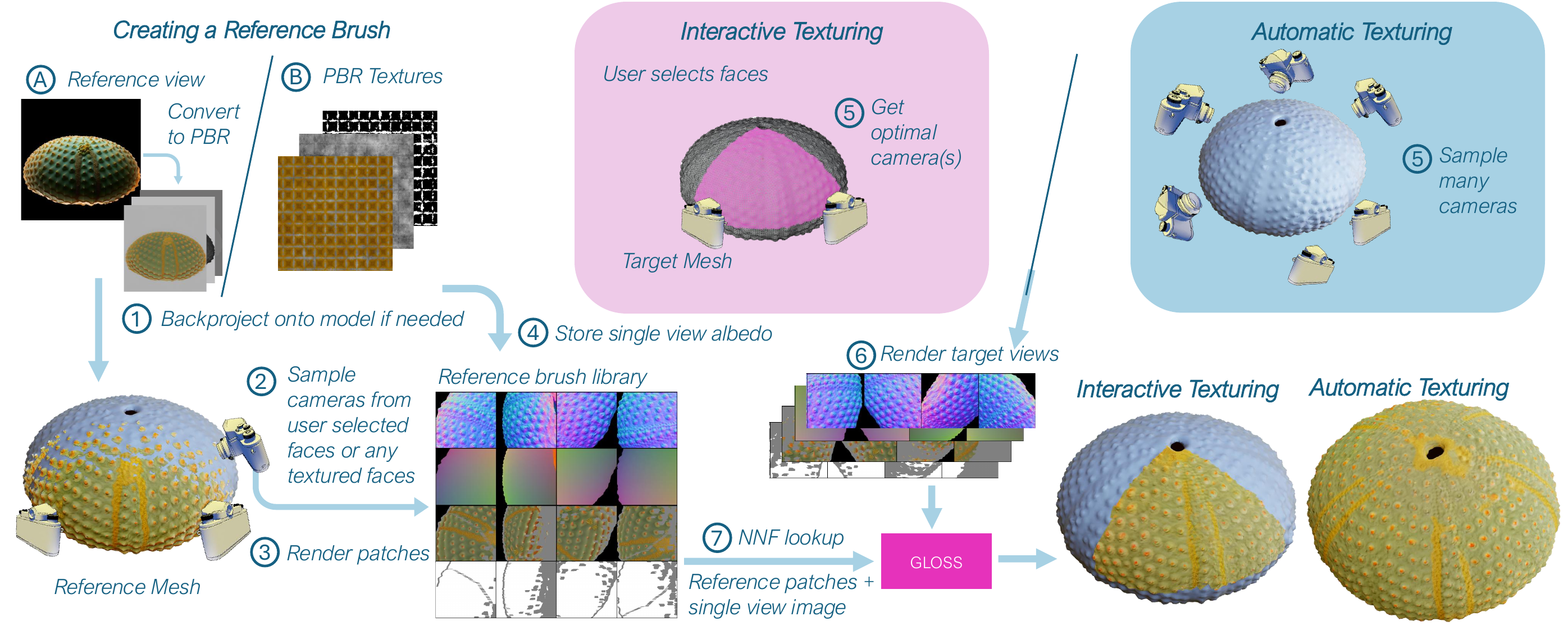}
    }
    \subfloat[Blender UI prototype\label{fig:ui}]{
		\includegraphics[width=0.3\linewidth]{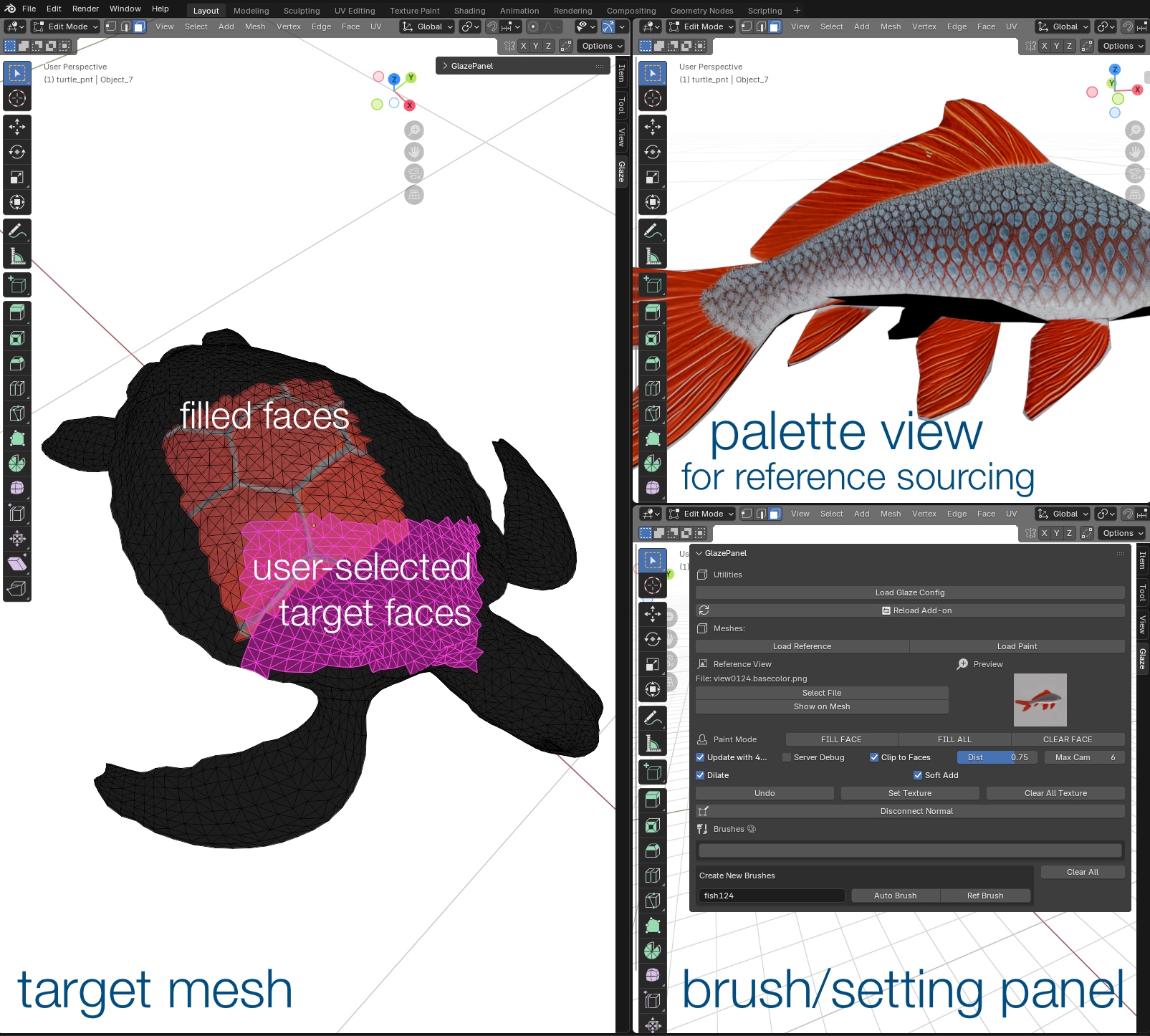}
    }
	\caption{\textbf{Model Inference Applications:} Reference brush creation from (A) a single view image or (B) any PBR texture map. The model can be used interactively to complete select faces, or fully automatically to complete the whole mesh.}\label{fig:inference_and_ui}
\end{figure*}
\section{TEXTURING APPLICATIONS}\label{sec:applications}

Once trained, $\cG_M$ opens many applications for texturing the source mesh $M$. We explored interactive texture fill and editing (\S\ref{ssec:app:interactive}), automatic texture generation (\S\ref{ssec:app:auto}), and extension to PBR material authoring (\S\ref{ssec:app:pbr}),
all guided by patch-level references (\S\ref{ssec:app:brushes}).

\subsection{Reference Brushes}\label{ssec:app:brushes}

The batch multi-attention design (\S\ref{ssec:model}) enables fine-grained control over output texture via input patches $\bI_{\rho}$, allowing the model to retrieve relevant context from known regions. This enables us to construct reference ``brushes'' during inference.
Given target patches with empty or partial albedo $\bI^t = (\bI^t_{\rho}, \bI^t_{N}, \bI^t_{\mathrm{geo}}, \bI^t_{\alpha})$ and reference patches $\bI^* = (\bI^*_{\rho}, \bI^*_{N}, \bI^*_{\mathrm{geo}}, \bI^*_{\alpha})$, we jointly process $\bI^t,\bI^*$ with $\cG_M$ in one batch and retain only predictions for $\bI^t$. However, different regions may require different references due to geometric differences (e.g.\ fish fins vs. scales). Thus, we construct a reference library (we call ``brush'') $\cL$ containing texture-geometry patch sets $(I^*_{\rho}, I^*_{N}, I^*_{\mathrm{geo}}, I^*_{\alpha})$ for a \emph{single particular texture}.
For each target batch, we select references via Nearest Neighbor Feature Matching (NNFM)~\cite{zhang2022arf} on VGG features of normal renderings, and filter out candidates with large color mismatch to existing target albedo.

To build a brush $\cL$, users can explore \textbf{single-view looks} for $M$ with generative models following our data-generation pipeline (\S\ref{ssec:data}), creating references through patch rendering. 
We experiment with two modes for sampling $\cL$.
In automatic mode, we sample from known texture areas similarly to training, relying on NNFM to select fitting references. 
In user-guided mode, the user can restrict face sampling to selected areas of interest. Our interactive prototype implements both modes.
We experiment with other ways to create $\cL$, without a single view rendering of $M$. For example, we apply a 
\textbf{PBR material} to a square mesh, and sample reference patches from it in a similar way, leveraging aligned
normals and albedo textures to create meaningful references (Fig.~\ref{fig:inference} \textcircled{\footnotesize B}). Alternatively, we can use existing techniques
such as \cite{DiffusionRenderer,zeng2024rgb,alhaija2023aimaterials} to convert \textbf{any image} into PBR material maps
or images with depth, albedo and normals, containing enough information to render references. Finally, we can
use a \textbf{single view or existing texture of another mesh $M'$} to sample reference patches for texture transfer. 
In all cases, we simply provide a single CLIP-encoded albedo for the unCLIP image embedding (\S\ref{ssec:model}).
See Video for examples of all the above brush construction strategies.

\subsection{Interactive Texture Fill}\label{ssec:app:interactive}

{\ourmodel} enables novel interactive texturing applications using $\cG_M$, including reference-guided texture fill, supporting seamless inpainting and blending of different textures. This workflow allows 3D artists to more easily integrate references into the interactive texturing workflow. 
Given any single view $\tilde{I}$ of interest, we construct a reference library $\cL$ using the strategy in \S\ref{ssec:app:brushes}, and then apply $\cG_M$ by sampling local views $\hat{c}_i$ on $M$, inpainting them and back-projecting inpainted albedo patches $I^*_{\rho}$ into the texture map $T_{\rho}$ using Eq.~\ref{eq:R_inv} while using $T_{\alpha}$ for blending with texture already present. Given any reference brush $\cL$ constructed in \S\ref{ssec:app:brushes}, the artist can specify a set of target faces to be filled. Using pre-computed local cameras and face-camera visibility matrix, we greedily sample local views $\hat{c}_i$ on $M$ covering the target area and generate target patches. We implement additional camera sampling in areas where precomputed local cameras fail to cover due to complex mesh topology or occlusion (e.g. small faces in between the dragon scalp and horn). We enable both small area and large area filling by creating multiple target cameras. Naively filling patches iteratively on a large selected area where multiple cameras are sampled will cause fine variation and inconsistency due to the generative nature of the model. To improve consistency, we first apply a synchronized diffusion strategy \cite{liu2024text} for the first 8/20 diffusion steps. Then, we add noise for 18/20 diffusion steps to this globally consistent yet over-smoothed texture and denoise from there. Using the same views $\hat{c}_i$, we add noise to the per-camera rendering of this guidance texture, and run the remaining diffusion steps in render space. 
The user can mix-and-match multiple references to texture from scratch or edit an existing or automatically generated texture of $M$. We implement all of these capabilities into our Blender
prototype (Fig.~\ref{fig:ui}).

\subsection{PBR Material Fill and Transfer}\label{ssec:app:pbr}

Although $\cG_M$ is only trained to inpaint albedo, preliminary results reveal that
it generalizes to other PBR channels, such as metallic and roughness concatenated with zeros
into an RGB image. We use the strategies in \S\ref{ssec:app:brushes} to create 
brushes $\cL$ containing material, not albedo, references. This extends prior applications to PBR material
generation and interactive editing as well. Materials and albedo 
are painted independently, but both are conditioned on underlying geometry, which
results in partial but not full consistency.

\subsection{Automatic Texture Generation}\label{ssec:app:auto}

GLOSS also enables automatic texture generation (Fig.~\ref{fig:inference}), based on a novel single view $\tilde{I}$ of $M$ using $\cG_M$. 
For the automatic setting, we sample batches of cameras over the entire mesh greedily, selecting those that contain at least 20\% overlap with fully completed texture
(for consistency) and would update the most texture. We construct batches of 8 target and 8 reference patches. 
We also enforce non-overlapping patches by applying rejection sampling, ensuring that independently filled patches within a batch
do not conflict with each other. We also apply the synchronized diffusion strategy to enforce global consistency. Inpainted patch albedo is back-projected into the still empty areas of the texture $T_{\rho}$, updating
$T_{\alpha}$ accordingly. We post-process the texture with dilation to fill the small seams due to the imprecision of backprojection.






%
\section{RESULTS AND EVALUATION}\label{sec:results}
\begin{figure*}[h!tbp]
	\centering

        \begin{tabular}{c}
        \includegraphics[trim={1cm 1cm 3cm 2cm}, clip, width=0.85\linewidth, valign=m]{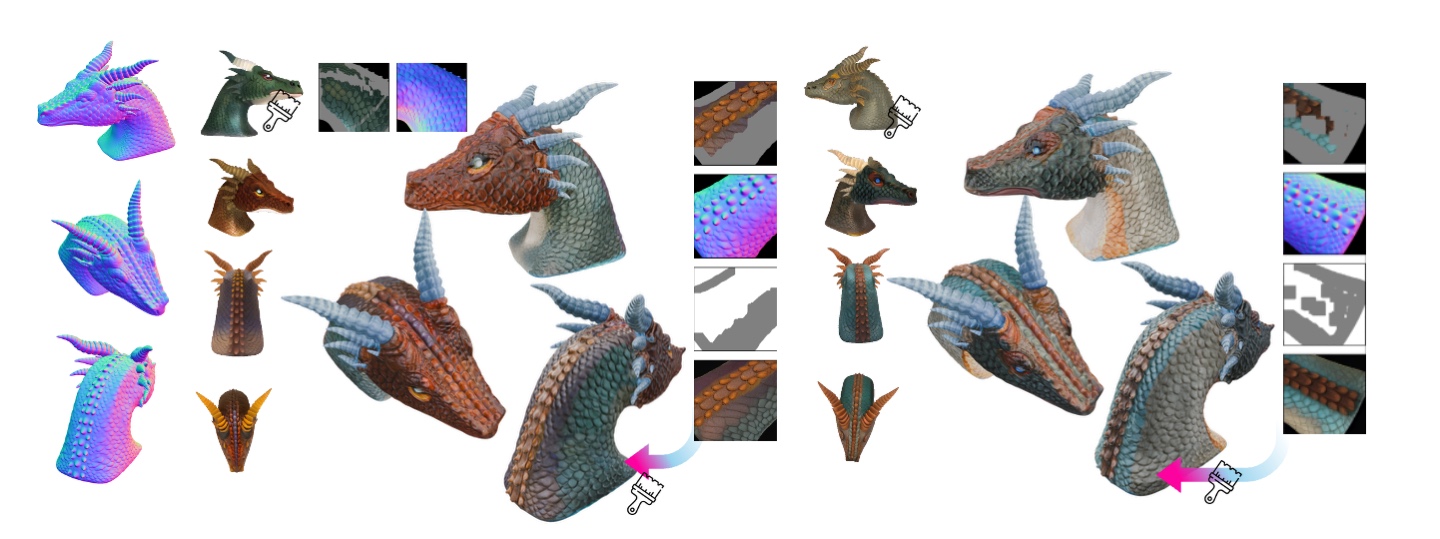}
        
        \end{tabular}

	\caption{\textbf{Interactive texturing} with multiple source single views on various parts of the object. We show the reference texture patch from the single view along with the normal, and where the texture brush created with that reference was applied to inpaint on the existing painted texture.}
	\label{fig:interactive_small}
\end{figure*}

\begin{figure*}[h!tbp]
	\centering
    \subfloat[Interactive texture completion by combining multiple texture in single views. In the lizard example, one can source texture from the head and transfer the texture onto the body.\label{fig:gallery:interative}]
    {
        \begin{tabular}{c}
        \includegraphics[trim={0cm 2cm 0cm 2cm}, clip, width=0.85\linewidth, valign=m]{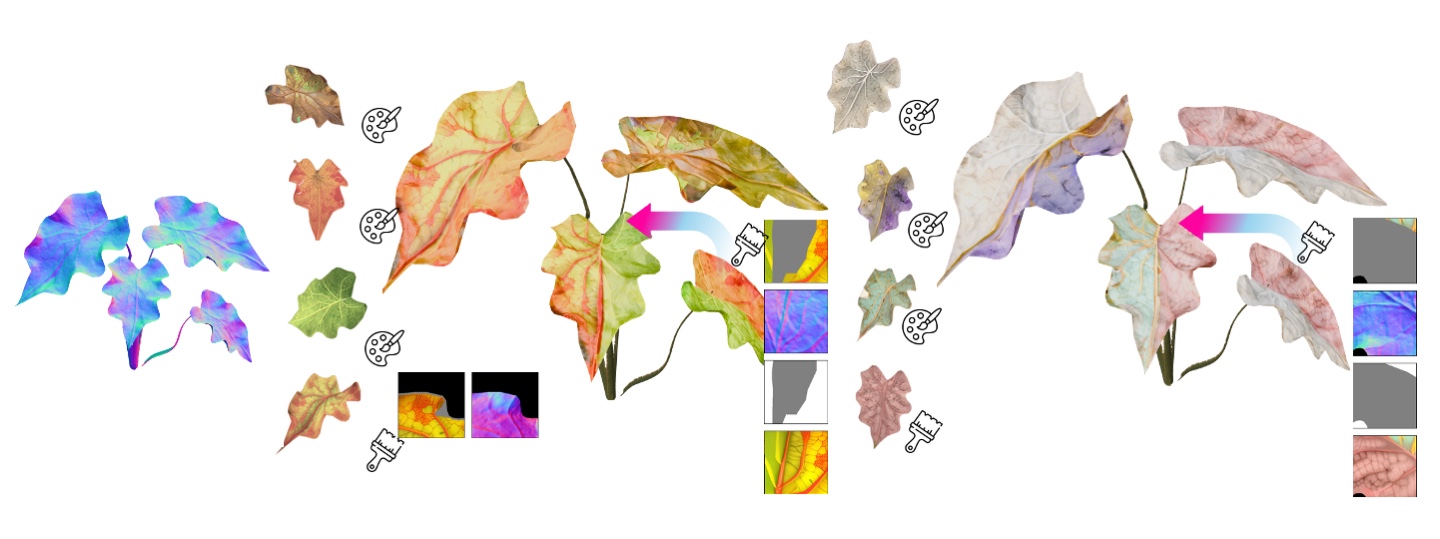} \\
        \includegraphics[trim={0cm 0cm 0cm 0cm}, clip, width=0.85\linewidth, valign=m]{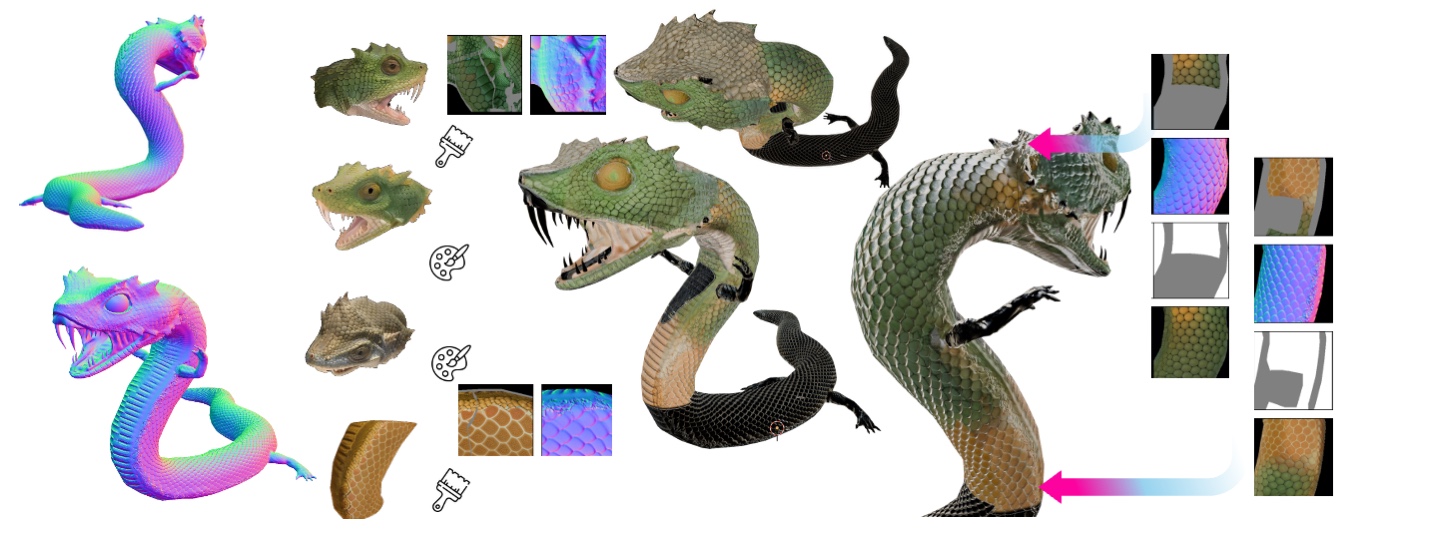} \\
        
        \end{tabular}
    } 

    \subfloat[Texture filling given various reference source and geometry condition. \label{fig:gallery:inpaint}]
    {
        \begin{tabular}{ccc}
        \includegraphics[trim={12cm 0cm 12cm 0cm}, clip, width=0.3\linewidth, valign=m]{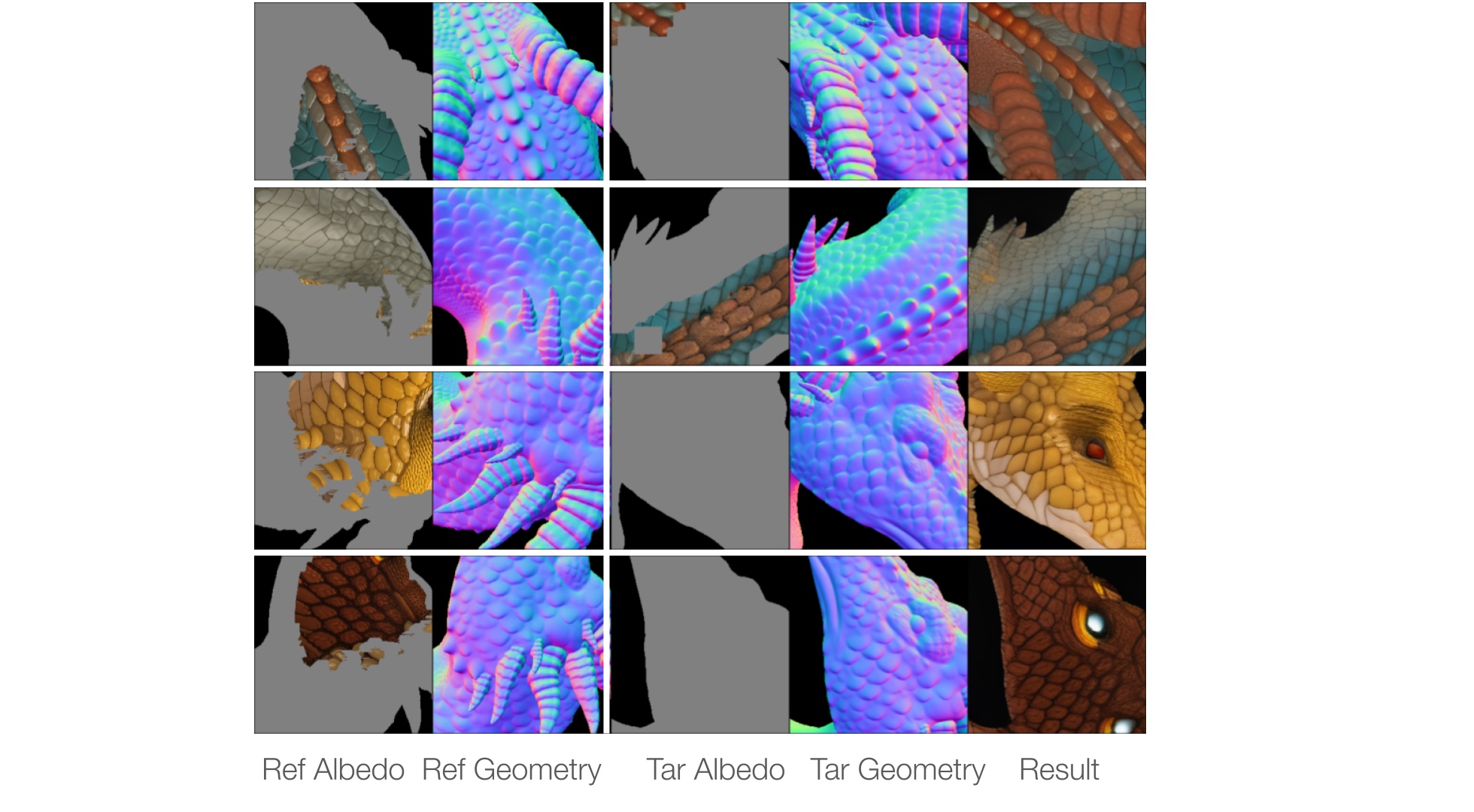} &
         \includegraphics[trim={12cm 0cm 12cm 0cm}, clip, width=0.3\linewidth, valign=m]{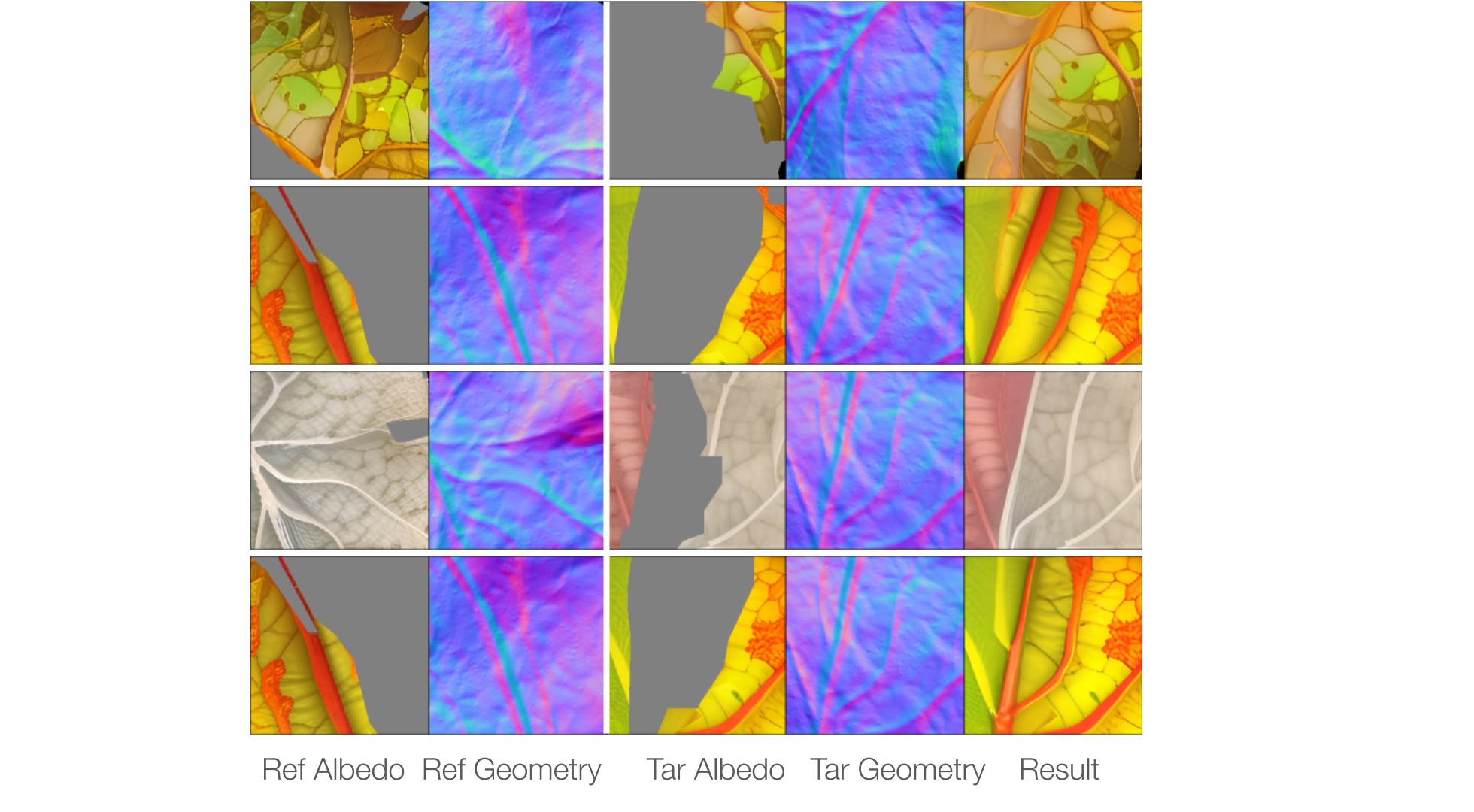} 
         &
          \includegraphics[trim={12cm 0cm 12cm 0cm}, clip, width=0.3\linewidth, valign=m]{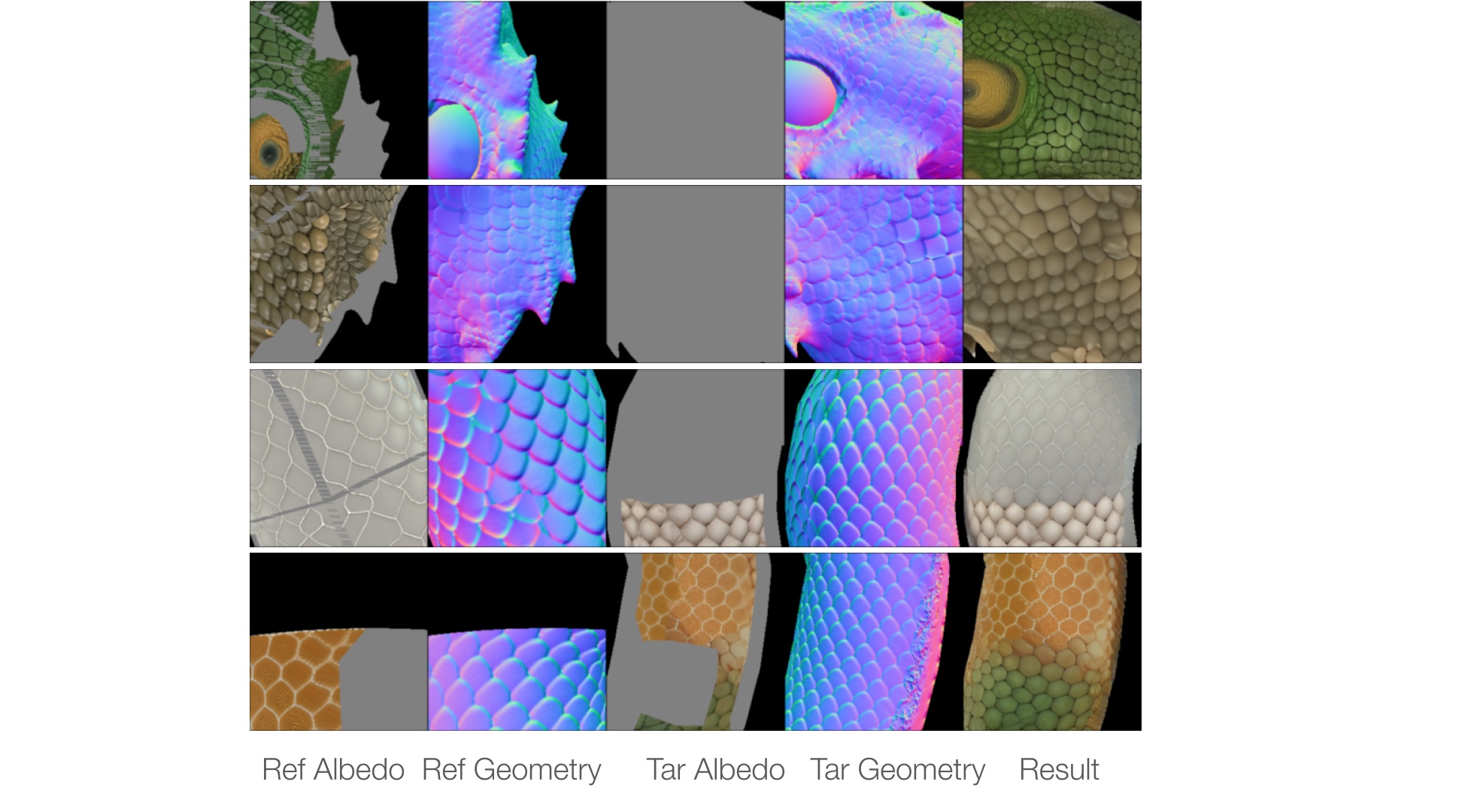}
        \\
        
        \end{tabular}
    } 
    
    \subfloat[Comparison to automatic generation methods.\label{fig:gallery:comp_baseline}]
    {
        \begin{tabular}{cc}
        \includegraphics[trim={5cm 0cm 5cm 0cm}, clip, width=0.45\linewidth, valign=m]{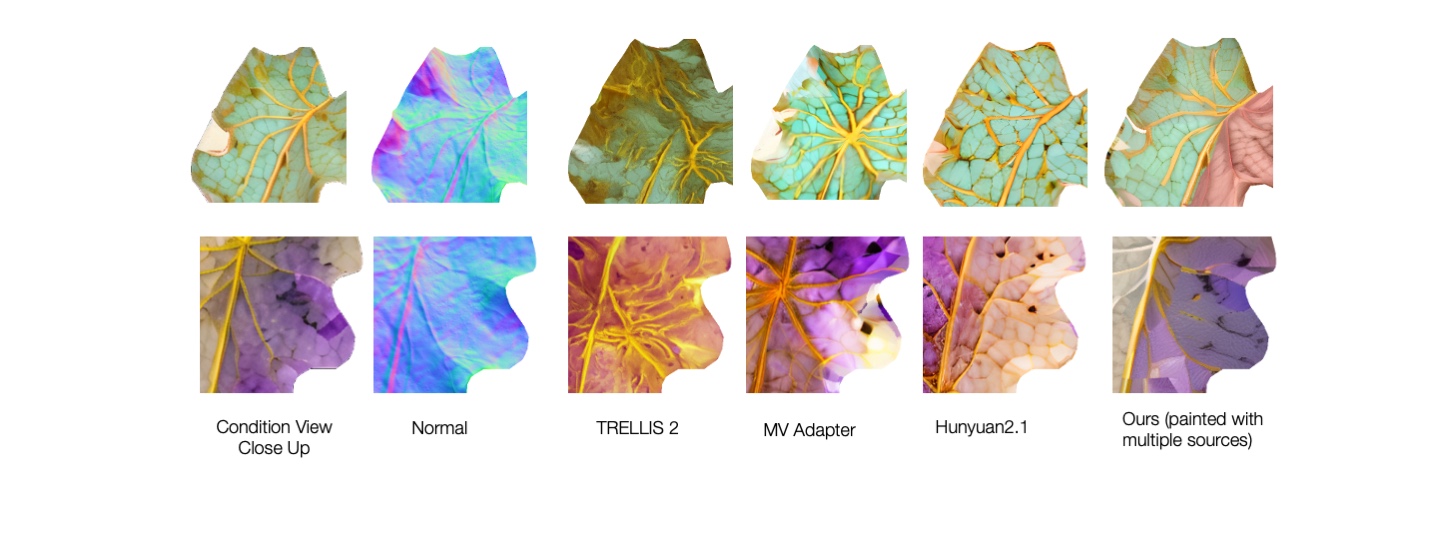} &
        
          \includegraphics[trim={5cm 0cm 5cm 0cm}, clip, width=0.45\linewidth, valign=m]{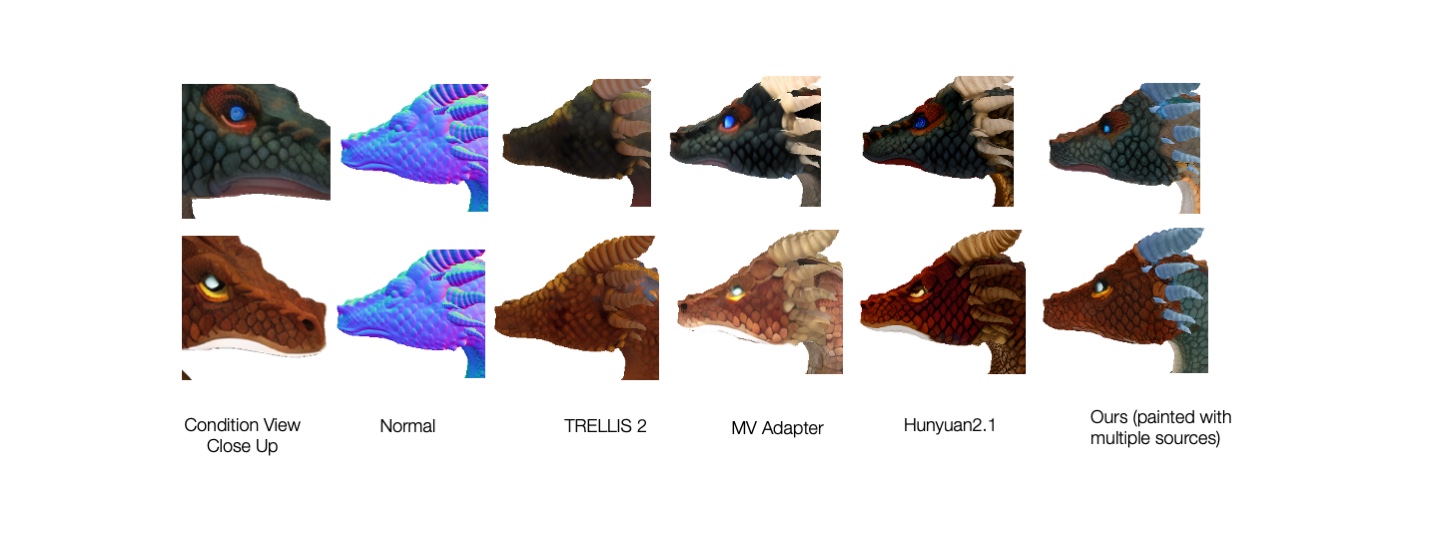}
        \\
        
        \end{tabular}
    } 
    
	\caption{\textbf{Interactive texturing}: our model supports faithful and controllable texture fills from multiple sources.}
	\label{fig:interactive_gallery}
\end{figure*}

In \S\ref{ssec:eval:interactive}, we show novel
interactive capabilities, texture transfer and generalization results, and a pilot user study.  We evaluate our method on automatic texture generation in \S\ref{ssec:eval:auto}, with
competitive performance against strong generative baselines. We present ablations in \S\ref{ssec:ablations}. We include more evaluation details, experimental result and attention analysis in Appendix~\ref{app:eval}.

\subsection{Experimental Settings}\label{ssec:exp:settings_data}


We select the following baselines. \textbf{TEXGen} \cite{yu2024texgen}, a representative feed-forward method working directly in texture UV space,
\textbf{Hunyuan 2.1} (Hunyuan3D-Paint model) \cite{hunyuan3d2025hunyuan3d}, a state-of-the-art method predicting multi-view consistent albedo
renderings which are fused into the final 3D shape texture, and \textbf{Paint3D} \cite{zeng2024paint3d}, a coarse-to-fine approach combining
both multi-view diffusion and learning in UV space. \textbf{TRELLIS 2} \cite{xiang2025native} is a state-of-the-art method that leverages a structured sparse 3D latent space for image to 3D generation. \textbf{MV-Adapter} \cite{huang2025mv} is a consistent multi-view generation method that supports texture generation from an image. All of these models support image conditioned texture generation. For the from-scratch experiment setting, we use AdamW optimizer with learning rate $1e-4$ with batch size of 32 for 80k steps and EMA decay of 0.999. For the finetune experiment setting, we use 100 single views, batch size 16, and 20k steps. We use the finetuned model for interactive applications.

\subsection{Interactive Applications}\label{ssec:eval:interactive}

We demonstrate interactive affordances of {\ourmodel} for artist-guided fill, inpainting and blending (\S\ref{ssec:app:interactive})
of textures using reference brushes constructed using varied methods \S\ref{ssec:app:brushes}, compare with another generative interactive technique (\S\ref{ssec:eval:dtp}) and pilot our prototype with several 3D professionals (\S\ref{ssec:ustudy}). We show that with \ourmodel, artists can interactively paint, and can create texture with close adherence to the reference texture and to the underlying geometry in Fig.~\ref{fig:gallery:interative} and Fig.~\ref{fig:gallery:inpaint}. We compare our method to other automatic methods, which fail to respond to the artist intent of interaction, and produce bad alignment to reference and geometry in Fig.~\ref{fig:gallery:comp_baseline}.

\subsubsection{Interactive Texture Fill}\label{ssec:eval:fill}

Many references are used in an interactive workflow to generate the final result. Using our interface, artists fill target areas of the mesh using references generated via diverse methods. For example, in Fig.~\ref{fig:teaser},
single-view references of the dragon were used to paint textures that are geometrically consistent with its surface.
We show more interactive examples in Fig.~\ref{fig:interactive_gallery}. In both the dragon head and foliage examples, ours is faithful to reference texture and geometry, while other methods show saturated color and fail to align with local dragon scales in Fig.~\ref{fig:gallery:comp_baseline}. Further,
we show that we can source brushes from images and PBR materials in Fig.~\ref{fig:gallery:brushes}, and show preliminary results of $\cG_M$
transferring to PBR material painting. For example, in Fig.~\ref{fig:gallery:brushes}, painting the urchin with references from the waffled metal PBR material
causes the distressed material artifacts to appear on the sea urchin bumps by attending to normal map similarities. This achieves a much more natural texture transfer than na\"ive application of the material.

\subsubsection{Generalization}\label{ssec:app:general}

While the data generation and pre-training of $\cG_M$ for a specific mesh $M$ are automated, 
such pre-processing limits the practical application of our method. To probe our model's capacity
for generalization, we apply it \textit{zero-shot} to novel meshes. 
We show that a model $\cG_{M}$ trained on $M$ can also texture other similar shapes in Fig.~\ref{fig:gallery:transfer1}. While these are preliminary results, future work
could extend the generalization of our method across shapes.

\subsubsection{Comparison with Diffusion Texture Painting}\label{ssec:eval:dtp}

\begin{figure*}[h!tbp]
	\centering
    \subfloat[\textbf{Blending} multiple textures seamlessly: results of interactive application with references shown in insets.\label{fig:gallery:blend3}]
    {
        \begin{tabular}{ccc}
        \includegraphics[trim={0cm 9cm 0cm 8cm}, clip, width=0.32\linewidth, valign=m]{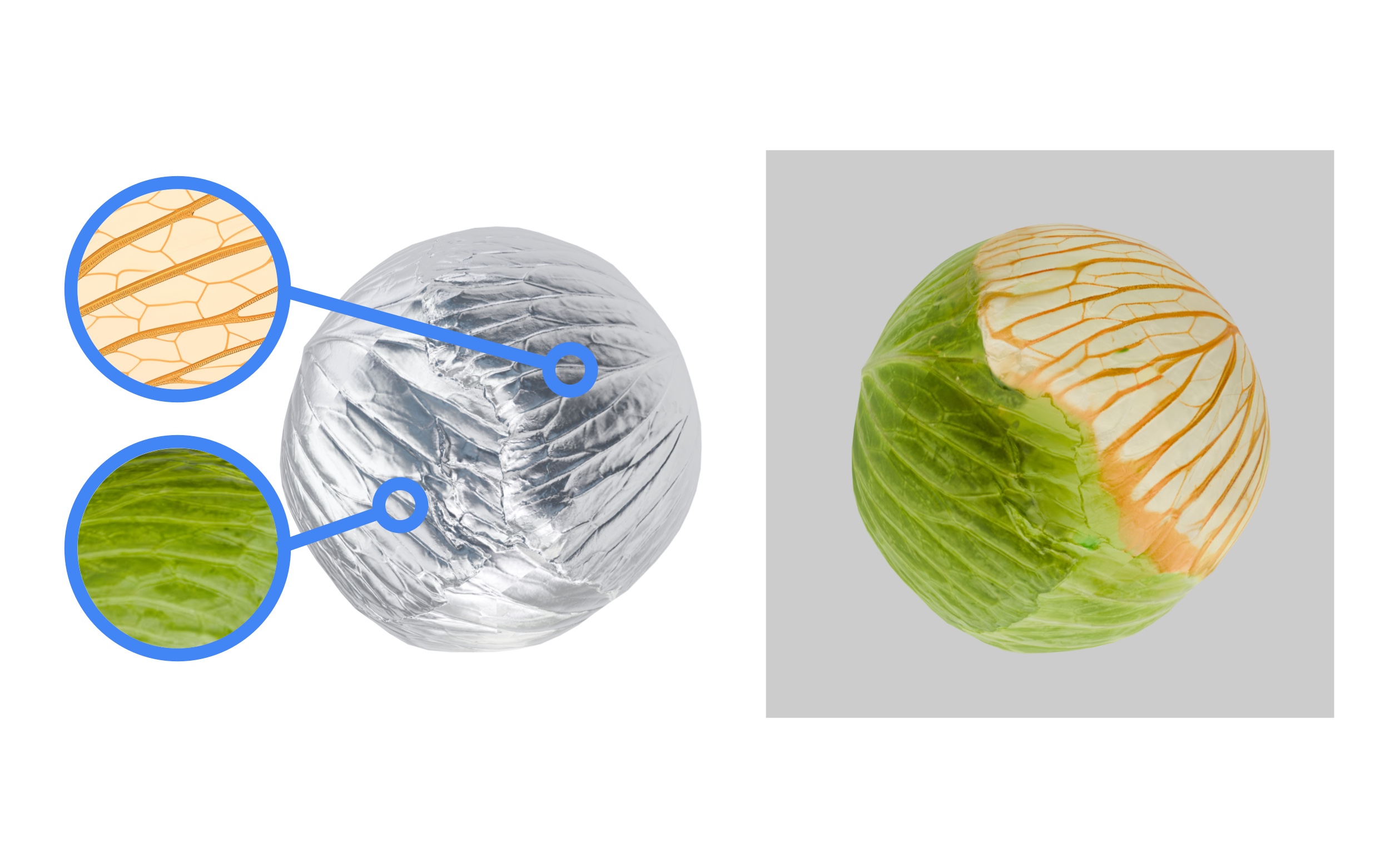} &
        \includegraphics[trim={0cm 9cm 0cm 8cm}, clip, width=0.32\linewidth, valign=m]{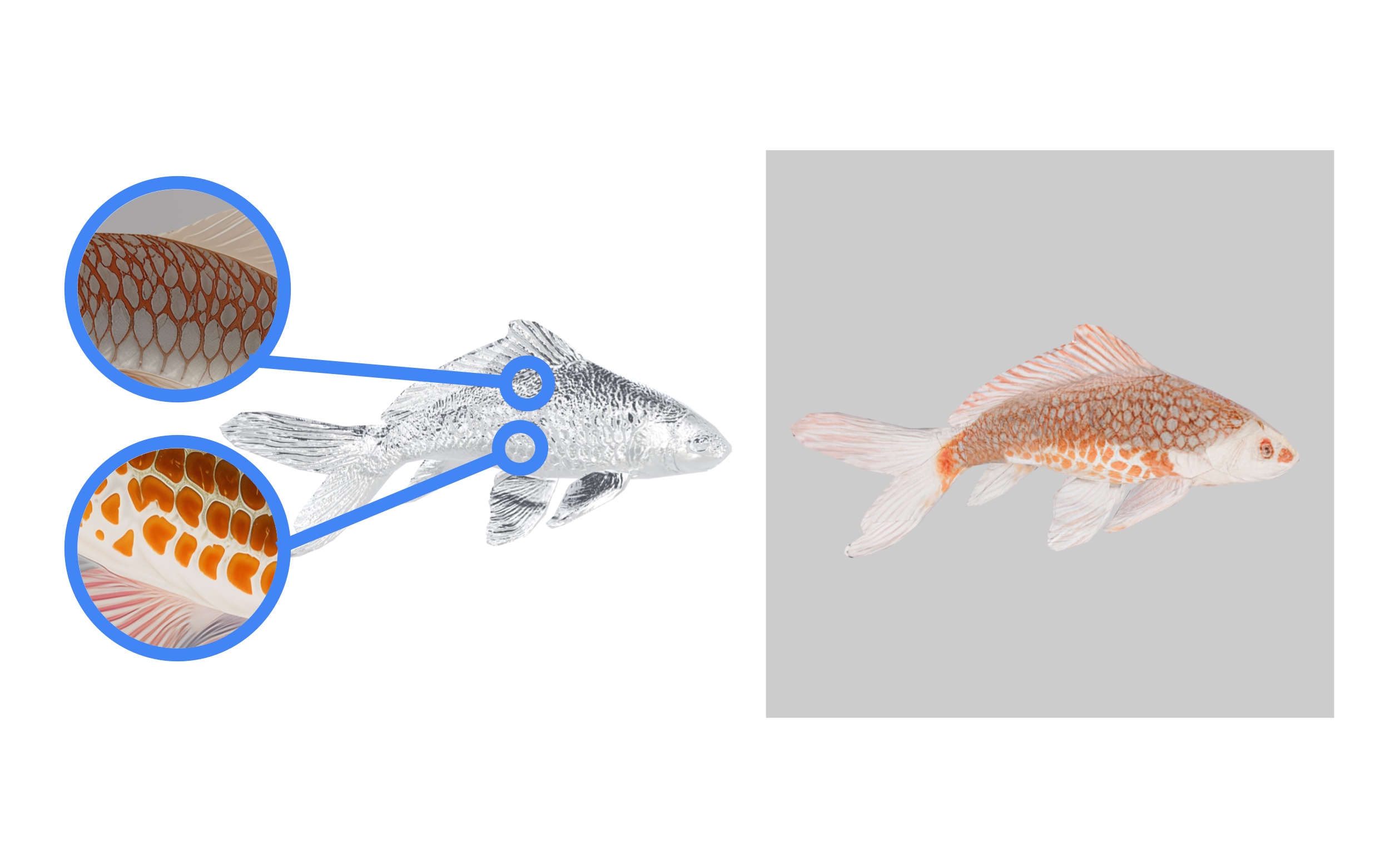} &
        \includegraphics[ trim={0cm 9cm 0cm 8cm}, clip, width=0.32\linewidth, valign=m]{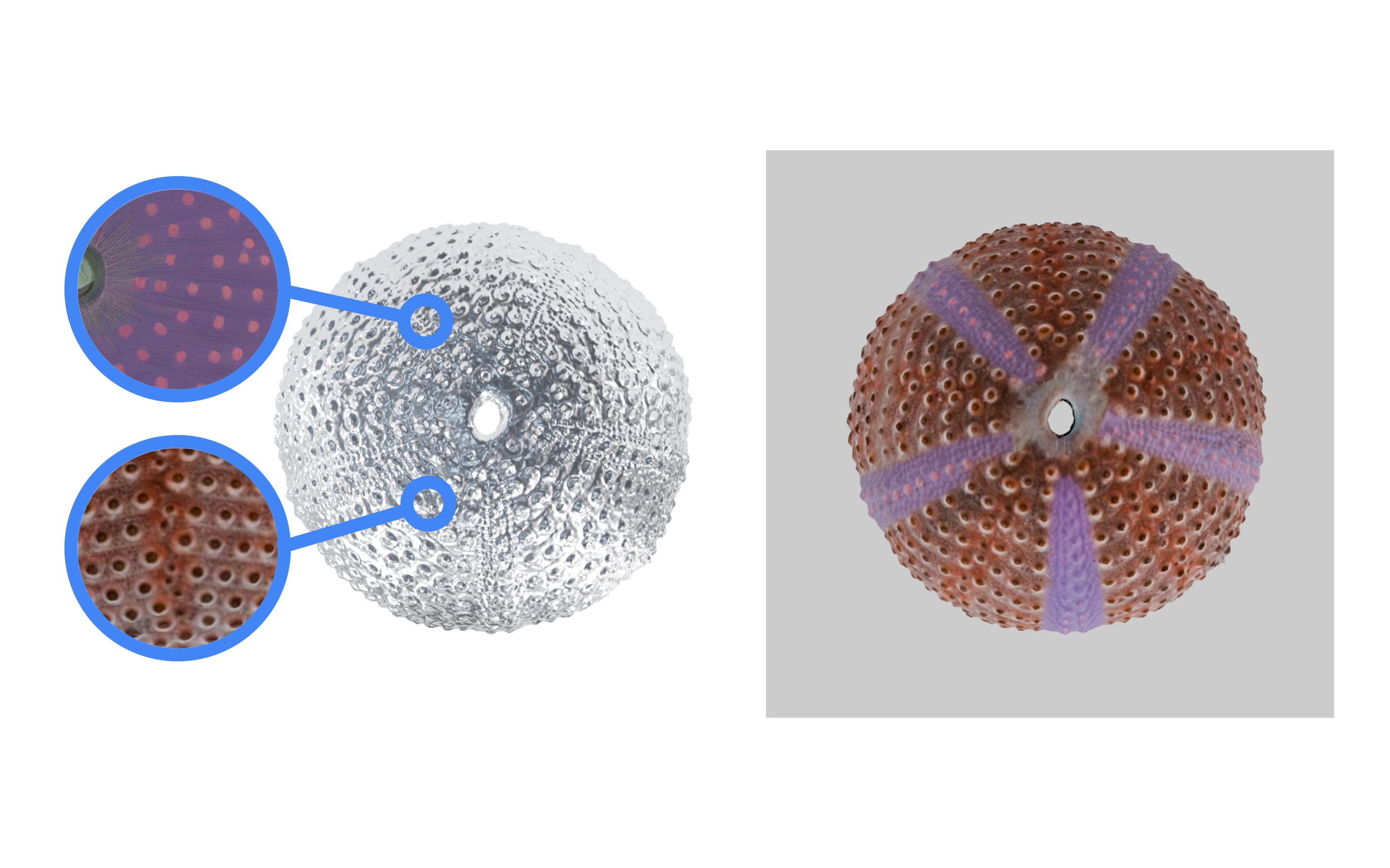}
        \end{tabular}
    } \\
    \subfloat[\textbf{Generalization:} a model $\cG_{M}$ trained on the mesh $M$ (left), can be applied to texture similar unseen shape $N$ (right), with references
    sourced from either, facilitating texture transfer and other applications (\S\ref{ssec:app:general}). \label{fig:gallery:transfer1}]
    {
        \begin{tabular}{cc}
        \includegraphics[trim={0cm 0cm 0cm 0cm}, clip, width=0.5\linewidth, valign=m]{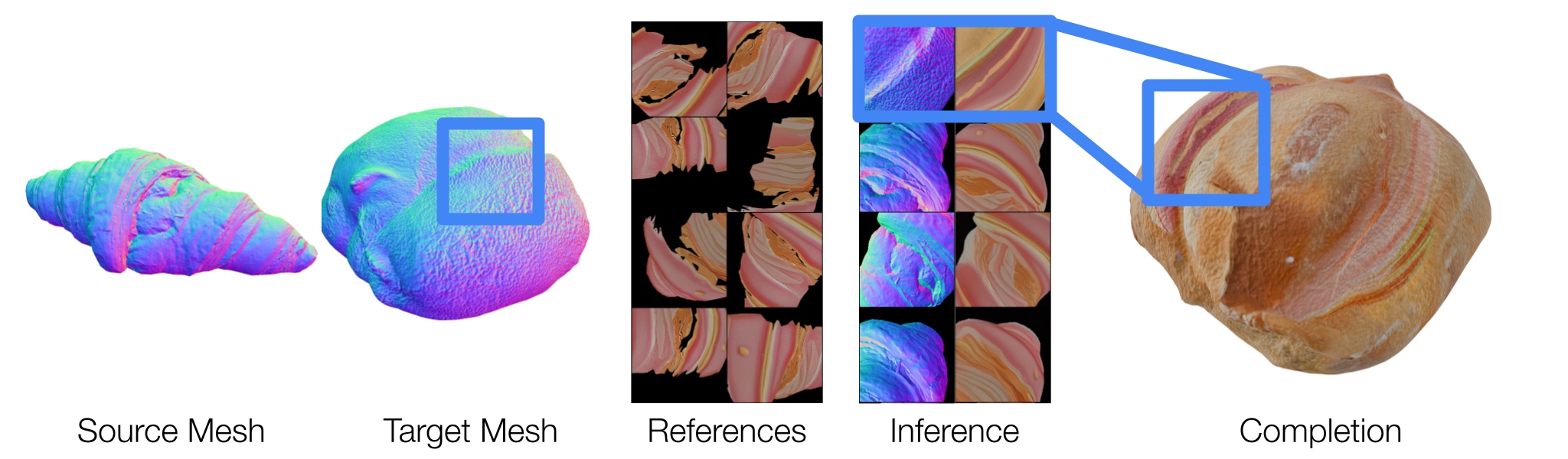} &
        \includegraphics[trim={0cm 0cm 0cm 0cm}, clip, width=0.5\linewidth, valign=m]{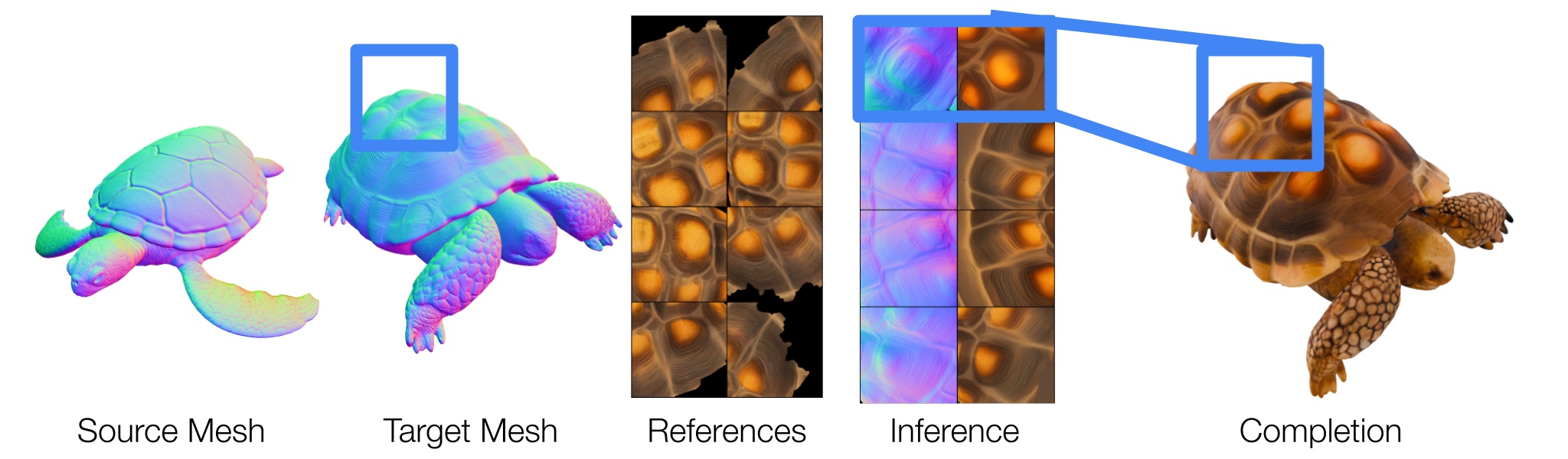} 
        \end{tabular}
    } \\
    \subfloat[\textbf{Comparison to Diffusion Texture Painting} on painting a stroke (left) and fill with random cameras (right) (see \S\ref{ssec:eval:dtp}). \label{fig:dpt_compare}]
    {
        \includegraphics[width=0.95\linewidth]{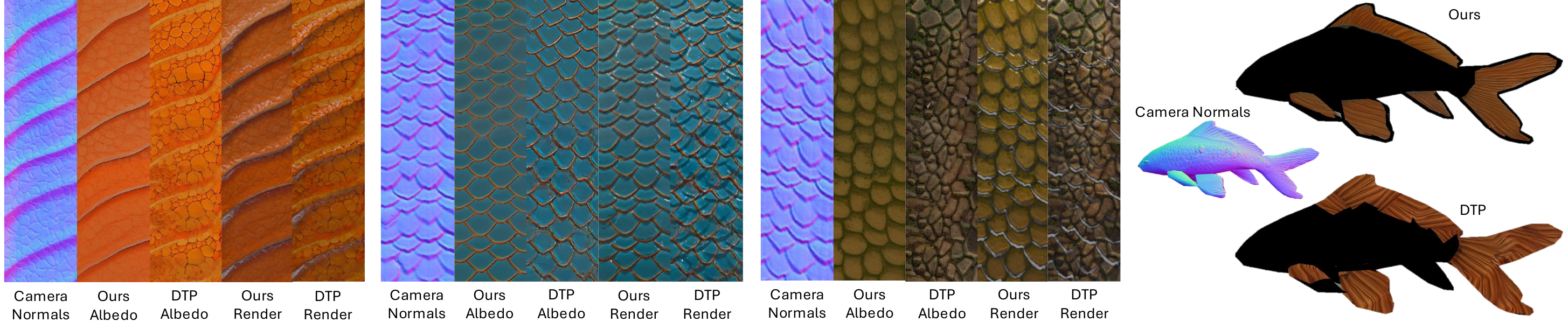}
    } \\
    \subfloat[\textbf{Brush Creation Methods:} generating a brush from an RGB image (left), and from a PBR material (right). See \S\ref{ssec:app:brushes}, \S\ref{ssec:eval:fill}. \label{fig:gallery:brushes}]
    {
       \begin{tabular}{cc}
        \includegraphics[trim={0cm 2cm 0cm 2cm}, clip, width=0.55\linewidth, valign=m]{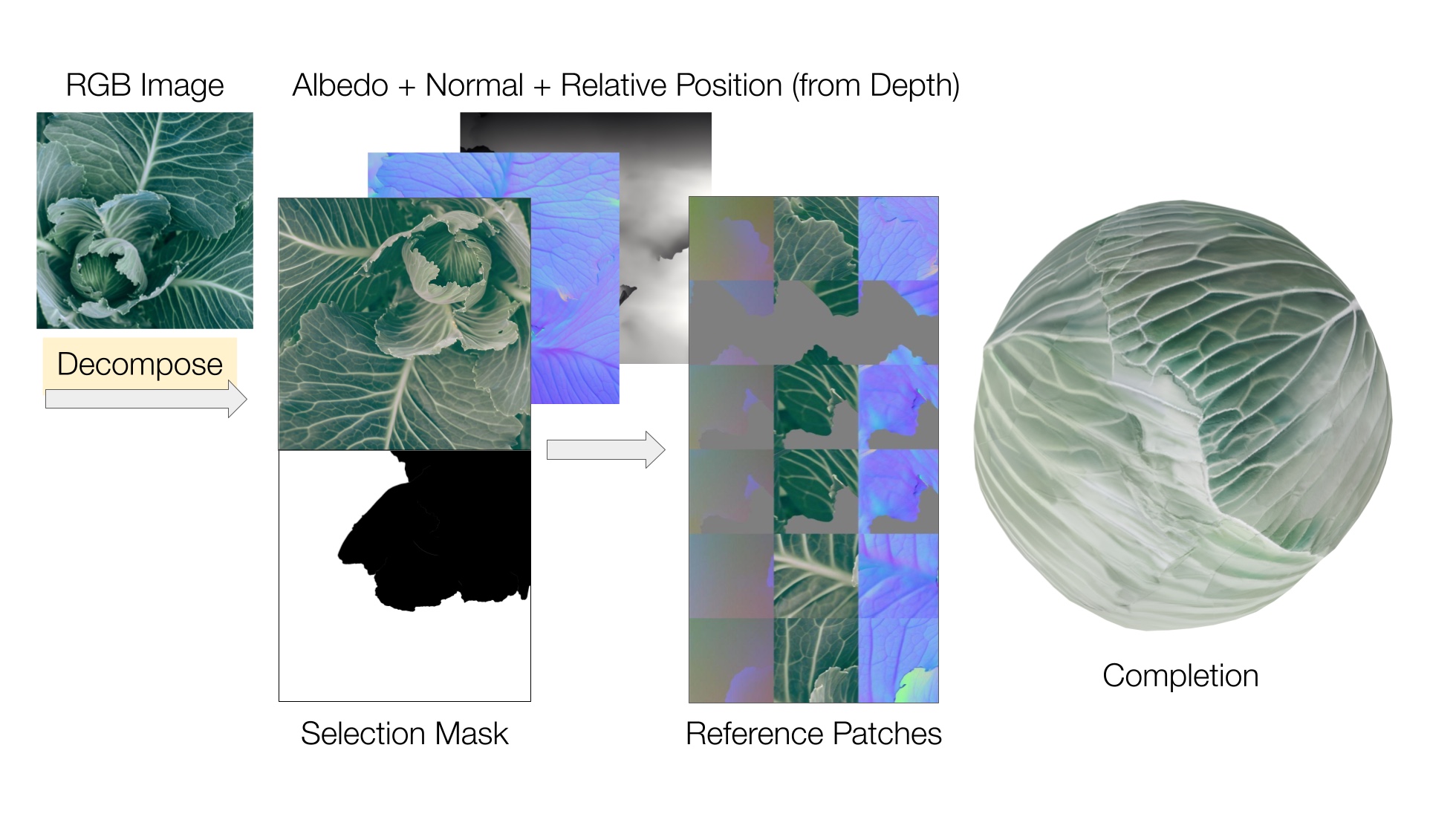} &
        \includegraphics[trim={0cm 1cm 0cm 1cm}, clip, width=0.45\linewidth, valign=m]{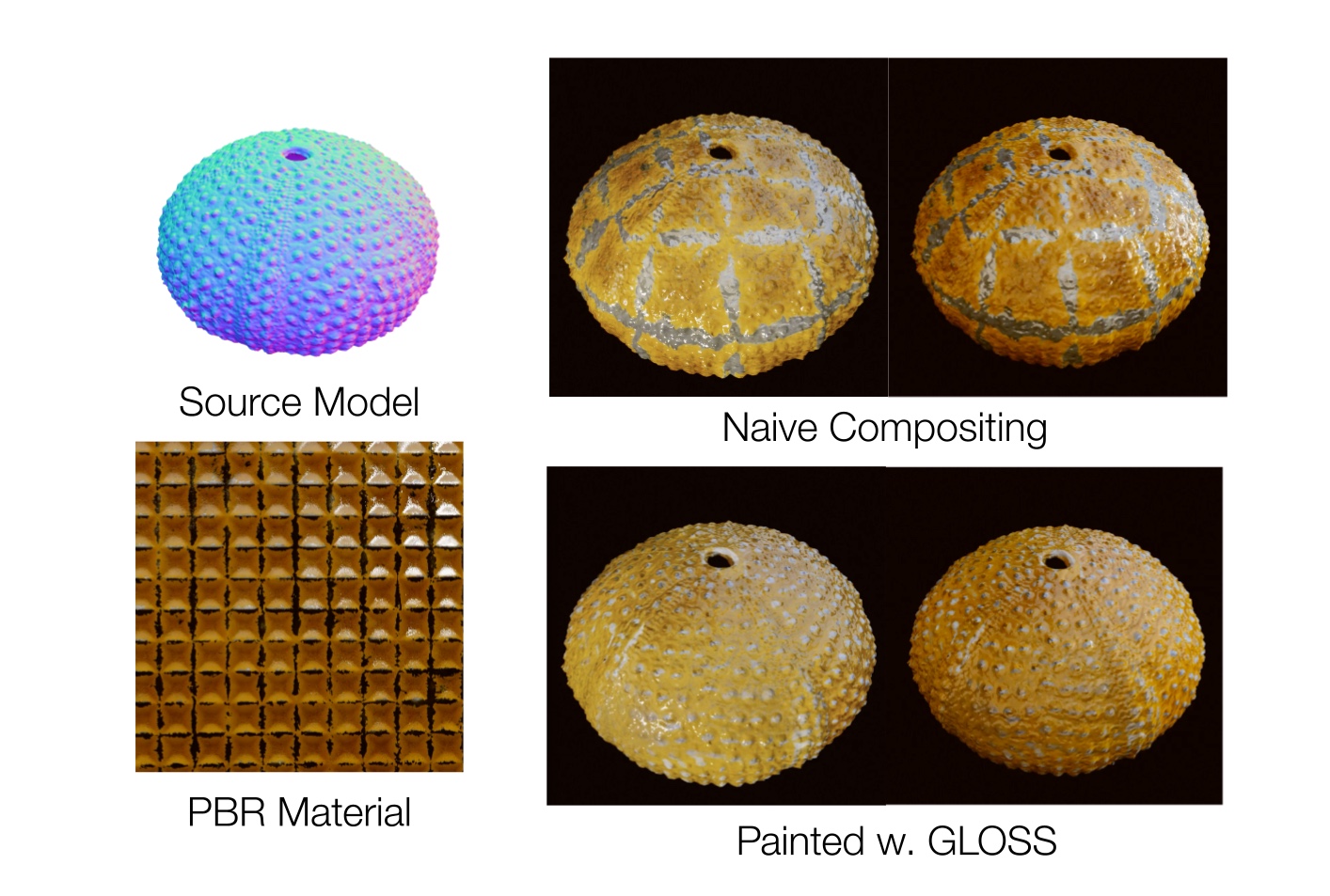} 
        \end{tabular}
    } \\
    \subfloat[\textbf{PBR Material Completion:} {\ourmodel} can generalize to metallic and roughness channels (also previous line, right). See \S\ref{ssec:app:pbr}, \S\ref{ssec:eval:fill}. \label{fig:gallery:pbrgen}]
    {
       \begin{tabular}{cc}
        \includegraphics[trim={0cm 5cm 0cm 5cm}, clip, width=0.50\linewidth, valign=m]{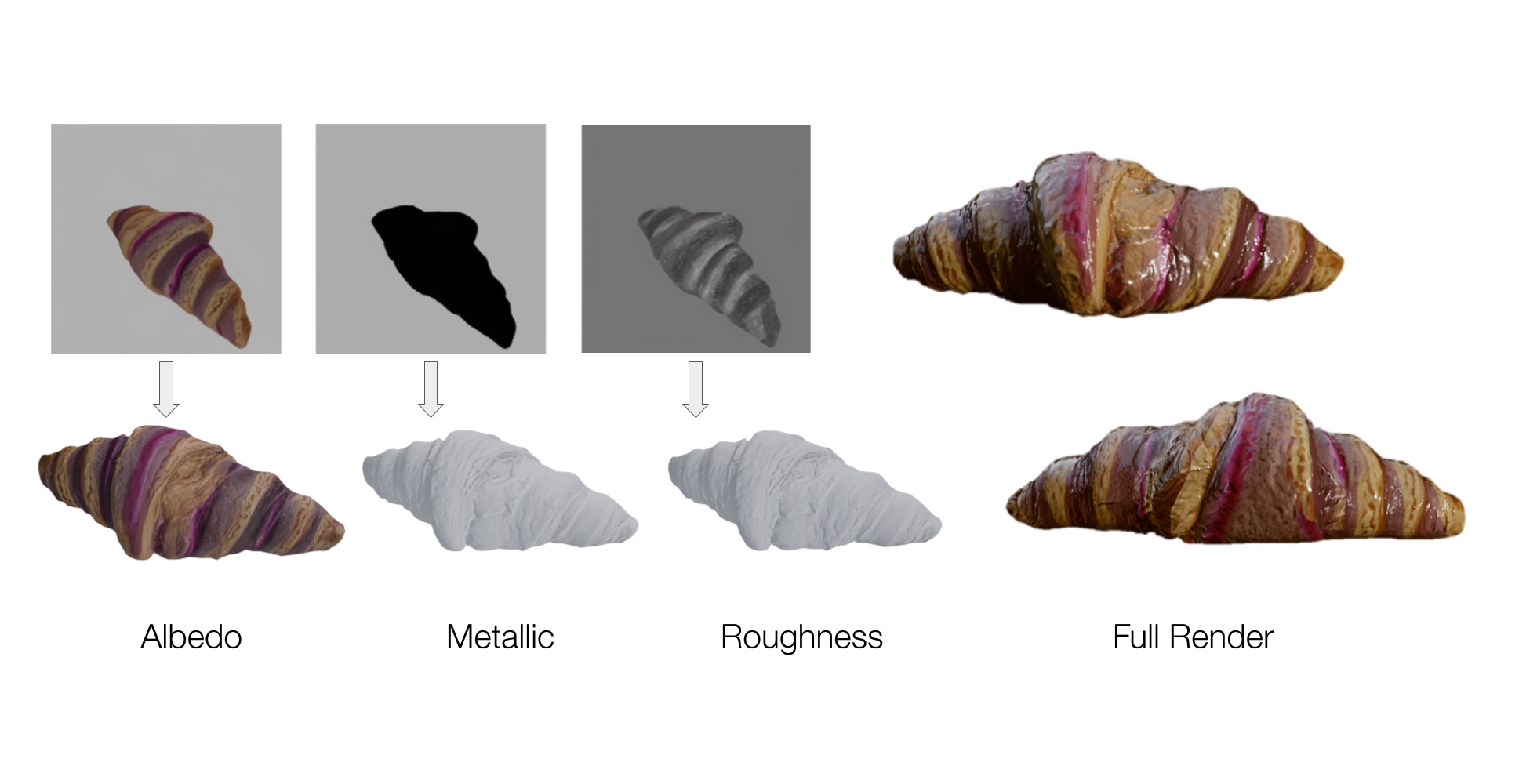} &
        \includegraphics[trim={0cm 5cm 0cm 5cm}, clip, width=0.50\linewidth, valign=m]{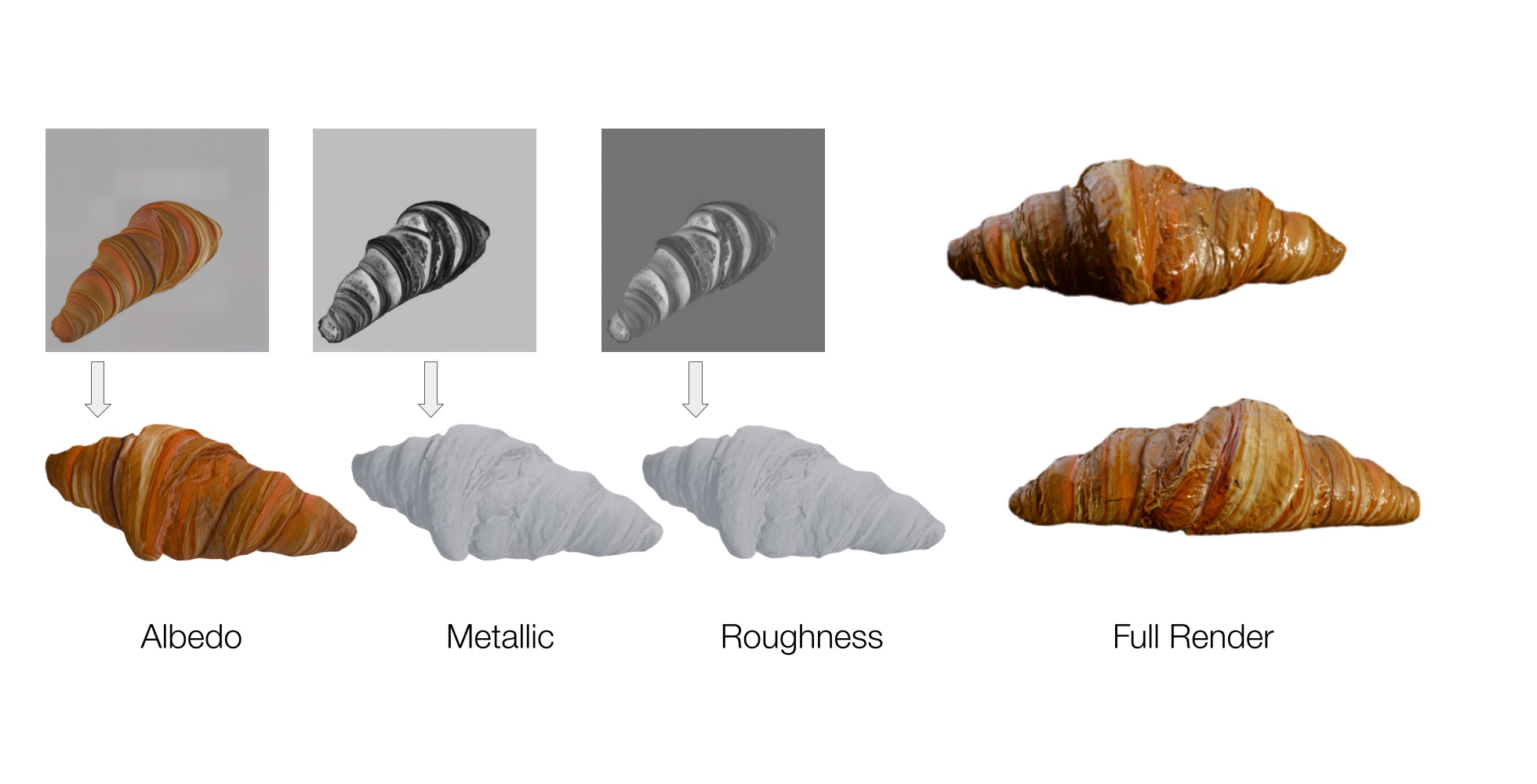} 
        \end{tabular}
    } \\
	\caption{\textbf{Interactive applications, transfer, comparisons}: our model supports a wide range of texturing applications.}
	\label{fig:gallery}
\end{figure*}

Recent work, Diffusion Texture Painting (DTP) \cite{hu2024diffusion}, demonstrated the utility of generative inpainting models for local texture painting on meshes, a similar setting. For progressively filling strokes one camera at a time (original DTP setting), both methods produce seamless strokes. In Fig.~\ref{fig:dpt_compare} left, DTP ignores geometry and produces only minor variations from the single reference patch, while our method adapts multiple references to match geometric details (see rendered strokes for texture/geometry misalignment). This allows DTP to paint with references that have no relation to underlying geometry (rocks over scales), whereas our method produces geometrically consistent texture transfer (rocks turn into scales), making the two approaches useful for orthogonal use cases. For texture fill with unordered local cameras (Fig.~\ref{fig:dpt_compare}, right), DTP produces chaotic results due to its reliance on stroke orientation, while our method uses local geometry to create consistent fills. 

\newcommand{\transfernormalswidth}{0.2\linewidth}
\newcommand{\transferviewwidth}{0.2\linewidth}
\newcommand{\transferreswidth}{0.2\linewidth}

\begingroup
\setlength{\tabcolsep}{3pt}
\renewcommand{\arraystretch}{0}

\begin{figure}[tb]
  \centering
  \begin{tabular}{cccc}
  \small Geometry & \small Single View & \small View 1 & \small View 2 \\


  \includegraphics[width=\transfernormalswidth, valign=m]{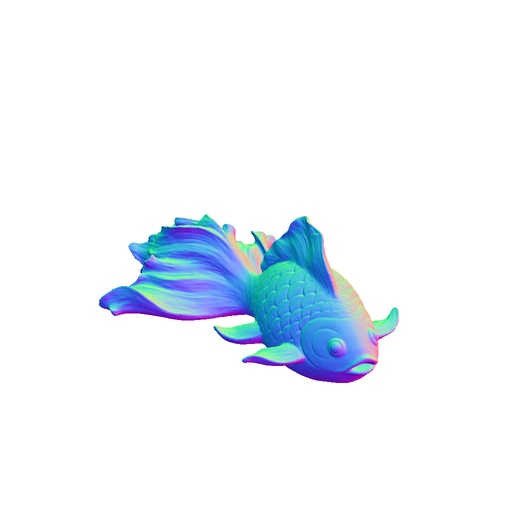} & \includegraphics[width=\transferviewwidth, valign=m]{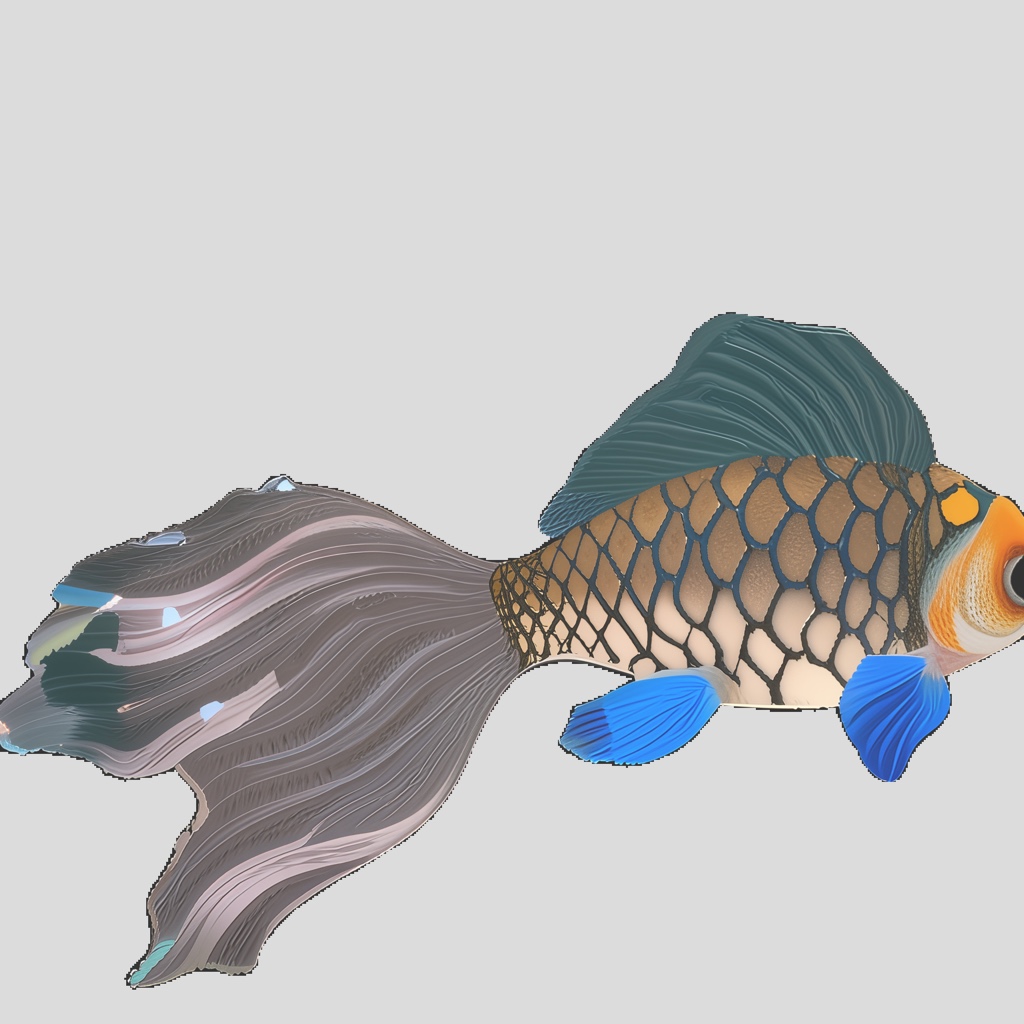} & \includegraphics[trim={10mm 10mm 10mm 10mm}, clip, width=\transferreswidth, valign=m]{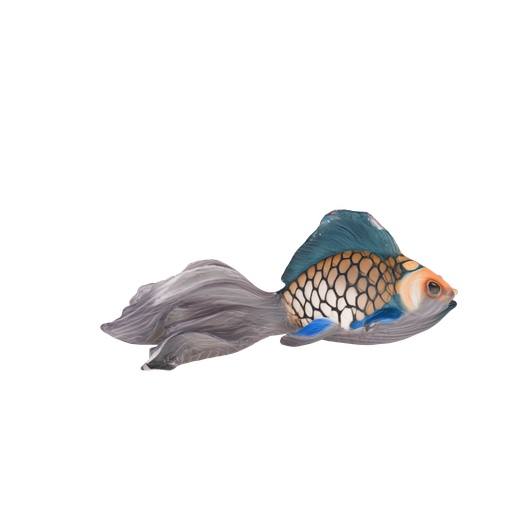} & \includegraphics[trim={10mm 10mm 10mm 10mm}, clip, width=\transferreswidth, valign=m]{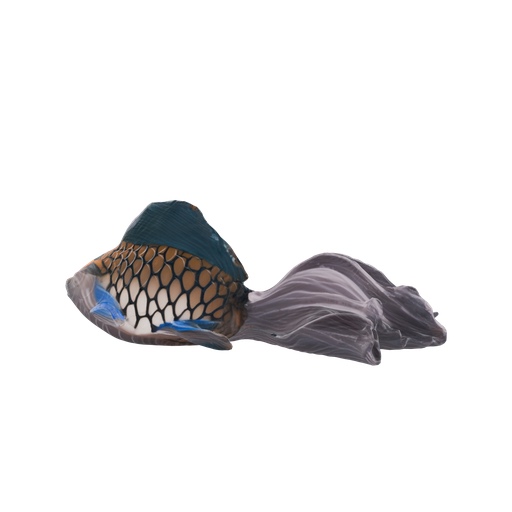} \\

  \includegraphics[width=\transfernormalswidth, valign=m]{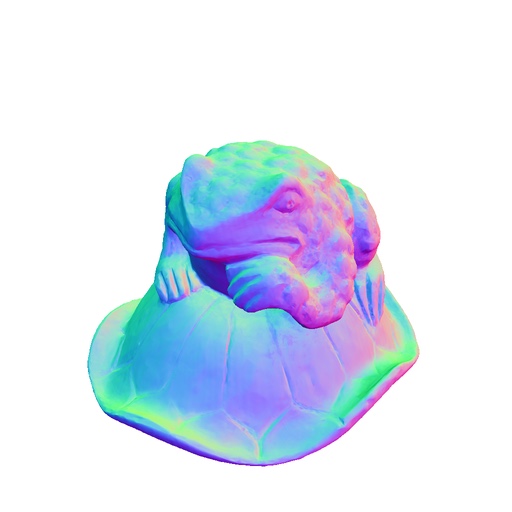} & \includegraphics[width=\transferviewwidth, valign=m]{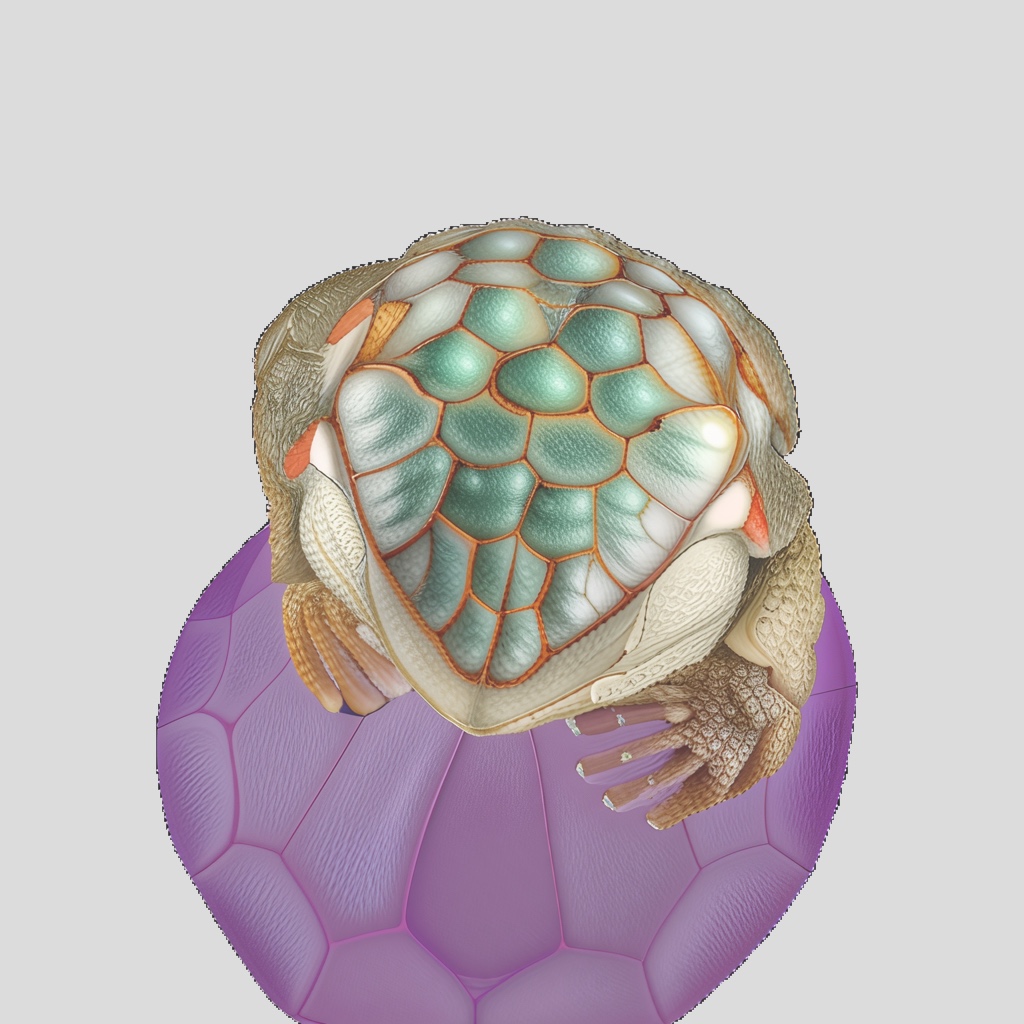} & \includegraphics[trim={10mm 10mm 10mm 10mm}, clip, width=\transferreswidth, valign=m]{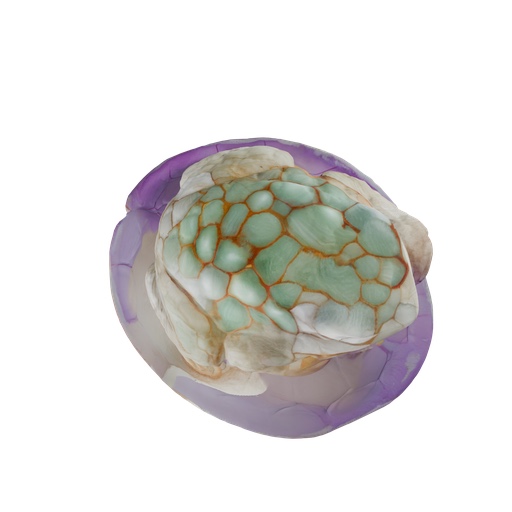} & \includegraphics[trim={10mm 10mm 10mm 10mm}, clip, width=\transferreswidth, valign=m]{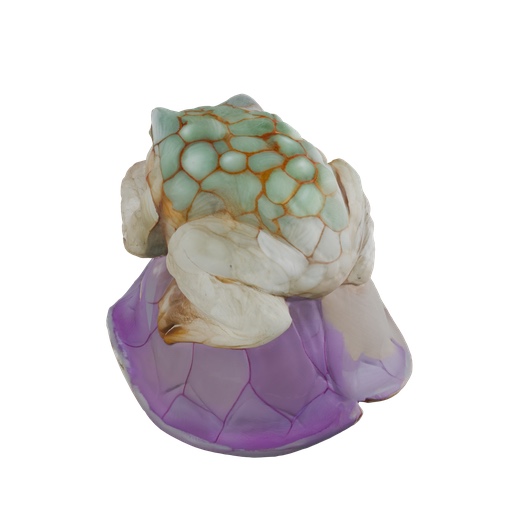} \\

  \end{tabular}
  \caption{\textbf{Automatic texturing} with fine-tuned models for the fish and reptile categories.}
  \label{fig:transfer_small}
\end{figure}

\endgroup
\newcommand{\shortnormalswidth}{0.12\linewidth}
\newcommand{\shortviewwidth}{0.12\linewidth}
\newcommand{\shortreswidth}{0.12\linewidth}

\begingroup
\setlength{\tabcolsep}{3pt}
\renewcommand{\arraystretch}{0}

\begin{figure}[tb]
  \centering
  \begin{tabular}{cccccc}
  \small Geometry & \small Single View & \small MV-Adapter & \small Hunyuan2.1 & Trellis & Ours \\

  \includegraphics[width=\shortnormalswidth, valign=m]{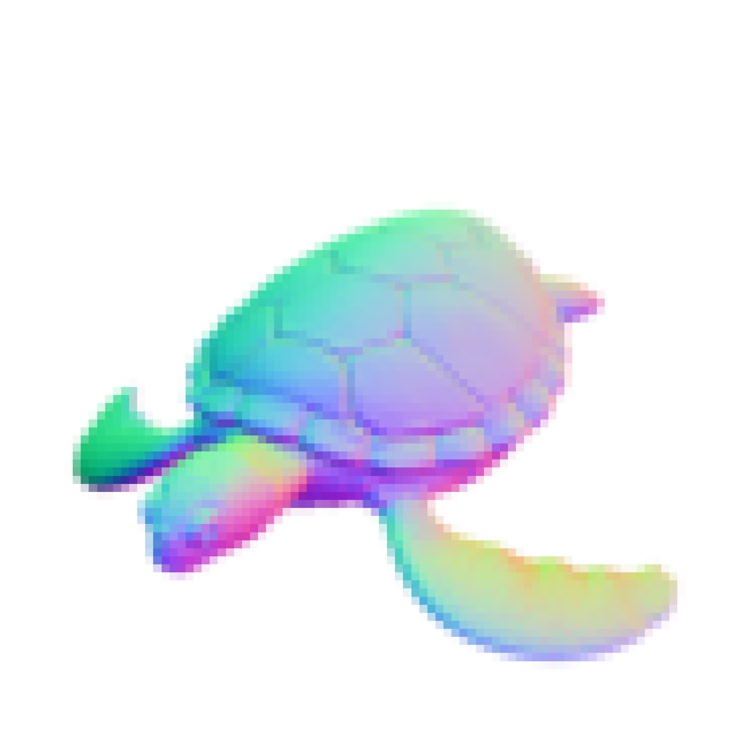} &
  \includegraphics[width=\shortviewwidth, valign=m]{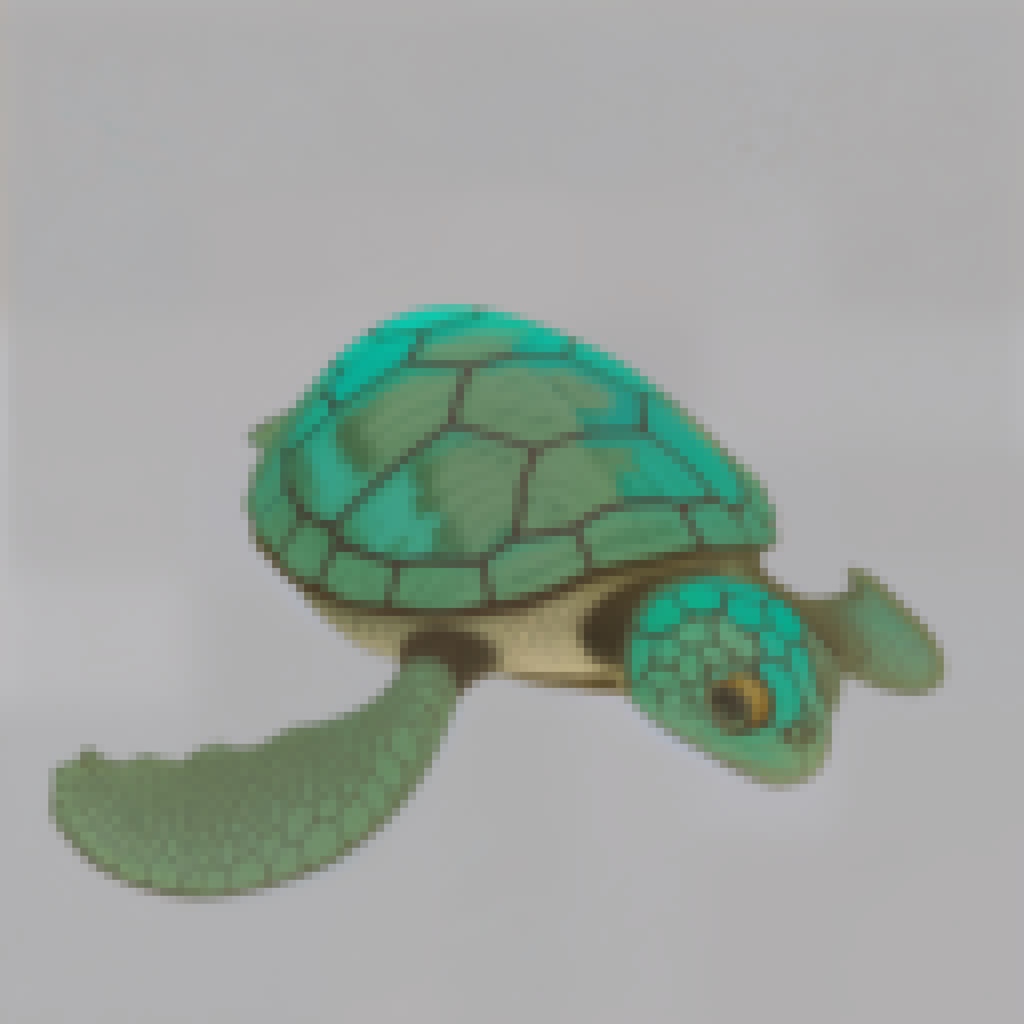} &

  \includegraphics[width=\shortreswidth, valign=m]{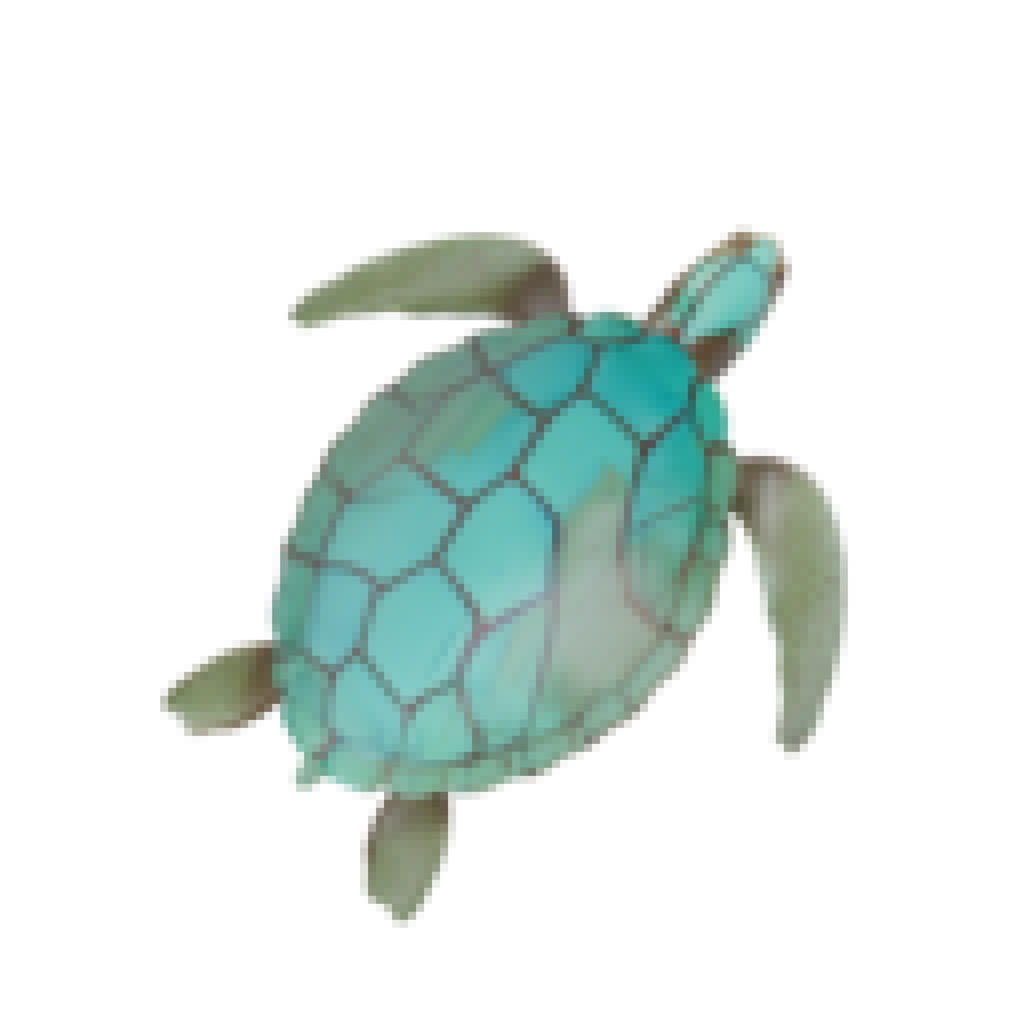} &
  \includegraphics[width=\shortreswidth, valign=m]{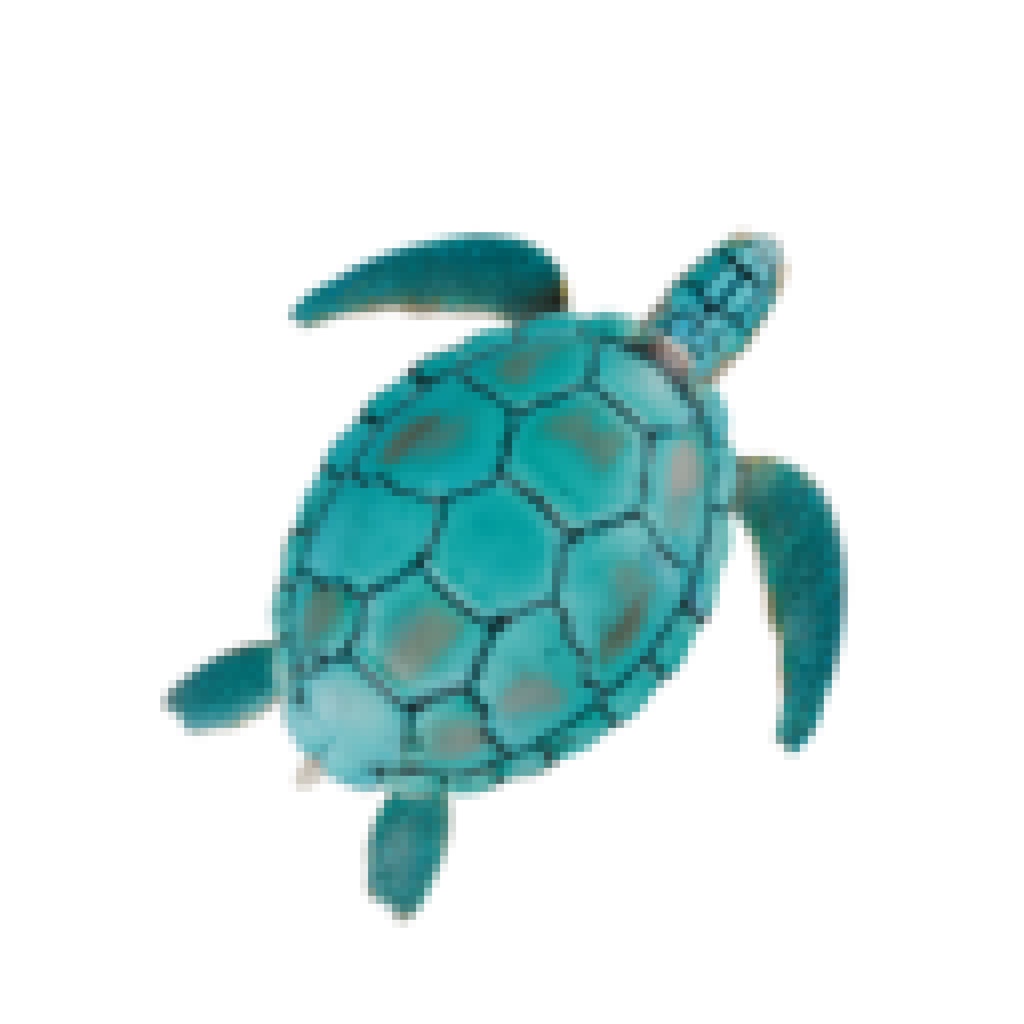} &
  \includegraphics[width=\shortreswidth, valign=m]{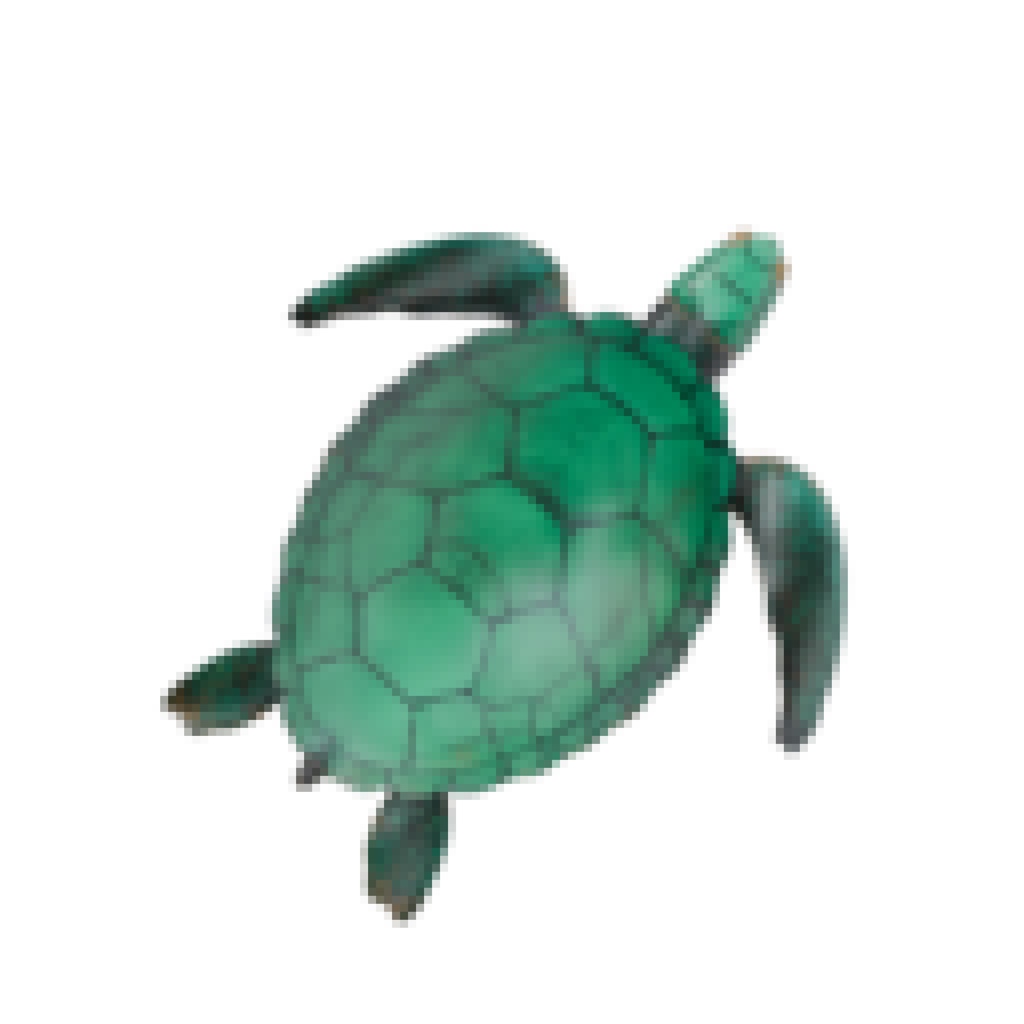} &
  \includegraphics[width=\shortreswidth, valign=m]{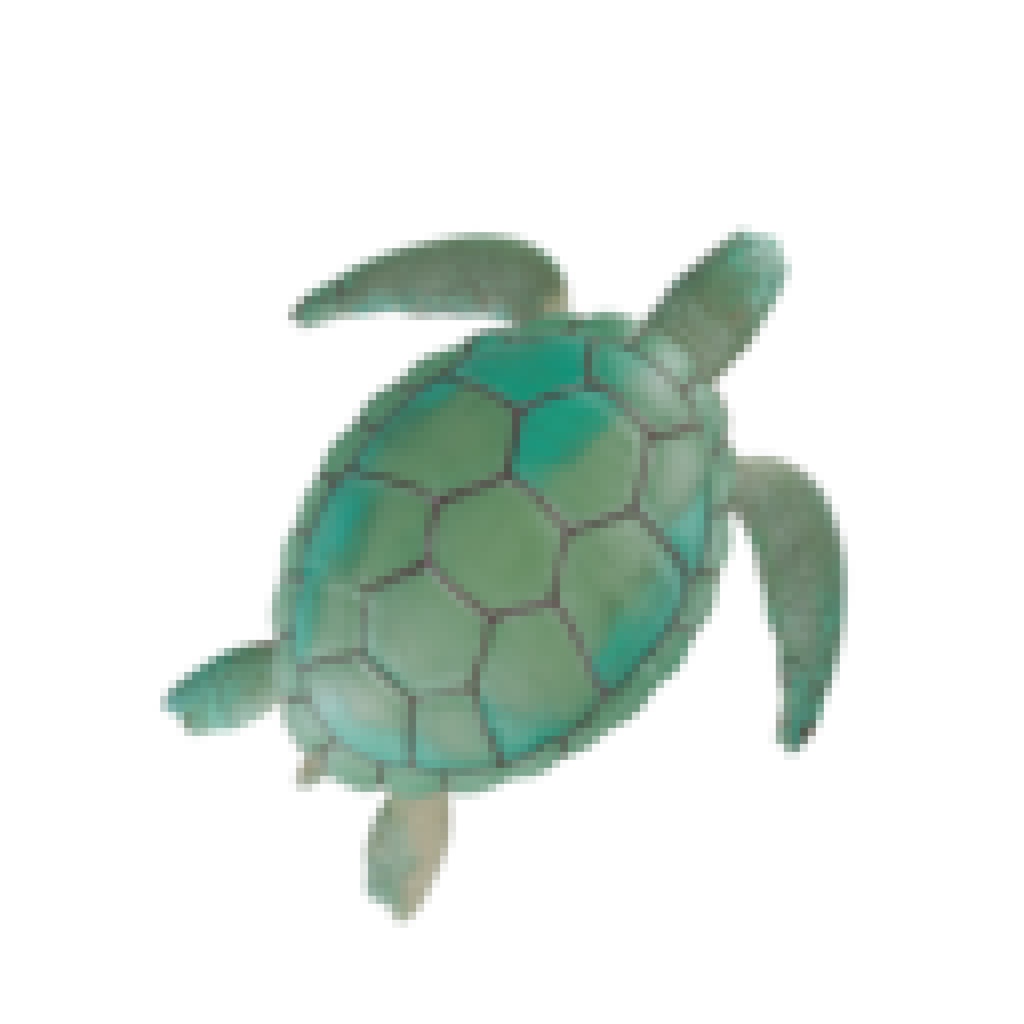} \\
  \includegraphics[width=\shortnormalswidth, valign=m]{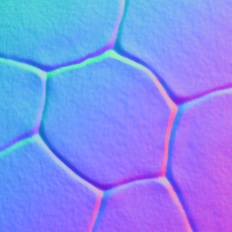} &
  \includegraphics[width=\shortviewwidth, valign=m]{img/basecolor/turtle/view0021.basecolor.jpg} &

  \includegraphics[width=\shortreswidth, valign=m]{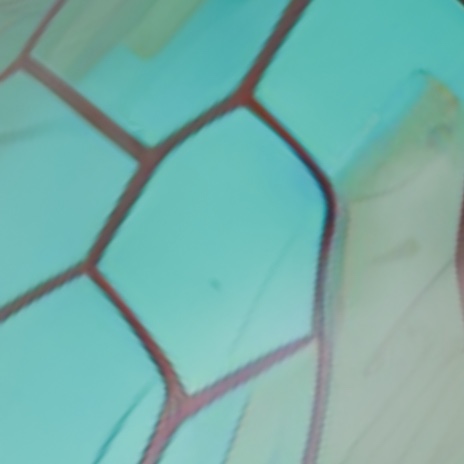} &
  \includegraphics[width=\shortreswidth, valign=m]{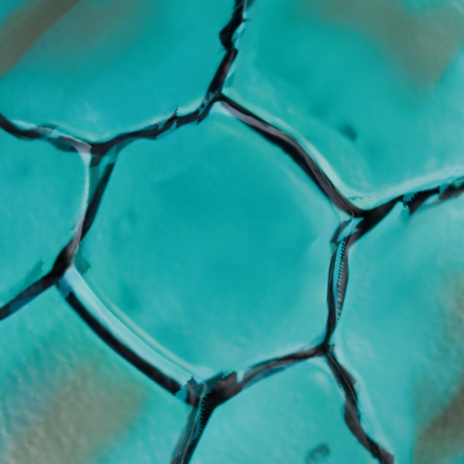} &
  \includegraphics[width=\shortreswidth, valign=m]{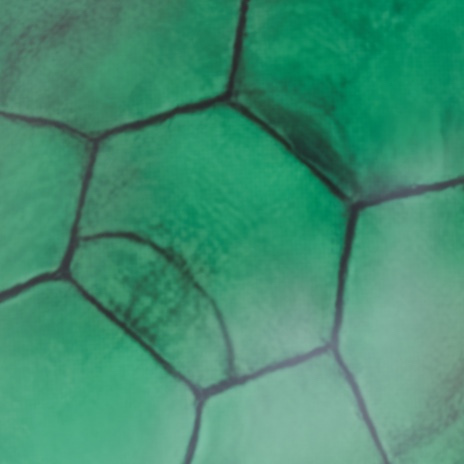} &
  \includegraphics[width=\shortreswidth, valign=m]{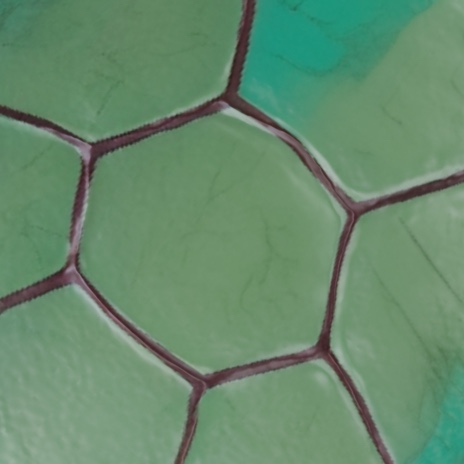} \\

  \end{tabular}
  \caption{An example of \textbf{Automatic texturing} visual comparisons.}
  \label{fig:comparison_small}
\end{figure}

\endgroup
\begin{table}[t]
\centering

{
\begingroup
\setlength{\tabcolsep}{2pt} 

\footnotesize
\renewcommand{\arraystretch}{1}

\resizebox{\linewidth}{!} {
\begin{tabular}{@{}l|cccc|cccc@{}}
\Xhline{1.0pt}
\multirow{2}{*}{\textbf{Method}} & \multicolumn{4}{c|}{\textbf{Per-mesh}} & \multicolumn{4}{c}{\textbf{Fine-tune}} \\
& LPIPS$\downarrow$ & FID$\downarrow$ & DS$\downarrow$ & CMMD$\downarrow$ & LPIPS$\downarrow$ & FID$\downarrow$ & DS$\downarrow$ & CMMD$\downarrow$ \\
\Xhline{1.0pt}
TexGen &
- &
106.917 &
0.436 &
0.883 & 
- &
124.460 & 
0.538 & 
1.143 \\

MV-Adapter &
0.358 &
76.98 & 
0.370 & 
0.666 & 
0.366 & 
122.865 & 
0.473 & 
0.961 \\

Hunyuan2.1 &
0.383 & 
\textbf{73.702} & 
0.437 & 
0.494 & 
0.299 & 
\textbf{93.045} & 
0.356 & 
\textbf{0.570} \\

Trellis2 & 
0.356 & 
102.22 & 
0.411 & 
0.545 & 
0.307 & 
110.773 & 
0.371 & 
0.777 \\

\hline

\textbf{Ours} &
\textbf{0.302} & 
80.258 & 
\textbf{0.344} & 
\textbf{0.478} & 
\textbf{0.282} & 
104.108 & 
\textbf{0.352} & 
0.694 \\

\Xhline{1.0pt}
\end{tabular}
}
\endgroup
}

\caption{\textbf{Quantitative Results:} Overall (DreamSim) and patch-level (LPIPS, FID, CMMD) metrics on automatic texture generation for per-mesh and fine-tuned models. Best in bold.}
\label{tab:quant_overall}
\end{table}

\subsection{Automatic Texture Generation}\label{ssec:eval:auto}
We showed that our model is powerful for local texture completion. Because this modality is difficult to compare quantitatively, we also compare the per-mesh and fine-tuned models we trained against large scale generative models on the full texture generation task. For our selected meshes, our data-limited texture model still shows competitive behavior.

\subsubsection{Task and Metrics} 
We evaluate our model on from-scratch and finetune settings over 10 base meshes and 9 transfer meshes. To assess overall consistency and adherence to $\tilde{I}$, we render each mesh from the opposite side of the test view and compute \textbf{DreamSim} \cite{fu2023dreamsim}, a perceptual metric aligned with human visual similarity.
For local texture quality, we estimate distributional metrics per-shape and per-view between these two patch samples using Fréchet Inception Distance (\textbf{FID}) \cite{heusel2017gans} and CLIP-based Maximum Mean Discrepancy (\textbf{CMMD}) \cite{jayasumana2024rethinking}, a more recent metric that is robust for assessing image quality. Finally, we also compute \textbf{LPIPS} \cite{zhang2018unreasonable} between patches $I_{\rho}$ rendered from known texture region using original backprojected view and generated textures, respectively. This metric is \emph{omitted for TEXGen}, where the texture in this region is the same as the back-projection. 


\subsubsection{Results} Quantitative results in Tb.~\ref{tab:quant_overall} reveal that our method achieves
the best overall score in 3/4 metrics. Our method remains competitive when transferring to 9 other meshes with shorter finetuning steps, suggesting that a single texturing model could be learned for a class of shapes for more practical artist use cases. We report full per shape metrics and visual comparisons in Supplement.
Qualitative results in Fig.~\ref{fig:comparison_small} 
(and video) 
likewise show competitive performance against state-of-the-art methods. Our method has limited generalization across shapes but shows the surprising power of local reference-guided texture generation. While {\ourmodel} lacks global context during generation and sometimes has trouble enforcing globally consistent texturing, its reference-guided design ensures close adherence to the target $\tilde{I}_{\rho}$, which is critical for user-guided applications.  



\subsubsection{User Study}\label{ssec:ustudy}

To get early feedback on the interactive potential of {\ourmodel} (\S\ref{ssec:app:brushes}, \S\ref{ssec:app:interactive}), we piloted our Blender prototype (Fig.~\ref{fig:ui}) with five professional 3D artists. Participants experimented with interactive fill and reference brush selection, and also tried fully automated texturing using the Hunyuan2.1 baseline. See Appendix~\ref{app:ustudy} for more details. 

All participants acknowledged the value of our iterative workflow, with 4/5 artists reporting that it would make their work more efficient (2 strongly agree, 2 agree, 1 neutral on a 5-point Likert scale) and all reporting that they would integrate the interactive reference and geometry-guided fill feature into their texturing workflow (2 strongly agree, 3 agree). Artist 1 noted that it “opens up a lot of creative possibilities, like creatively texturing different areas from multiple texture style images” and found selecting reference brushes ``very intuitive and convenient.'' Artist 2 remarked that the tool “blends well with [sic] multiple textures that [they] want to paint.” After comparing with automatic Hunyuan2.1 texturing, all participants reported that automatic and interactive texturing approaches are complementary, quoting Artist 3: ``the two tools could be combined to make a more powerful and flexible tool, giving creatives both a speed-up and greater creative and quality control.'' This suggests that the finer local and reference control of {\ourmodel} complements existing texture generation solutions. Participants also identified key areas for improvement, including speed and PBR integration. We further iterated on the prototype by speeding up inference time and dilating the inpainting area to avoid model's over-inpainting over geometric cues.

\subsection{Ablations}\label{ssec:additional} \label{ssec:ablations}


We conduct ablation studies to validate design choices of our model. In Tb.~\ref{tb:seg_ablations}, we compare our proposed method with the following variations with the same metrics.
\textbf{No batch-wise attention}: standard self-attention layers within $\cG_M$.
\textbf{No geometry conditioning}: No relative position and normal conditioning, $\bI_{\mathrm{geo}}$, which serves as a DTP-style baseline where the texture guidance formulation ignores local geometry.
\textbf{No image loss}: The auxiliary image-space loss, $\cL_{\mathrm{image}}$, is omitted.
\textbf{No fine-tuning}: The model is trained without initializing from the pretrained network weights. The batch-wise attention and geometry conditioning have the greatest effect on texture generation quality.

\begin{table}[t!]
\centering
\begingroup
\footnotesize
\setlength{\tabcolsep}{2pt} 
\begin{tabular}{@{}l|cc|cc|cc|cc@{}}
\Xhline{1.0pt}
\multirow{2}{*}{\textbf{Variant}} & \multicolumn{2}{c|}{\textbf{Croissant}} 
& \multicolumn{2}{c|}{\textbf{Koi Fish}} 
& \multicolumn{2}{c|}{\textbf{Sea Shell}} 
& \multicolumn{2}{c}{\textbf{Overall}} \\
& LPIPS$\downarrow$ & FID$\downarrow$ 
& LPIPS$\downarrow$ & FID$\downarrow$  
& LPIPS$\downarrow$ & FID$\downarrow$  
& LPIPS$\downarrow$ & FID$\downarrow$ \\ 
\Xhline{1.0pt}
No batch-attn. & 0.4277 & 100.93 & 0.2248 & 57.09 & 0.2923 & 81.06 & 0.3262 & 79.70 \\
No geom.\ cond. & 0.3838 & 64.62  & 0.2512 & 79.73 & 0.2810 & 90.48 & 0.3175 & 78.27 \\
No img.\ loss & 0.3006 & 62.31  & 0.2027 & 53.25 & 0.2628 & \textbf{76.76} & 0.2517 & 64.11 \\
No fine-tuning  & 0.3020 & 59.85  & 0.2011 & \textbf{51.77} & \textbf{0.2623} & 78.53 & 0.2518 & \textbf{63.38} \\
\hline
\textbf{Full model} & \textbf{0.2830} & \textbf{59.45}  & \textbf{0.2009} & 55.54 & 0.2673 & 77.11 & \textbf{0.2420} & 64.03 \\
\hline
\end{tabular}
\endgroup
\caption{\textbf{Ablations} on 3 shapes and a truncated training run using our validation set suggest that batch-attention and geometric conditioning have the greatest effect on local texture quality.}
\label{tb:seg_ablations}

\end{table}

\section{CONCLUSION}

{\ourmodel} integrates generative models into artist workflows while preserving control and flexibility. Our data pipeline and training strategy require no additional 3D data, yet maintain geometric consistency and high-quality outputs. We demonstrate that {\ourmodel} is competitive with state-of-the-art automatic methods while also advancing the frontier of interactive texturing.

\textit{Limitations and Future Work.} Operating in the small-data regime is a key strength, but it limits generalization and increases cost per shape. Future work could further develop {\ourmodel} to improve scalability and generalization across shapes. Another limitation is that our model leverages a local geometry-texture joint prior; when there is little such correlation (e.g. semantic texture, weak geometry assets), our method would not be applicable. We present such failure cases in the supplemental material. We also acknowledge that PBR generation is preliminary as the given material channels are not jointly trained; thus consistency across channels is not guaranteed and remains an important direction for future work.

\begin{acks}
We would like to thank members of the PIXL group for their feedback on the project.
\end{acks}

\bibliographystyle{ACM-Reference-Format}
\bibliography{main}

\clearpage


\clearpage
\appendix
\section{Data Generation}\label{app:datagen}
\paragraph{Prompt generation} We use OpenAI's o5 model to generate detailed prompts for single-view image generation. The system prompt used is ``Generate concise, diverse and vivid prompts ($\sim$15-25 words) for objects. Output *one prompt per line*, no numbering.'' This is followed by the user prompt: ``Generate 10 prompts for different kinds of **{subject}** with different appearances in a studio setting. Use vivid short object centric adjectives instead of long prose'' where ``**subject**'' is replaced by a short description of the mesh object e.g.\ ``sea urchin shell''. If satisfied with the response, the model is prompted to continue generating prompts with the following user prompt: ``Great! Now generate 500 more prompts for **{subject}** with different appearances just like that.'' 
Here is an example of a prompt generated for the sea urchin: ``A sea urchin shell in smoky purple, its texture lightly powdered, rows of tiny spines etched as shallow lumps.''

\paragraph{Single View Generation}
 We apply CLIP-based aesthetic score filtering to remove low-quality ControlNet-generated views. We use DiffusionRenderer~\cite{DiffusionRenderer} for intrinsic decomposition. In our experiments, the predicted albedo generally has sufficient quality for training, and we do not observe severe artifacts propagating to the final textured results. 

\section{Additional Model Details}\label{app:model}

\begin{figure}
\centering
    \centering
    \includegraphics[ width=\linewidth]{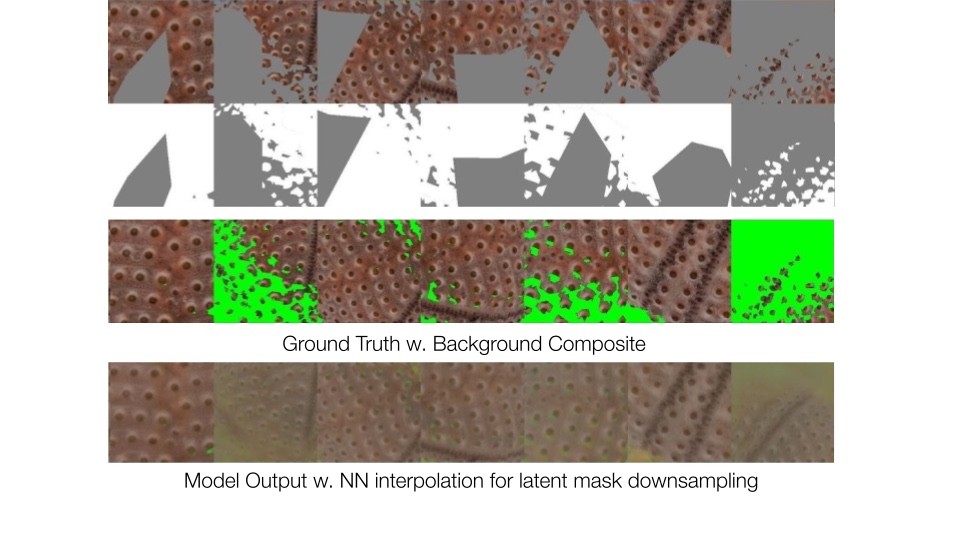}
    \caption{Color leak effect from training model using nearest neighbor interpolation for latent mask downsampling. Top two rows show masked albedo and corresponding mask as model inputs; the third row is ground truth composited with green; the last row shows the model output. }
    \label{fig:colourleak}
\end{figure}

\paragraph{Inpainting mask down-sampling} We apply a min-pooling operation to downsample masks to the latent dimension instead of nearest neighbor interpolation used in stable diffusion inpainting. This ensures that a latent pixel is masked if any image pixel in that region is masked. This is especially important during training for masking the latent space loss as nearest neighbor interpolation will leak invalid untextured regions into the loss.  
As shown in Fig.~\ref{fig:colourleak}, the model trained with a nearest neighbor interpolation mask shows signs of green from the invalid texture regions. With min-pooling, the model does not show signs of colour leak.

\section{Additional Training Details}\label{app:training}

\paragraph{Diffusion loss formulation} We adopt the noise schedule and loss formulation from Stable Diffusion 2.1. Our noisy input, $z_t$, is computed using the DDPM formulation:
\begin{equation}
z_t = \sqrt{\bar{\alpha}_t} z_0 + \sqrt{1 - \bar{\alpha}_t} \, \epsilon, \quad \epsilon \sim \mathcal{N}(0, \mathbf{I}).
\end{equation}
The parameters $\alpha_t$ are computed via the scaled linear DDPM schedule. Specifically, $\beta_{\min}=8.5\times10^{-4}$ and $\beta_{\max}=1.2\times10^{-2}$.
Define
\begin{equation}
\beta_t
= \left(
\sqrt{\beta_{\min}}
+ \frac{t-1}{T-1}
\left(\sqrt{\beta_{\max}}-\sqrt{\beta_{\min}}\right)
\right)^2,
\qquad t=1,\dots,T.
\end{equation}
Then set $\alpha_t = 1-\beta_t$ and
$\bar{\alpha}_t = \prod_{s=1}^t \alpha_s$.

For the velocity prediction~\cite{salimans2022progressive}, we compute,
\begin{equation}
v_t = \sqrt{\bar{\alpha}_t}\,\epsilon
      - \sqrt{1-\bar{\alpha}_t}\,z_0.
\end{equation}
\noindent This is the target for our trained model $\cG_M$.

\paragraph{Hyperparameters} For the from-scratch experiment setting, we use AdamW optimizer with learning rate $1e-4$ with batch size of 32 for 80k steps and EMA decay of 0.999. The training for a single object takes around 40 hours to train on 4 NVIDIA A100 GPUs. For the finetune experiment setting, we use 100 single views, batch size 16, and 20k steps on 2 NVIDIA L40 GPUs for 8hr. We use the finetuned model for interactive applications for the foliage, the dragon head and the lizard creature.

\paragraph{Training Data Generation}


To train a shape-specific {\ourmodel} model on each mesh,
we generate 550 single-view renderings (512$\times$512) using Realistic Vision v5.1 \cite{realisticvisioncivitai}, a community photorealistic variant of Stable Diffusion v1.5 \cite{rombach2022high}, augmented with ControlNets \cite{zhang2023controlnet} for normals, depth, and canny edges. To extract albedo, we use \citet{liang2025diffrend} and upscale it to 2048$\times$2048 using \citet{yue2025invsr}. Each single-view dataset is split into 50 \textbf{eval}, 50 \textbf{test} and 450 \textbf{train} views. 
To finetune a new shape-specific model from a base model we generate 200 single views, 100 for training, 50 for \textbf{eval}, 50 for \textbf{test}. We choose these based on ablations of the number of single views, and the number of finetuning steps in Tb.~\ref{tb:ablation_step} and Tb.~\ref{tb:ablation_views}.

\section{Evaluation Details}\label{app:eval}
\begin{table*}[t]
\centering

{
\begingroup
\setlength{\tabcolsep}{4.5pt} 

\footnotesize
\renewcommand{\arraystretch}{1}

\begin{tabular}{l|l|c||ccccccccc c}
\Xhline{1.0pt}
 \multirow{2}{*}{\textbf{Metric}} & \multirow{2}{*}{\textbf{Method}} & \multirow{2}{*}{\textbf{Overall}} &
\multicolumn{10}{c}{\textbf{by Example}} \\
\cline{4-13}
& & & Brick & Cabbage & Croissant & Tire & Fire Hydrant &
Gourd & Koi Fish & Barrel & Sea Shell & Turtle  \\
\Xhline{2.0pt}
\multirow{3}{*}{LPIPS$\downarrow$}
& Paint3D
& 0.378 & 0.580 & 0.501 & 0.370 & 0.379
& 0.265 & 0.432 & 0.337 & 0.328 & 0.348 & 0.242  \\
& MV-Adapter
& 0.358 & 0.650 & 0.485 & 0.340 & 0.352 & 0.226 & 0.429 &
  0.286 & 0.295 & 0.302 & 0.211 \\
& Hunyuan2.1
& 0.383 & 0.577 & 0.498 & 0.352 & 0.370
& 0.272 & 0.604 & 0.303 & 0.314 & 0.328 & 0.214  \\
& Trellis2.0
& 0.356 & 0.547 & 0.482 & 0.341 & 0.358 & 0.283 & 0.407 &
  0.299 & 0.307 & 0.327 & 0.213 \\
& \textbf{Ours} 
& \textbf{0.302} & \textbf{0.462} & \textbf{0.386} & \textbf{0.256} & \textbf{0.319}
& \textbf{0.208} & \textbf{0.361} & \textbf{0.281} & \textbf{0.283} & \textbf{0.283} & \textbf{0.181} 
\\
\Xhline{1.0pt}

\multirow{4}{*}{FID$\downarrow$}

& Paint3D
& 98.765 & 124.633 & 131.087 & 83.139 & 149.776
& 82.467 & 97.246 & 84.359 & 76.715 & 79.076 & 79.149 \\ 
& TexGen
& 106.917 & 224.421 & 130.891 & 60.657 & 120.995
& 105.353 & 80.442 & 80.190 & 52.466 & 109.557 & 104.197 \\
& MV-Adapter
& 76.98 & 166.61 & 83.54 & 48.58 & \textbf{97.36} & 85.45 & \textbf{51.04} &
  51.10 & 53.52 & 64.99 & 67.60 \\
& Hunyuan2.1
& \textbf{73.702} & \textbf{93.570} & \textbf{88.890} & 63.040 & 99.270
& 87.750 & 73.760 & 52.510 & \textbf{47.840} & \textbf{51.560} & 78.830 \\
& Trellis2.0
& 102.22 & 185.81 & 153.14 & 75.48 & 108.47 & 106.83 & 96.52 &
   61.30 & 51.99 & 96.71 & 85.91 \\
& \textbf{Ours} 
& 80.258 & 130.522 & 95.320 & \textbf{50.175} & 119.024
& \textbf{63.553} & \textbf{59.799} & 60.633 & 75.219 & 88.413 & \textbf{59.925} \\
\Xhline{1.0pt}

\multirow{4}{*}{DreamSIM$\downarrow$}

& Paint3D
& 0.412 & 0.480 & 0.368 & 0.360 & 0.543
& 0.396 & 0.348 & 0.368 & 0.507 & 0.387 & 0.359 \\
& TexGen
& 0.436 & 0.626 & 0.391 & 0.353 & 0.526
& 0.377 & 0.398 & 0.361 & 0.547 & 0.422 & 0.361 \\
& MV-Adapter
& 0.370 & 0.629 & 0.301 & 0.314 & 0.430 & 0.356 & 0.301 &
  0.300 & 0.458 & 0.318 & 0.293 \\
& Hunyuan2.1
& 0.437 & 0.503 & 0.372 & 0.394 & 0.493
& 0.438 & 0.509 & 0.395 & 0.525 & 0.380 & 0.362 \\
& Trellis2.0 
& 0.411 & 0.484 & 0.354 & 0.395 & 0.453 & 0.464 & 0.358 &
  0.381 & 0.497 & 0.373 & 0.352 \\
& \textbf{Ours}
& \textbf{0.344} & \textbf{0.396} & \textbf{0.296} & \textbf{0.289} & \textbf{0.439}
& \textbf{0.340} & \textbf{0.278} & \textbf{0.331} & \textbf{0.486} & \textbf{0.316} & \textbf{0.270} \\
\Xhline{1.0pt}

\multirow{4}{*}{CMMD$\downarrow$}
& Paint3D
& 0.728 & \textbf{0.597} & 1.198 & 0.709 & 0.836
& 0.827 & 0.706 & 0.923 & 0.274 & 0.607 & 0.604 \\
& TexGen
& 0.883 & 1.538 & 1.262 & 0.743 & 0.942
& 0.800 & 0.897 & 0.803 & \textbf{0.257} & 0.845 & 0.738 \\
& MV-Adapter
& 0.666 & 1.424 & 0.896 & 0.816 & 0.507 & 0.678 & 0.521 &
  0.570 & 0.224 & 0.476 & 0.546 \\
& Hunyuan2.1
& 0.494 & 0.806 & 0.664 & 0.476 & 0.591
& 0.470 & \textbf{0.345} &\textbf{0.407} & 0.312 & \textbf{0.417} & 0.456 \\
& Trellis2.0
& 0.545 & 0.913 & 0.812 & 0.619 & 0.475 & 0.685 & 0.512 &
  0.370 & 0.179 & 0.361 & 0.521 \\
& \textbf{Ours}
& \textbf{0.478} & 0.888 & \textbf{0.430} & \textbf{0.384} & \textbf{0.573}
& \textbf{0.415} & 0.365 & 0.506 & 0.291 & 0.518 & \textbf{0.405} \\
\Xhline{1.0pt}
\end{tabular}

\endgroup
}

\caption{\textbf{Quantitative Results:} Overall (DreamSim) and patch-level (LPIPS, FID, CMMD) metrics on texture generation task conditioned on a single view, across 10 shapes with 50 views each, reveal competitive performance of our method, despite using no textured 3D training data for training. }
\label{tab:quantitative}
\end{table*}

\begin{table*}[t]
\centering

{
\begingroup
\setlength{\tabcolsep}{4.5pt}

\footnotesize
\renewcommand{\arraystretch}{1}

\begin{tabular}{l|l|c||ccccccccc}
\Xhline{1.0pt}
 \multirow{2}{*}{\textbf{Metric}} & \multirow{2}{*}{\textbf{Method}} & \multirow{2}{*}{\textbf{Overall}} &
\multicolumn{9}{c}{\textbf{by Example}} \\
\cline{4-12}
& & & Cabbage1 & Cabbage3 & Cantaloup & Fish1 & Fish3 & Fish4 & Toad & Tortoise & Turtle  \\
\Xhline{2.0pt}
\multirow{5}{*}{LPIPS$\downarrow$}
& MV-Adapter
& 0.366 & 0.319 & 0.299 & 0.598 & 0.301 & 0.409 & 0.247 & 0.399 & 0.387 & 0.334 \\
& Hunyuan2.1
& 0.299 & 0.287 & 0.266 & 0.440 & 0.253 & 0.335 & 0.187 & 0.327 & 0.318 & 0.279 \\
& Trellis2.0
& 0.307 & 0.284 & 0.279 & 0.471 & 0.268 & 0.379 & 0.188 & 0.321 & 0.306 & 0.269 \\
& \textbf{Ours}
& \textbf{0.282} & \textbf{0.278} & \textbf{0.255} & \textbf{0.407} & \textbf{0.248} & \textbf{0.308} & \textbf{0.172} & \textbf{0.315} & \textbf{0.287} & \textbf{0.268} \\
\Xhline{1.0pt}

\multirow{5}{*}{FID$\downarrow$}
& TexGen
& 124.460 & \textbf{116.061} & 84.111 & 290.178 & 130.762 & 120.826 & \textbf{107.264} & 52.956 & 113.996 & 103.989 \\
& MV-Adapter
& 122.865 & 202.740 & 117.259 & 175.604 & 106.143 & 113.812 & 146.131 & 63.382 & 84.237 & 96.476 \\
& Hunyuan2.1
& \textbf{93.045} & 196.583 & \textbf{60.828} & \textbf{96.139} & \textbf{89.571} & \textbf{84.585} & 119.341 & 51.564 & \textbf{70.519} & \textbf{68.278} \\
& Trellis2.0
& 110.773 & 191.301 & 99.255 & 134.525 & 113.500 & 116.402 & 120.003 & 62.463 & 77.913 & 81.596 \\
& \textbf{Ours}
& 104.108 & 123.885 & 83.397 & 198.546 & 98.445 & 101.567 & 110.265 & \textbf{50.494} & 87.696 & 82.681 \\
\Xhline{1.0pt}

\multirow{5}{*}{DreamSim$\downarrow$}
& TexGen
& 0.538 & 0.459 & 0.499 & 0.493 & 0.564 & 0.509 & 0.571 & 0.595 & 0.599 & 0.552 \\
& MV-Adapter
& 0.473 & 0.318 & 0.361 & 0.540 & 0.507 & 0.487 & 0.481 & 0.525 & 0.525 & 0.512 \\
& Hunyuan2.1
& 0.356 & \textbf{0.264} & \textbf{0.317} & 0.280 & 0.425 & \textbf{0.317} & 0.355 & \textbf{0.415} & 0.441 & 0.387 \\
& Trellis2.0
& 0.371 & 0.293 & 0.341 & \textbf{0.263} & 0.427 & 0.367 & 0.372 & 0.458 & 0.417 & 0.401 \\
& \textbf{Ours}
& \textbf{0.352} & 0.265 & 0.342 & 0.319 & \textbf{0.407} & 0.340 & \textbf{0.352} & 0.427 & \textbf{0.392} & \textbf{0.326} \\
\Xhline{1.0pt}

\multirow{5}{*}{CMMD$\downarrow$}
& TexGen
& 1.143 & 1.346 & 1.041 & 2.980 & 0.990 & 0.883 & 0.584 & 0.585 & 0.912 & 0.969 \\
& MV-Adapter
& 0.961 & 1.376 & 0.873 & 1.798 & 0.904 & 0.942 & 0.867 & 0.482 & 0.590 & 0.821 \\
& Hunyuan2.1
& \textbf{0.570} & 1.455 & \textbf{0.546} & \textbf{0.491} & \textbf{0.497} & \textbf{0.444} & \textbf{0.540} & \textbf{0.309} & \textbf{0.380} & \textbf{0.471} \\
& Trellis2.0
& 0.777 & 2.058 & 0.835 & 0.793 & 0.571 & 0.626 & 0.602 & 0.437 & 0.492 & 0.577 \\
& \textbf{Ours}
& 0.694 & \textbf{1.336} & 0.721 & 0.905 & 0.640 & 0.609 & 0.598 & 0.383 & 0.548 & 0.510 \\
\Xhline{1.0pt}
\end{tabular}

\endgroup
}

\caption{\textbf{Quantitative Results on Transfer Meshes:} Overall and per-example LPIPS, FID, DreamSim, and CMMD on the texture-transfer evaluation set (9 sub-meshes spanning 3 base shapes). Best per column in bold. }
\label{tab:quantitative_transfer}
\end{table*}

\subsection{Baseline Details}
For fair comparison, we set the text conditioning for all methods as the object name, as ours takes in an empty string and relies solely on the single view conditioning. Although Paint3D includes intrinsic image decomposition (de-lighting) in its pipeline, for fair comparison, we bypass this step and instead directly provide the albedo extracted from the single input image as supervision.



\subsection{Metrics Details}
For all patch-level experiments, we render patches at a resolution of 256 by 256. We fix the camera–mesh distance to 0.25 and vary the field of view from 0.4 to 0.8 to introduce variations in perceived viewing distance. For LPIPS evaluation, we sample camera views on faces with existing textures to enable comparison against ground-truth textures. For FID and CMMD, we select views that are entirely untextured.

\subsection{Model Ablations}
\begin{figure}
\centering
    \hfill
    \subfloat[ablations]{
    \begin{tabular}{c}
        \includegraphics[width=\linewidth]{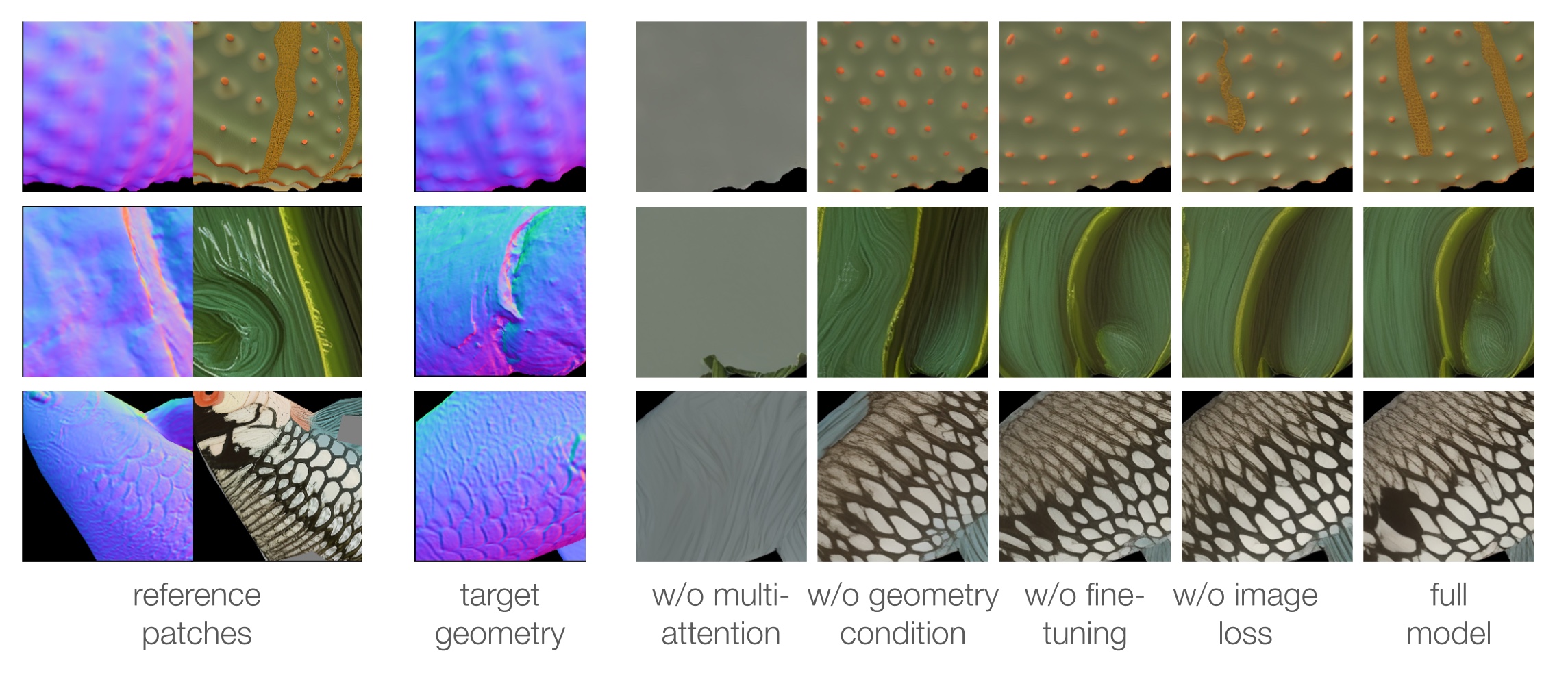} \\
    \end{tabular}
    }
    \caption{We show a visual comparison of our model ablation experiments. We show reference albedo and geometry patches on the left and each model variant's output.}
    \label{fig:ablations}
\end{figure}

We present visual comparisons to analyze key architectural and training design choices. Removing multi-attention causes the model to completely fail at texture generation. Without geometry conditioning, the model relies primarily on the inpainting mask shape and texture priors, leading to geometry-inconsistent results: in the croissant examples, surface creases are misaligned with the underlying geometry, and in the fish example, fin textures incorrectly extend onto the body. (See Fig.~\ref{fig:ablations})

The variant without the image-space loss performs competitively with the full model but occasionally loses fine-grained texture details due to the absence of image-level supervision. Similarly, the model without fine-tuning produces competitive results overall, but sometimes fails to capture subtle texture variations, as illustrated by the urchin shell example.

\subsection{Completion Ablations}
We report quantitative metrics for completion results produced by different variants of our automatic texturing pipeline (Tb.~\ref{fig:completion-ablation}). Our results show that setting SyncMVD latent guidance strength to 2 yields the highest quantitative texture quality, whereas increasing the guidance strength to 10 degrades performance. All ablated variants underperform the original completion setting, indicating that each component of the pipeline is essential for GLOSS to achieve high-quality automatic texture generation.
\begin{table}[t!]
\centering
\begingroup
\footnotesize
\setlength{\tabcolsep}{4pt} 
\begin{tabular}{@{}|l||c|c|c|@{}}
\Xhline{2.0pt}
\textbf{Experiment} & \textbf{LPIPS$\downarrow$} & \textbf{DreamSim$\downarrow$} & \textbf{FID$\downarrow$} \\ 
\Xhline{2.0pt}
guidance\_2 (final) & \textbf{0.3031} & 0.3441 & 80.24 \\
wo\_syncmvd & 0.3040 & \textbf{0.3393} & \textbf{74.81} \\
guidance\_4 & 0.3044 & 0.3514 & 86.64 \\
guidance\_10 & 0.3169 & 0.3829 & 120.97 \\
overlap\_0.4 & 0.3099 & 0.3644 & 97.85 \\
wo\_color\_correct & 0.3044 & 0.3497 & 84.41 \\
wo\_custom\_weight & 0.3070 & 0.3493 & 82.12 \\
wo\_nnfm & 0.3090 & 0.3507 & 84.79 \\
\Xhline{2.0pt}
\end{tabular}
\endgroup
\caption{Quantitative evaluation of textures generated with variations of final completion experiment settings. We report average LPIPS, DreamSim, and FID metrics over 10 objects.}
\label{fig:completion-ablation}
\end{table}

\subsection{Finetune Ablations}
We conduct finetune experiment ablation on single view numbers and training steps.
\begin{table}[t!]
\centering
\begingroup
\footnotesize
\setlength{\tabcolsep}{2pt}
\begin{tabular}{@{}l|c|c|c|c|c|c|c|c@{}}
\Xhline{1.0pt}
\multirow{2}{*}{} & \multicolumn{2}{c|}{Cabbages} & \multicolumn{2}{c|}{Fishes} & \multicolumn{2}{c|}{Turtles}
& \multicolumn{2}{c}{Overall} \\
\cline{2-9}
& \textbf{LPIPS$\downarrow$} & \textbf{FID$\downarrow$} & \textbf{LPIPS$\downarrow$} & \textbf{FID$\downarrow$} & \textbf{LPIPS$\downarrow$} & \textbf{FID$\downarrow$} & \textbf{LPIPS$\downarrow$} & \textbf{FID$\downarrow$} \\
\Xhline{2.0pt}
v100--10k & 0.260 & \textbf{75.87} & 0.176 & 111.95 & 0.263 & \textbf{76.32} & 0.233 & \textbf{88.05} \\
v100--20k & \textbf{0.253} & 80.78 & \textbf{0.169} & 109.63 & \textbf{0.253} & 76.33 & \textbf{0.225} & 88.91 \\
v100--50k & 0.255 & 80.16 & 0.170 & \textbf{104.27} & 0.268 & 88.24 & 0.231 & 90.89 \\
\hline
\end{tabular}
\endgroup
\caption{\textbf{Step-count ablation} on the 100 view variant. LPIPS / FID at 10k, 20k, and 50k fine-tuning steps. We show that finetuning results are best at an early stage, enabling faster convergence.}
\label{tb:ablation_step}
\end{table}

\begin{table}[t!]
\centering
\begingroup
\footnotesize
\setlength{\tabcolsep}{2pt}
\begin{tabular}{@{}l|c|c|c|c|c|c|c|c@{}}
\Xhline{1.0pt}
\multirow{2}{*}{} & \multicolumn{2}{c|}{Cabbages} & \multicolumn{2}{c|}{Fishes} & \multicolumn{2}{c|}{Turtles}
& \multicolumn{2}{c}{Overall} \\
\cline{2-9}
& \textbf{LPIPS$\downarrow$} & \textbf{FID$\downarrow$} & \textbf{LPIPS$\downarrow$} & \textbf{FID$\downarrow$} & \textbf{LPIPS$\downarrow$} & \textbf{FID$\downarrow$} & \textbf{LPIPS$\downarrow$} & \textbf{FID$\downarrow$} \\
\Xhline{2.0pt}
50 views & 0.255 & \textbf{72.80} & 0.170 & 108.66 & \textbf{0.257} & \textbf{78.77} & \textbf{0.227} & \textbf{86.74} \\
100 views & \textbf{0.255} & 80.16 & \textbf{0.170} & \textbf{104.27} & 0.268 & 88.24 & 0.231 & 90.89 \\
200 views & 0.255 & 83.40 & 0.172 & 110.26 & 0.268 & 82.68 & 0.232 & 92.11 \\
\hline
\end{tabular}
\endgroup
\caption{\textbf{View-count ablation} comparing v50, v100, and v200 fine-tunes. We show that we can finetune with a small number of views.}
\label{tb:ablation_views}
\end{table}

\subsection{Attention Analysis}

\begin{figure}
\centering
    \centering
    \includegraphics[width=\linewidth]{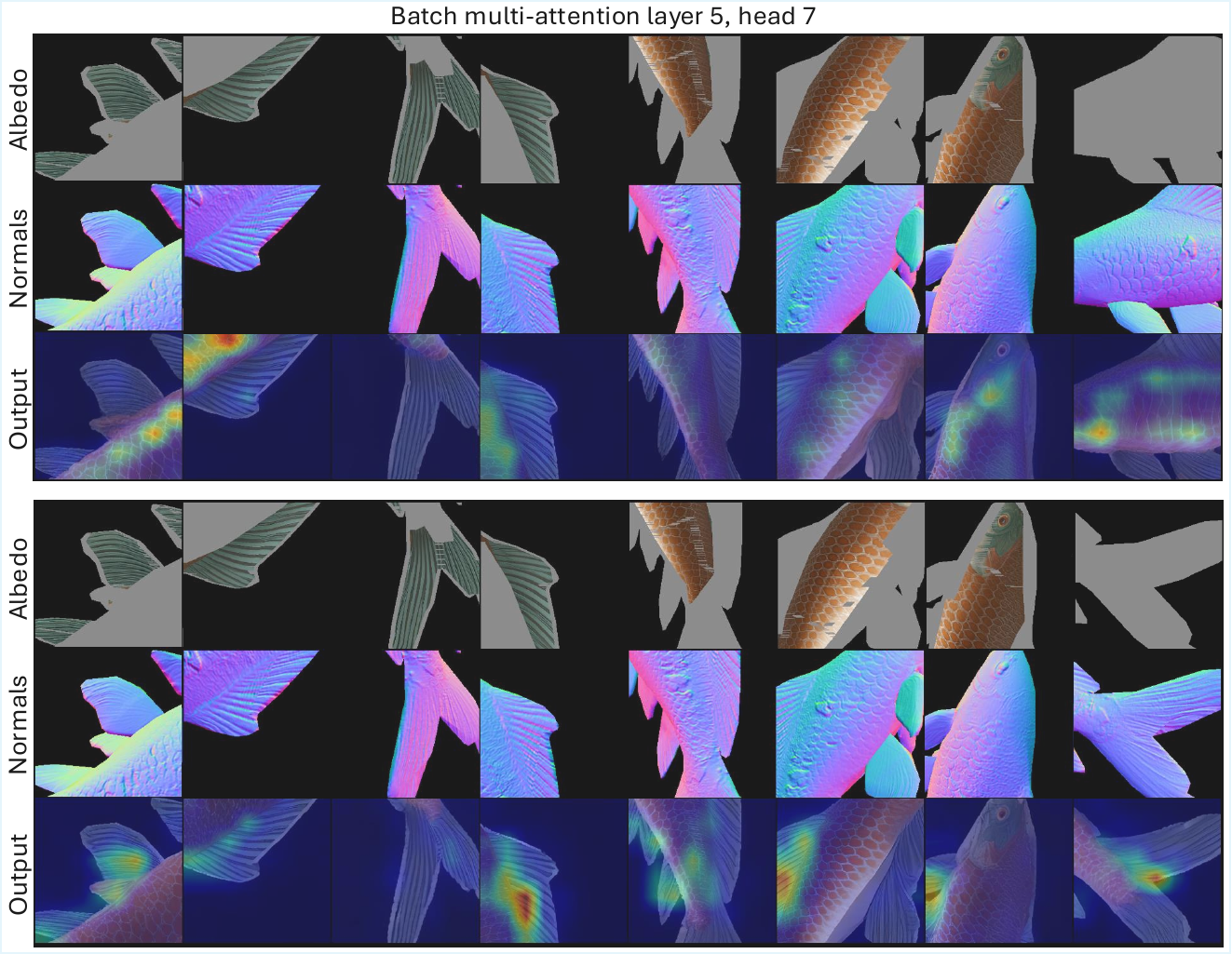}
    \caption{Batch multi-attention visualization comparing two batches with the same reference images but a different target image (last in the batch). The overlaid heatmap shows how the center patch token in the target image attends to all patch tokens across the batch.}
    \label{fig:attnviz}
\end{figure}

In Fig.~\ref{fig:attnviz} we visualize the batch multi-attention weights from a single attention head in the UNet. The heatmap overlaid on the model output is obtained using the attention weights from the center patch of the target image as the query. The attention weights are reshaped and upsampled to match the image resolution. The weights are then normalized across the batch before applying a colour map. 
While some attention heads may attend to similar parts of the image, some heads like the one shown in Fig.~\ref{fig:attnviz} show distinct differences based on the target image. We see that the first batch highlights regions with fish scales while the second batch highlights more fins. 


\section{Inference Settings and Timings}

We use a patch size of 256x256 at train and inference time, and 1K or 4K textures for our experiments (client-server architecture of our add-on makes 1K textures faster to work with). For interactive completion, typically 7 references and 1 target patch are used. 

\paragraph{Unoptimized timings} Rendering patches and NNFM reference matching takes around 1s. Model inference takes 1.4s. Backprojection from camera image space to texture space takes 1s for 1K texture resolution and almost 3s for 4K texture resolution. Total processing time is 3.4s for 1K texture painting and 5.3s for 4K textures. This is not including GLOSS server and Blender add-on communication time. Model inference and texture processing timings can be further optimized to near real-time speeds, as shown to be possible by \citet{hu2024diffusion}.

\section{Failure Cases}\label{app:failure}

\begin{figure}
\centering
    \centering
    \includegraphics[width=\linewidth]{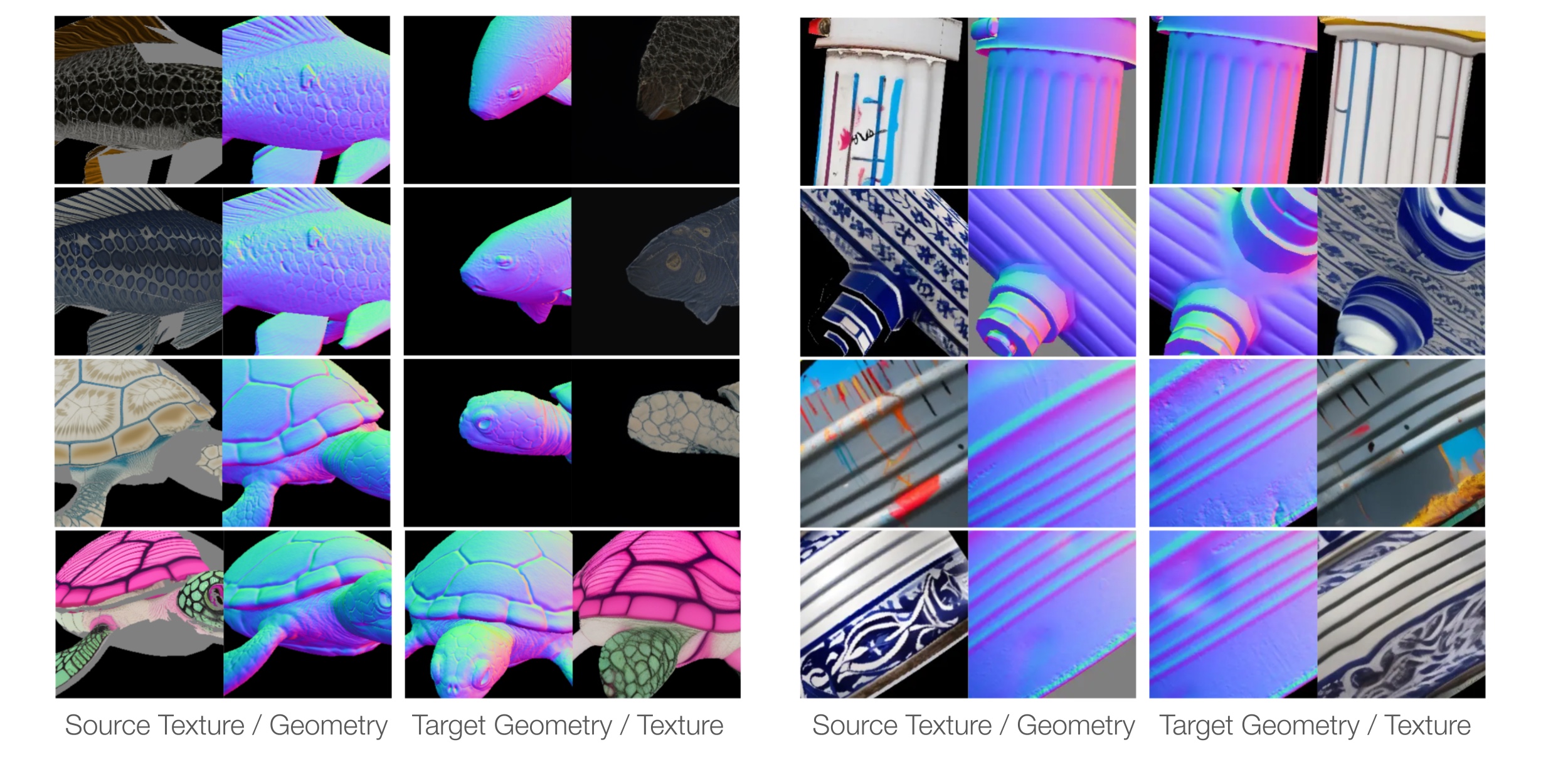}
    \caption{On the left we show examples of semantic texture failure. On the right we show examples of failure cases due to low geometry and texture correlation.}
    \label{fig:failure}

\end{figure}

Our model relies on learned local correlations between geometry and texture to texture an asset. While it excels at generating fine-grained details with high geometric and reference fidelity, its ability to infer semantic appearance from limited local reference context remains constrained. For example, the turtle’s face and mouth lack fine-grained semantic details, while the koi fish eyes are not rendered with sufficiently high-quality textures. We also examine failure cases in regions where the correlation between geometry and texture is weak. Although it can reproduce repetitive patterns to some extent, the fidelity and consistency of these patterns remain limited. The model may also fail to preserve small, stochastic, non-repetitive strokes that are weakly correlated with the local geometry. We show such cases in Fig.~\ref{fig:failure}.

\section{User Study Details}\label{app:ustudy}

To evaluate the potential of our interactive texture painting model, we conducted a pilot user study in which artists interacted with a Blender-based prototype built on our method. This section details the study design, participant background, experimental procedure, and collected feedback.

\subsection{Participants}
We recruited five participants with prior experience in 3D texturing and general familiarity with Blender~\cite{blender}. The study was advertised through online 3D artist communities. Applicants were screened based on their experience with 3D texture painting, and both students and professional artists were considered. Each study session lasted approximately 90 minutes. Participation was voluntary, and no compensation was provided.

\subsection{Methodology and Tasks}
The study consisted of three phases: a pre-study questionnaire, hands-on texturing tasks, and post-task interviews.
Prior to the tasks, participants completed a questionnaire covering their background in 3D texturing, typical workflows, tools used in practice, and familiarity with AI-assisted texturing methods. Participants were also asked to share their perspectives on usability and creative control in existing tools. Each session began with a brief introduction to generative AI concepts and conditional texture generation from rendered views, ensuring that participants shared a baseline technical understanding.
Participants then evaluated two texturing tools: (1) an automatic texturing baseline using the Hunyuan2.1 texture generation backbone~\cite{hunyuan3d2025hunyuan3d}, and (2) our interactive texture painting system. Both tools are implemented as Blender add-ons. The order of the tools was randomized across participants.
For both tools, participants selected one of three objects—a koi fish, a sea urchin, or a croissant—along with corresponding single-view reference images. For the automatic baseline, participants selected a single view of the mesh and triggered texture generation, after which they examined the resulting texture mapped onto the object in Blender.
For the interactive system, participants completed two tasks: (1) editing an existing partial texture map, and (2) painting a texture map from scratch. Before starting, participants were given a brief tutorial on the Blender add-on interface and were allowed to practice on a sample model. During the tasks, participants generated texture brushes from selected reference views and were instructed to follow a hypothetical art-director brief by combining visual elements from multiple views. Participants were asked to generate at least two texture brushes and to paint approximately 70\% of the mesh to gain sufficient hands-on experience with the system. Throughout the tasks, participants were encouraged to verbalize their thoughts and observations.

\subsection{Questions}
After completing each tool, participants answered the following questions:

\begin{enumerate}[label=Q\arabic*:]
  \item \textbf{Usefulness}: What features or capabilities of this tool did you find useful? Which aspects did not work well, and how could they be improved?
  \item \textbf{Controllability}: \\
  Rate from 1 to 5: I feel I have sufficient control over the final texture when using this tool.
  \item \textbf{Complementarity}: \\
  Rate from 1 to 5:
  \begin{itemize}
    \item[a.] I feel this tool complements the tools I use in my daily practice and I would love to use this feature as part of my texturing workflow.
    \item[b.] I would not like to use this system frequently.
    \item[c.] I feel I could work more efficiently if I used this system.
  \end{itemize}
\end{enumerate}
After completing both tools, participants were asked comparative questions:
\begin{enumerate}[label=Q\arabic*:]
  \item \textbf{Workflow Integration}: Which tool are you more likely to integrate into your texturing workflow, and why?
  \item \textbf{Texture Quality}: Which tool produced textures that were more faithful to the reference images? What differences did you observe between the two methods? Which tool better captured fine geometric details?
  \item \textbf{Creative Affordance}: Which tool better supports the creative process of texturing 3D objects, and why?
\end{enumerate}
Finally, participants answered feature-specific questions related to the interactive system:
\begin{enumerate}[label=Q\arabic*:]
  \item \textbf{Texture Quality}: When using entire views or selected regions as reference brushes, does the auto-brush tool produce textures aligned with your intent?
  \item \textbf{Controllability}: The interface supports undoing, regenerating, and clearing textures. Do you find this iterative process valuable compared to fully automatic texturing?
  \item \textbf{Creative Affordance}: Does the system enable experimentation with diverse ideas and combinations, and does this benefit your creative workflow?
\end{enumerate}

\subsection{Suggestions for Improvement}
Despite an overall positive response to our interactive texturing prototype, participants identified several limitations, with suggestions for implementation improvements and future work. Key areas for improvement include finer control over inpainting strength, real-time texture updates during brush strokes and painting with full material channels. Due to the fixed patch-based update scheme, participants found it difficult to make small, localized edits, as the inpainting model tends to overly smooth transitions and lose sharp details. In addition, the model occasionally struggles to produce high-fidelity results from certain camera angles, which can reduce perceived controllability and visual quality. With continued research and development, we aim to further advance interactive texturing techniques for high-quality, production-ready 3D assets.

\begin{figure*}[h!tbp]
	\centering
    \subfloat[More blending results. We blend multiple textures seamlessly on the same target mesh. \label{fig:supp_gallery:blend3}]
    {
        \begin{tabular}{ccc}
        
        \includegraphics[width=0.33\linewidth]{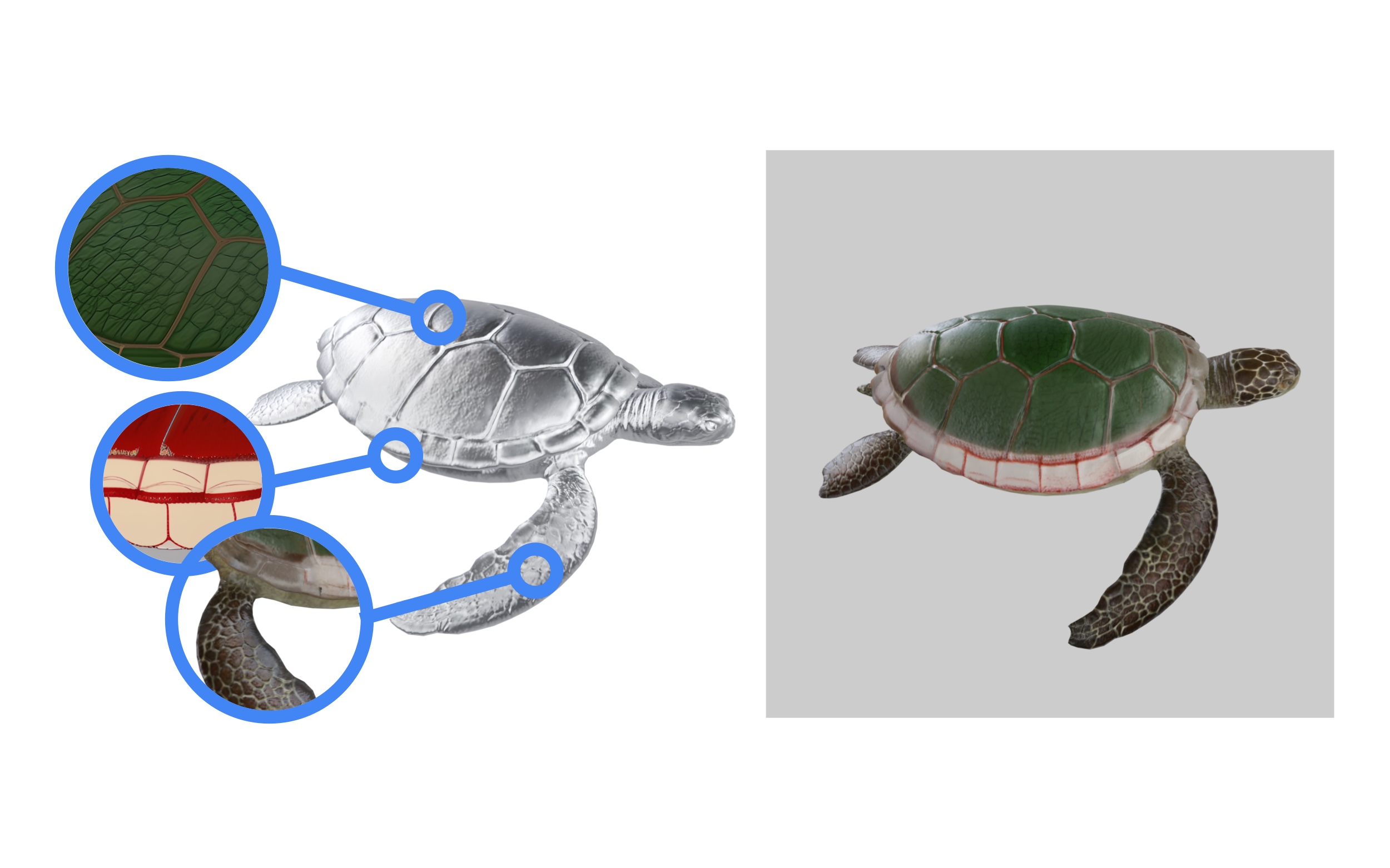} &
        \includegraphics[width=0.33\linewidth]{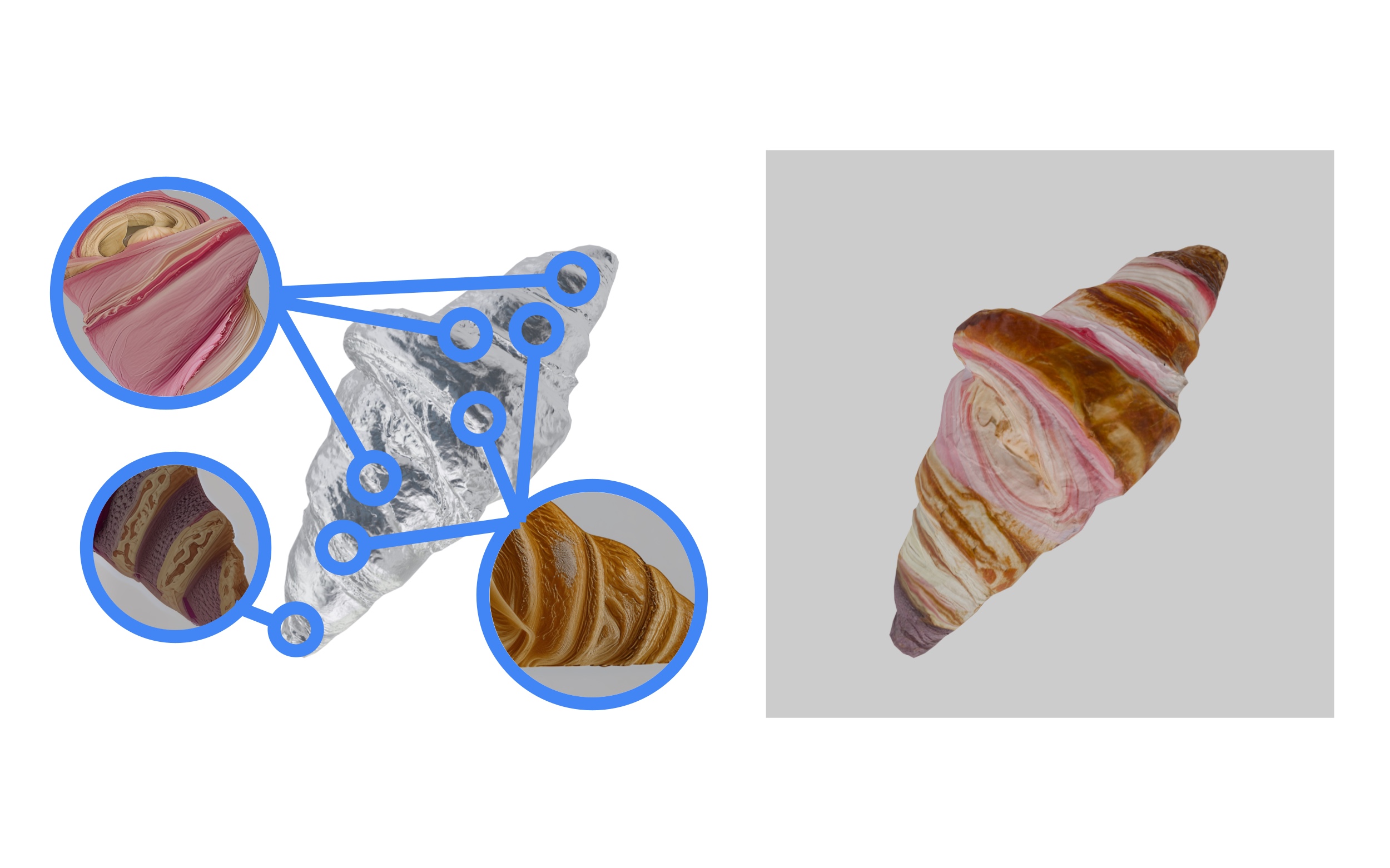} &
        \includegraphics[width=0.33\linewidth]{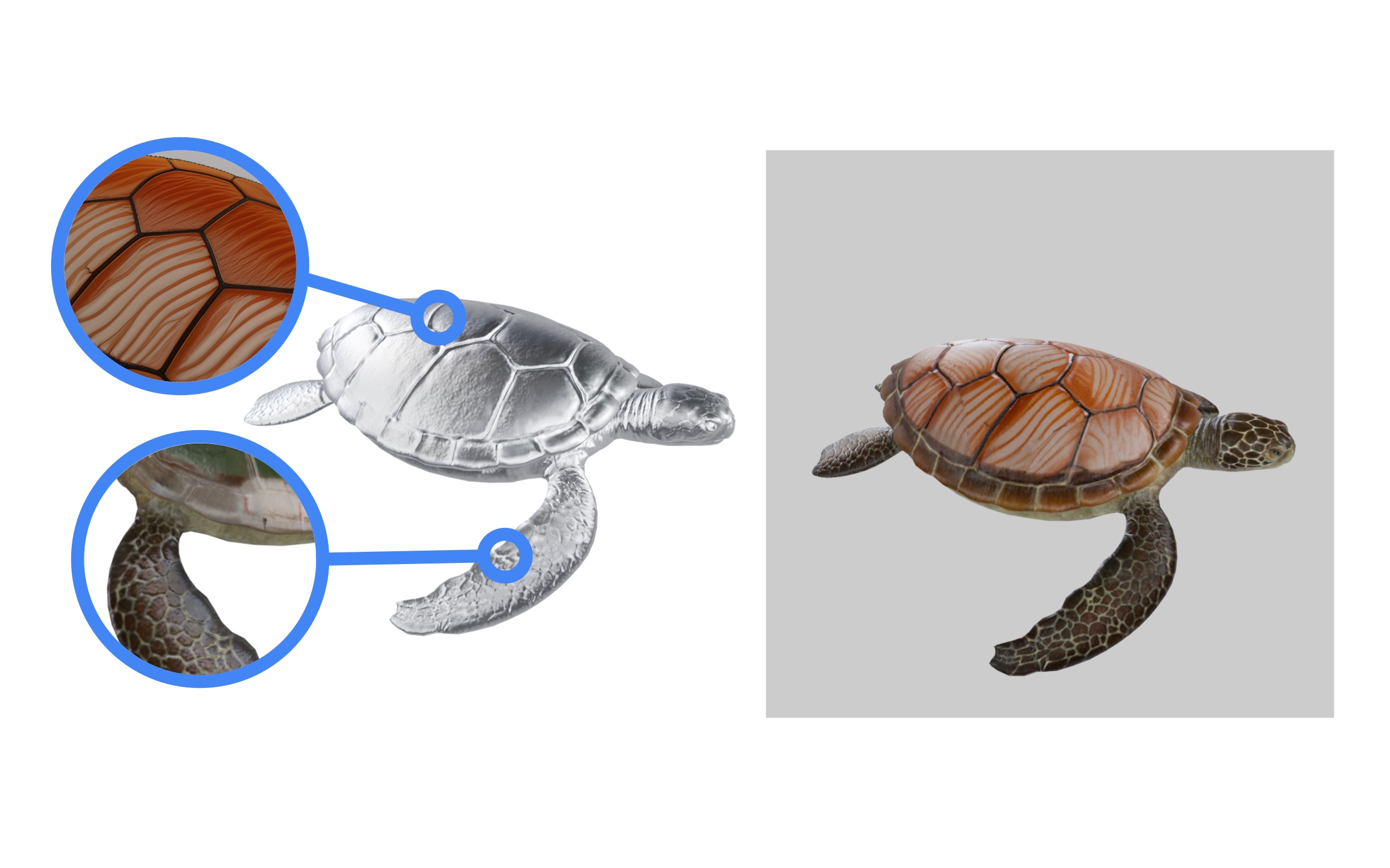} 
        \end{tabular}
    } \\

    \subfloat[More Results on Texture Transfer. Once we trained a GLOSS model on a source mesh, the same model can be used to texture new shapes. The transfer solely depends on the local patch-wise geometry and albedo between the target and the references, and generalizes to new geometry to a certain extent. Notice how we can transfer correlations between normal and urchin shell bump texture to the creases on a croissant. \label{fig:supp_gallery:transfer1}]
    {
        \begin{tabular}{cc}
        \includegraphics[width=0.5\linewidth]{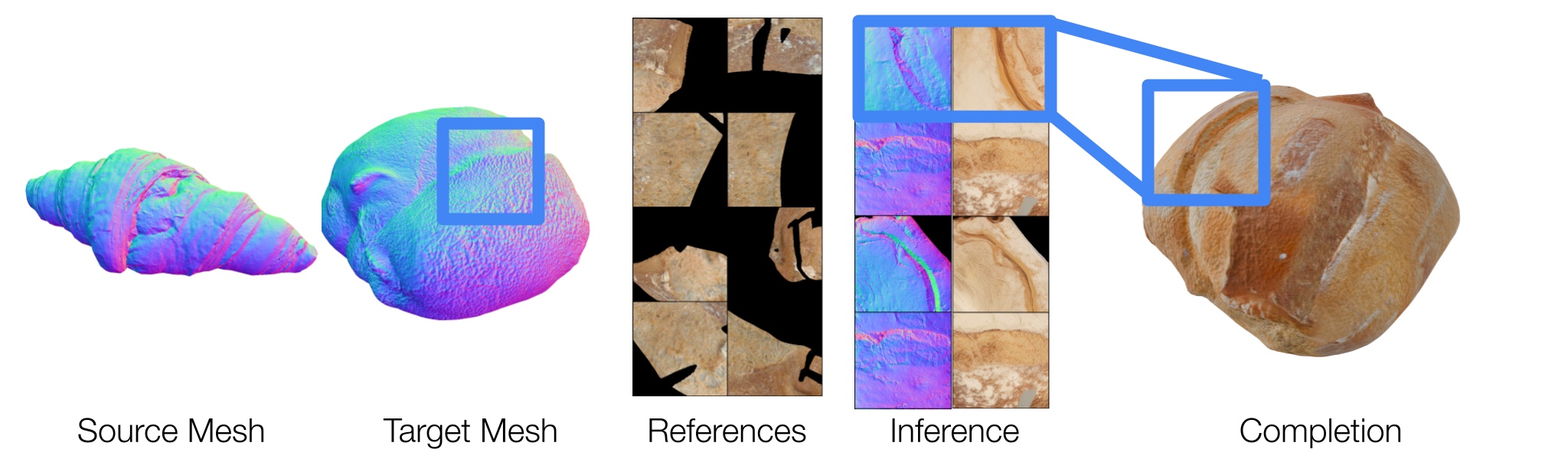} &
        \includegraphics[width=0.5\linewidth]{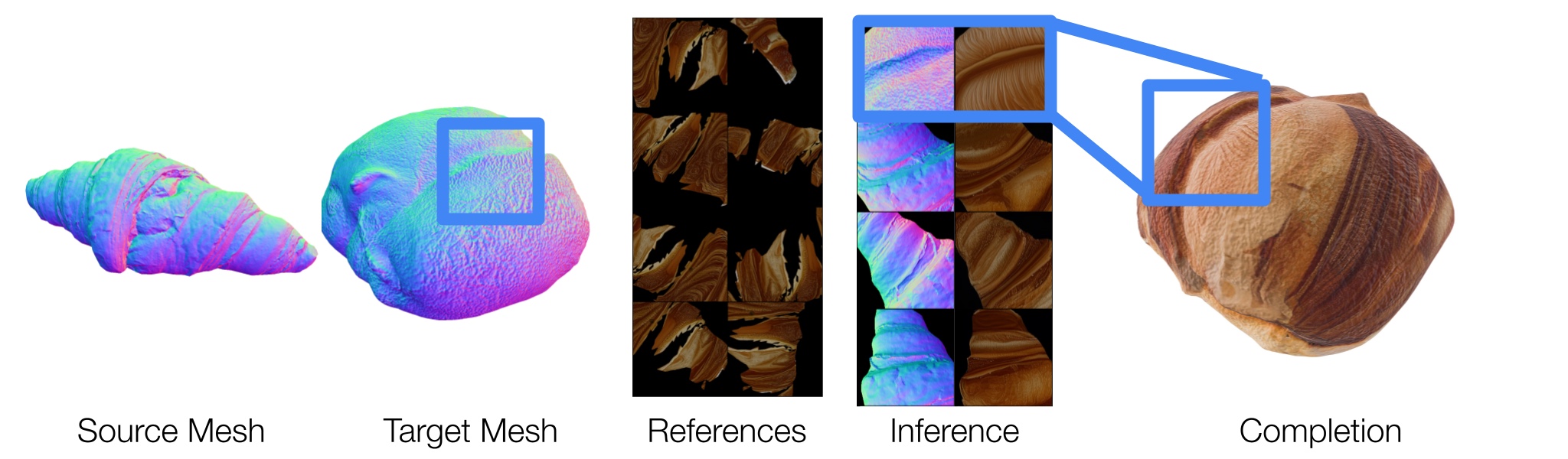} 
        \\
        \includegraphics[width=0.5\linewidth]{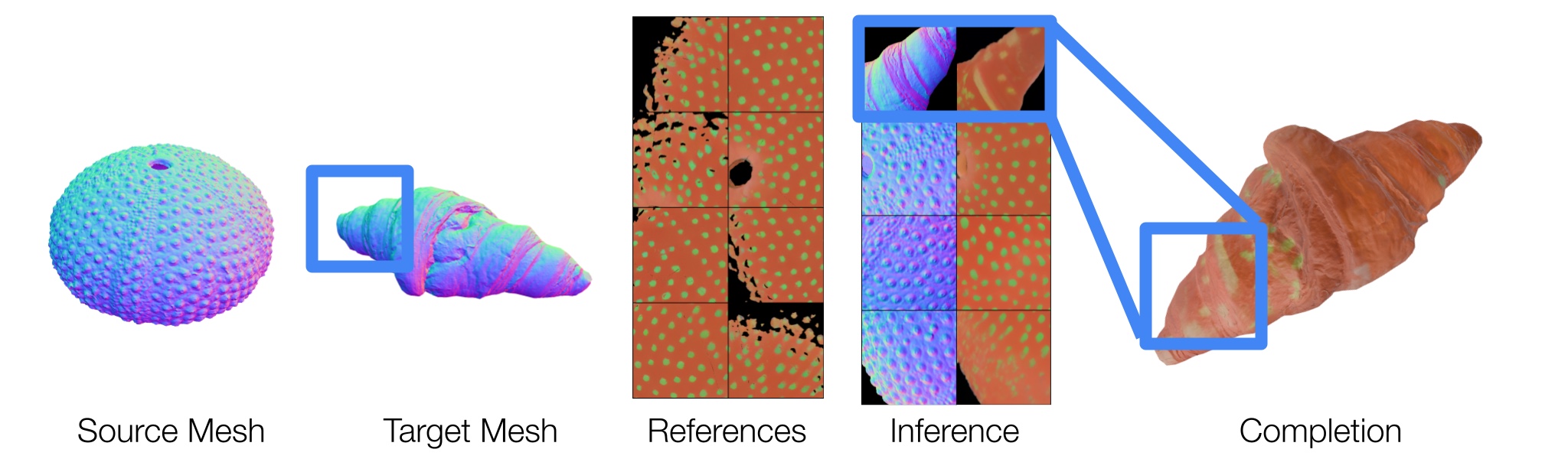} &
        \includegraphics[width=0.5\linewidth]{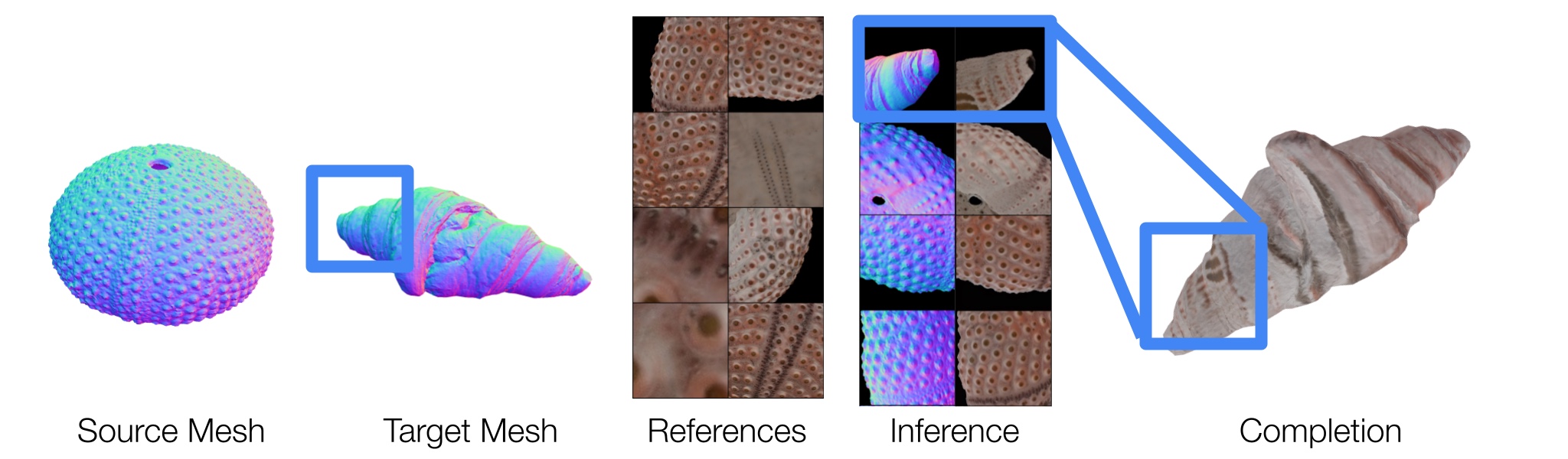}
        \end{tabular}
    }

    \subfloat[More Results on PBR Completion. Trained on albedo, GLOSS generalizes to completion in metallic and roughness channels. We show renders of albedo channel, metallic only (basecolor grey), roughness only (basecolor grey) as well as the full material channel renderings.\label{fig:supp_gallery:pbr}] {
        \begin{tabular}{cc}
        \includegraphics[trim={0cm 2cm 0cm 2cm}, clip, width=0.5\linewidth, valign=m]{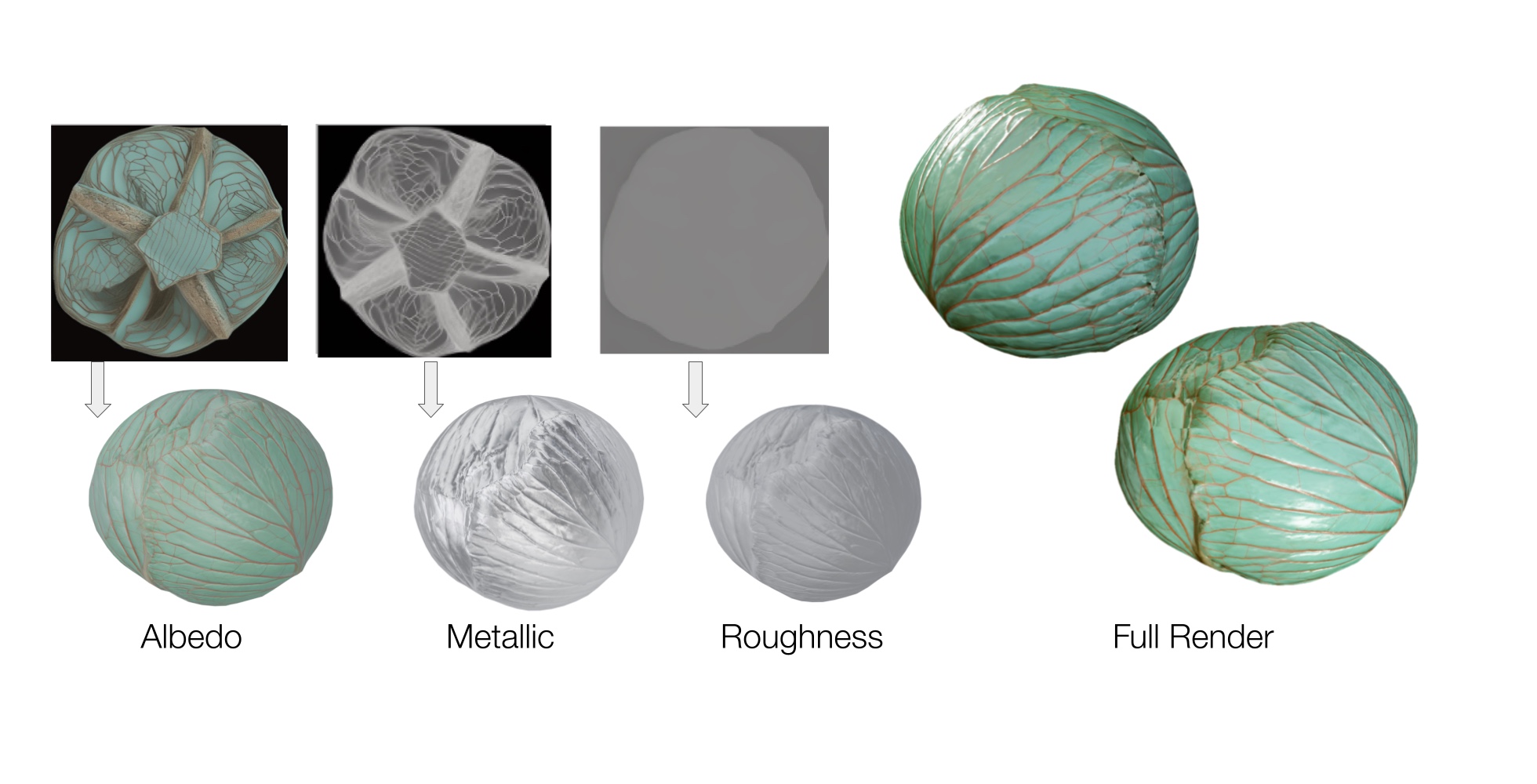} &
        \includegraphics[trim={0cm 2cm 0cm 2cm}, clip, width=0.5\linewidth, valign=m]{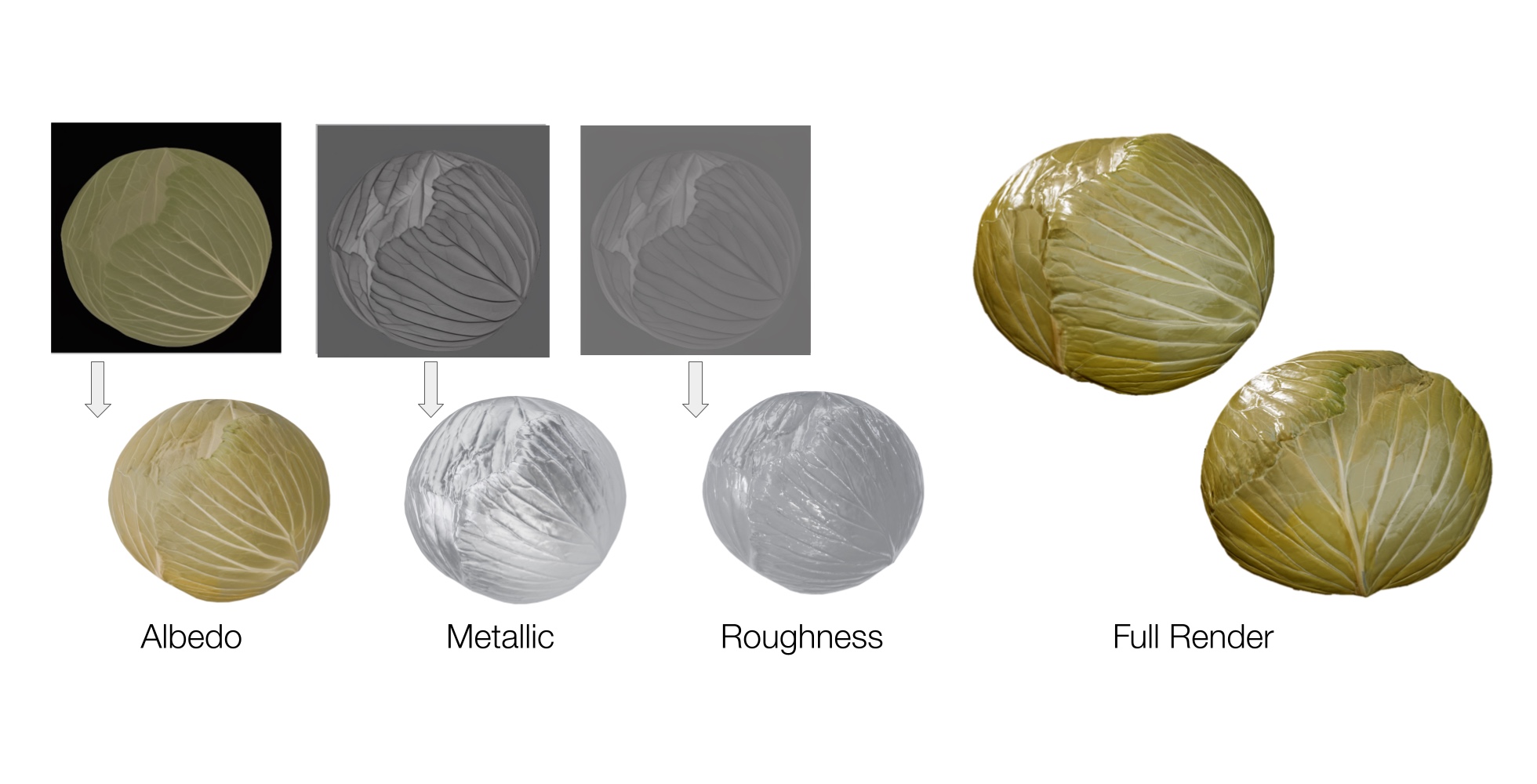} 
        \\
        \includegraphics[trim={0cm 2cm 0cm 2cm}, clip, width=0.5\linewidth, valign=m]{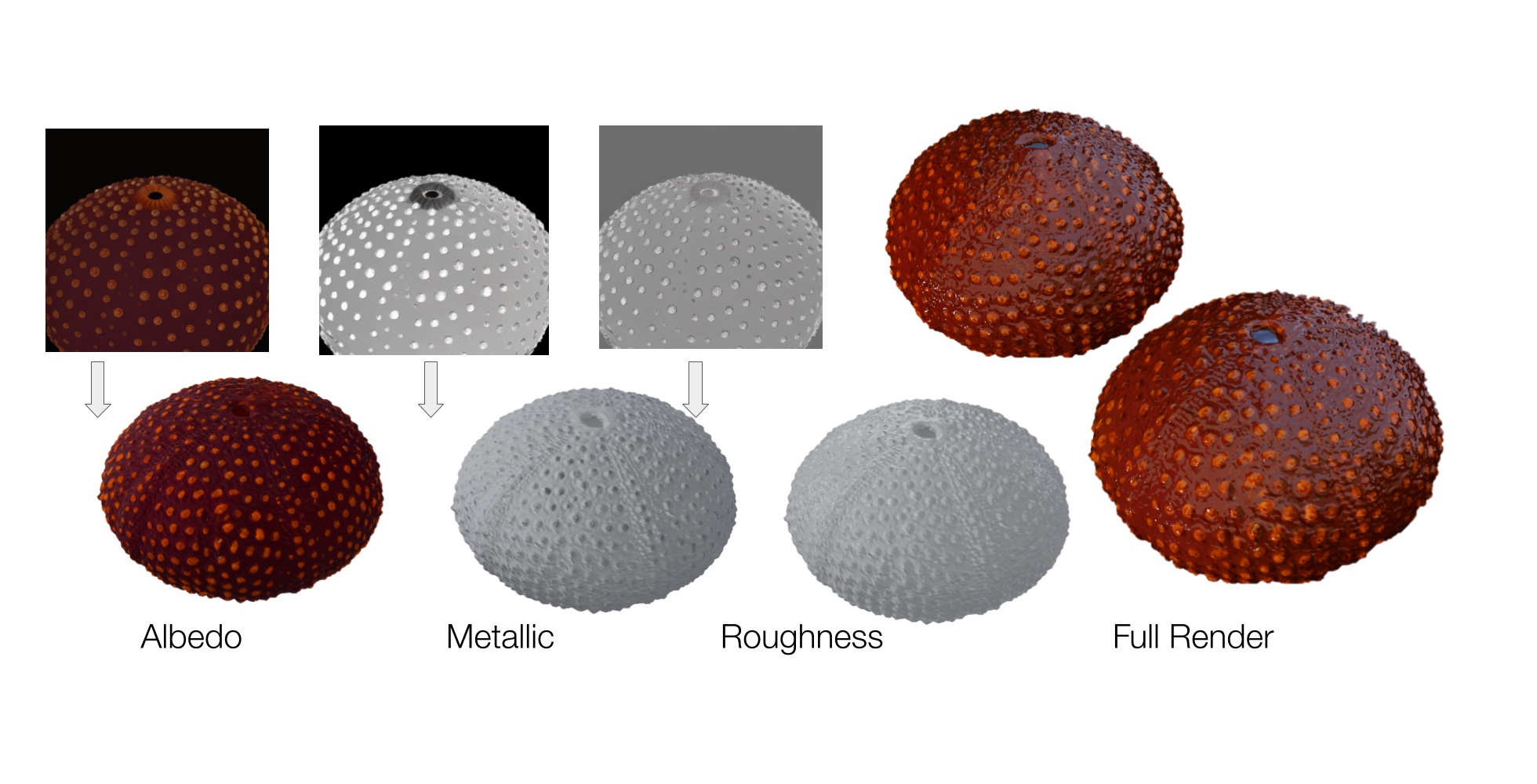} &
        \includegraphics[trim={0cm 2cm 0cm 2cm}, clip, width=0.5\linewidth, valign=m]{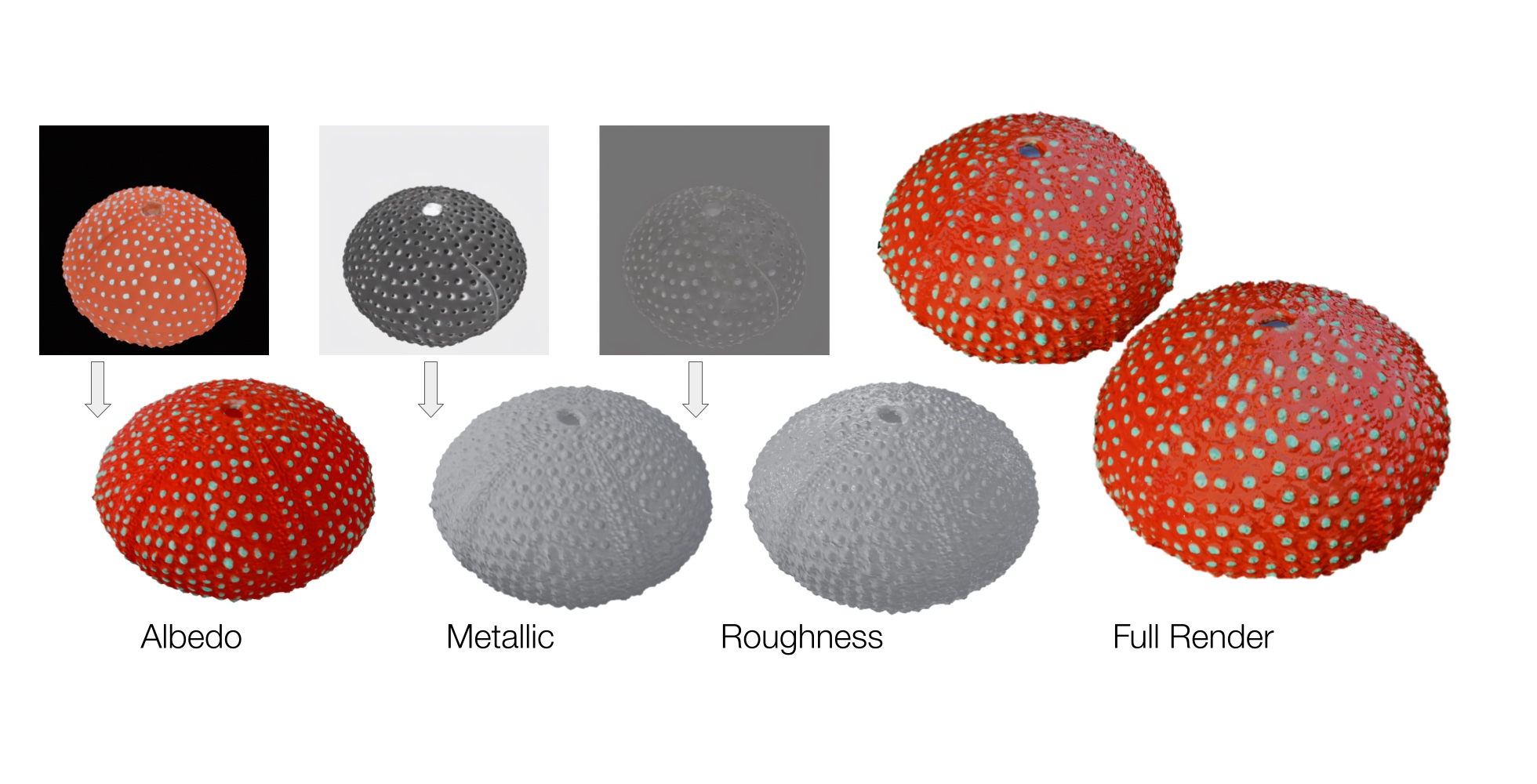}
        \end{tabular}
    }
	\caption{Additional result on texturing applications.}
	\label{fig:supp_gallery}
\end{figure*}
\newcommand{\transferrnormalswidth}{0.12\linewidth}
\newcommand{\transferrviewwidth}{0.12\linewidth}
\newcommand{\transferrreswidth}{0.12\linewidth}

\begingroup
\setlength{\tabcolsep}{3pt}
\renewcommand{\arraystretch}{0}

\begin{figure*}[h!tbp]
  \centering
  \begin{tabular}{ccccccc}
  \small Geometry & \small Single View & \small View 1 & \small View 2 & \small Single View & \small View 1 & \small View 2 \\

  \includegraphics[width=\transferrnormalswidth, valign=m]{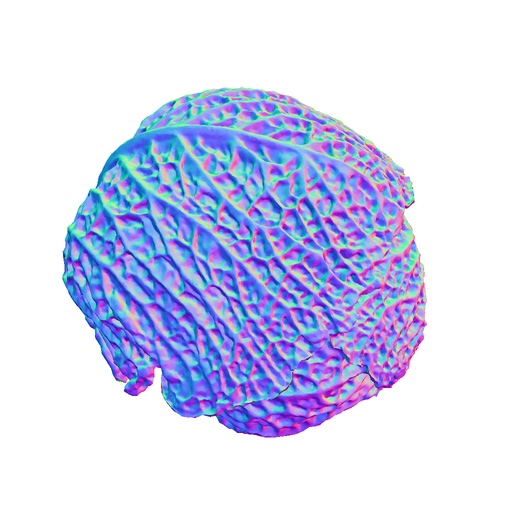} & \includegraphics[width=\transferrviewwidth, valign=m]{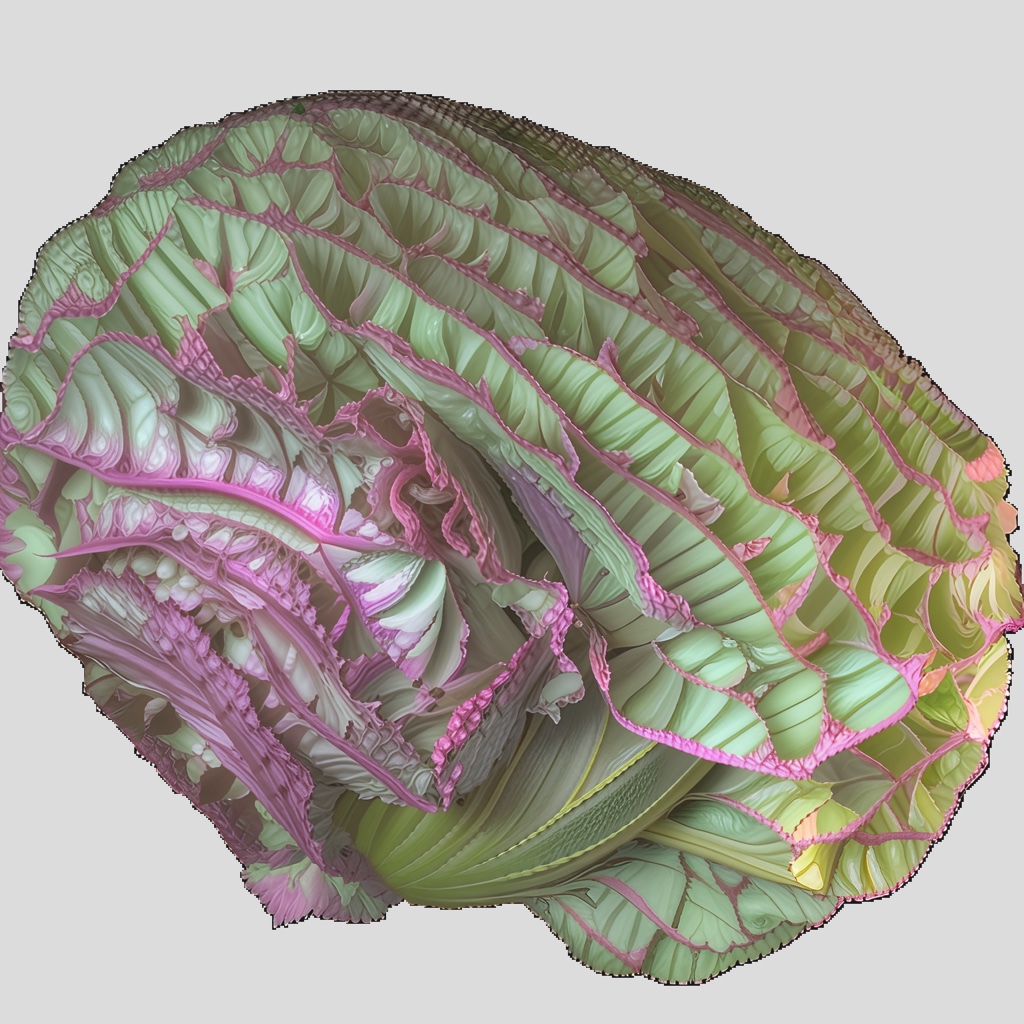} & \includegraphics[trim={5mm 5mm 5mm 5mm}, clip, width=\transferrreswidth, valign=m]{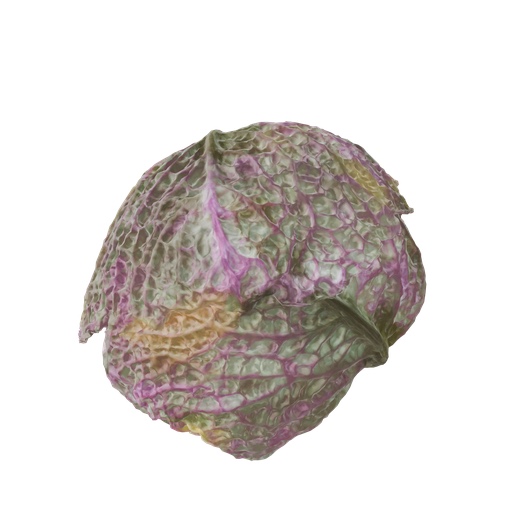} & \includegraphics[trim={5mm 5mm 5mm 5mm}, clip, width=\transferrreswidth, valign=m]{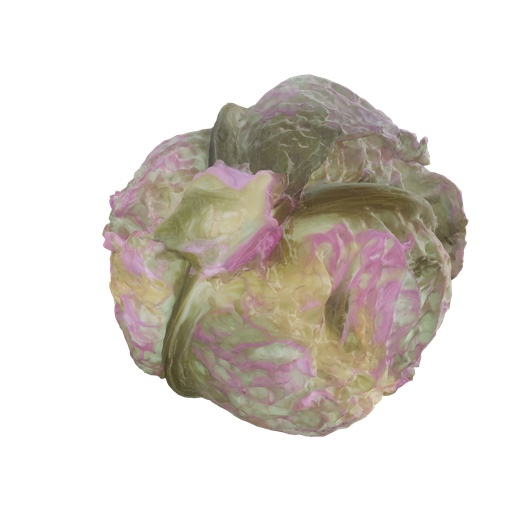} & \includegraphics[width=\transferrviewwidth, valign=m]{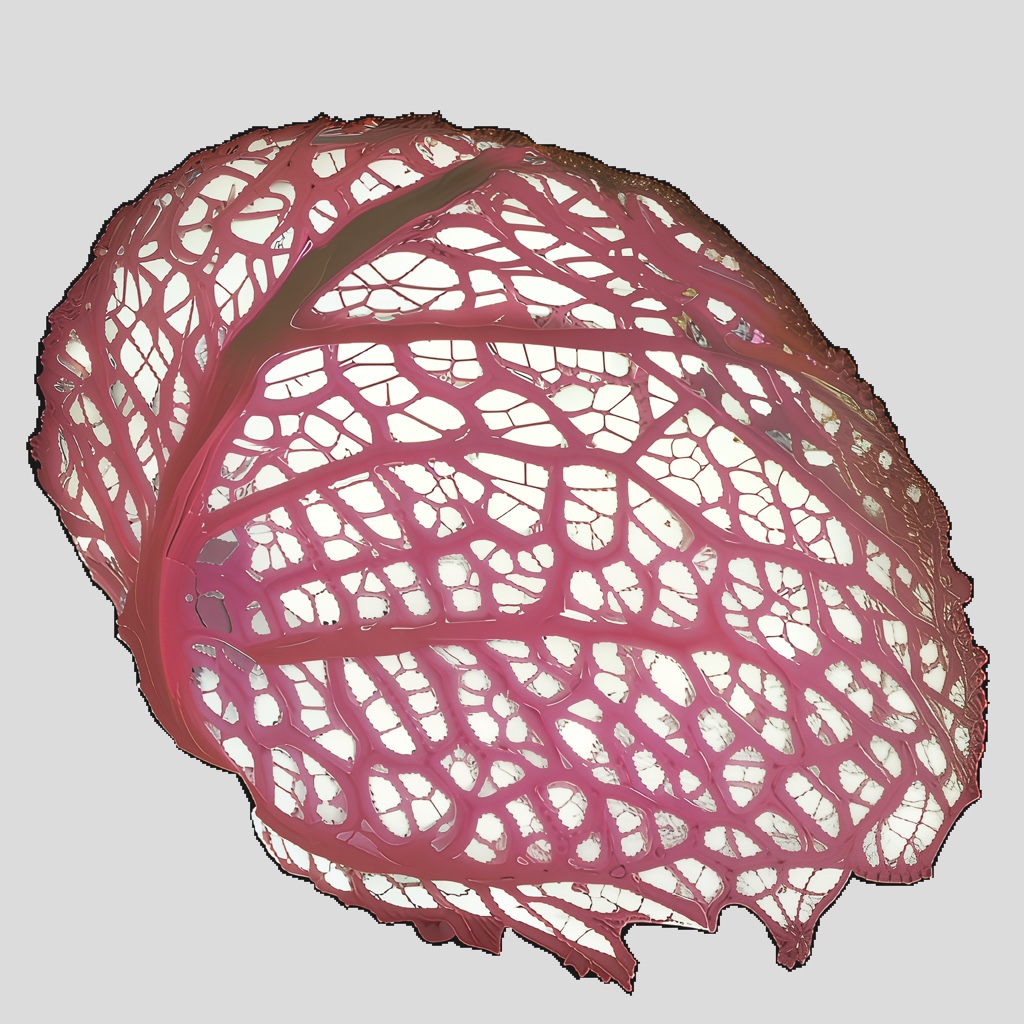} & \includegraphics[trim={5mm 5mm 5mm 5mm}, clip, width=\transferrreswidth, valign=m]{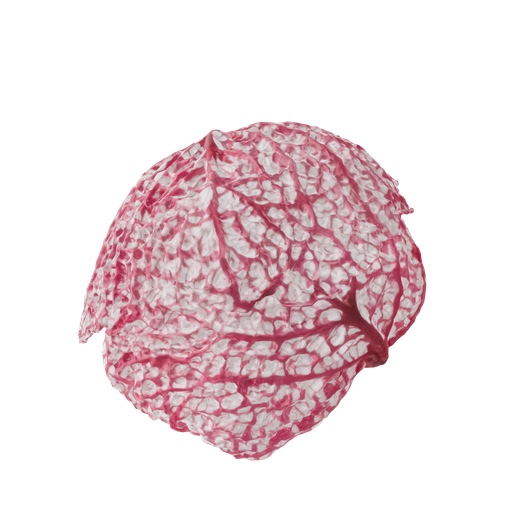} & \includegraphics[trim={5mm 5mm 5mm 5mm}, clip, width=\transferrreswidth, valign=m]{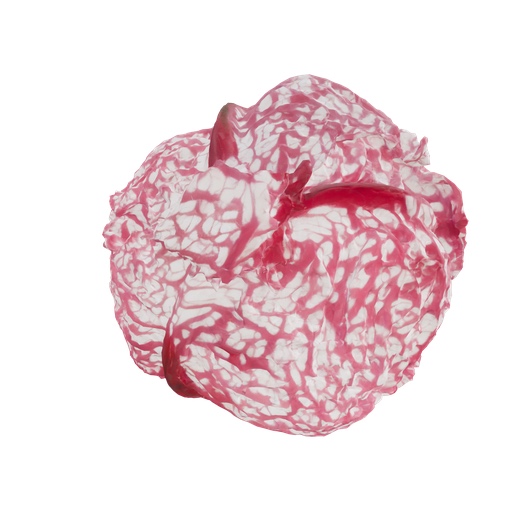} \\

  \includegraphics[width=\transferrnormalswidth, valign=m]{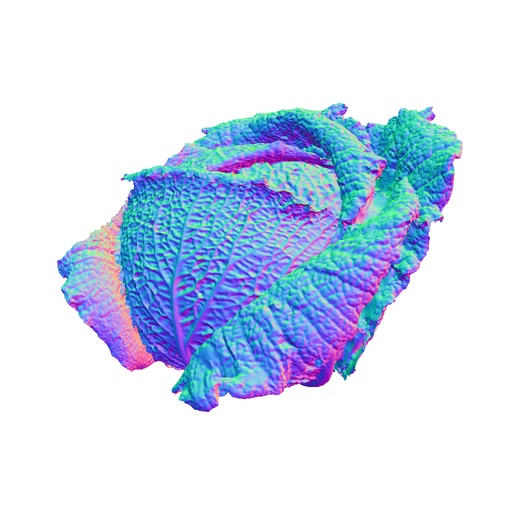} & \includegraphics[width=\transferrviewwidth, valign=m]{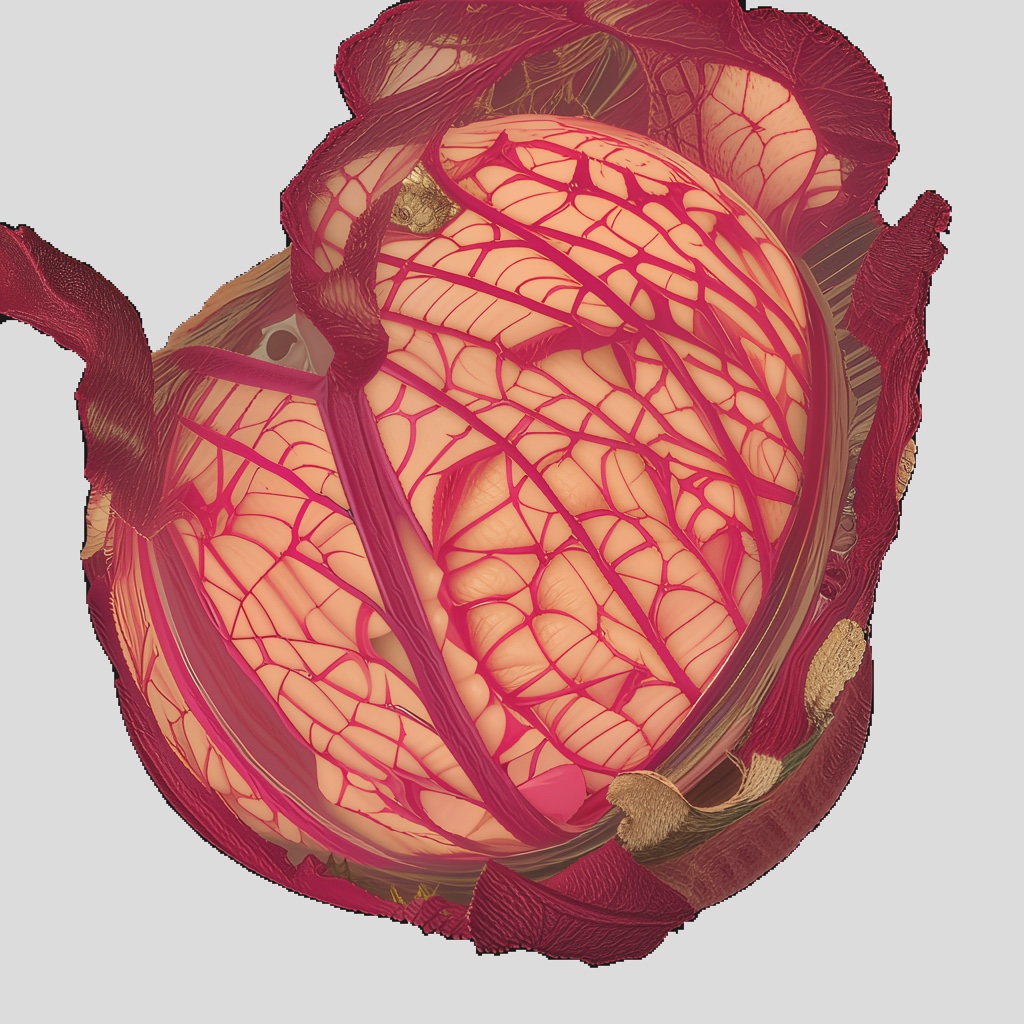} & \includegraphics[trim={5mm 5mm 5mm 5mm}, clip, width=\transferrreswidth, valign=m]{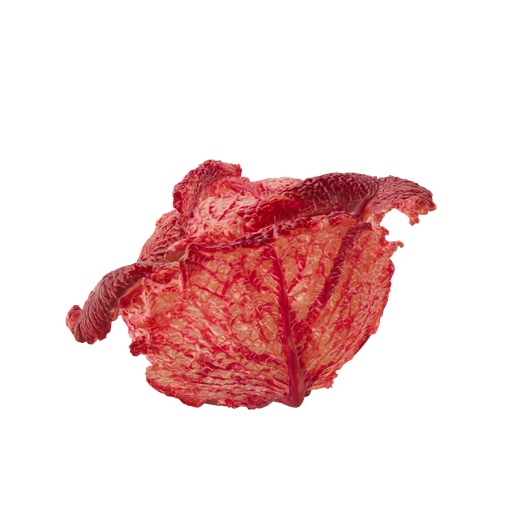} & \includegraphics[trim={5mm 5mm 5mm 5mm}, clip, width=\transferrreswidth, valign=m]{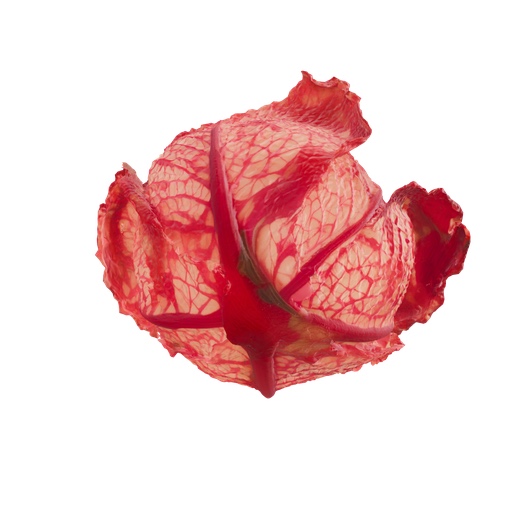} & \includegraphics[width=\transferrviewwidth, valign=m]{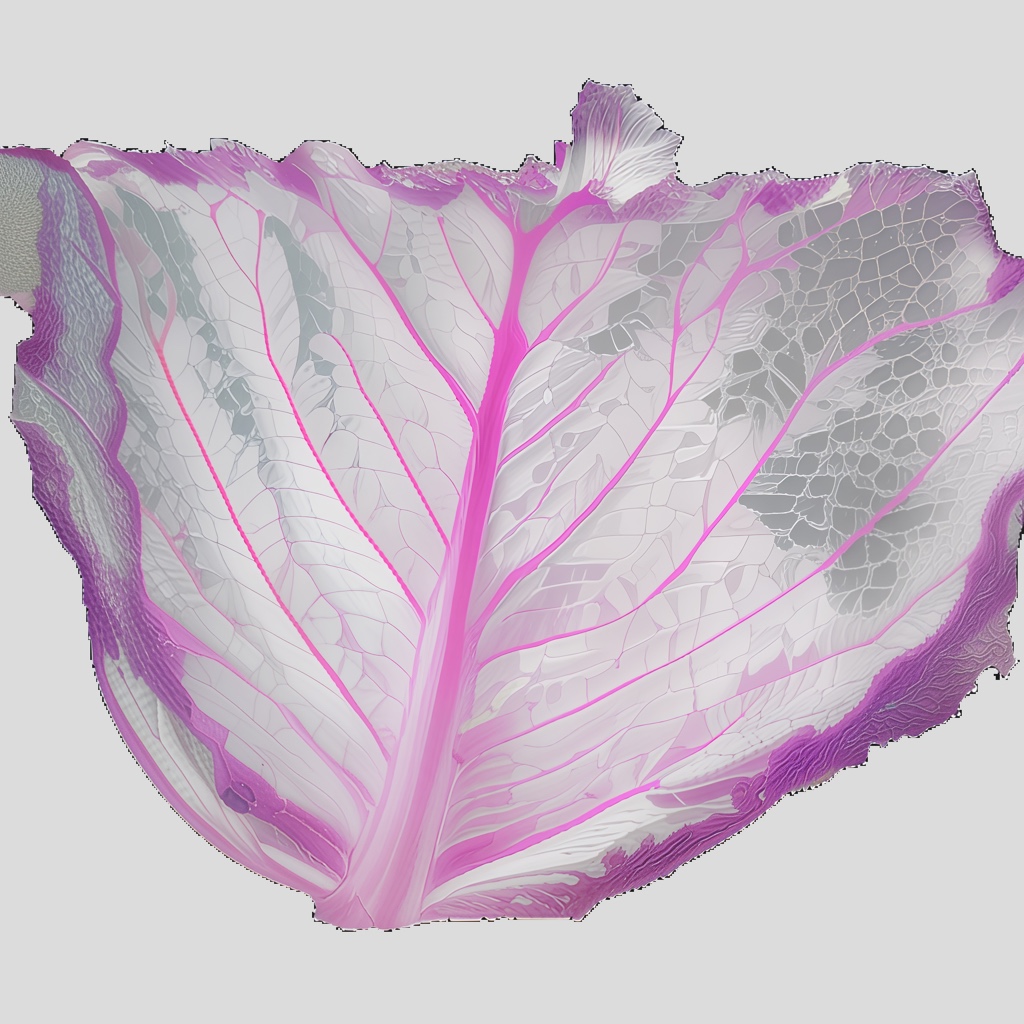} & \includegraphics[trim={5mm 5mm 5mm 5mm}, clip, width=\transferrreswidth, valign=m]{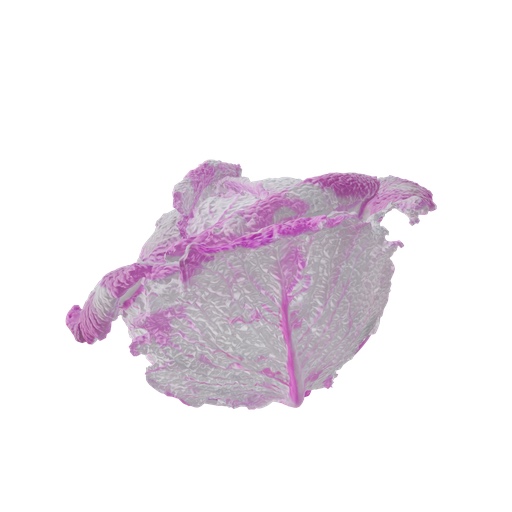} & \includegraphics[trim={5mm 5mm 5mm 5mm}, clip, width=\transferrreswidth, valign=m]{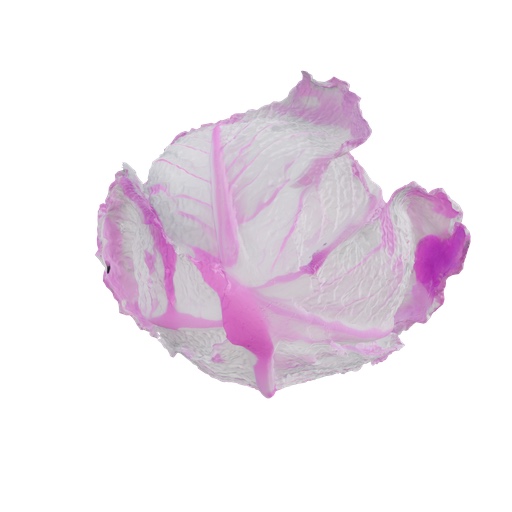} \\

  \includegraphics[width=\transferrnormalswidth, valign=m]{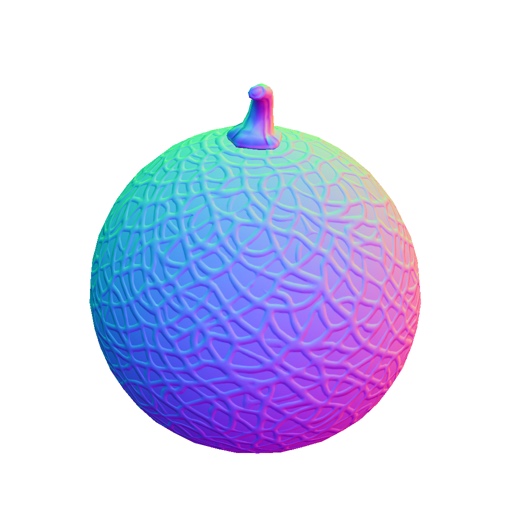} & \includegraphics[width=\transferrviewwidth, valign=m]{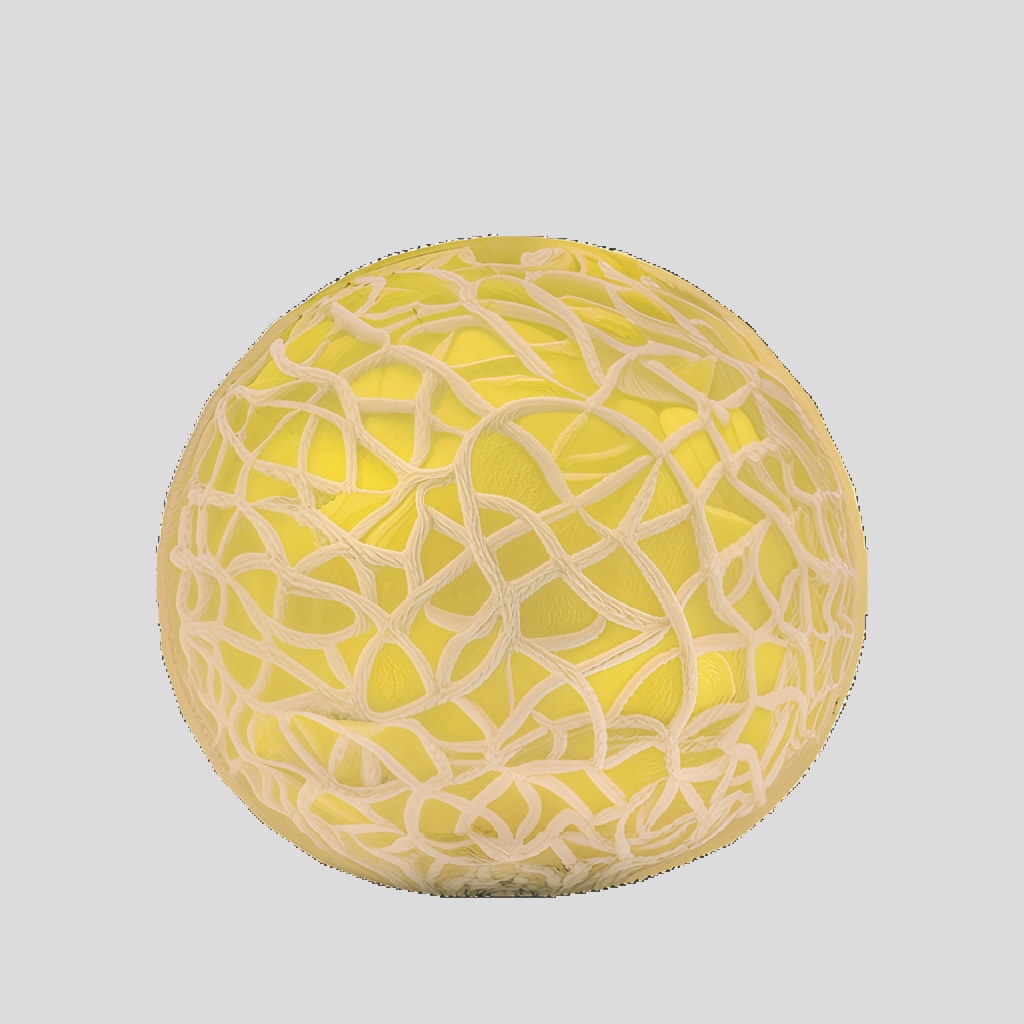} & \includegraphics[trim={5mm 5mm 5mm 5mm}, clip, width=\transferrreswidth, valign=m]{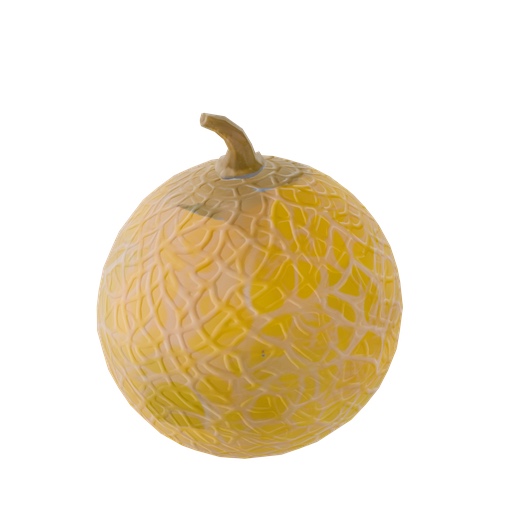} & \includegraphics[trim={5mm 5mm 5mm 5mm}, clip, width=\transferrreswidth, valign=m]{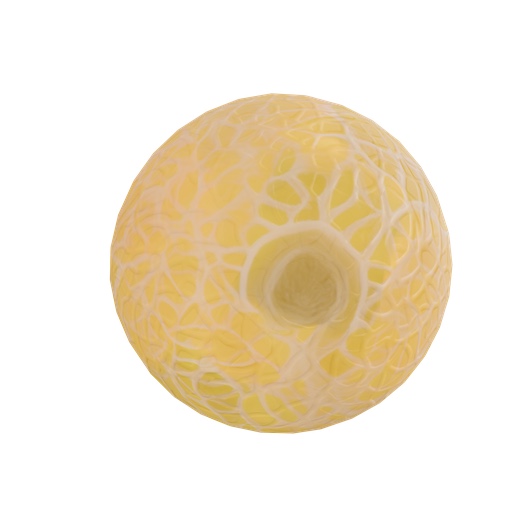} & \includegraphics[width=\transferrviewwidth, valign=m]{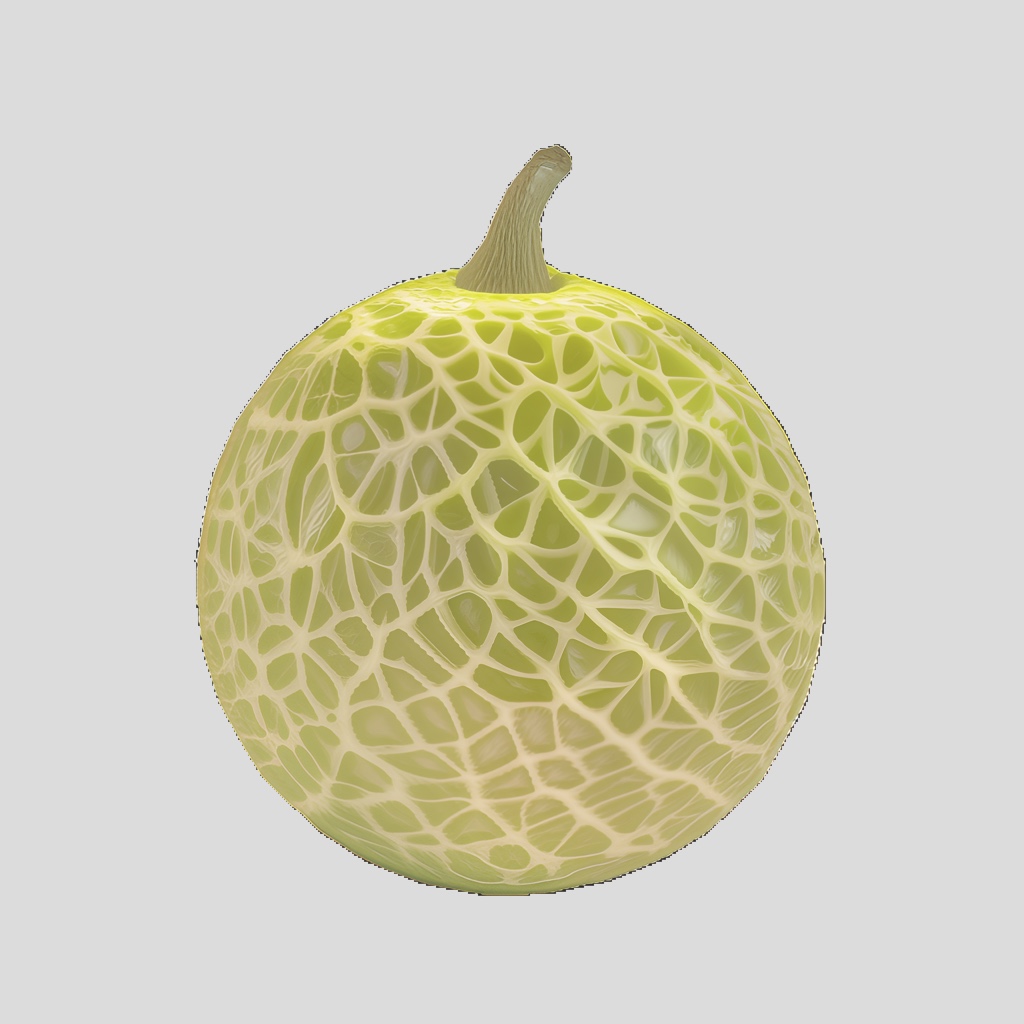} & \includegraphics[trim={5mm 5mm 5mm 5mm}, clip, width=\transferrreswidth, valign=m]{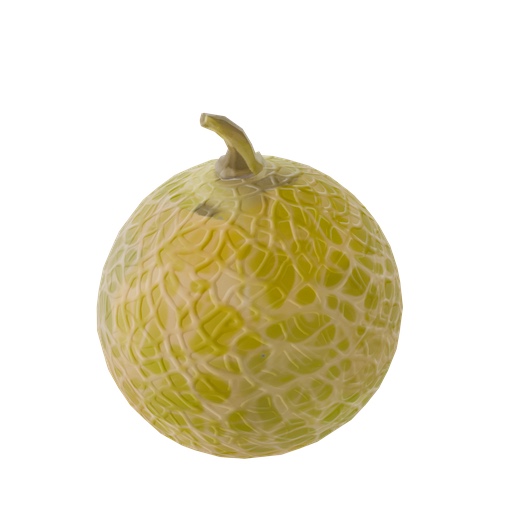} & \includegraphics[trim={5mm 5mm 5mm 5mm}, clip, width=\transferrreswidth, valign=m]{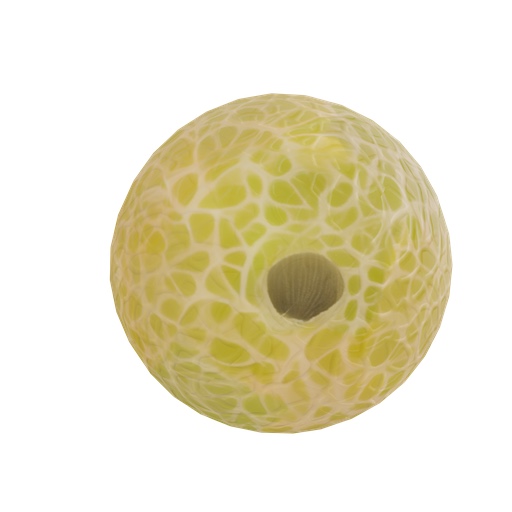} \\

  \includegraphics[width=\transferrnormalswidth, valign=m]{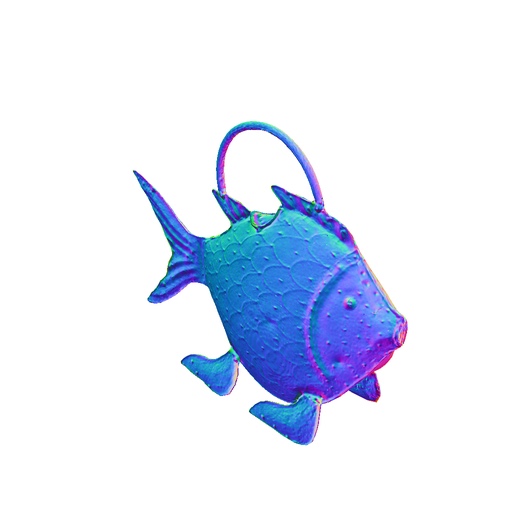} & \includegraphics[width=\transferrviewwidth, valign=m]{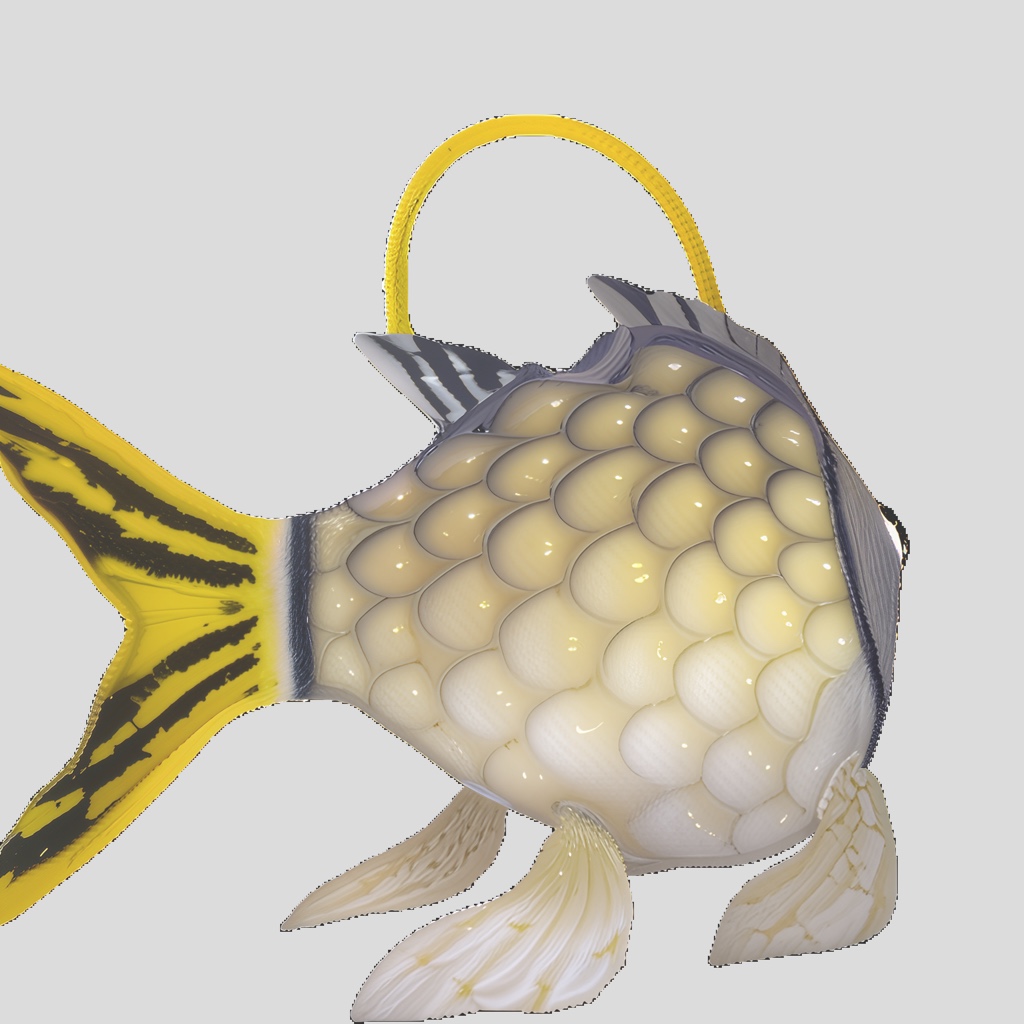} & \includegraphics[trim={5mm 5mm 5mm 5mm}, clip, width=\transferrreswidth, valign=m]{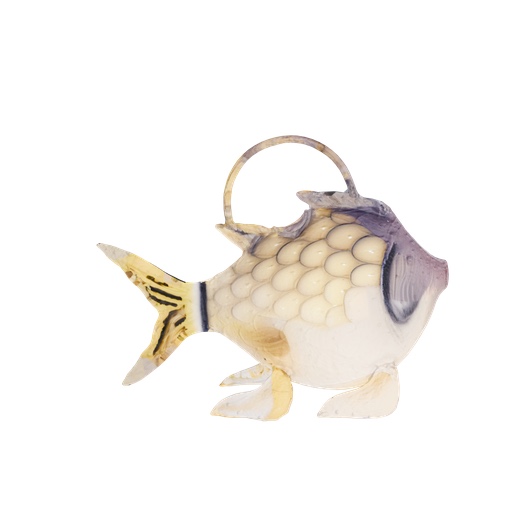} & \includegraphics[trim={5mm 5mm 5mm 5mm}, clip, width=\transferrreswidth, valign=m]{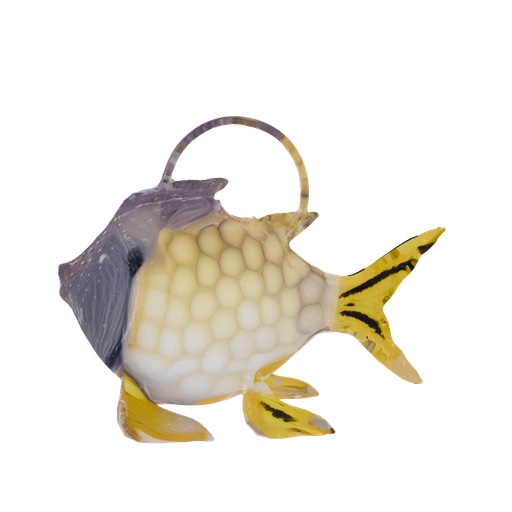} & \includegraphics[width=\transferrviewwidth, valign=m]{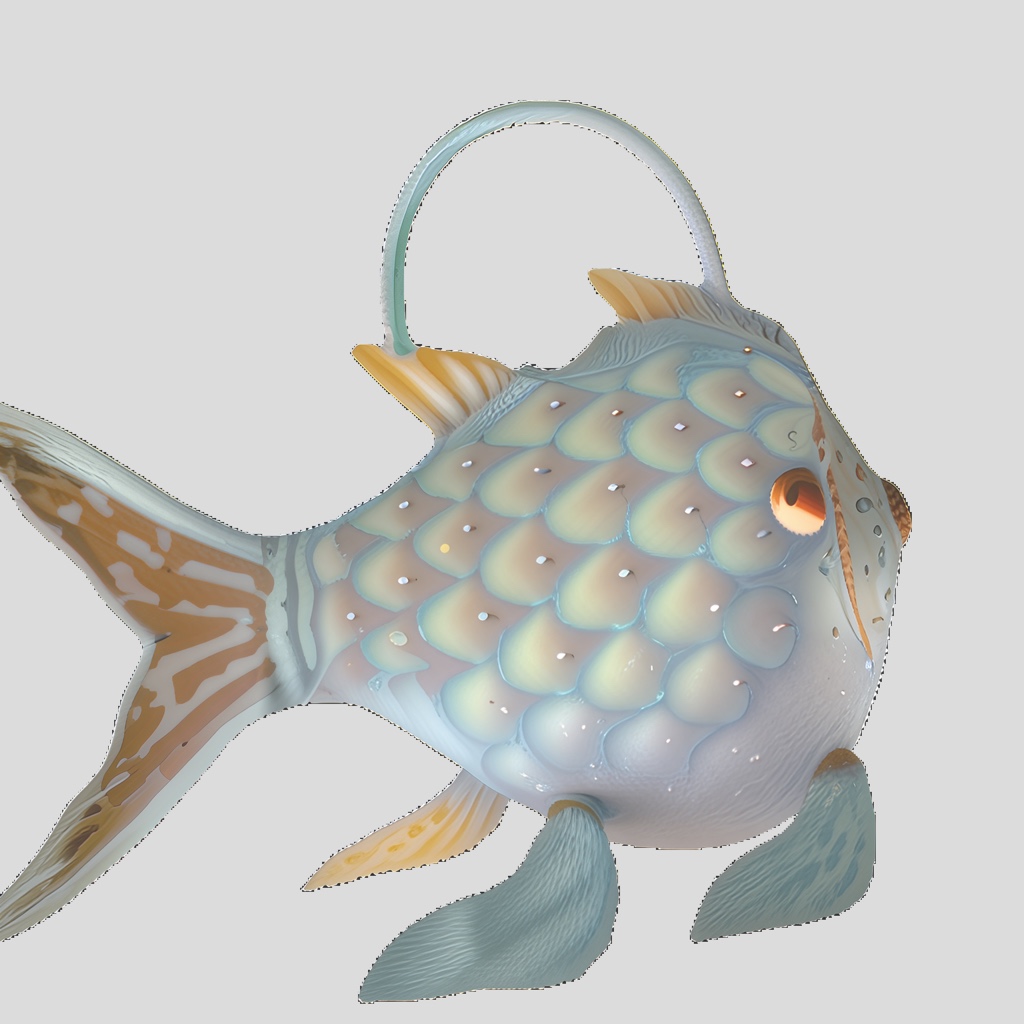} & \includegraphics[trim={5mm 5mm 5mm 5mm}, clip, width=\transferrreswidth, valign=m]{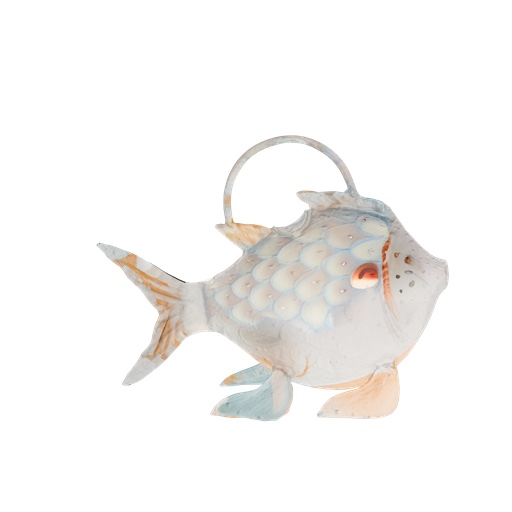} & \includegraphics[trim={5mm 5mm 5mm 5mm}, clip, width=\transferrreswidth, valign=m]{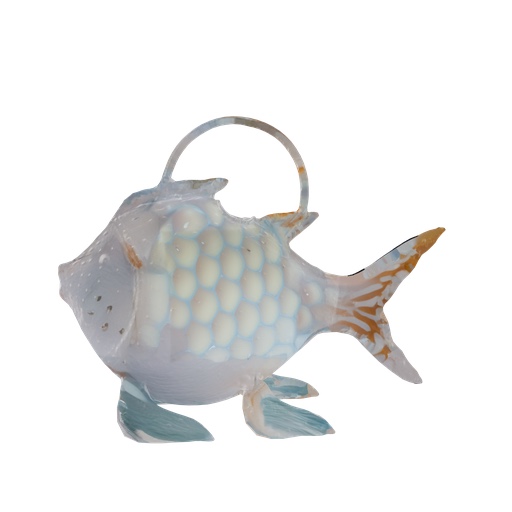} \\

  \includegraphics[width=\transferrnormalswidth, valign=m]{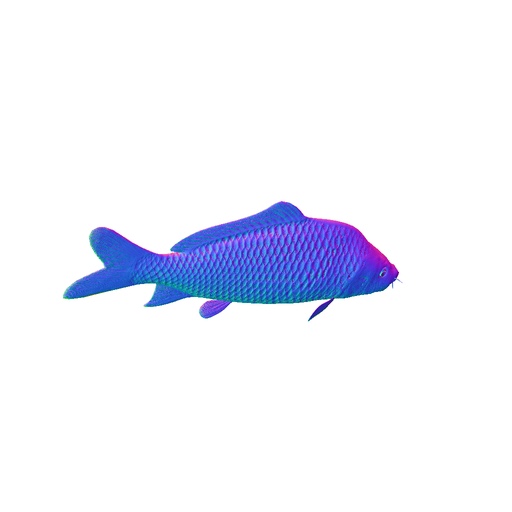} & \includegraphics[width=\transferrviewwidth, valign=m]{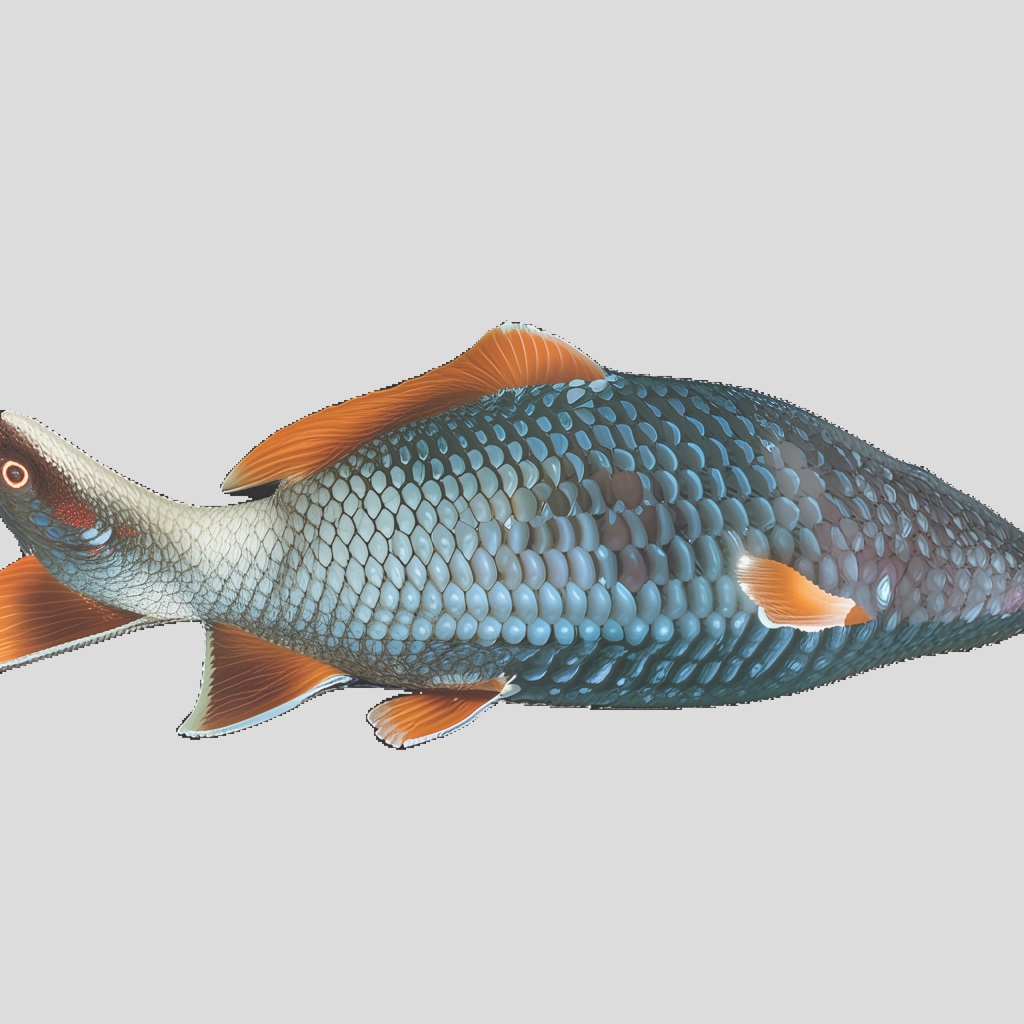} & \includegraphics[trim={5mm 5mm 5mm 5mm}, clip, width=\transferrreswidth, valign=m]{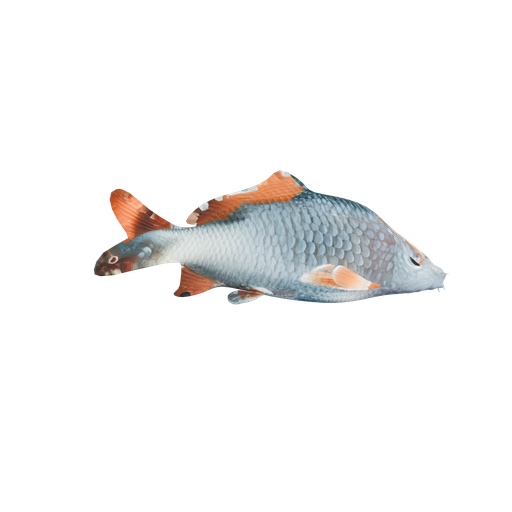} & \includegraphics[trim={5mm 5mm 5mm 5mm}, clip, width=\transferrreswidth, valign=m]{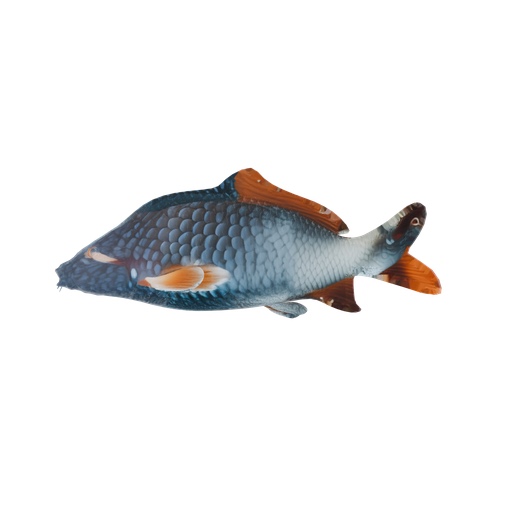} & \includegraphics[width=\transferrviewwidth, valign=m]{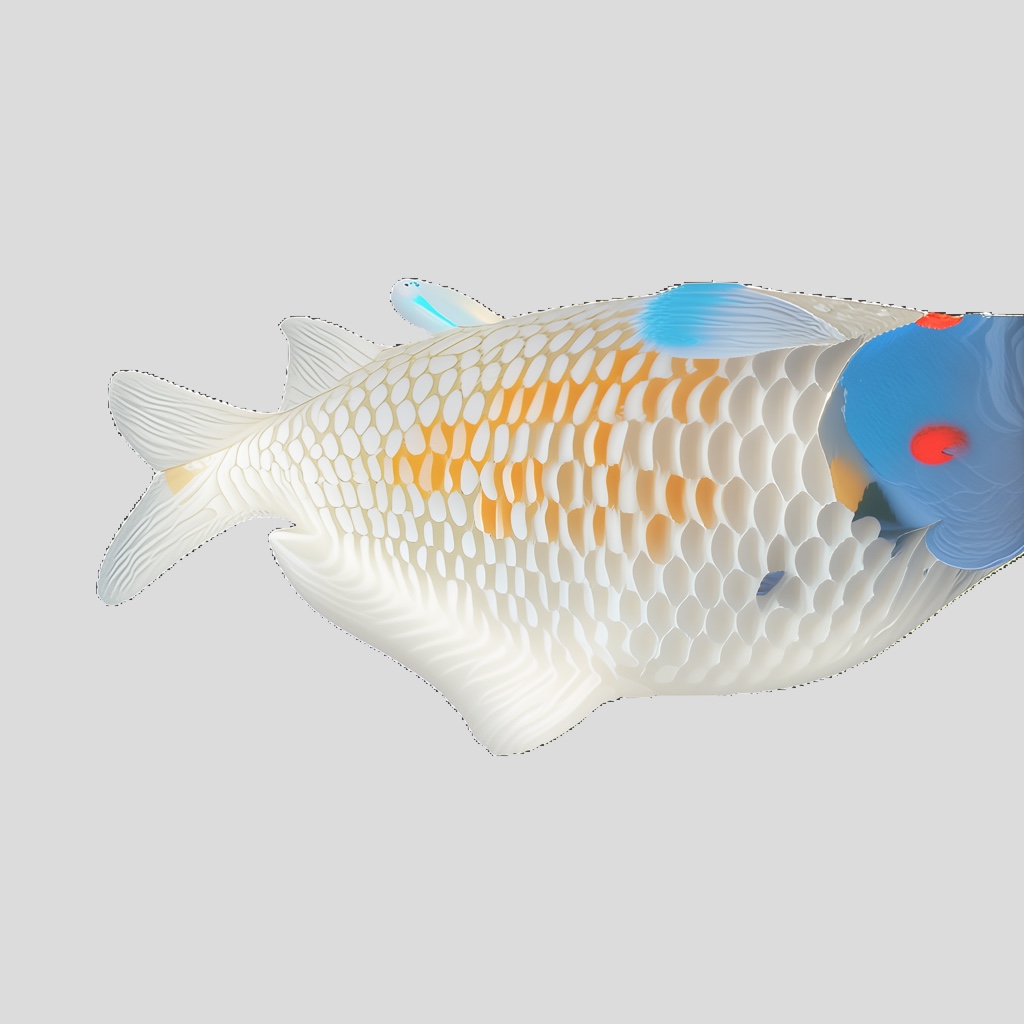} & \includegraphics[trim={5mm 5mm 5mm 5mm}, clip, width=\transferrreswidth, valign=m]{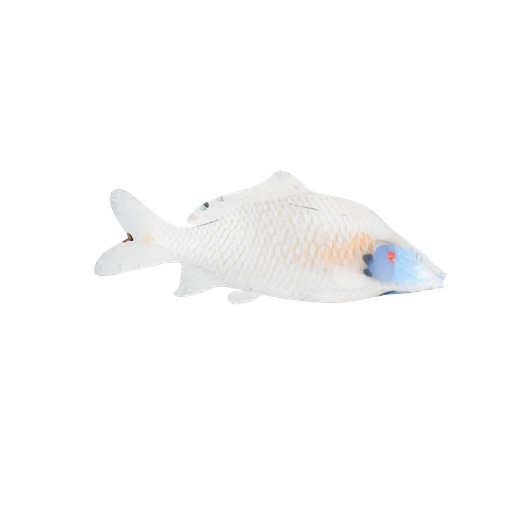} & \includegraphics[trim={5mm 5mm 5mm 5mm}, clip, width=\transferrreswidth, valign=m]{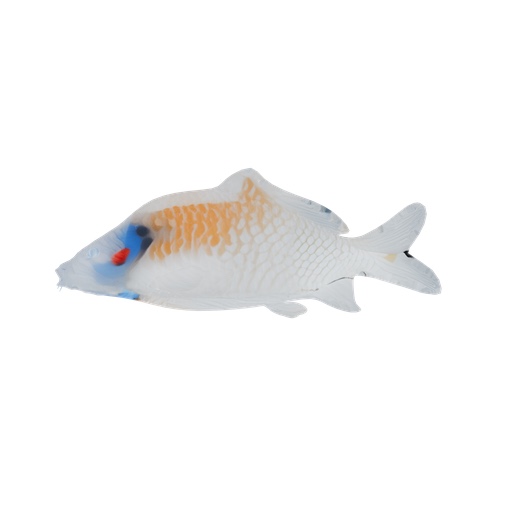} \\

  \includegraphics[width=\transferrnormalswidth, valign=m]{img/normals/fish_4.jpg} & \includegraphics[width=\transferrviewwidth, valign=m]{img/basecolor/fish_4/view0209.basecolor.jpg} & \includegraphics[trim={5mm 5mm 5mm 5mm}, clip, width=\transferrreswidth, valign=m]{img/render-keyframe/fish_4/view0209/view0209_glossft_diffuse_cam0004.jpg} & \includegraphics[trim={5mm 5mm 5mm 5mm}, clip, width=\transferrreswidth, valign=m]{img/render-keyframe/fish_4/view0209/view0209_glossft_diffuse_cam0003.jpg} & \includegraphics[width=\transferrviewwidth, valign=m]{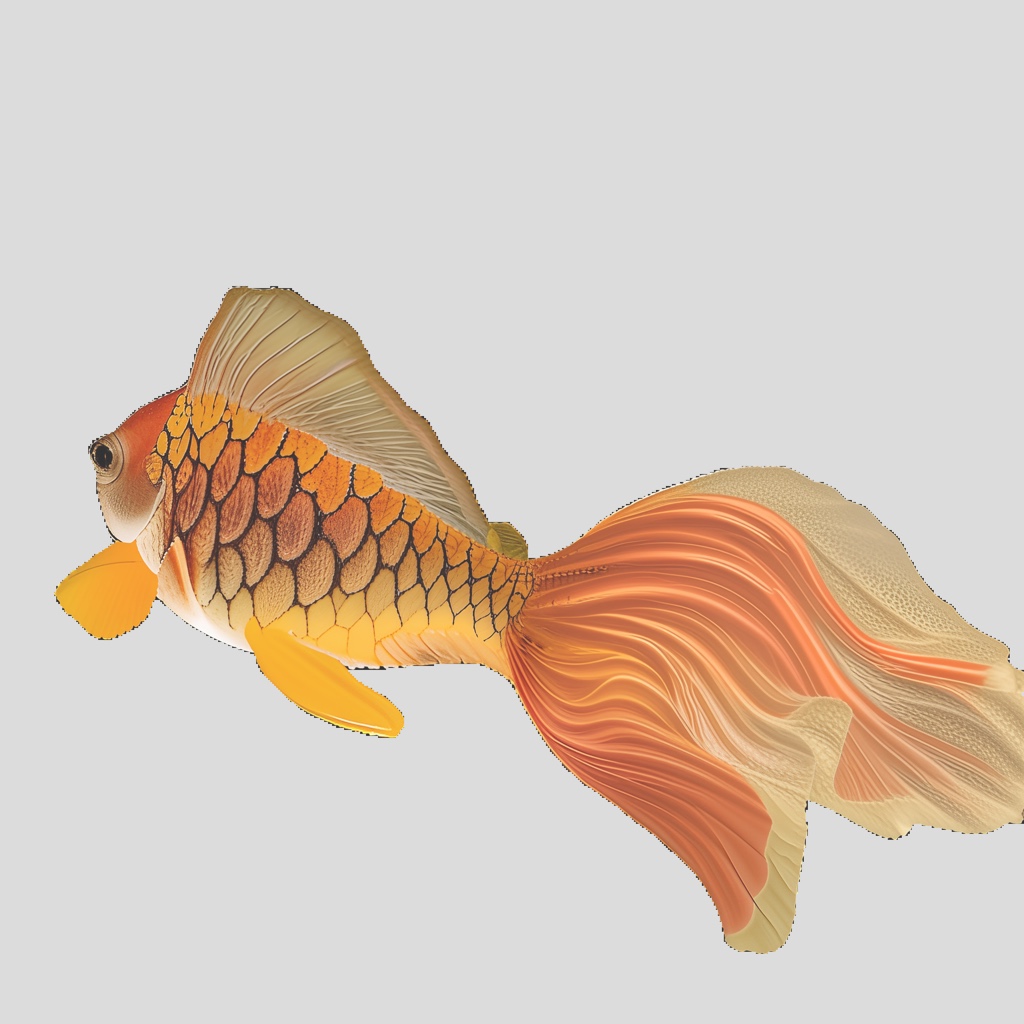} & \includegraphics[trim={5mm 5mm 5mm 5mm}, clip, width=\transferrreswidth, valign=m]{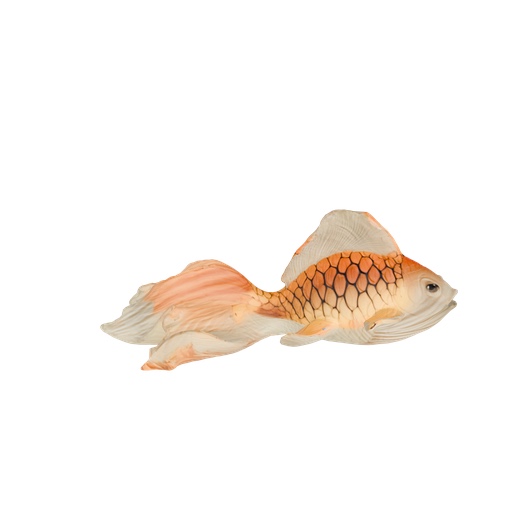} & \includegraphics[trim={5mm 5mm 5mm 5mm}, clip, width=\transferrreswidth, valign=m]{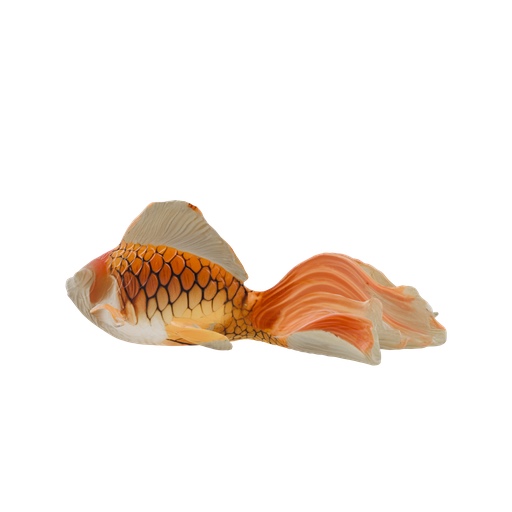} \\

  \includegraphics[width=\transferrnormalswidth, valign=m]{img/normals/toad_on_tortoise_shell_1.jpg} & \includegraphics[width=\transferrviewwidth, valign=m]{img/basecolor/toad_on_tortoise_shell_1/view0203.basecolor.jpg} & \includegraphics[trim={5mm 5mm 5mm 5mm}, clip, width=\transferrreswidth, valign=m]{img/render-keyframe/toad_on_tortoise_shell_1/view0203/view0203_glossft_diffuse_cam0003.jpg} & \includegraphics[trim={5mm 5mm 5mm 5mm}, clip, width=\transferrreswidth, valign=m]{img/render-keyframe/toad_on_tortoise_shell_1/view0203/view0203_glossft_diffuse_cam0002.jpg} & \includegraphics[width=\transferrviewwidth, valign=m]{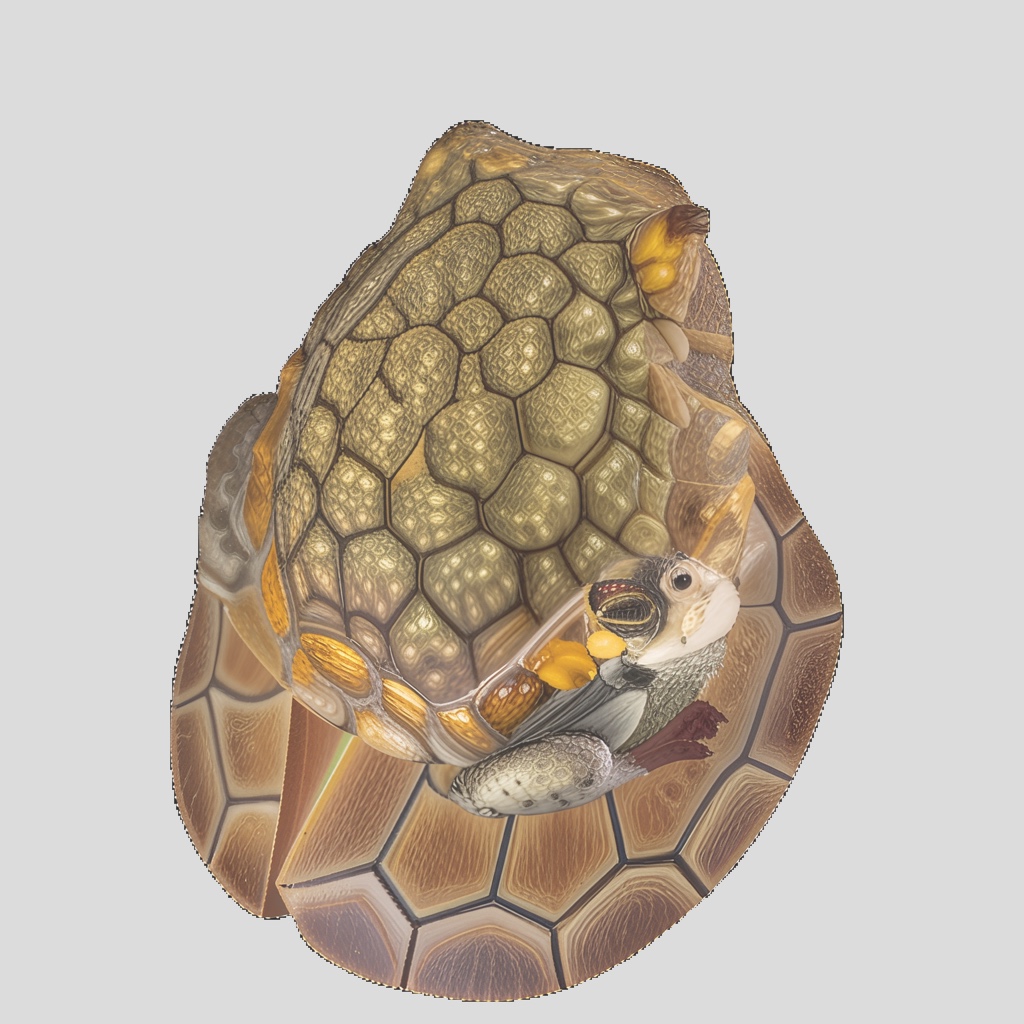} & \includegraphics[trim={5mm 5mm 5mm 5mm}, clip, width=\transferrreswidth, valign=m]{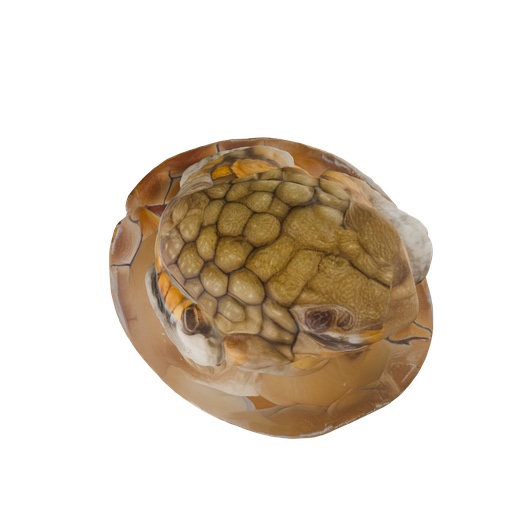} & \includegraphics[trim={5mm 5mm 5mm 5mm}, clip, width=\transferrreswidth, valign=m]{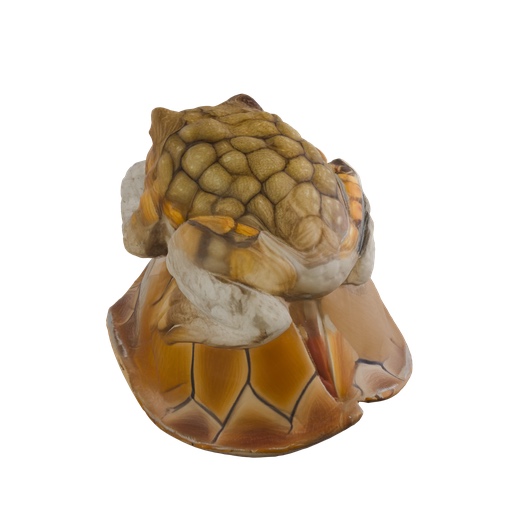} \\

  \includegraphics[width=\transferrnormalswidth, valign=m]{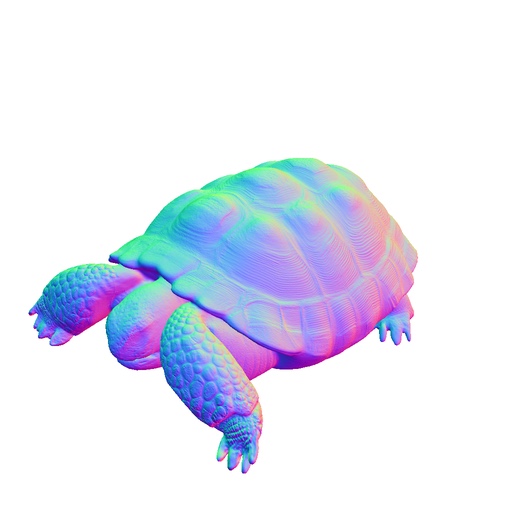} & \includegraphics[width=\transferrviewwidth, valign=m]{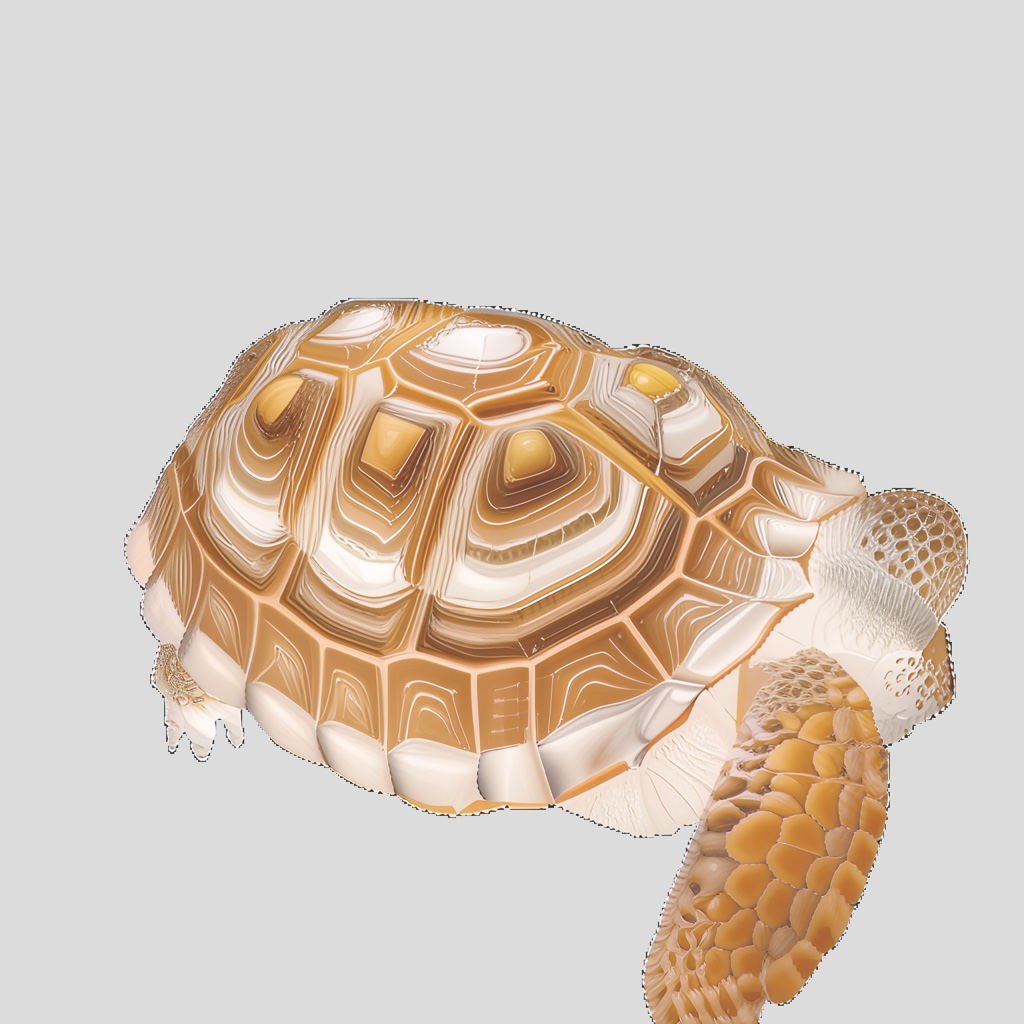} & \includegraphics[trim={5mm 5mm 5mm 5mm}, clip, width=\transferrreswidth, valign=m]{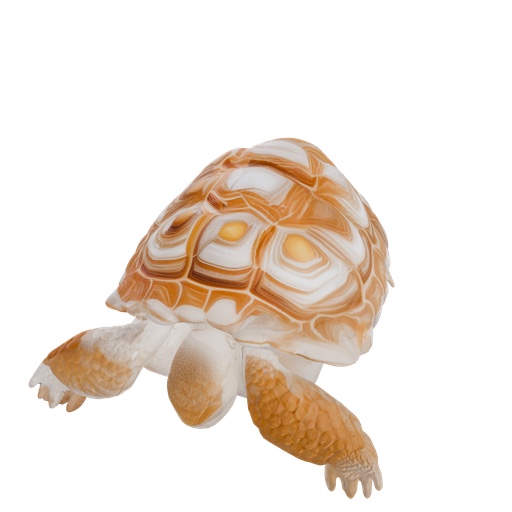} & \includegraphics[trim={5mm 5mm 5mm 5mm}, clip, width=\transferrreswidth, valign=m]{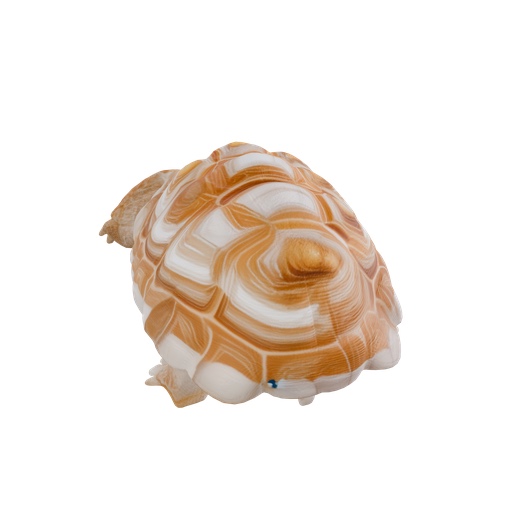} & \includegraphics[width=\transferrviewwidth, valign=m]{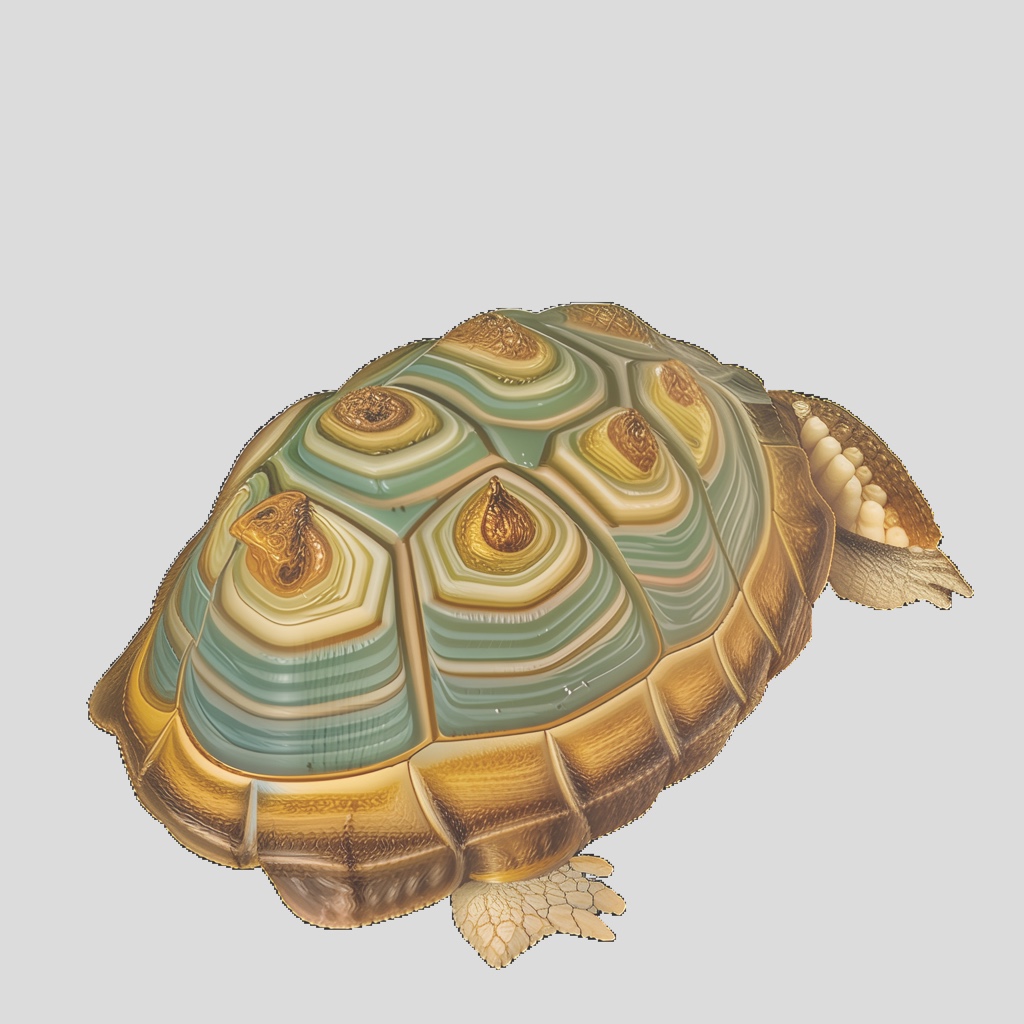} & \includegraphics[trim={5mm 5mm 5mm 5mm}, clip, width=\transferrreswidth, valign=m]{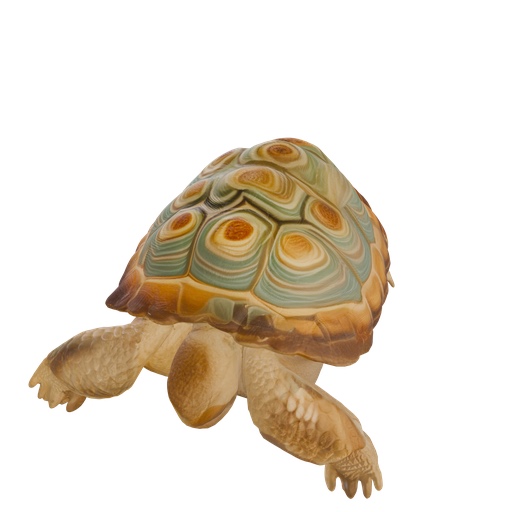} & \includegraphics[trim={5mm 5mm 5mm 5mm}, clip, width=\transferrreswidth, valign=m]{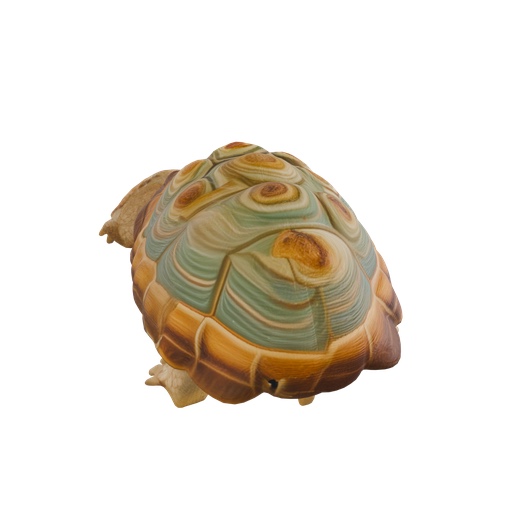} \\

  \end{tabular}
  \caption{Additional finetuning completion results.}
  \label{fig:supp_transferr_gallery_supp}
\end{figure*}

\endgroup
\providecommand{\shortsnormalswidth}{0.1\linewidth}
\providecommand{\shortsviewwidth}{0.1\linewidth}
\providecommand{\shortsreswidth}{0.1\linewidth}

\begingroup
\setlength{\tabcolsep}{2pt}
\renewcommand{\arraystretch}{0}

\begin{figure*}[h!tbp]
  \centering
  \begin{tabular}{cccccccc}
  \small Geometry & \small Single View & \small Paint 3D & \small TexGen & \small MV-Adapter & \small Hunyuan2.1 & \small Trellis & \small Ours  \\
  \includegraphics[width=\shortsnormalswidth, valign=m]{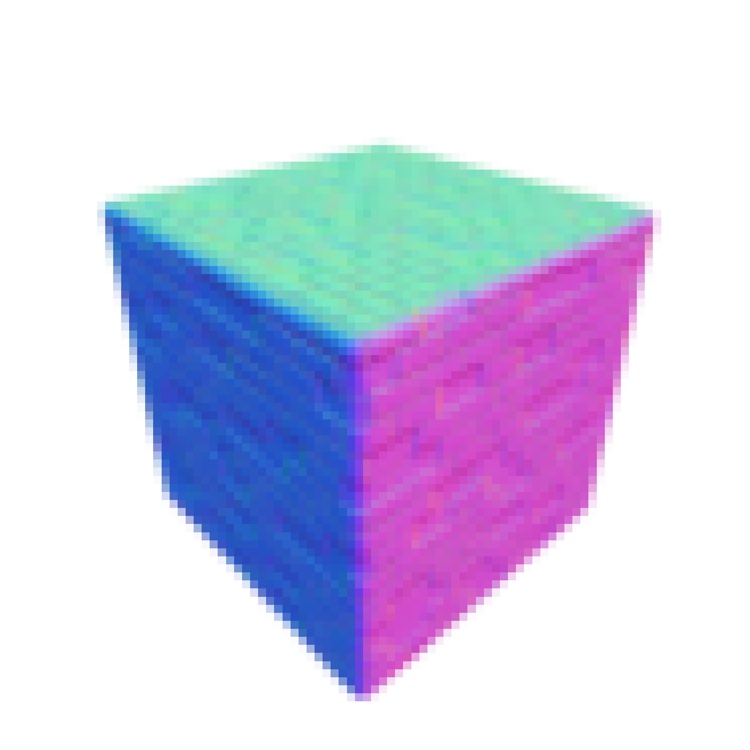} &
  \includegraphics[width=\shortsviewwidth, valign=m]{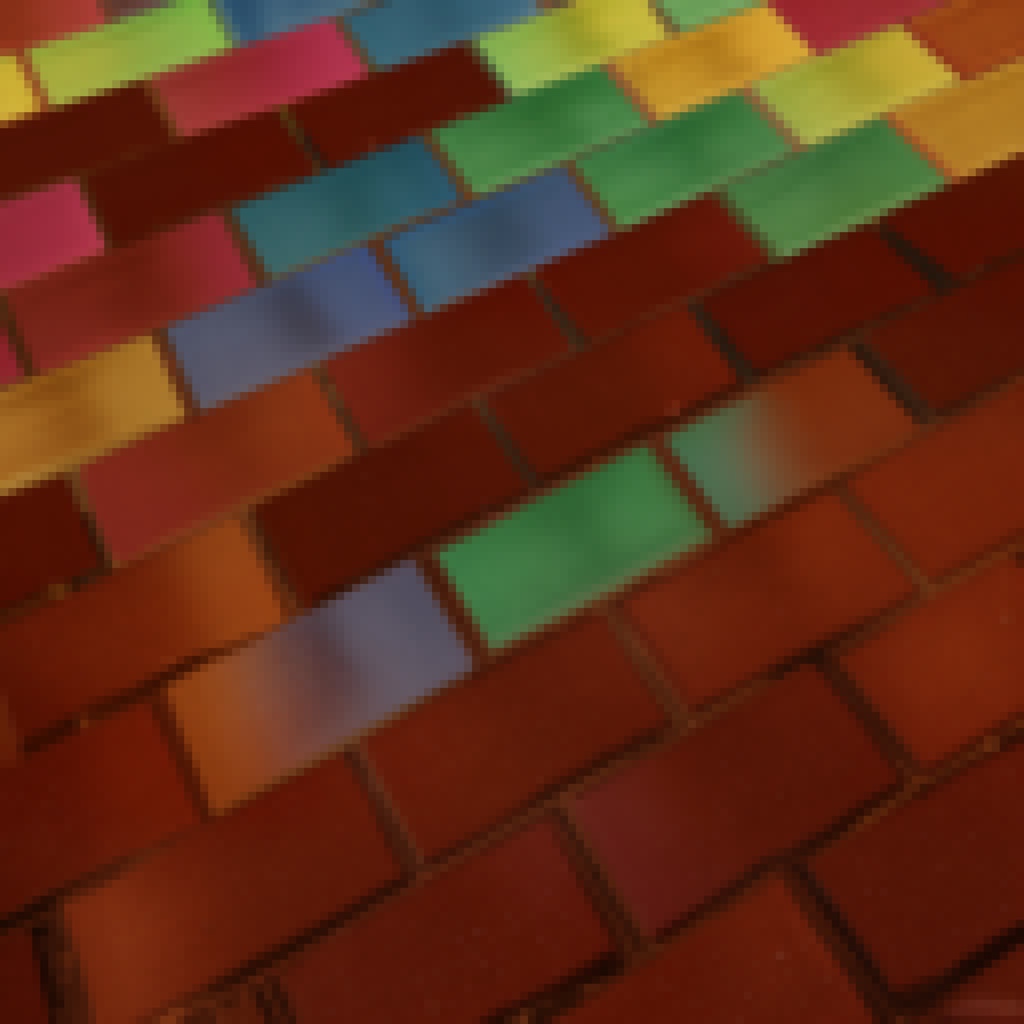} &
  \includegraphics[width=\shortsreswidth, valign=m]{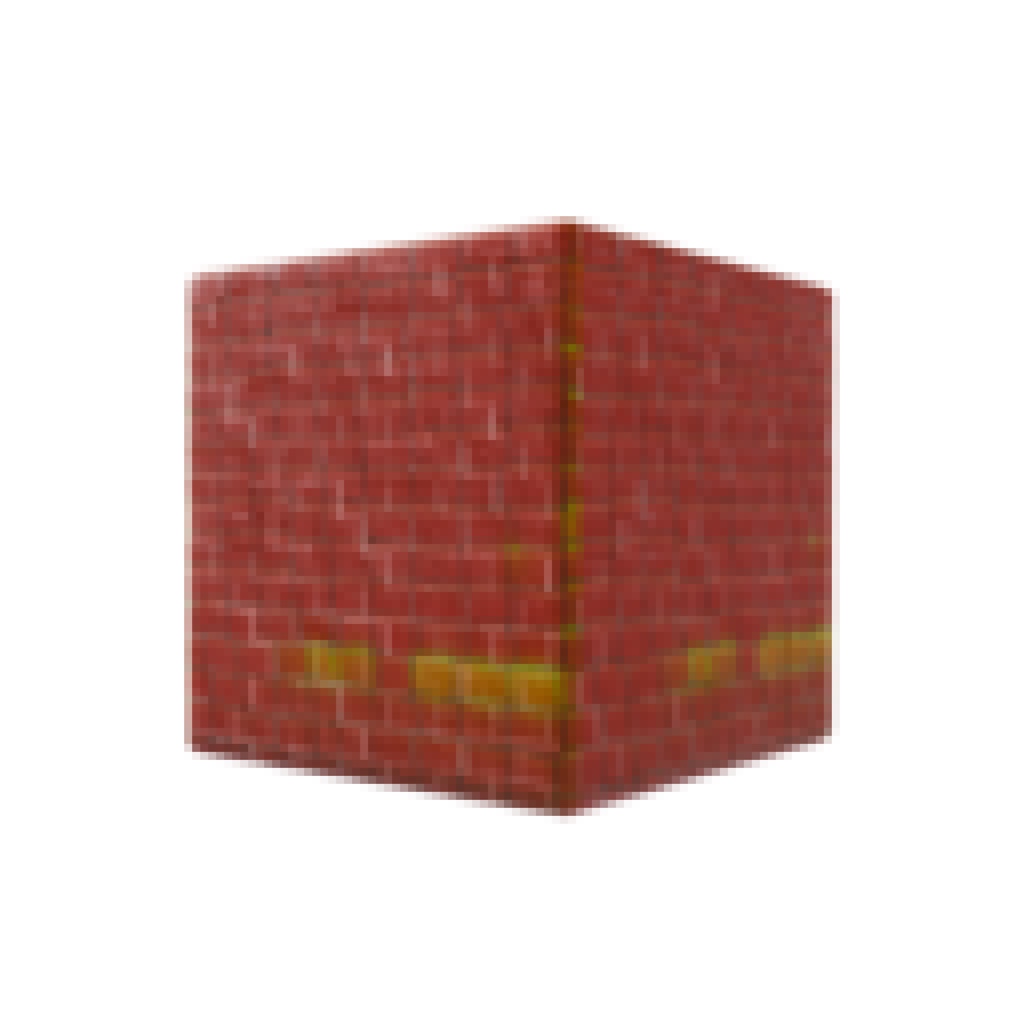} &
  \includegraphics[width=\shortsreswidth, valign=m]{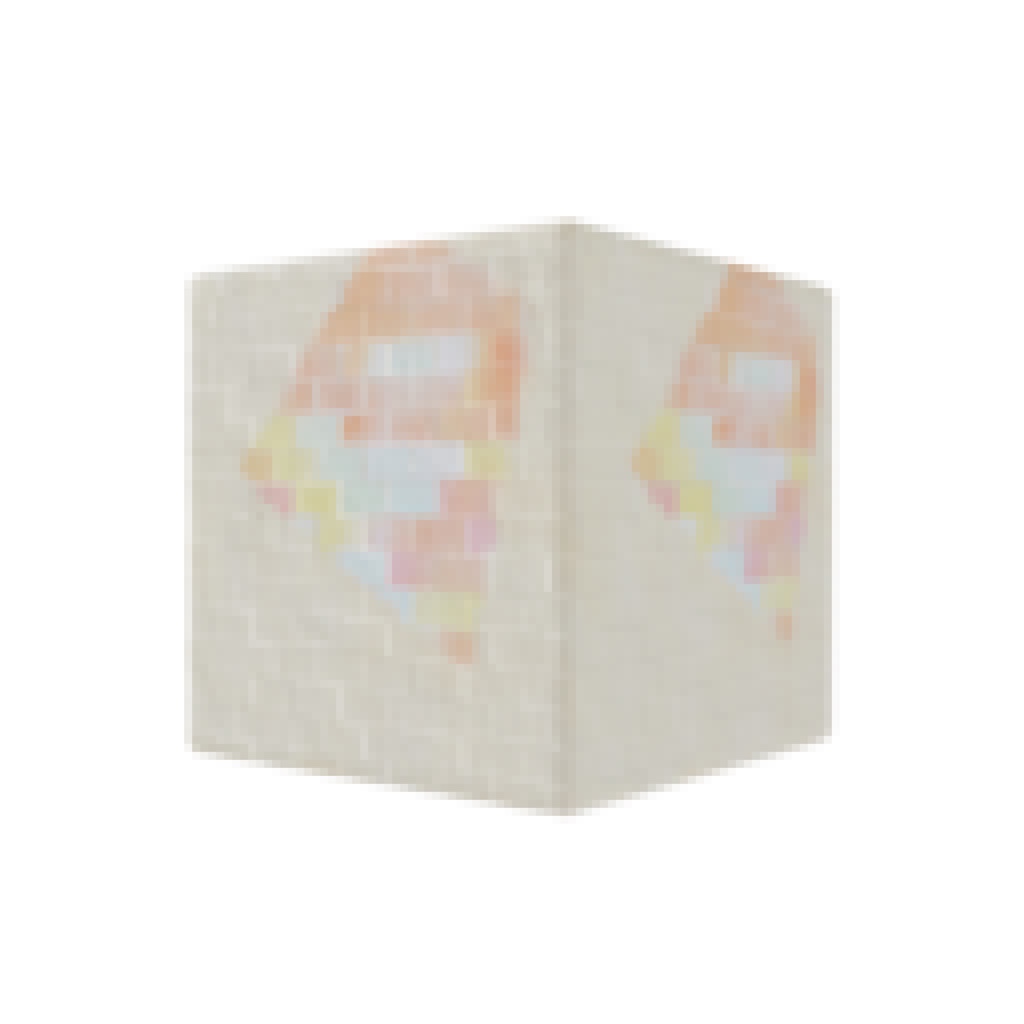} &
  \includegraphics[width=\shortsreswidth, valign=m]{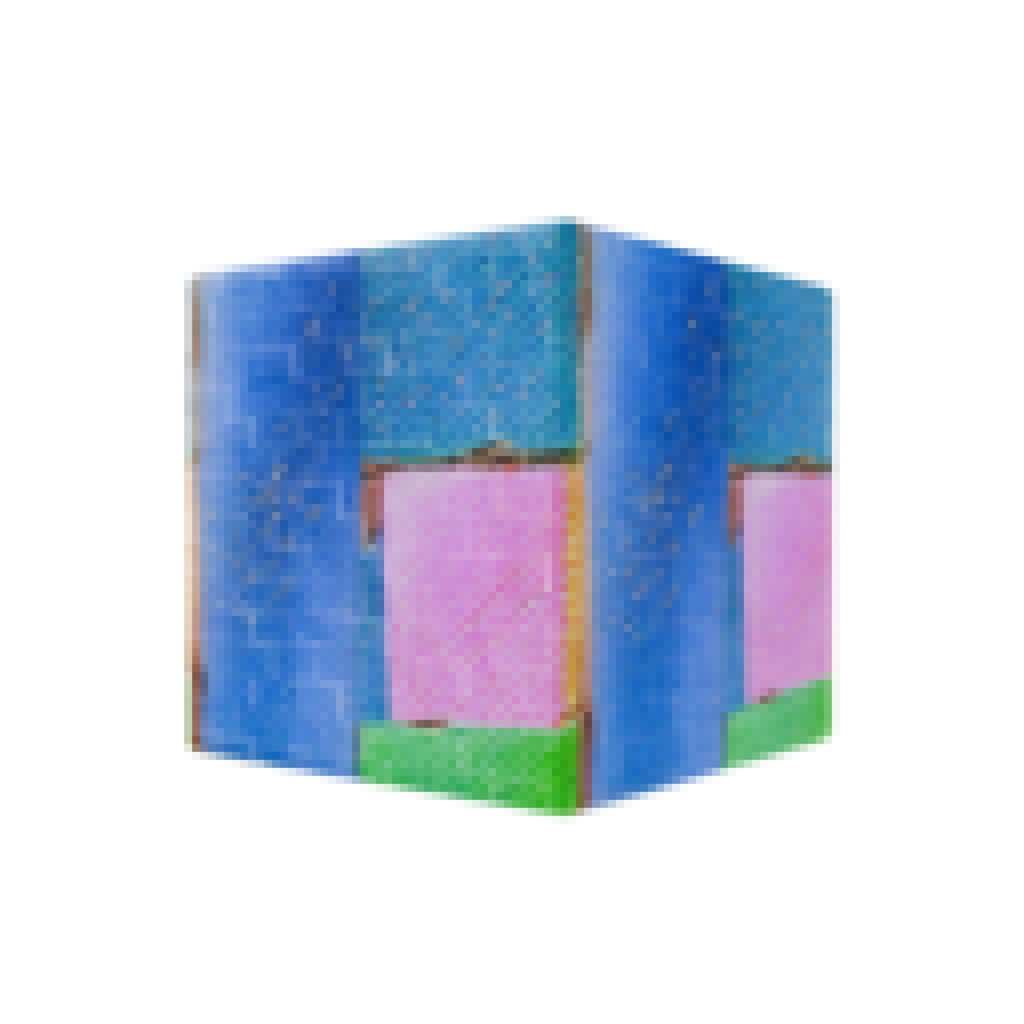} &
  \includegraphics[width=\shortsreswidth, valign=m]{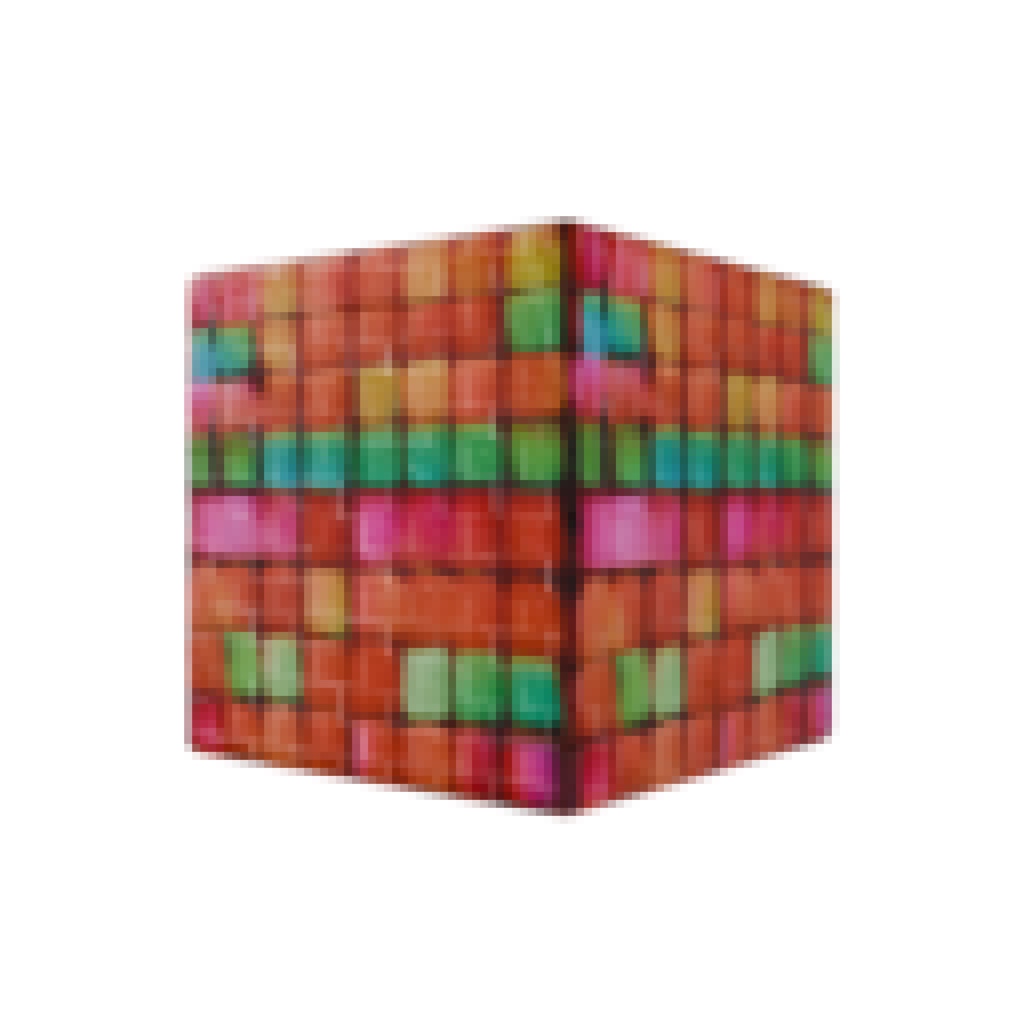} &
  \includegraphics[width=\shortsreswidth, valign=m]{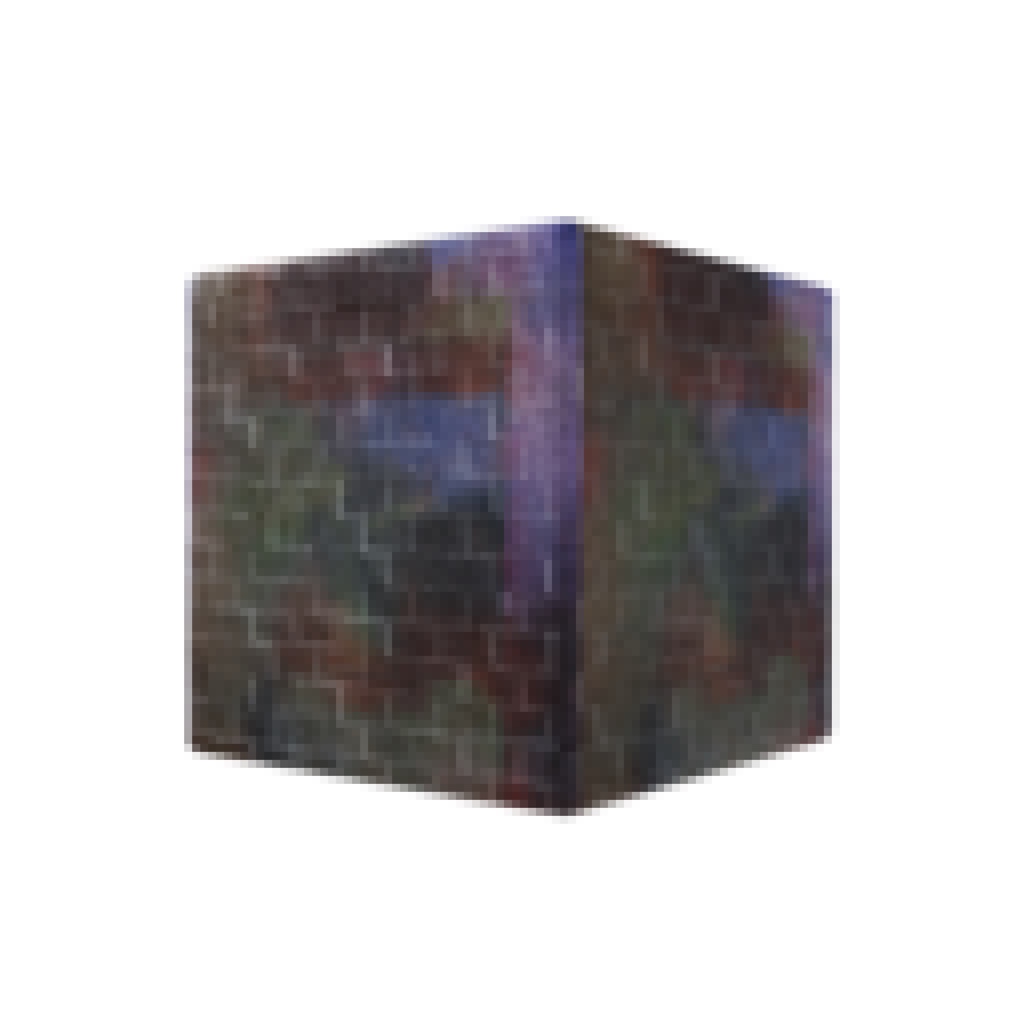} &
  \includegraphics[width=\shortsreswidth, valign=m]{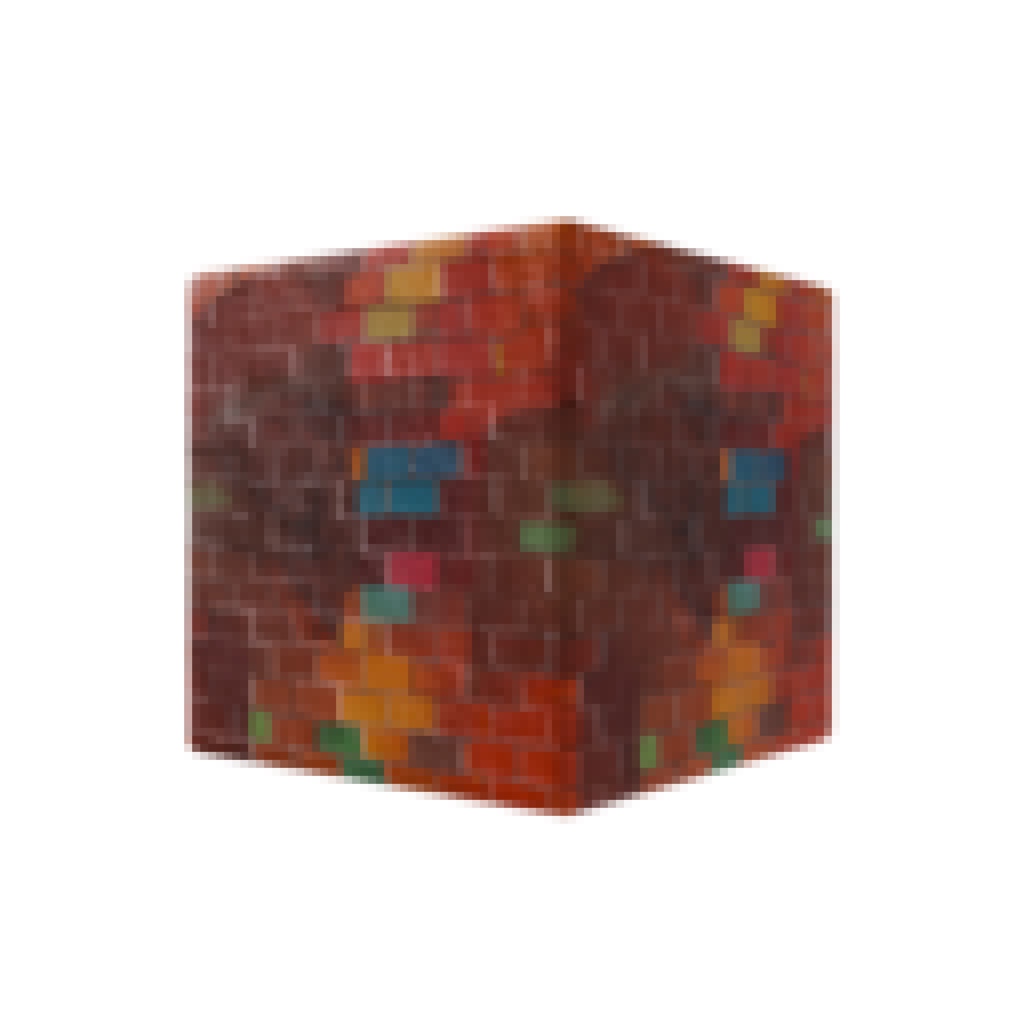}  \\
  \includegraphics[width=\shortsnormalswidth, valign=m]{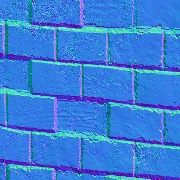} &
  \includegraphics[width=\shortsviewwidth, valign=m]{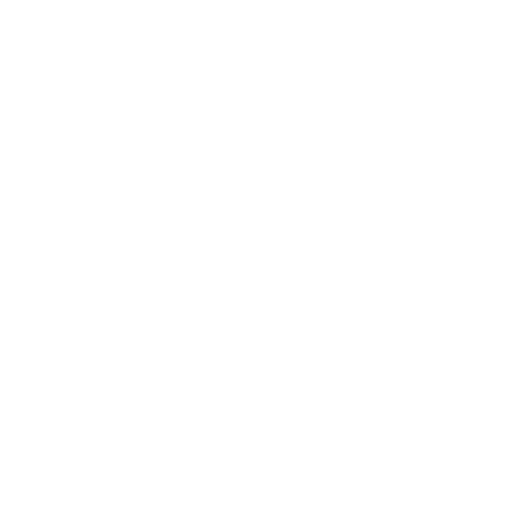} &
  \includegraphics[width=\shortsreswidth, valign=m]{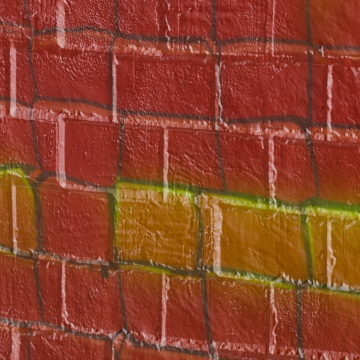} &
  \includegraphics[width=\shortsreswidth, valign=m]{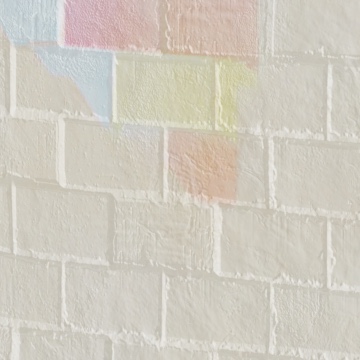} &
  \includegraphics[width=\shortsreswidth, valign=m]{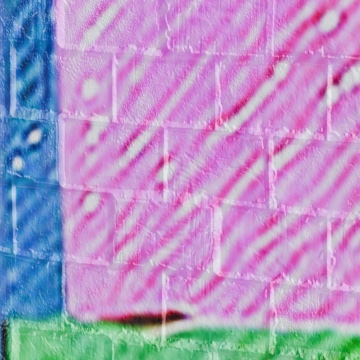} &
  \includegraphics[width=\shortsreswidth, valign=m]{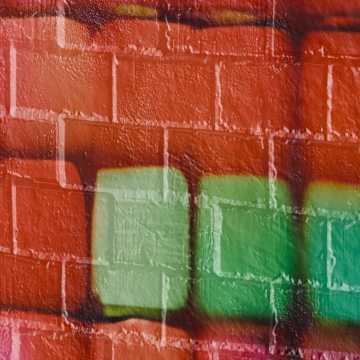} &
  \includegraphics[width=\shortsreswidth, valign=m]{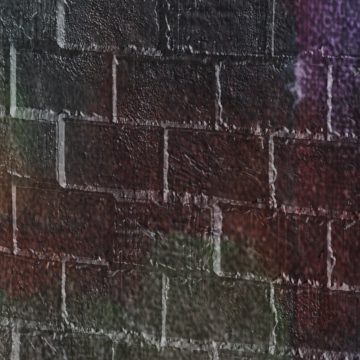} &
  \includegraphics[width=\shortsreswidth, valign=m]{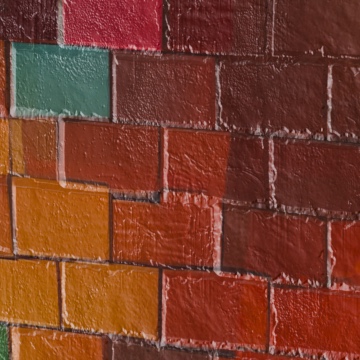}  \\
  \includegraphics[width=\shortsnormalswidth, valign=m]{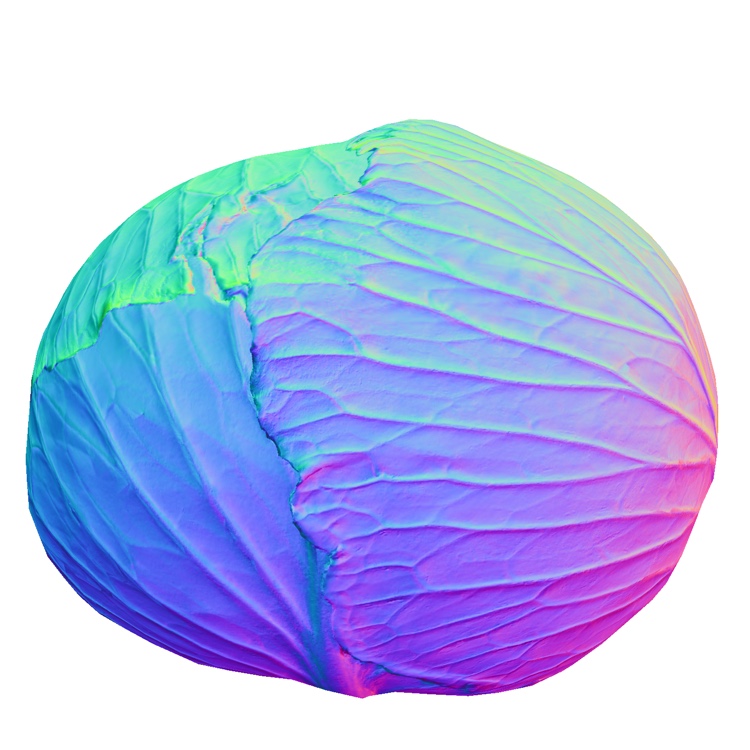} &
  \includegraphics[width=\shortsviewwidth, valign=m]{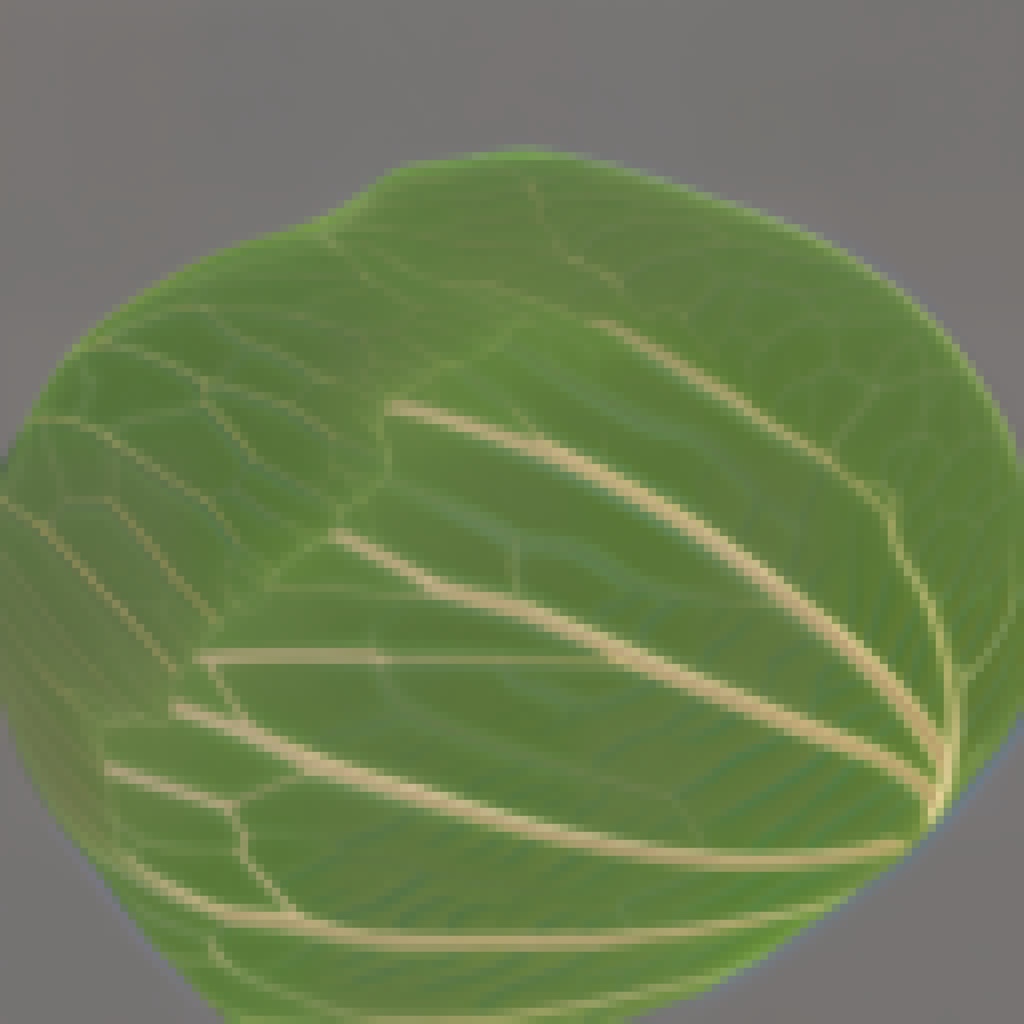} &
  \includegraphics[width=\shortsreswidth, valign=m]{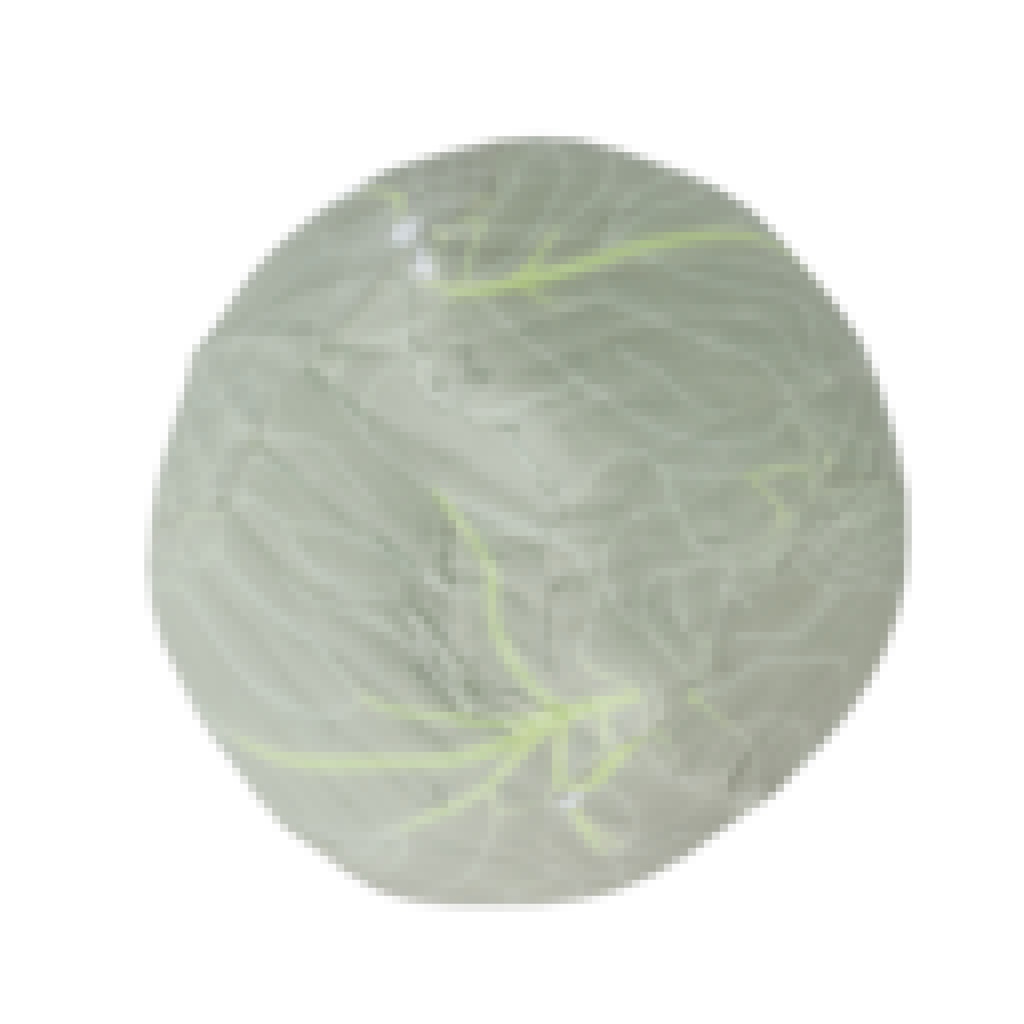} &
  \includegraphics[width=\shortsreswidth, valign=m]{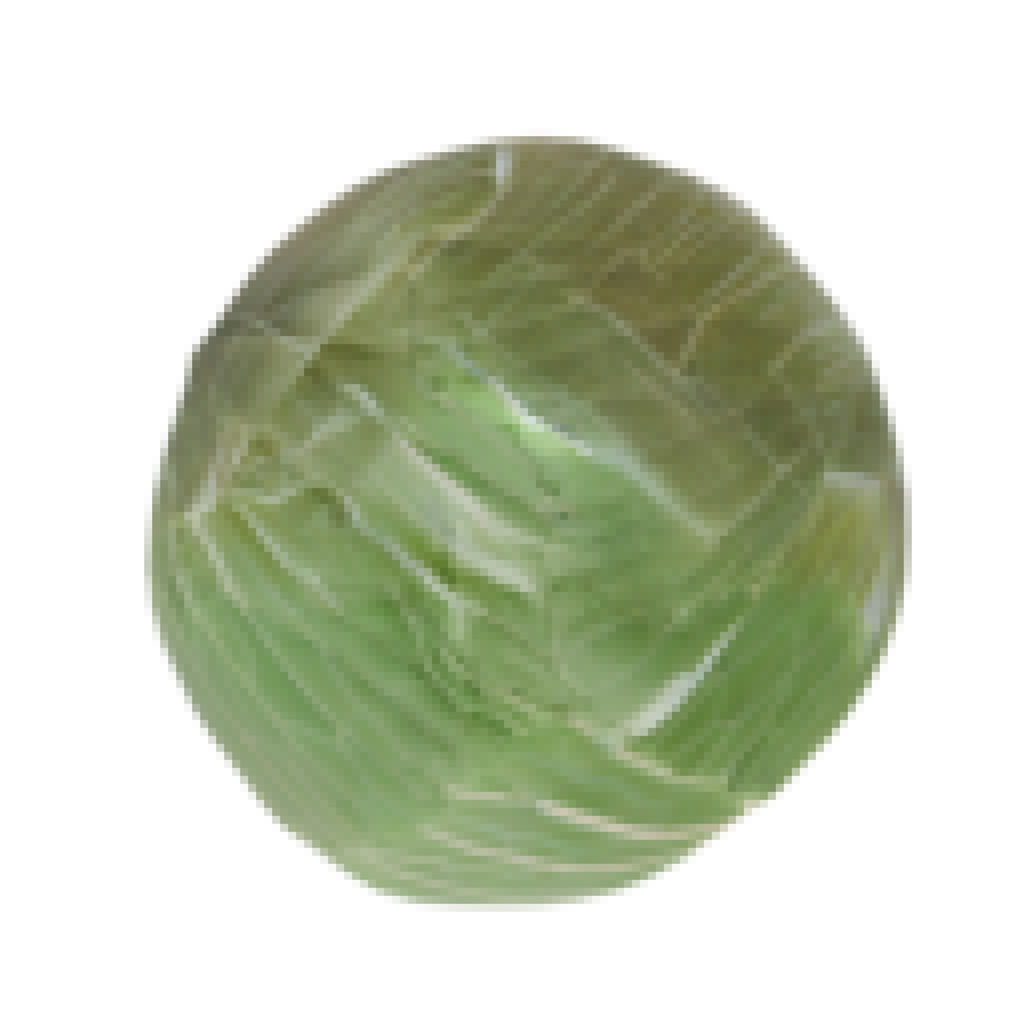} &
  \includegraphics[width=\shortsreswidth, valign=m]{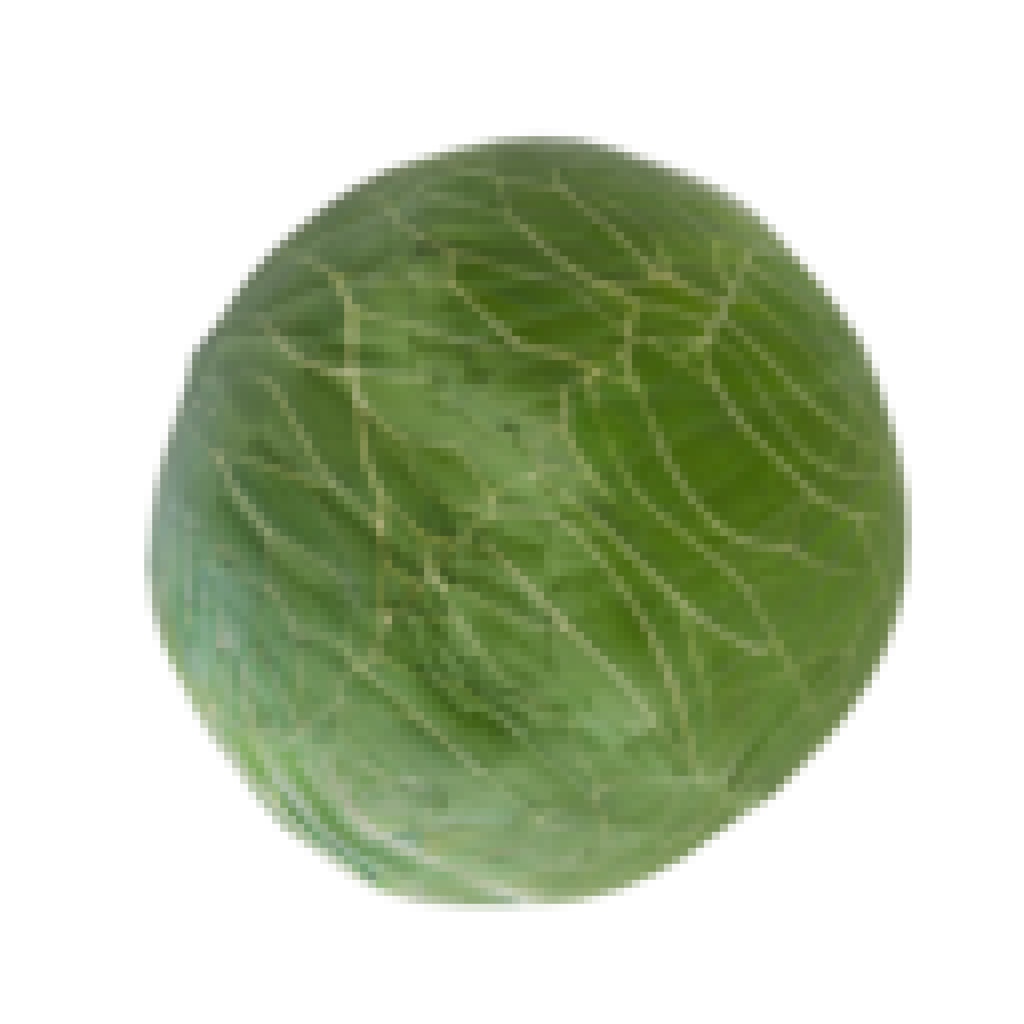} &
  \includegraphics[width=\shortsreswidth, valign=m]{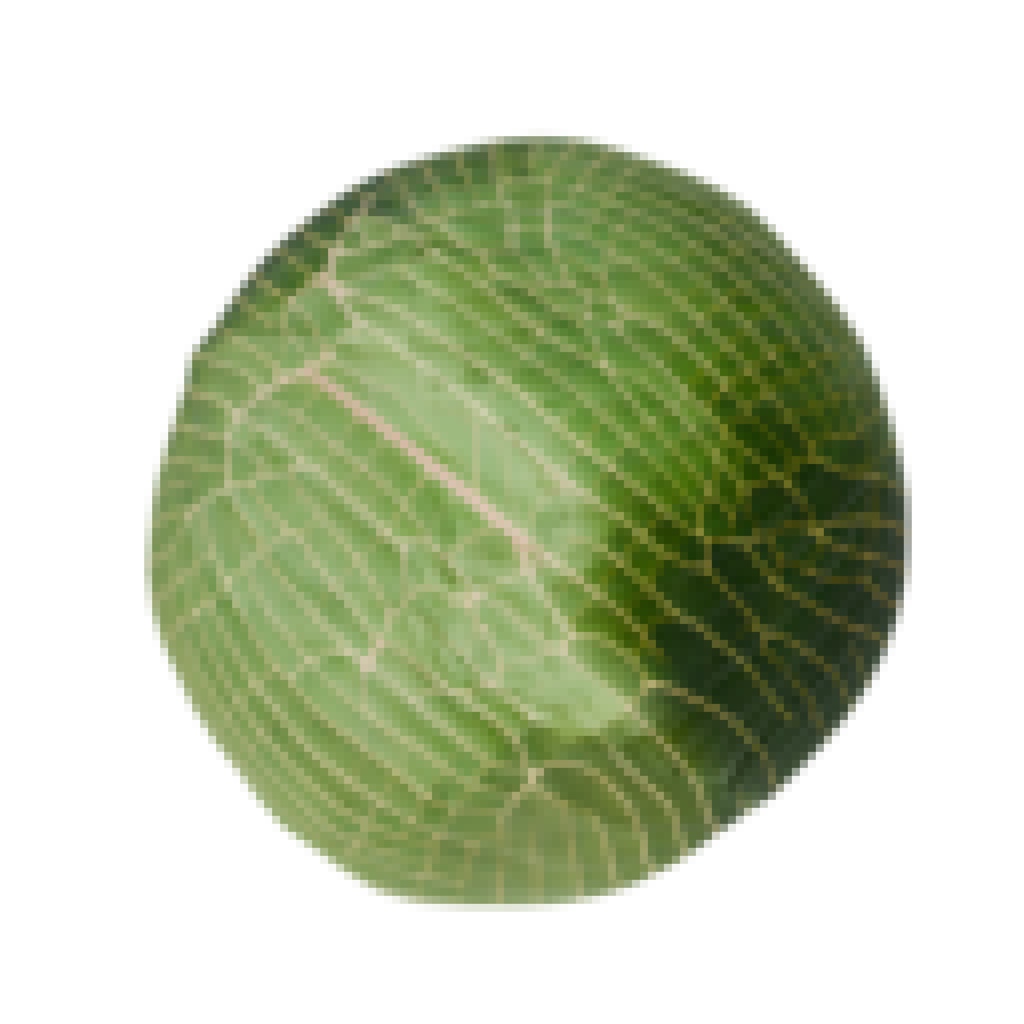} &
  \includegraphics[width=\shortsreswidth, valign=m]{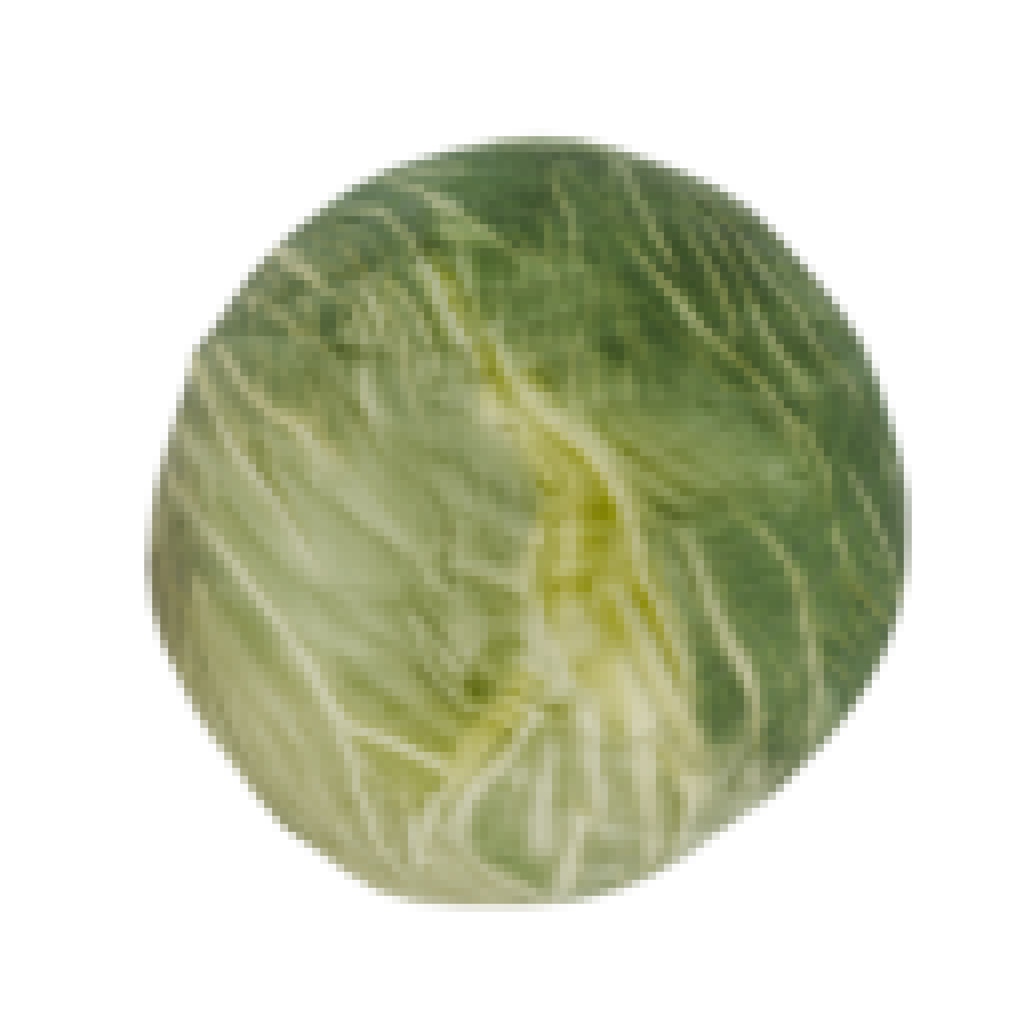} &
  \includegraphics[width=\shortsreswidth, valign=m]{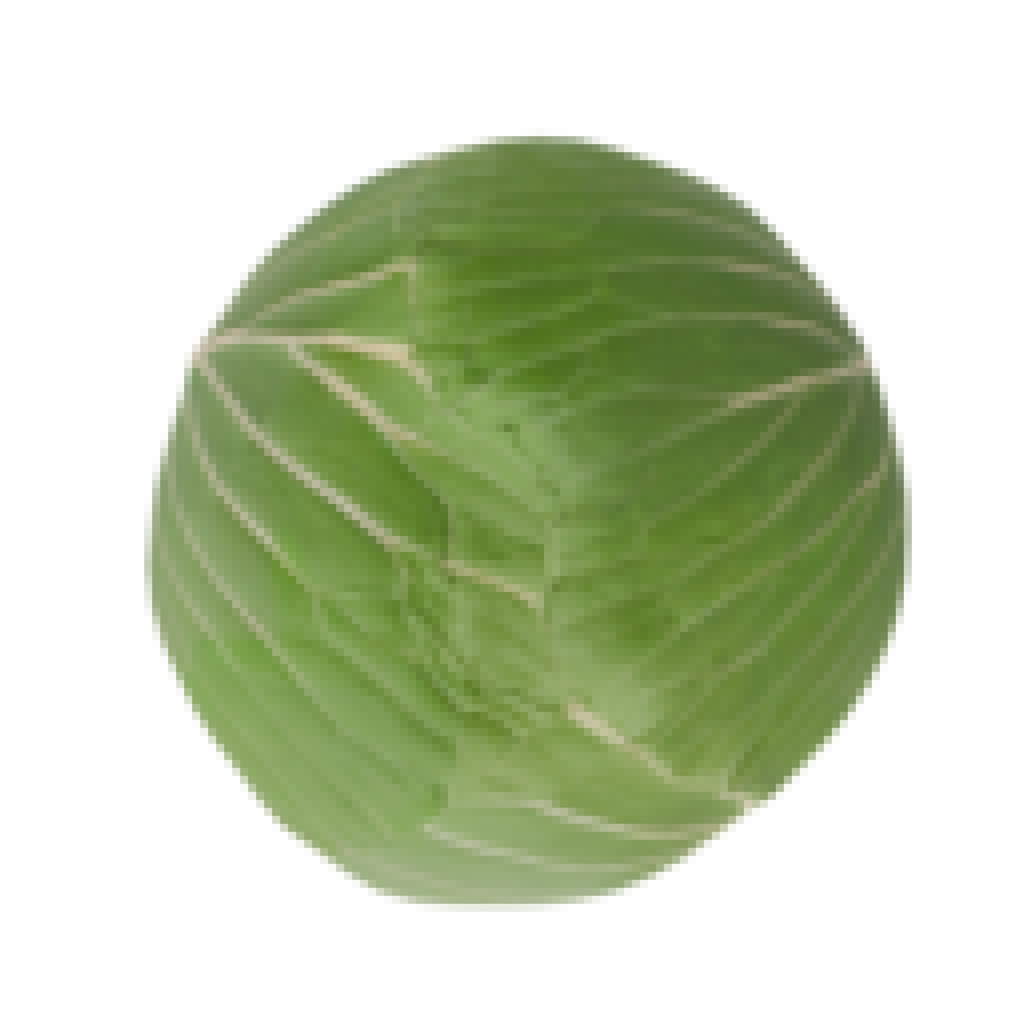}  \\
  \includegraphics[width=\shortsnormalswidth, valign=m]{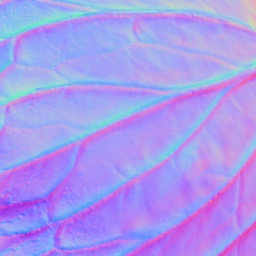} &
  \includegraphics[width=\shortsviewwidth, valign=m]{img/close-ups/_basecolor/white.jpg} &
  \includegraphics[width=\shortsreswidth, valign=m]{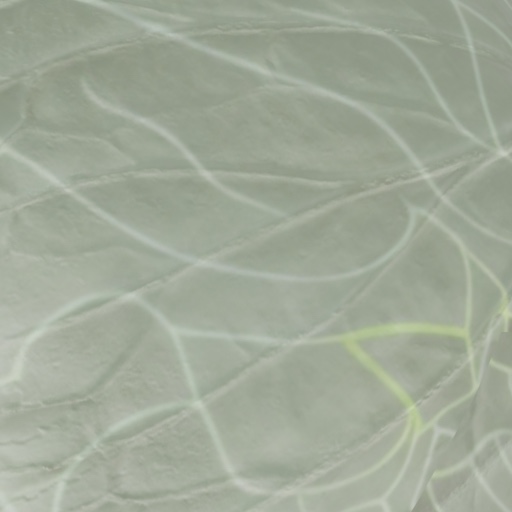} &
  \includegraphics[width=\shortsreswidth, valign=m]{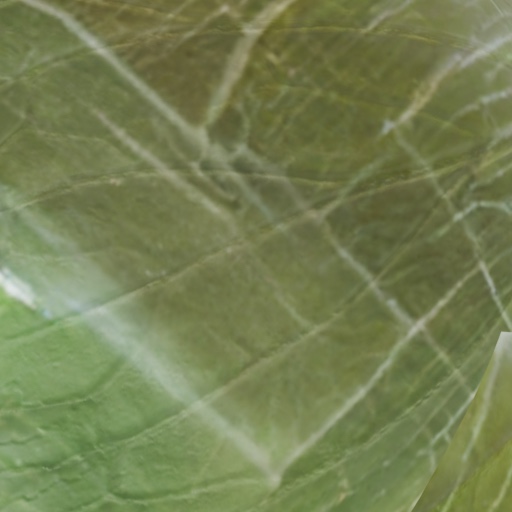} &
  \includegraphics[width=\shortsreswidth, valign=m]{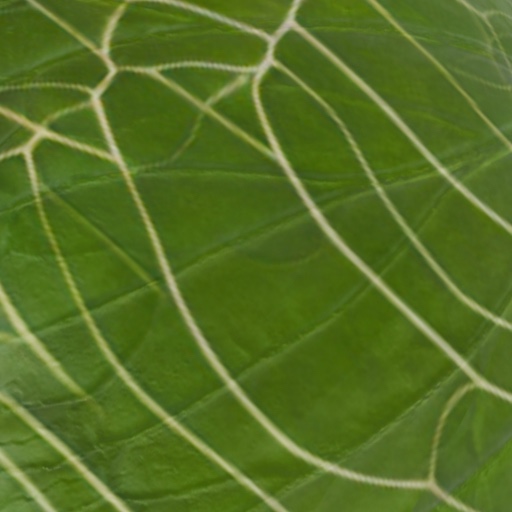} &
  \includegraphics[width=\shortsreswidth, valign=m]{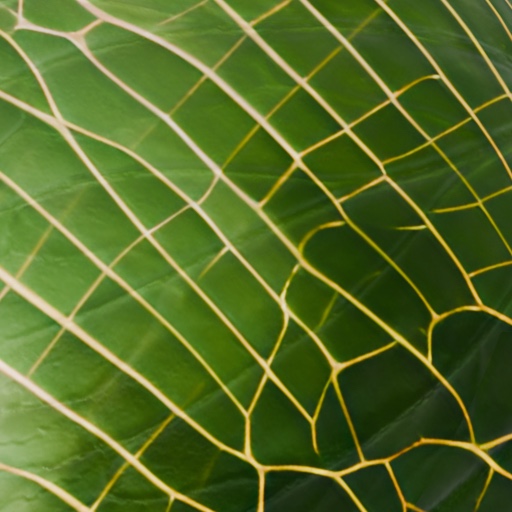} &
  \includegraphics[width=\shortsreswidth, valign=m]{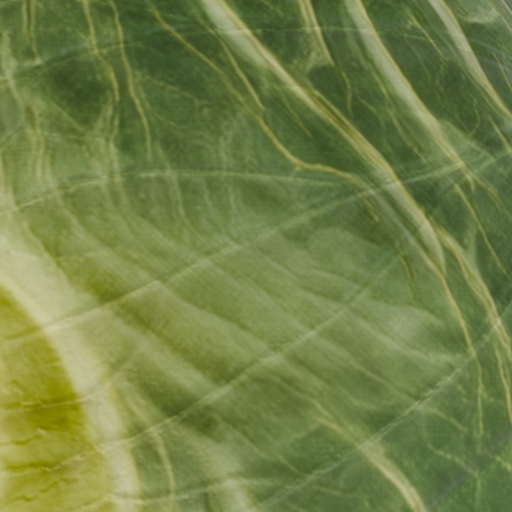} &
  \includegraphics[width=\shortsreswidth, valign=m]{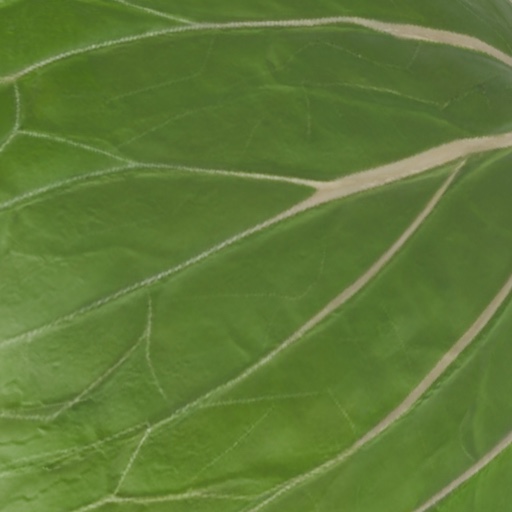}  \\
  \includegraphics[width=\shortsnormalswidth, valign=m]{img/normals/cabbage.jpg} &
  \includegraphics[width=\shortsviewwidth, valign=m]{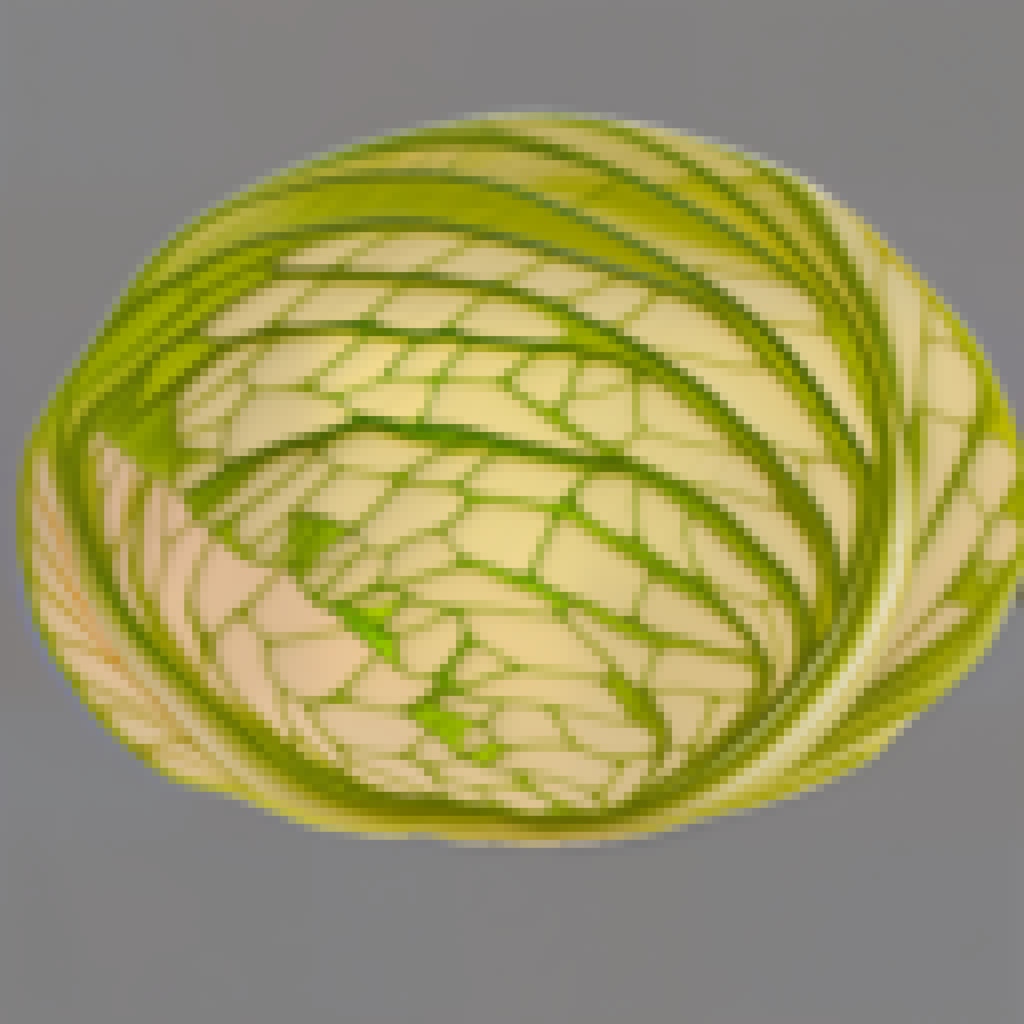} &
  \includegraphics[width=\shortsreswidth, valign=m]{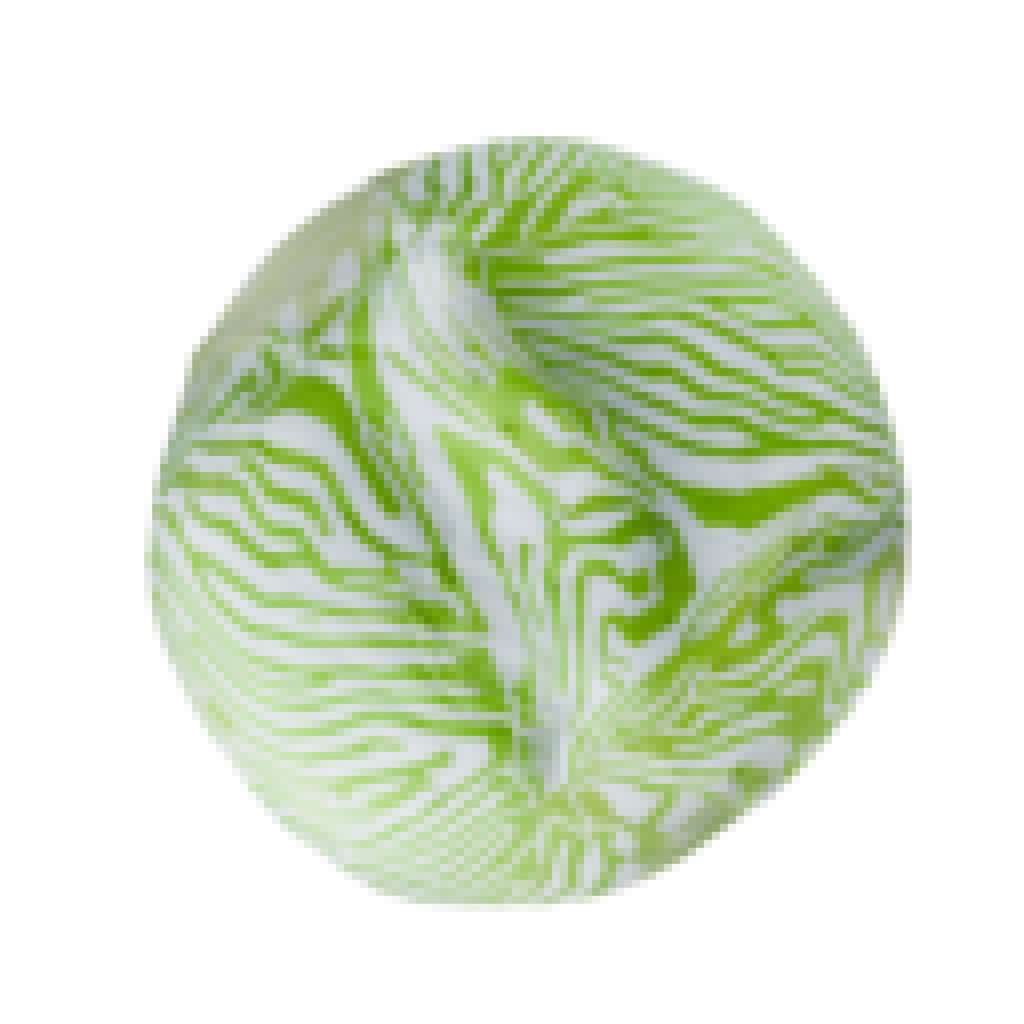} &
  \includegraphics[width=\shortsreswidth, valign=m]{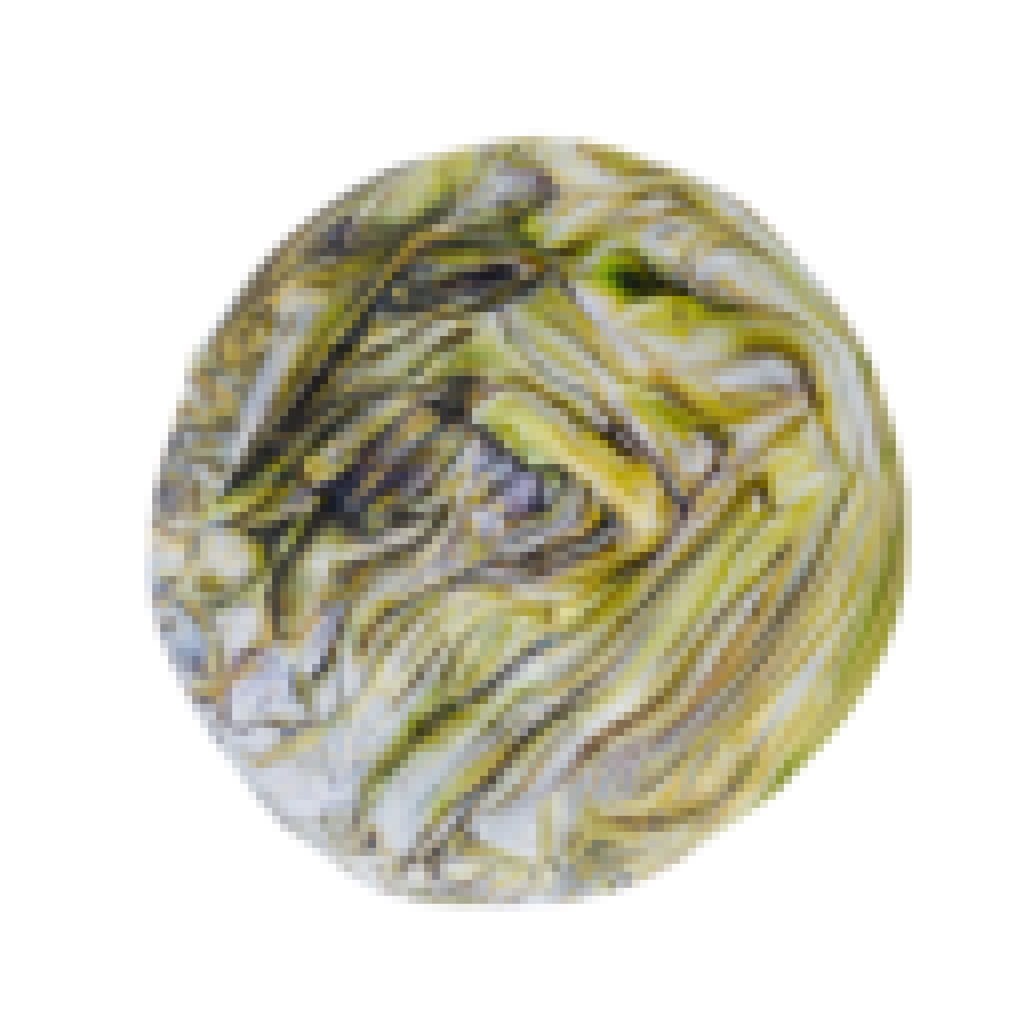} &
  \includegraphics[width=\shortsreswidth, valign=m]{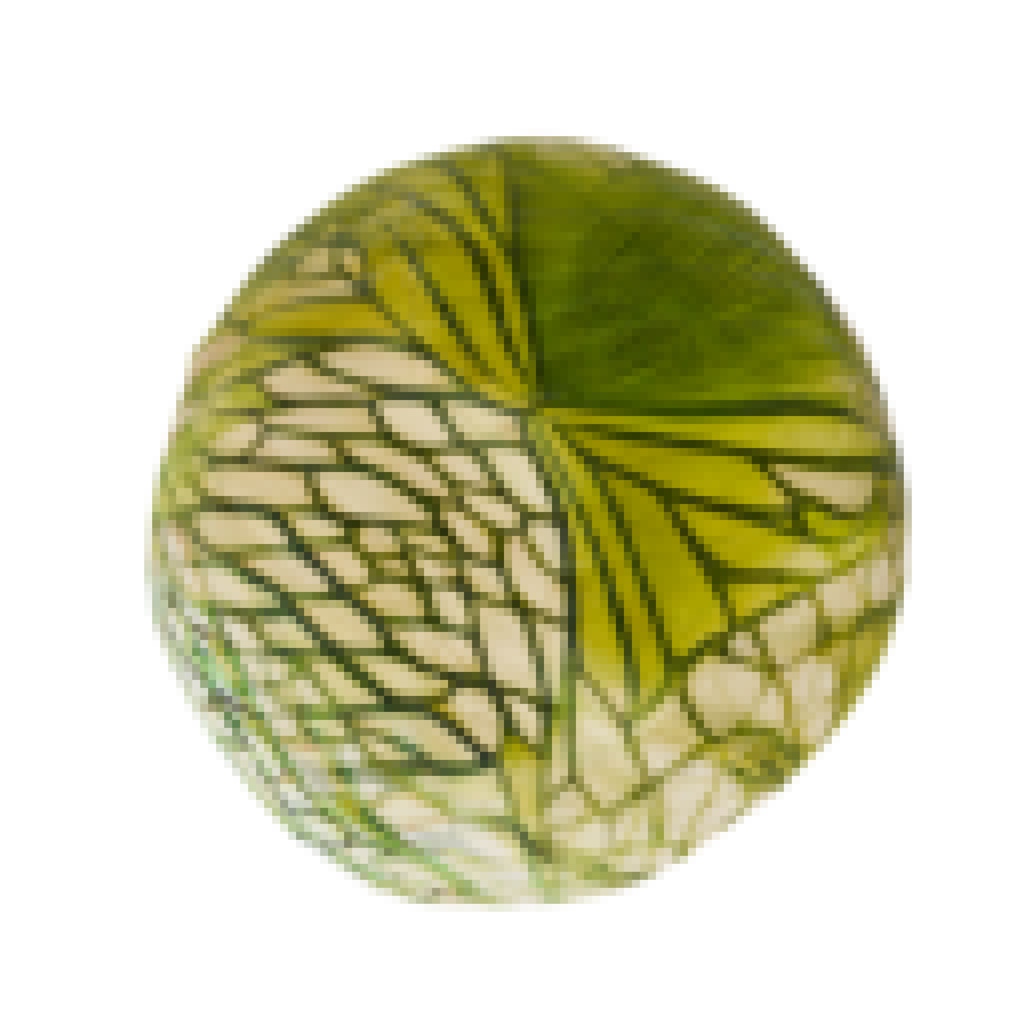} &
  \includegraphics[width=\shortsreswidth, valign=m]{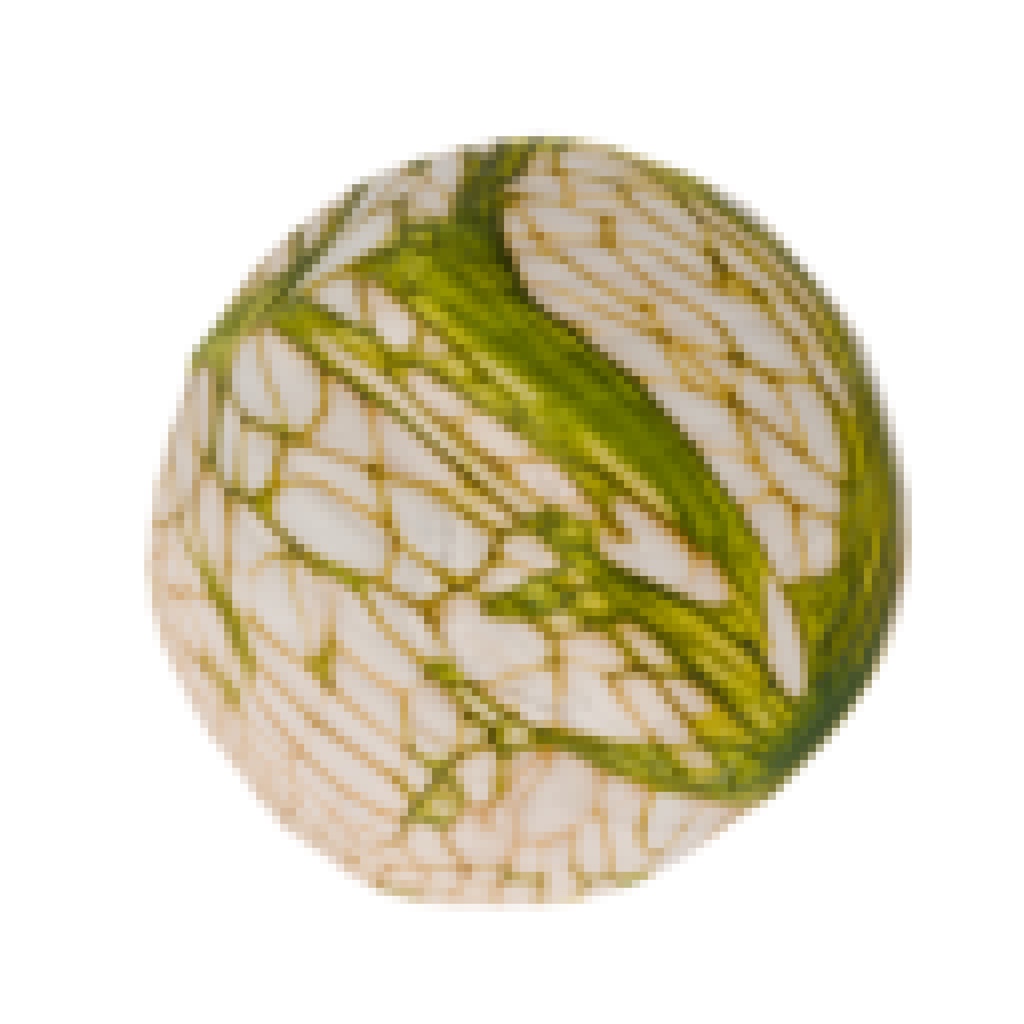} &
  \includegraphics[width=\shortsreswidth, valign=m]{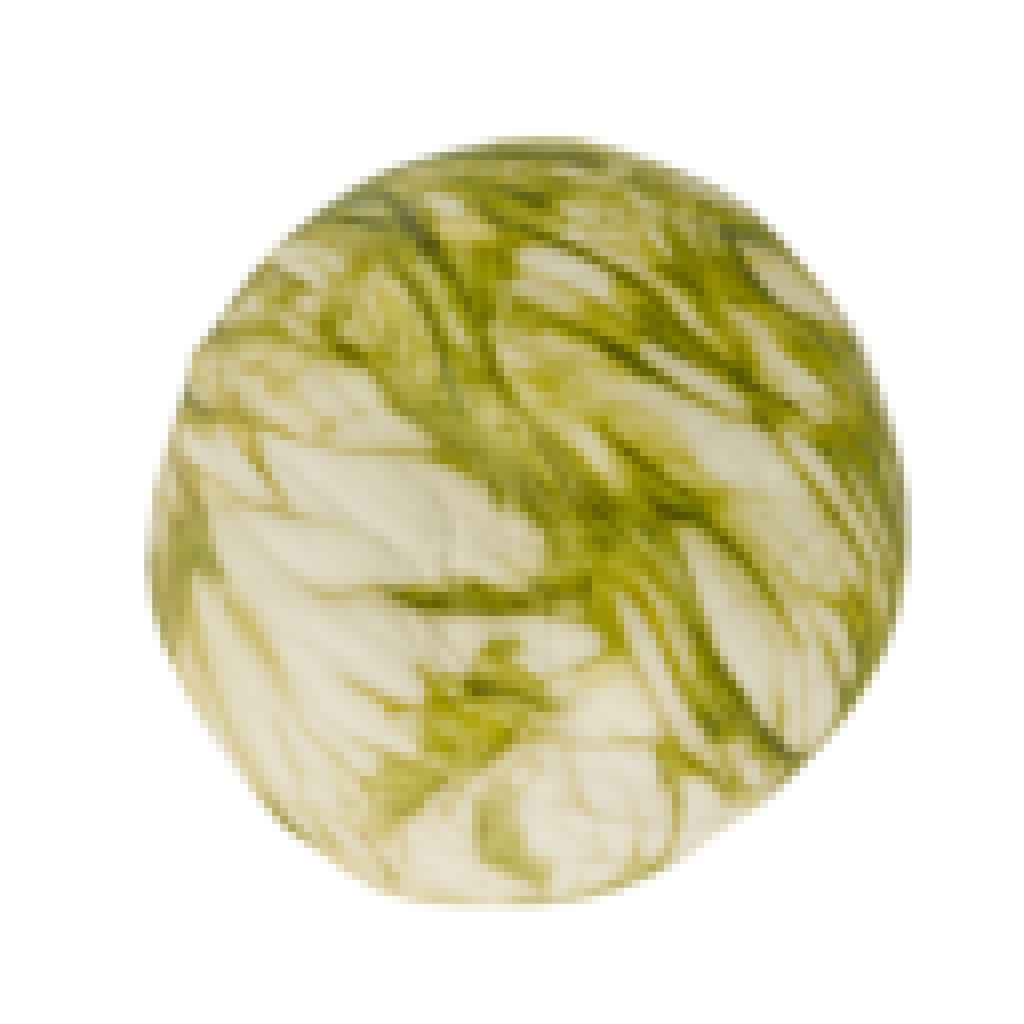} &
  \includegraphics[width=\shortsreswidth, valign=m]{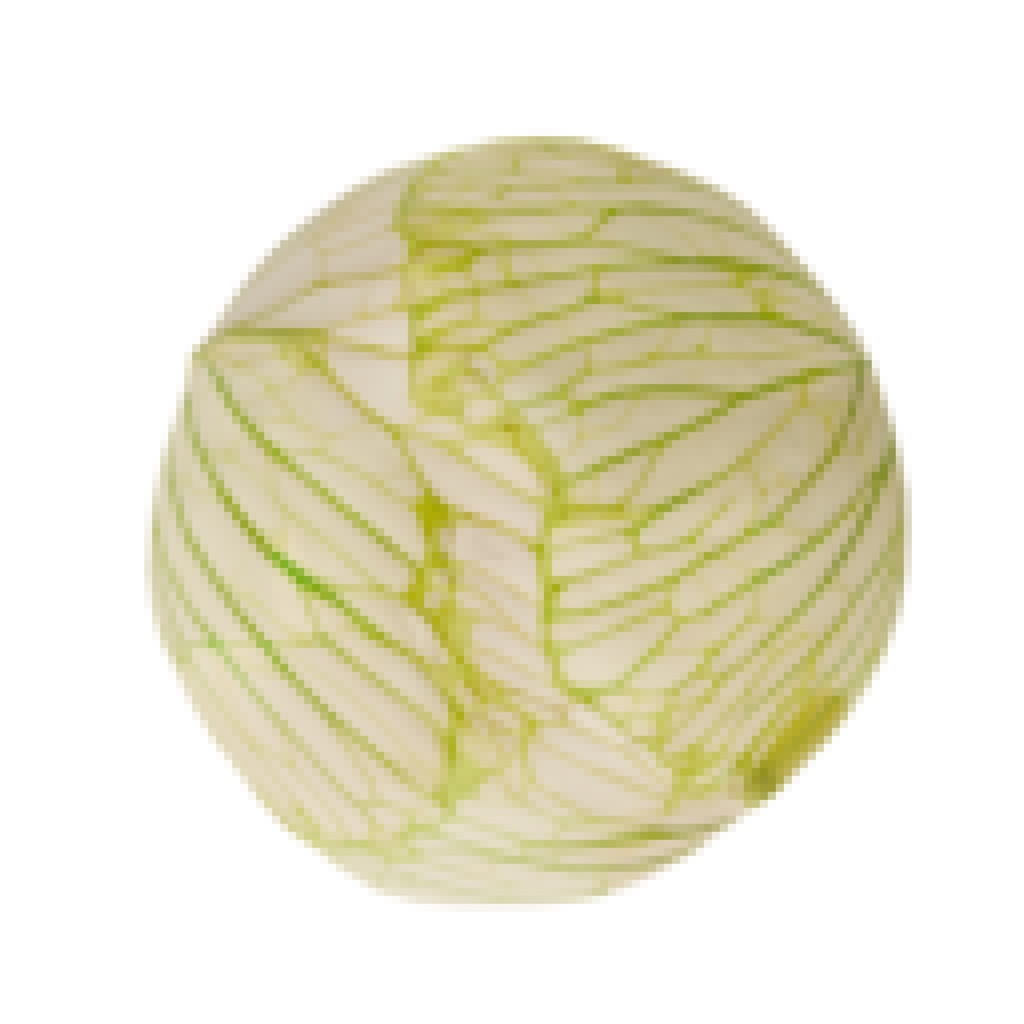}  \\
  \includegraphics[width=\shortsnormalswidth, valign=m]{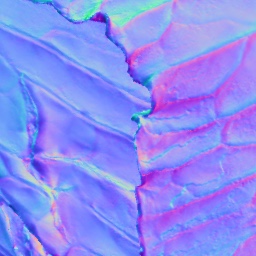} &
  \includegraphics[width=\shortsviewwidth, valign=m]{img/close-ups/_basecolor/white.jpg} &
  \includegraphics[width=\shortsreswidth, valign=m]{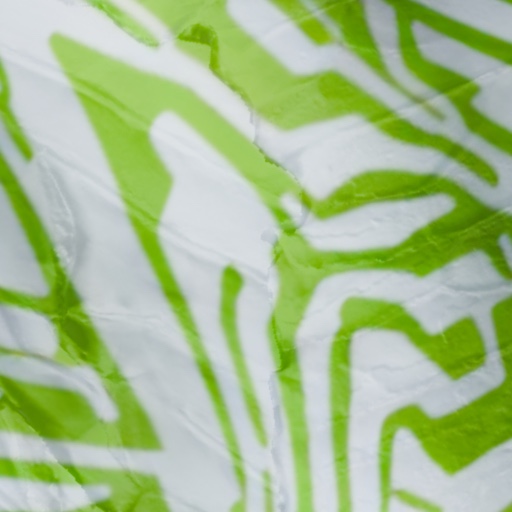} &
  \includegraphics[width=\shortsreswidth, valign=m]{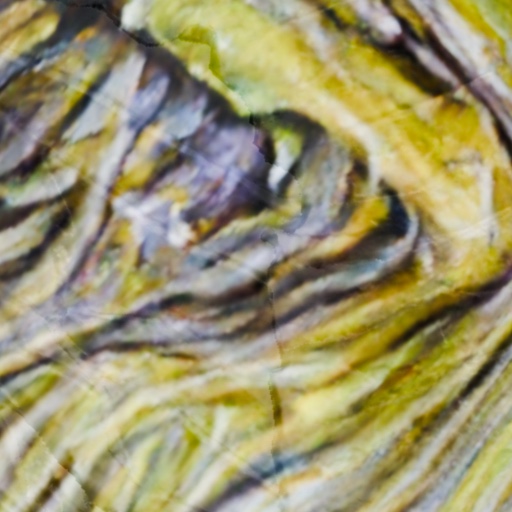} &
  \includegraphics[width=\shortsreswidth, valign=m]{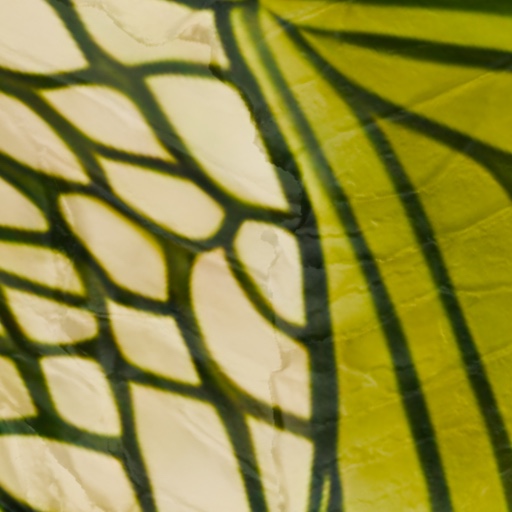} &
  \includegraphics[width=\shortsreswidth, valign=m]{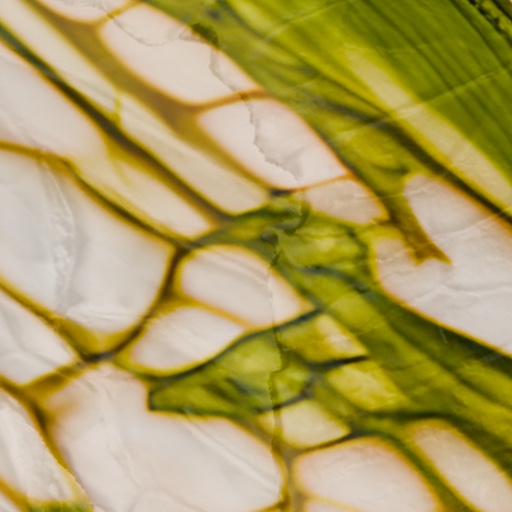} &
  \includegraphics[width=\shortsreswidth, valign=m]{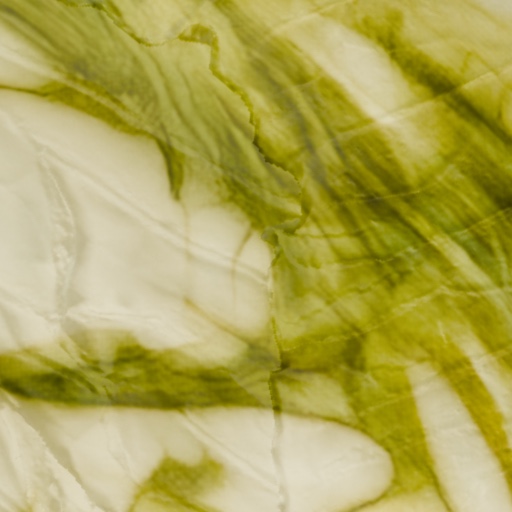} &
  \includegraphics[width=\shortsreswidth, valign=m]{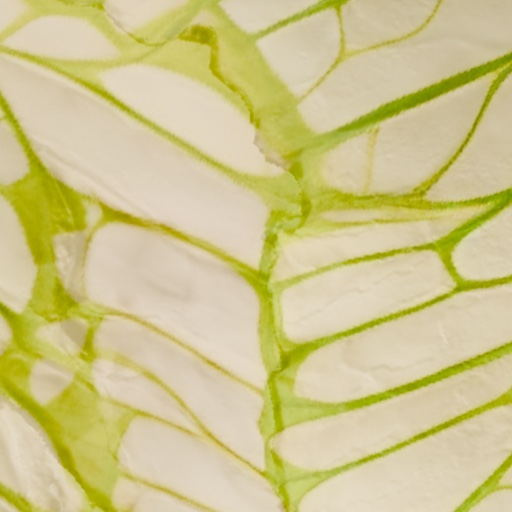}  \\
  \includegraphics[width=\shortsnormalswidth, valign=m]{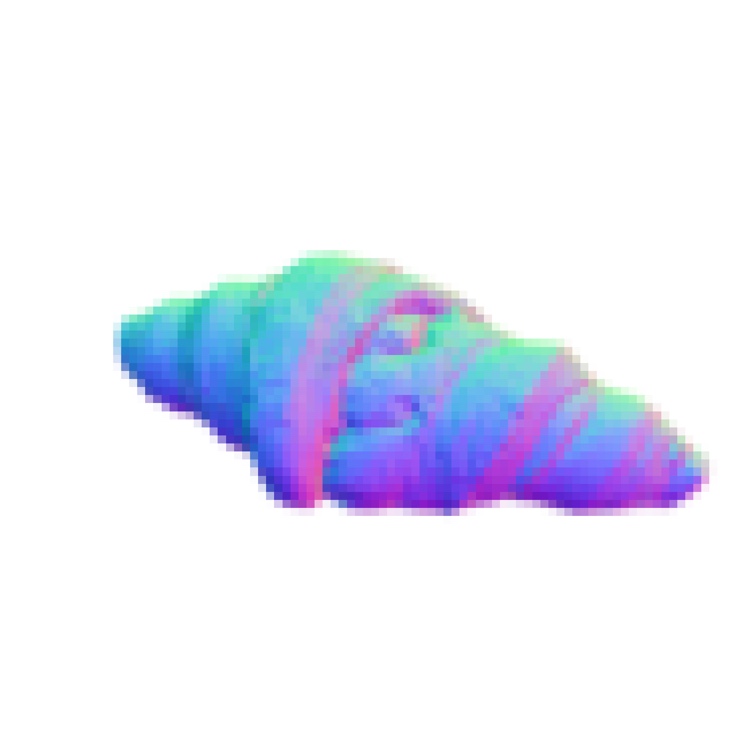} &
  \includegraphics[width=\shortsviewwidth, valign=m]{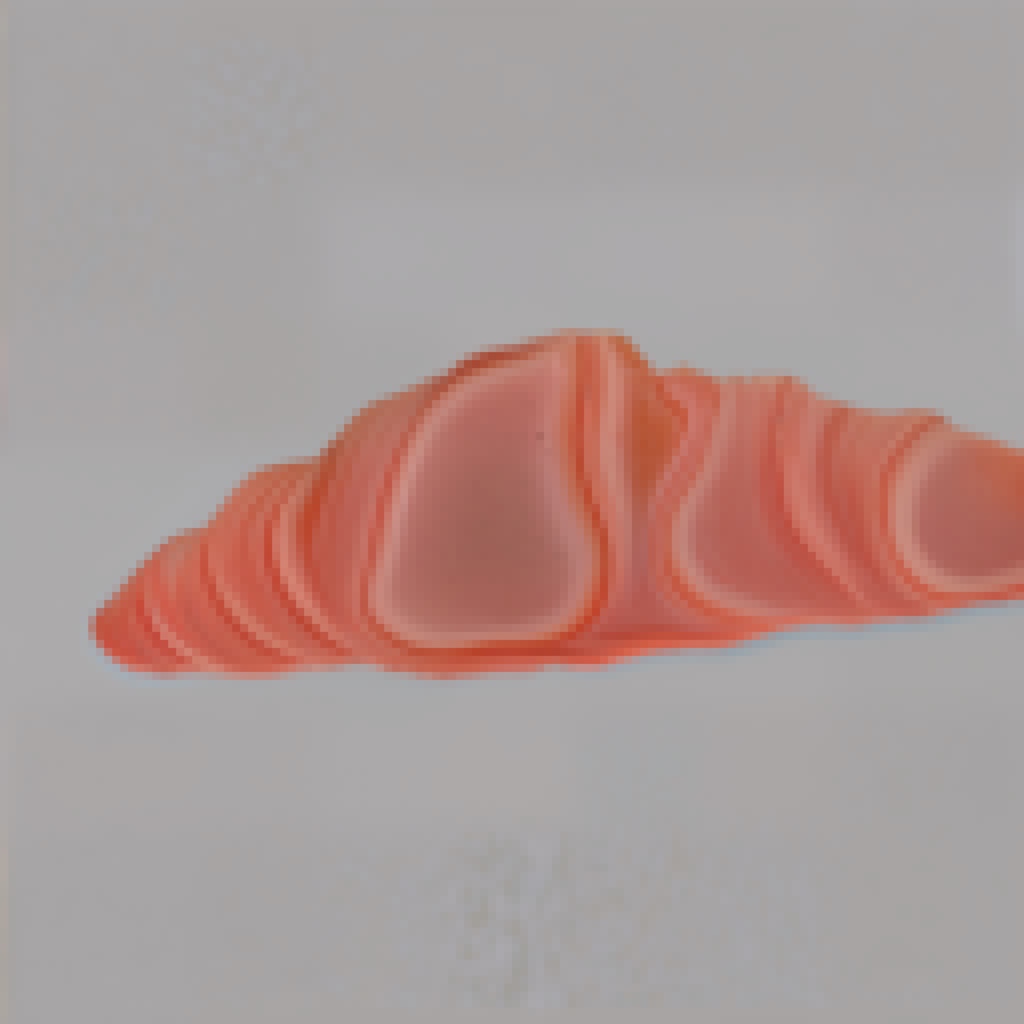} &
  \includegraphics[width=\shortsreswidth, valign=m]{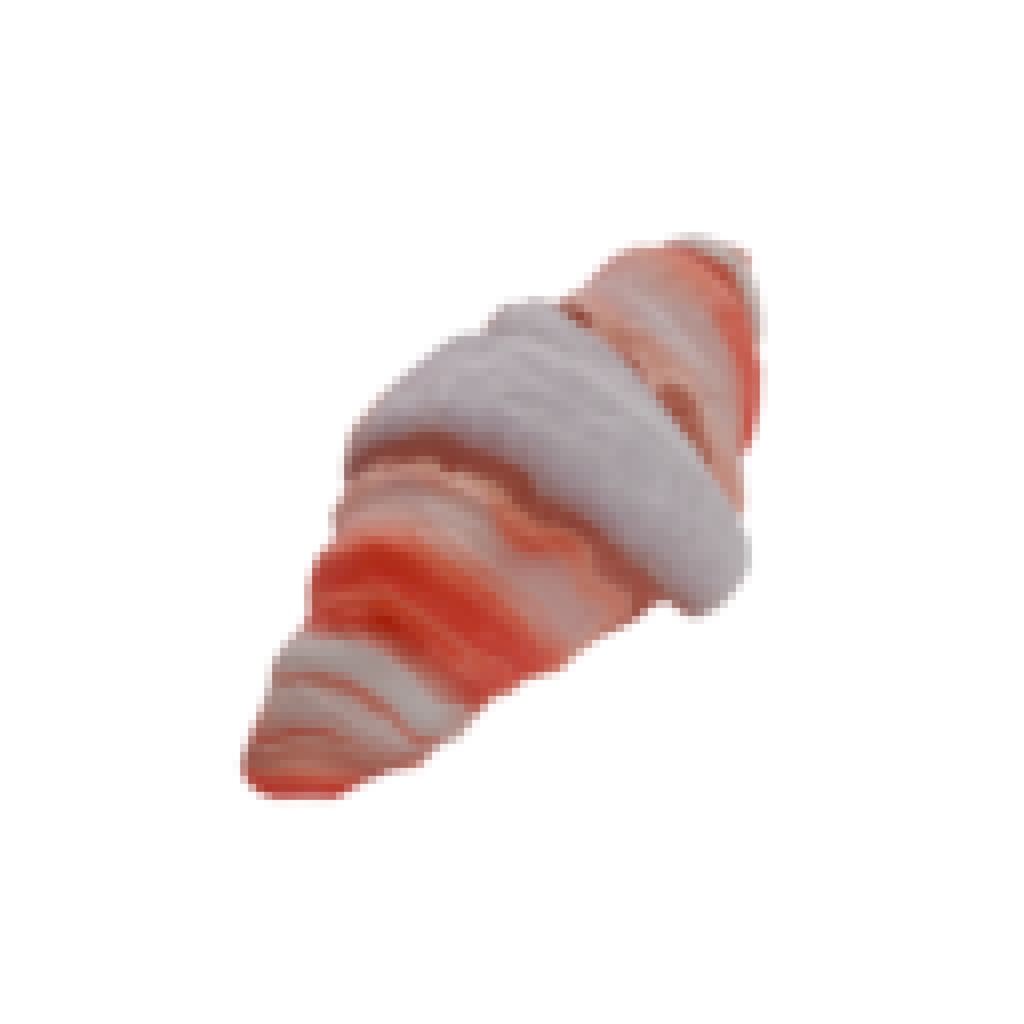} &
  \includegraphics[width=\shortsreswidth, valign=m]{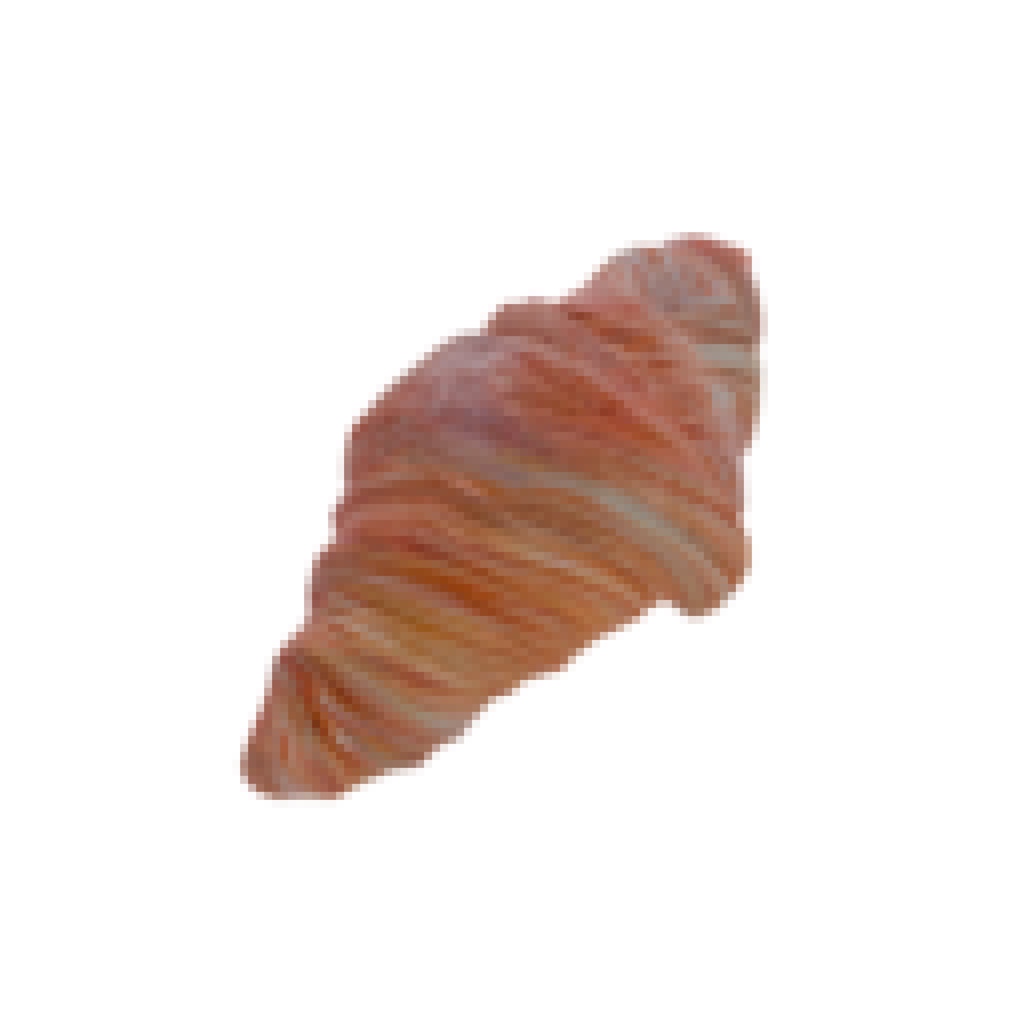} &
  \includegraphics[width=\shortsreswidth, valign=m]{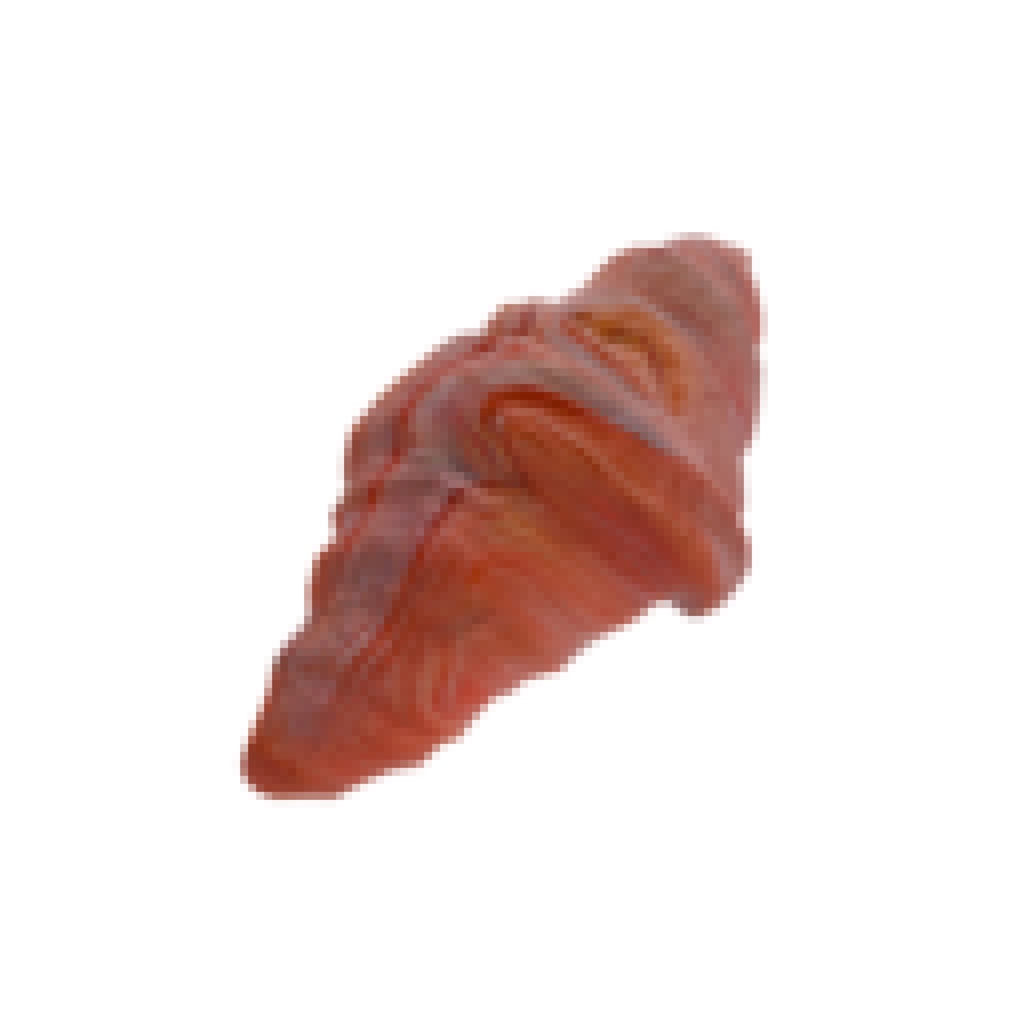} &
  \includegraphics[width=\shortsreswidth, valign=m]{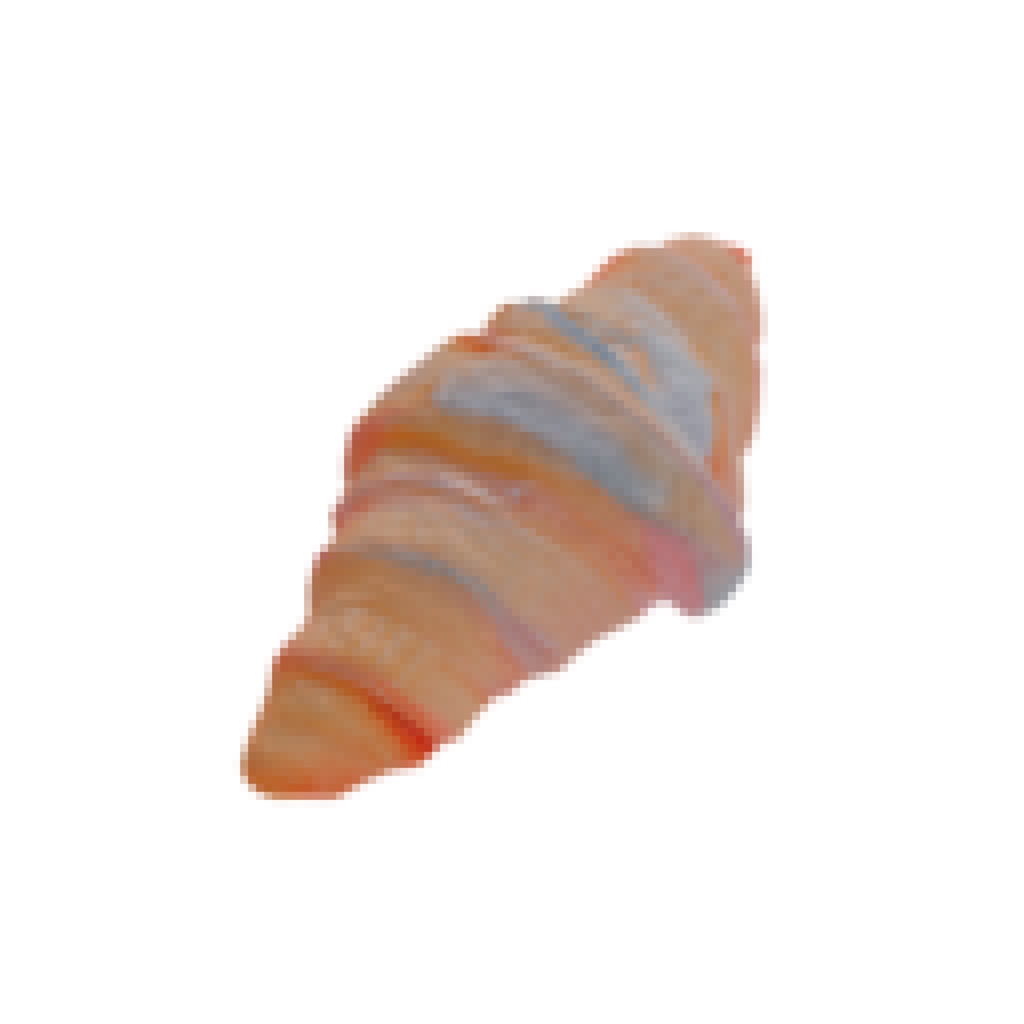} &
  \includegraphics[width=\shortsreswidth, valign=m]{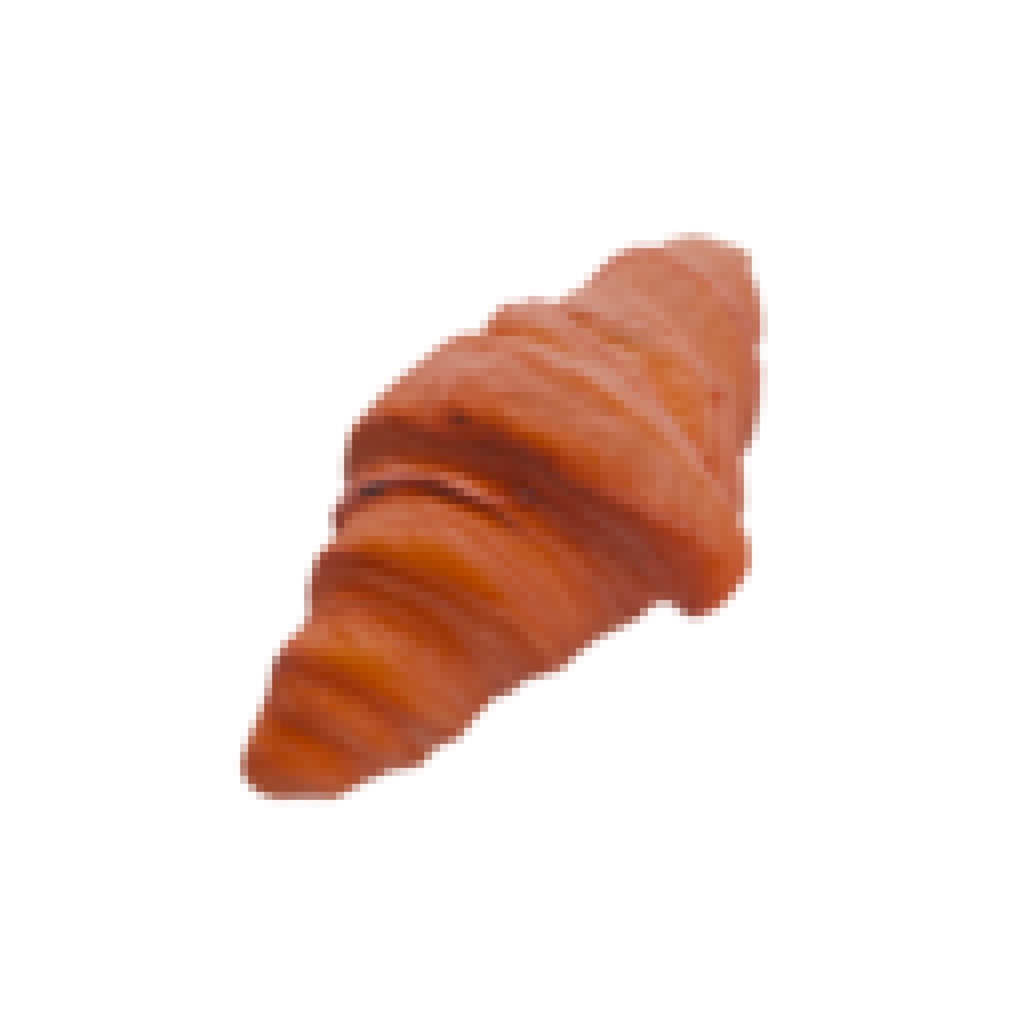} &
  \includegraphics[width=\shortsreswidth, valign=m]{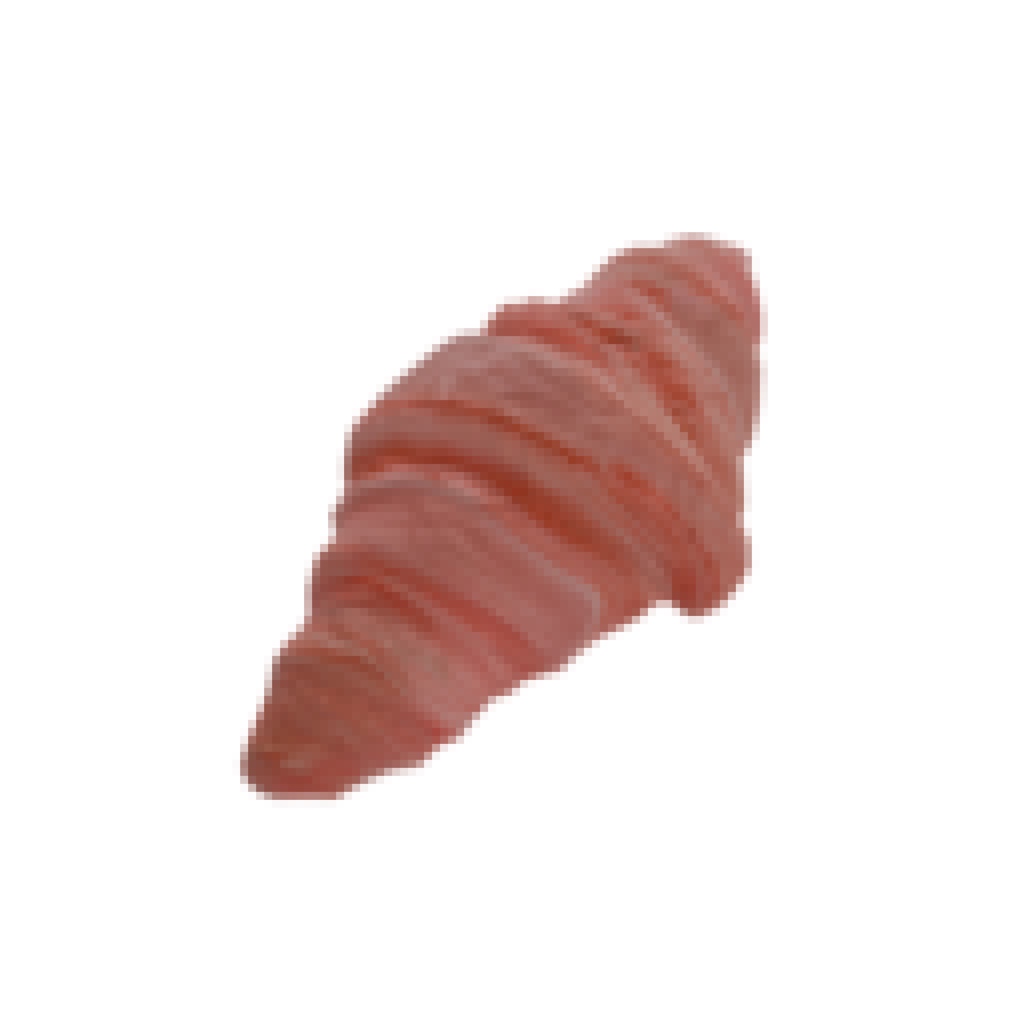}  \\
  \includegraphics[width=\shortsnormalswidth, valign=m]{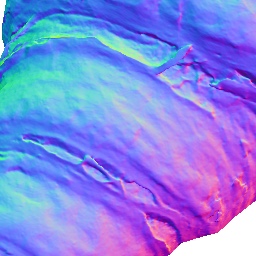} &
  \includegraphics[width=\shortsviewwidth, valign=m]{img/close-ups/_basecolor/white.jpg} &
  \includegraphics[width=\shortsreswidth, valign=m]{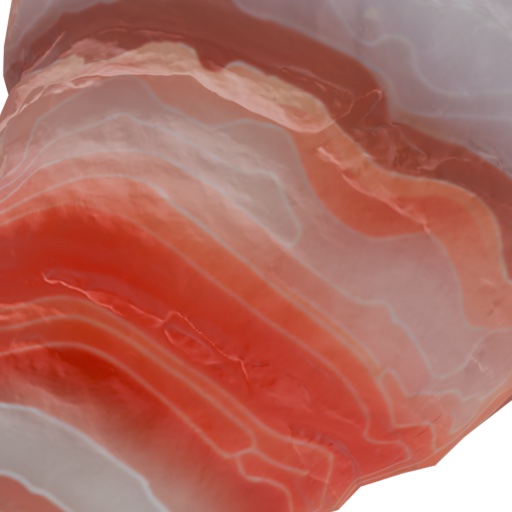} &
  \includegraphics[width=\shortsreswidth, valign=m]{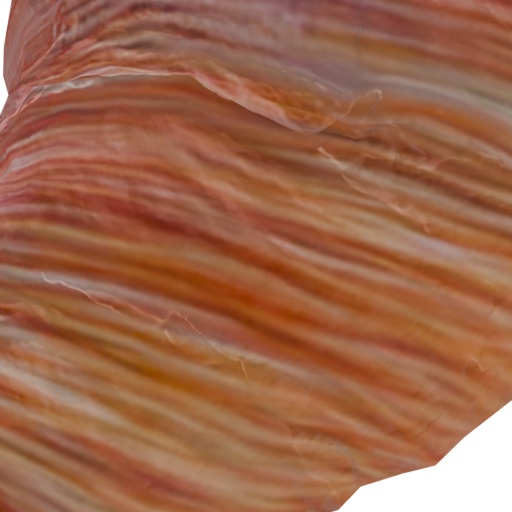} &
  \includegraphics[width=\shortsreswidth, valign=m]{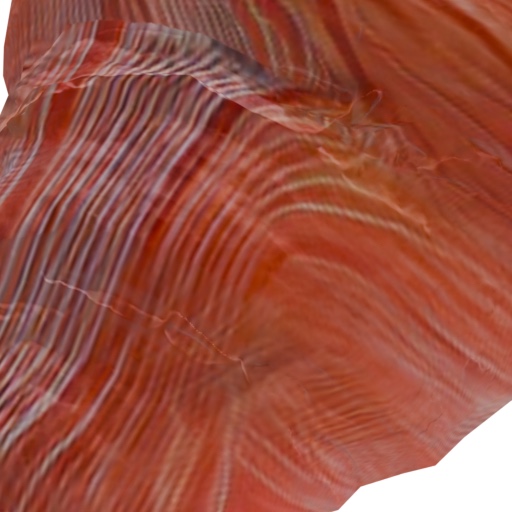} &
  \includegraphics[width=\shortsreswidth, valign=m]{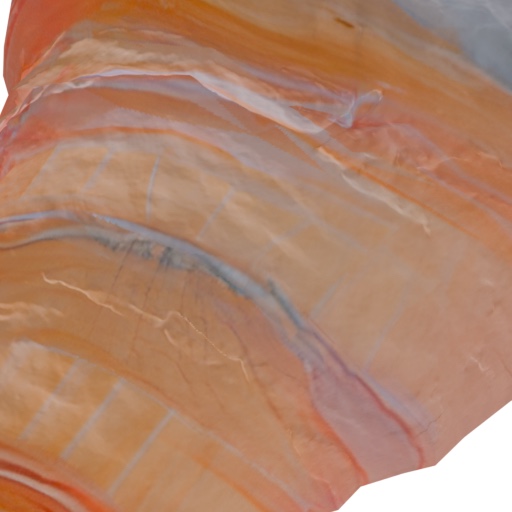} &
  \includegraphics[width=\shortsreswidth, valign=m]{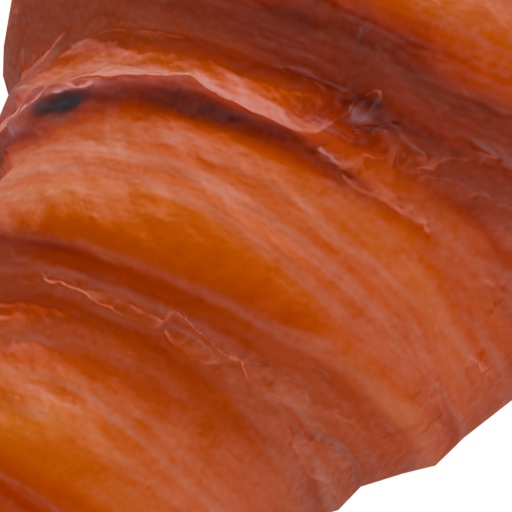} &
  \includegraphics[width=\shortsreswidth, valign=m]{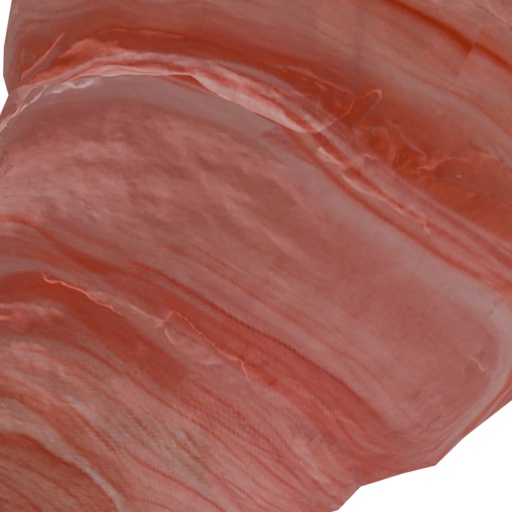}  \\
  \includegraphics[width=\shortsnormalswidth, valign=m]{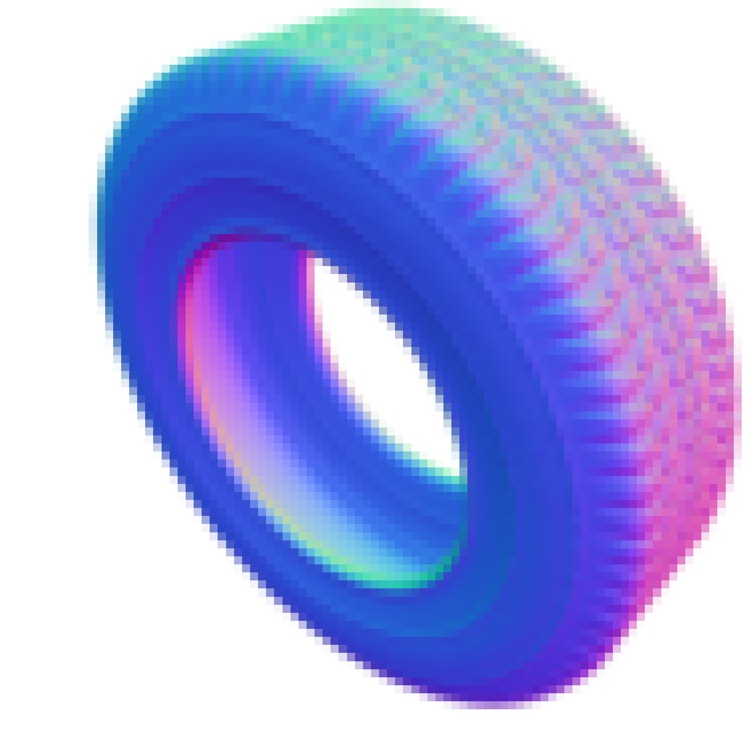} &
  \includegraphics[width=\shortsviewwidth, valign=m]{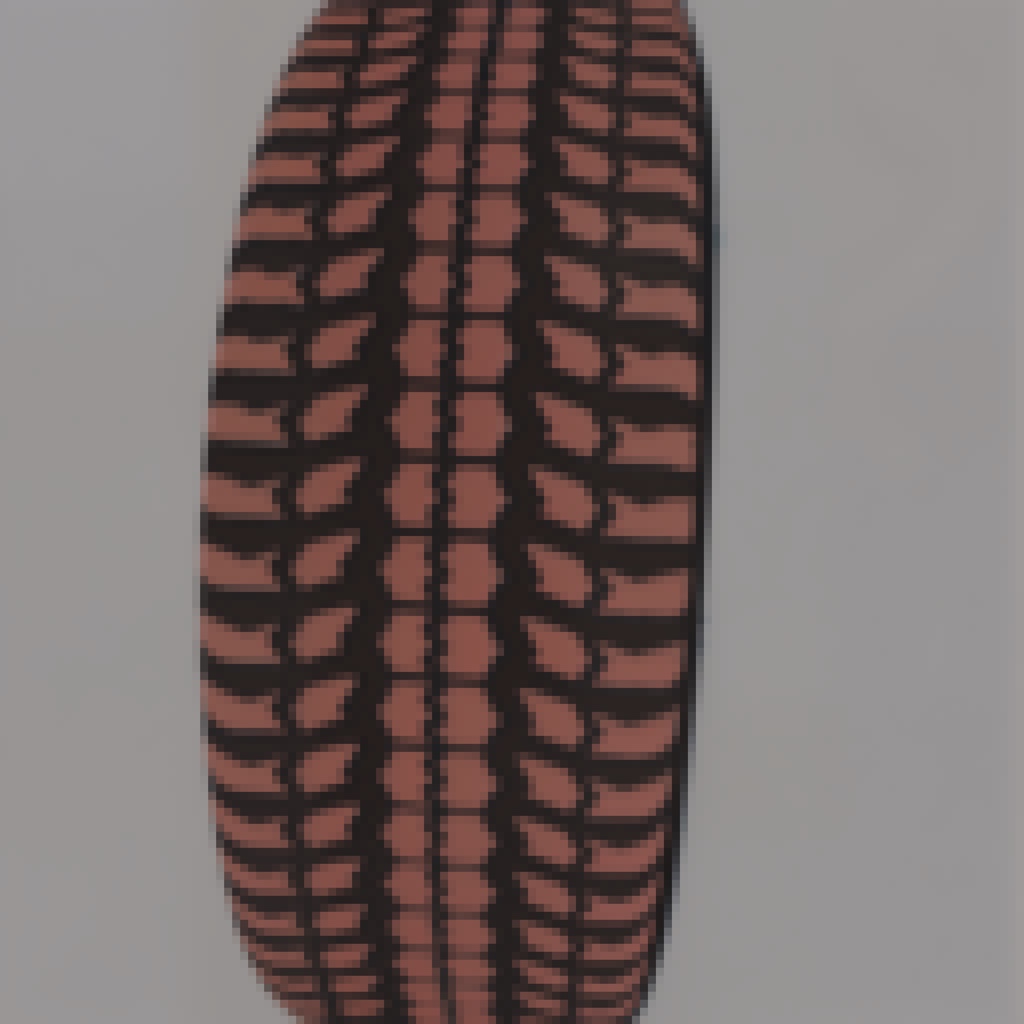} &
  \includegraphics[width=\shortsreswidth, valign=m]{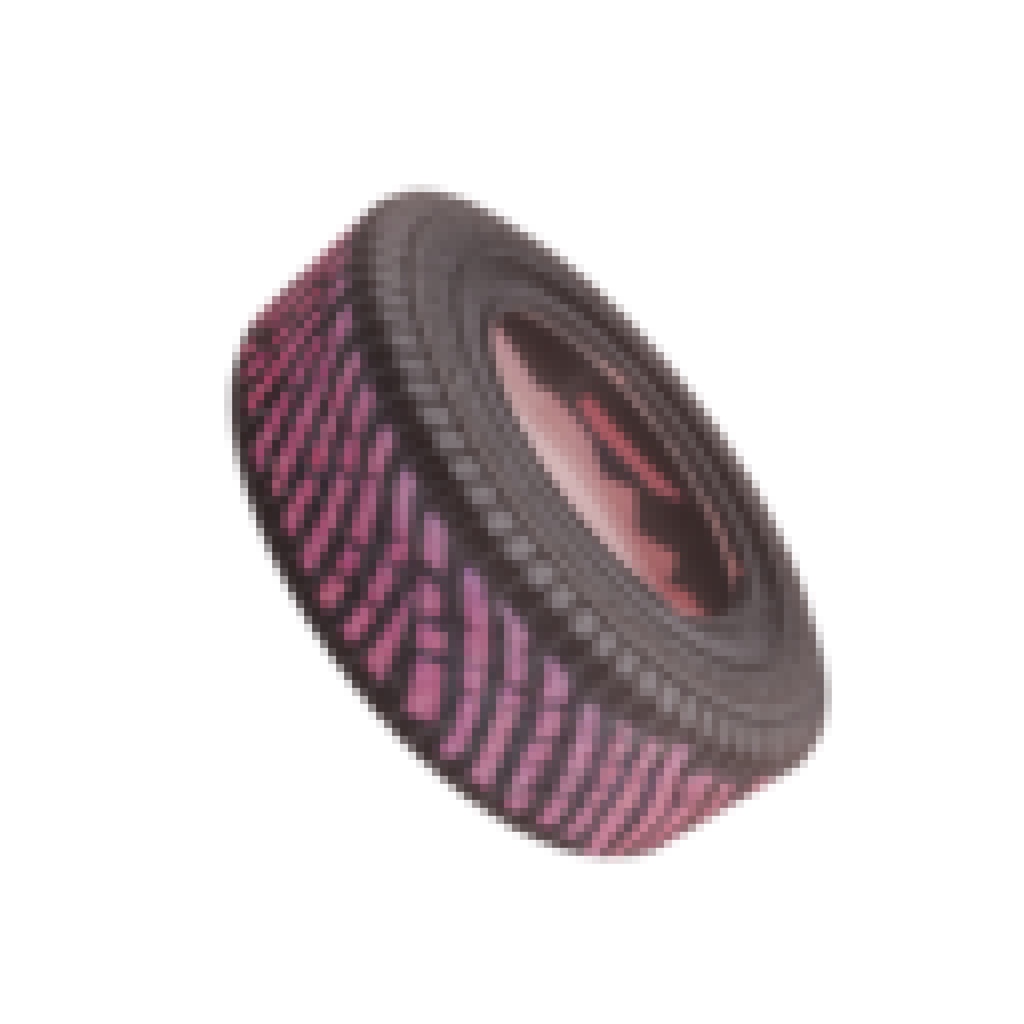} &
  \includegraphics[width=\shortsreswidth, valign=m]{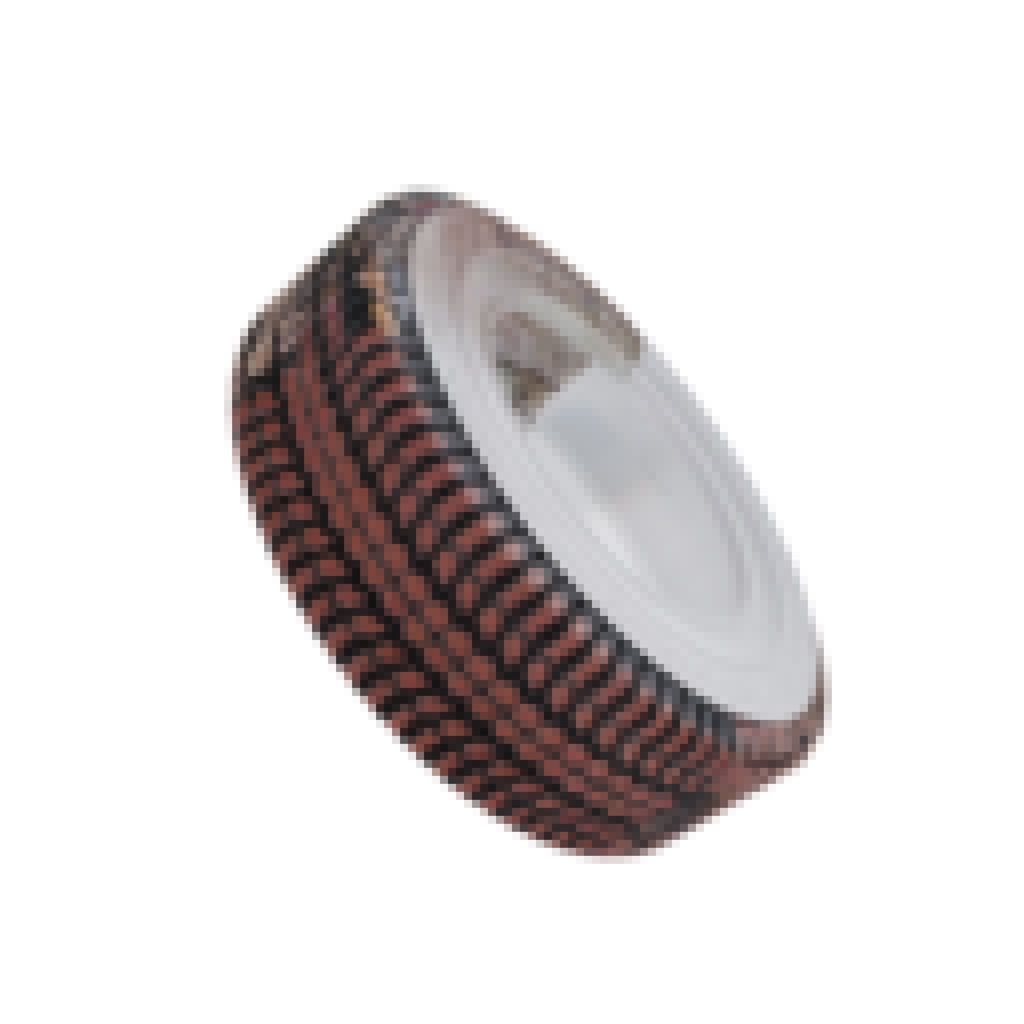} &
  \includegraphics[width=\shortsreswidth, valign=m]{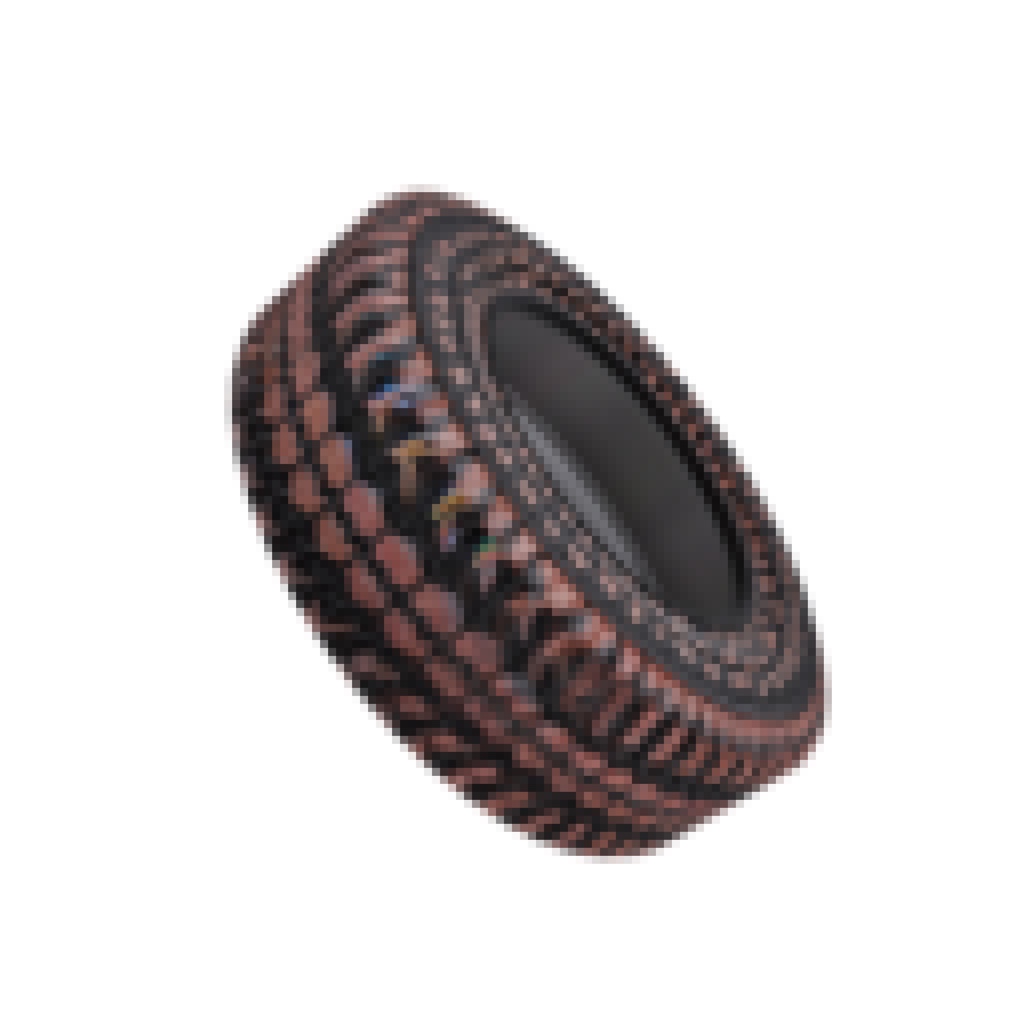} &
  \includegraphics[width=\shortsreswidth, valign=m]{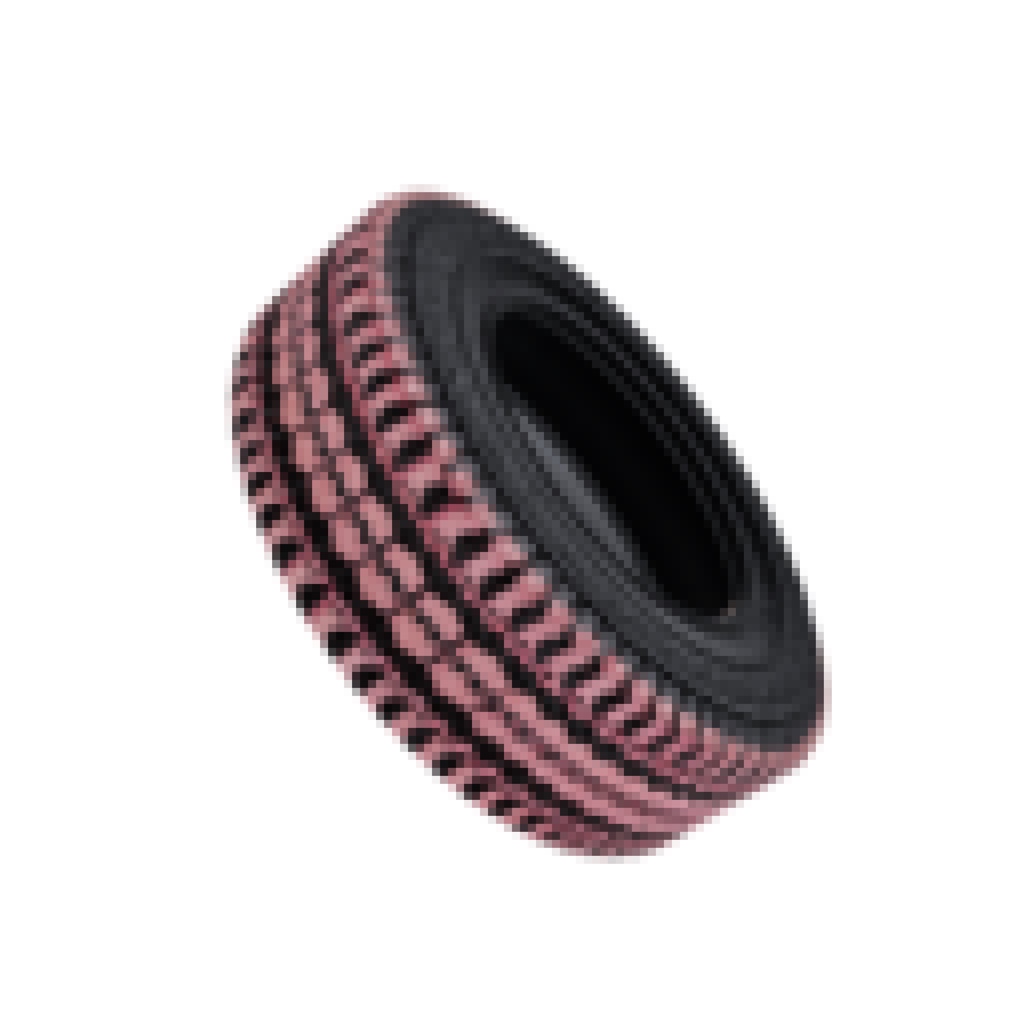} &
  \includegraphics[width=\shortsreswidth, valign=m]{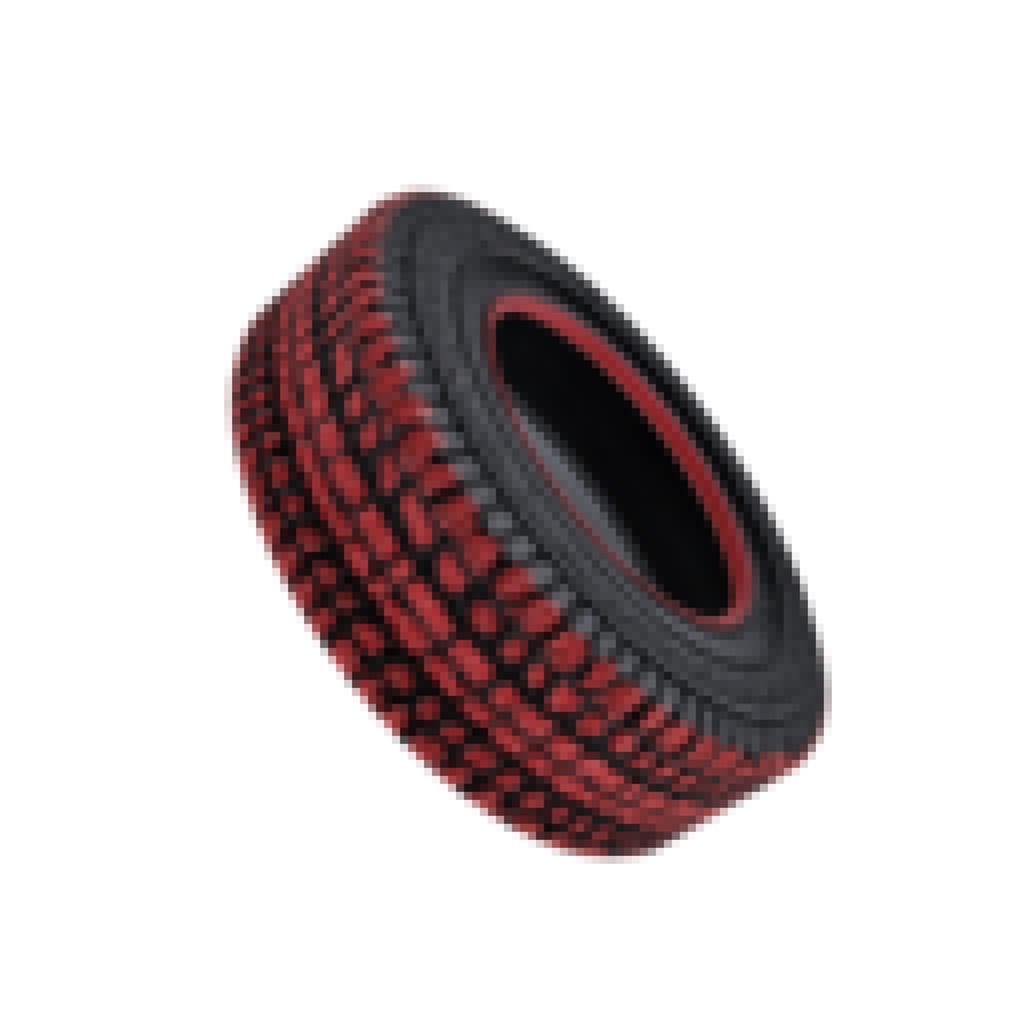} &
  \includegraphics[width=\shortsreswidth, valign=m]{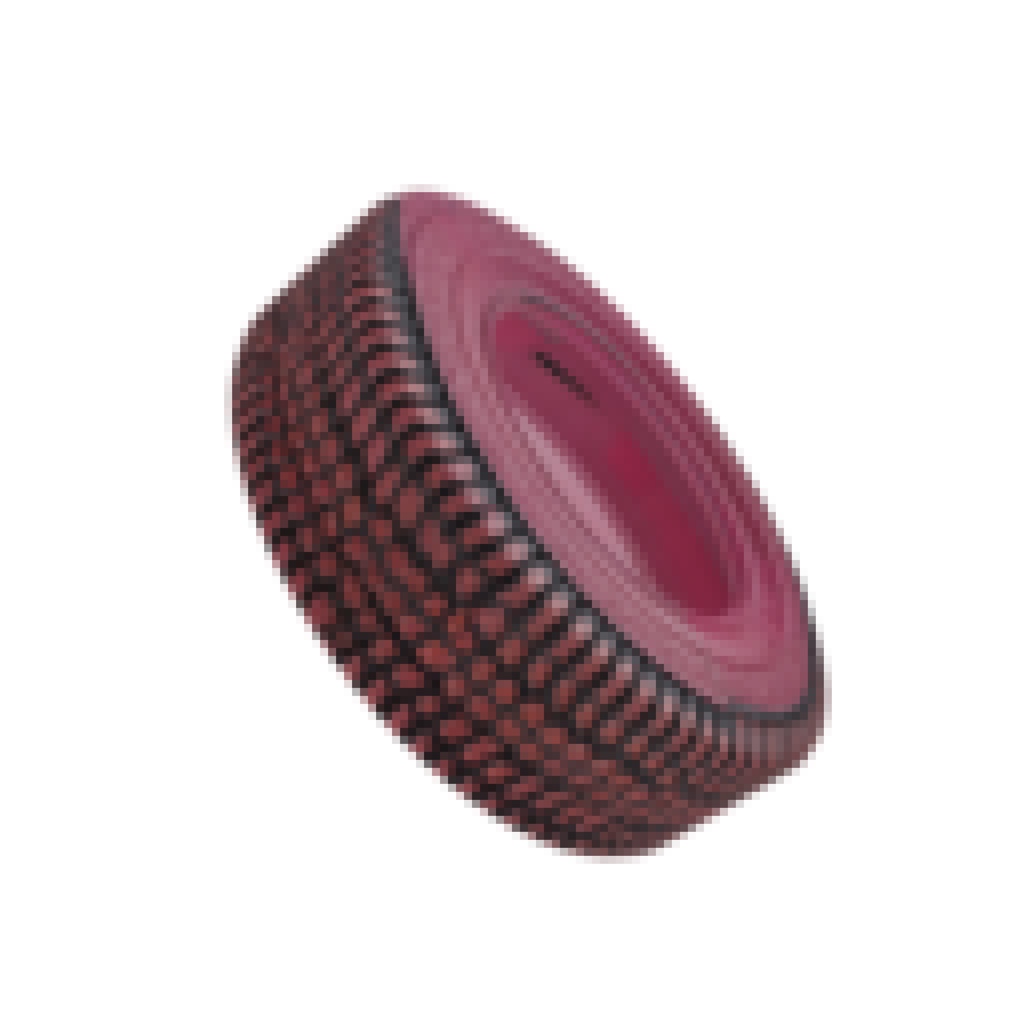}  \\
  \includegraphics[width=\shortsnormalswidth, valign=m]{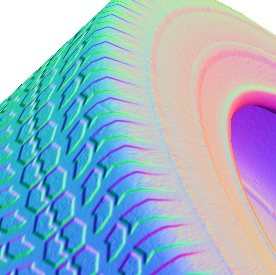} &
  \includegraphics[width=\shortsviewwidth, valign=m]{img/close-ups/_basecolor/white.jpg} &
  \includegraphics[width=\shortsreswidth, valign=m]{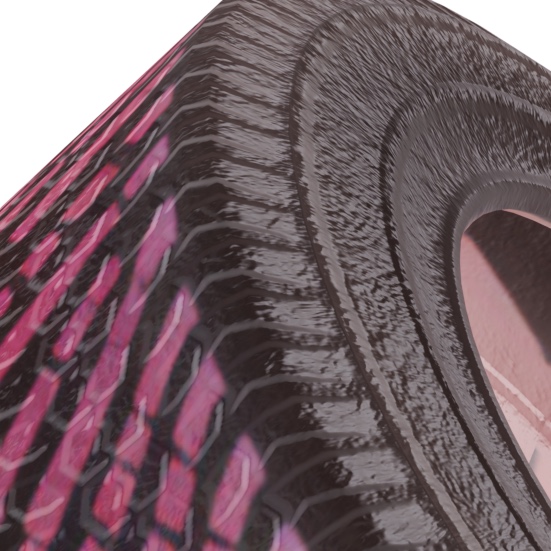} &
  \includegraphics[width=\shortsreswidth, valign=m]{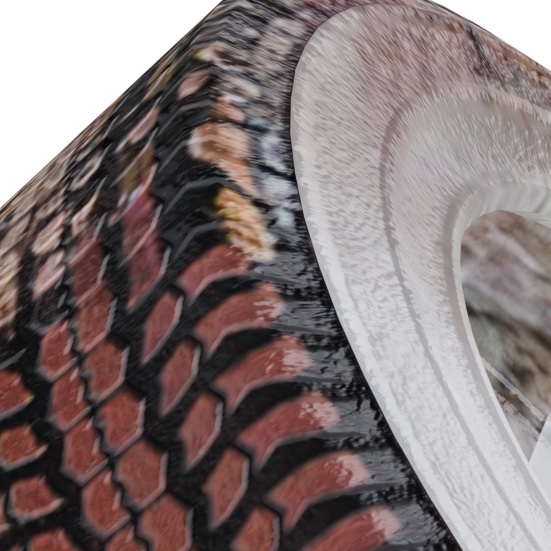} &
  \includegraphics[width=\shortsreswidth, valign=m]{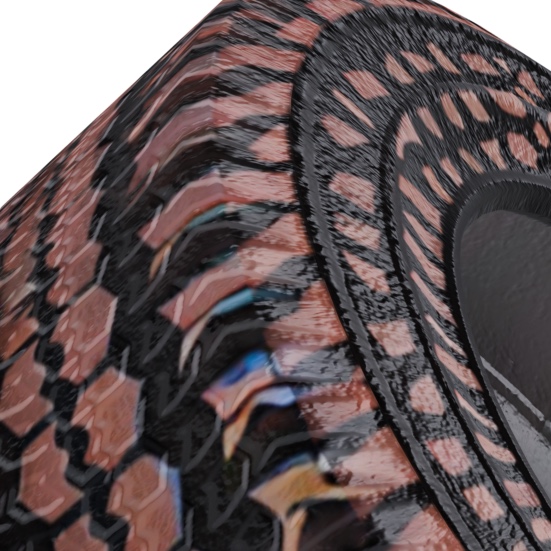} &
  \includegraphics[width=\shortsreswidth, valign=m]{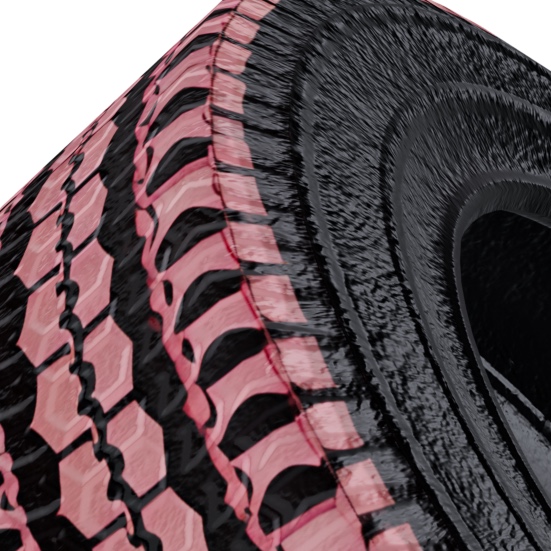} &
  \includegraphics[width=\shortsreswidth, valign=m]{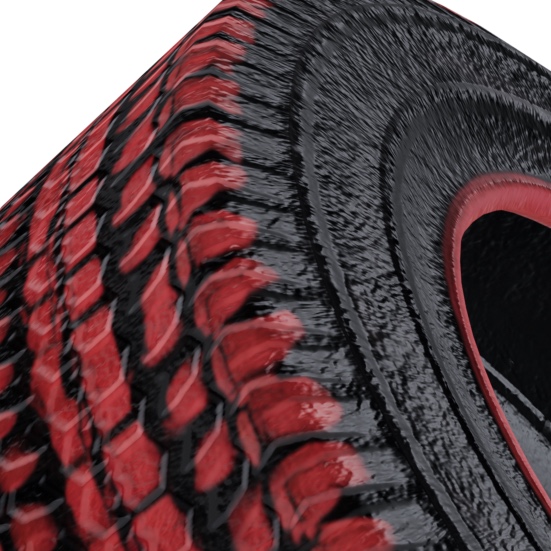} &
  \includegraphics[width=\shortsreswidth, valign=m]{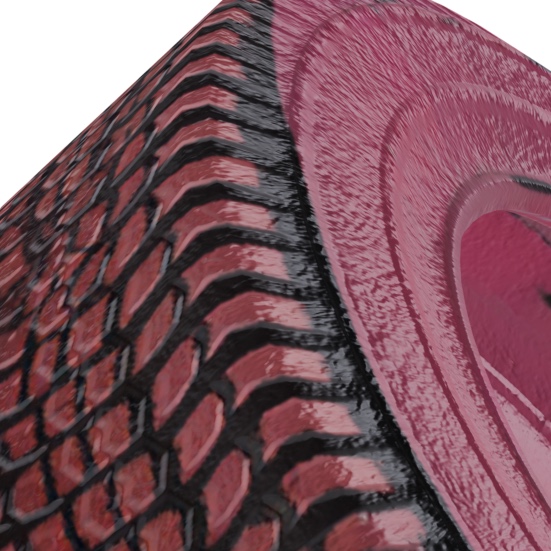}  \\
  \includegraphics[width=\shortsnormalswidth, valign=m]{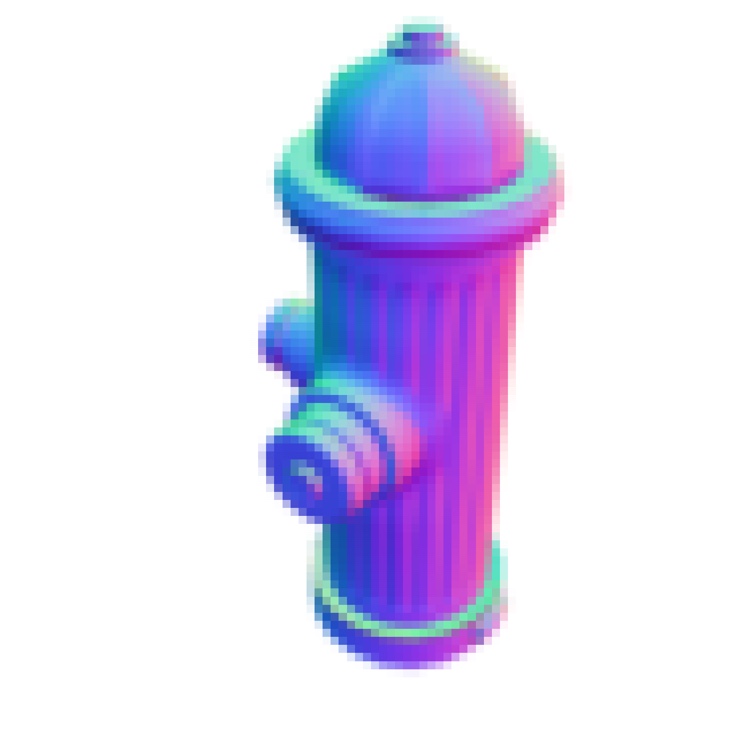} &
  \includegraphics[width=\shortsviewwidth, valign=m]{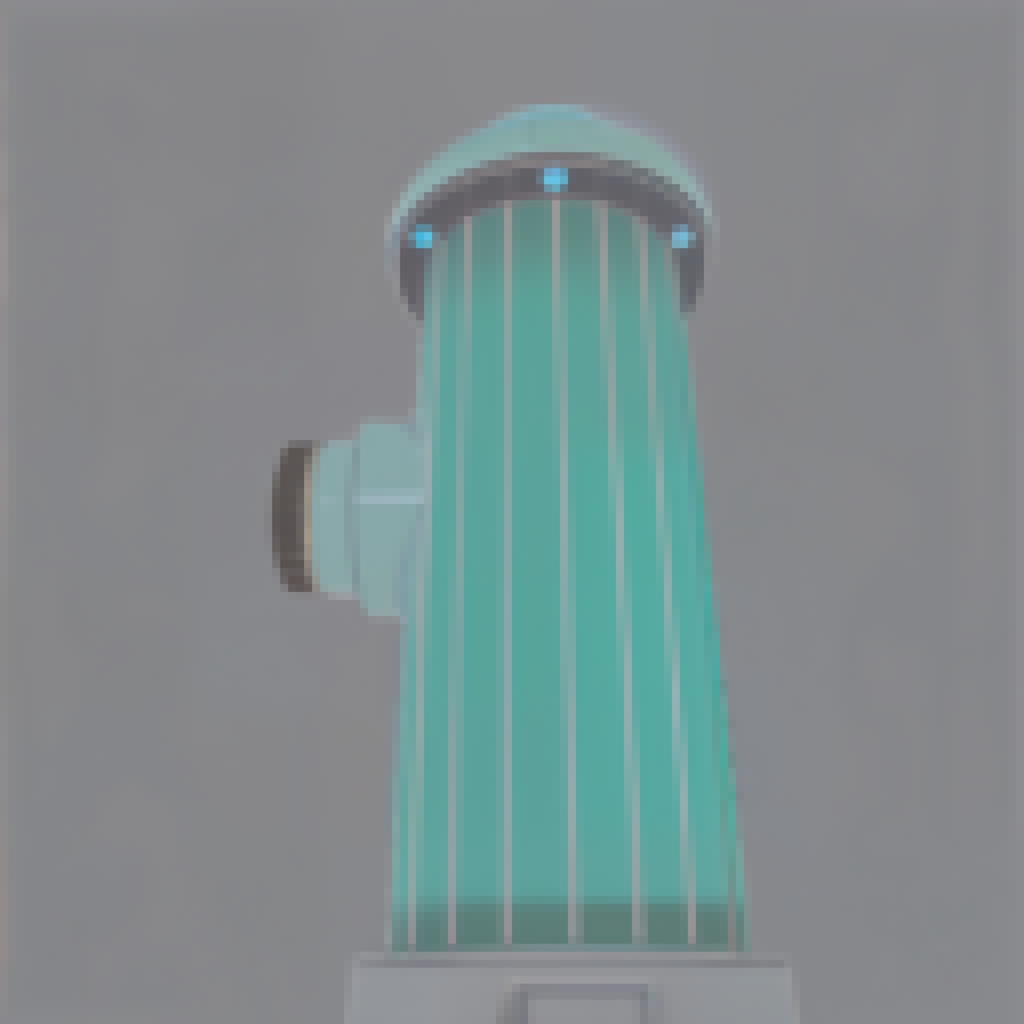} &
  \includegraphics[width=\shortsreswidth, valign=m]{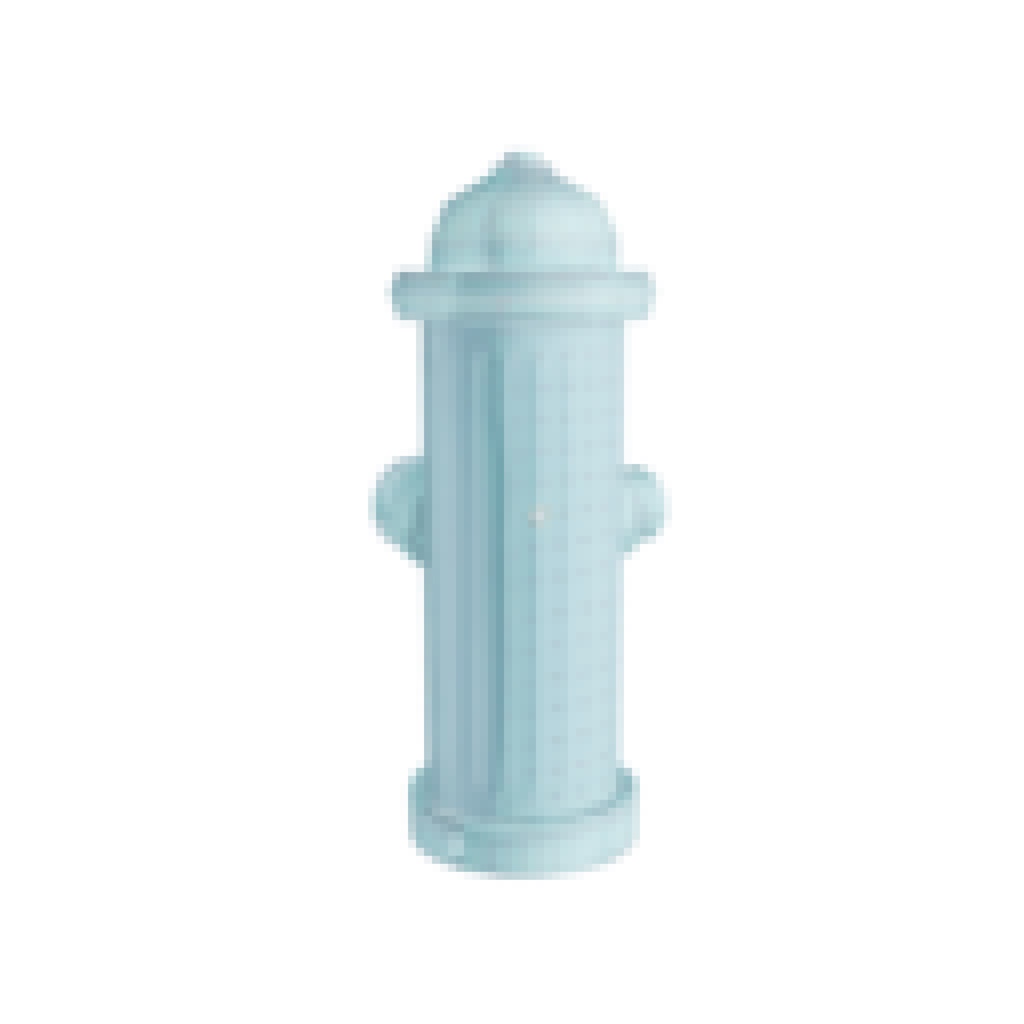} &
  \includegraphics[width=\shortsreswidth, valign=m]{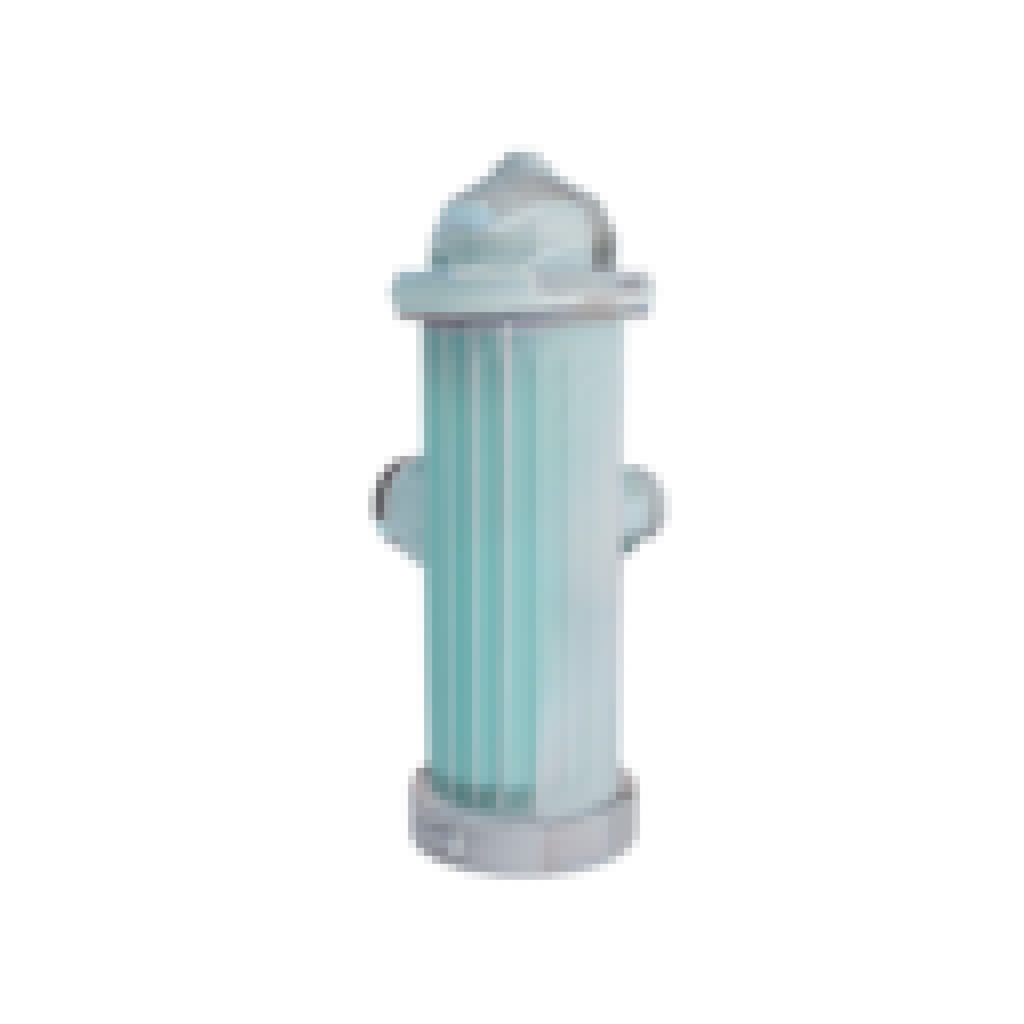} &
  \includegraphics[width=\shortsreswidth, valign=m]{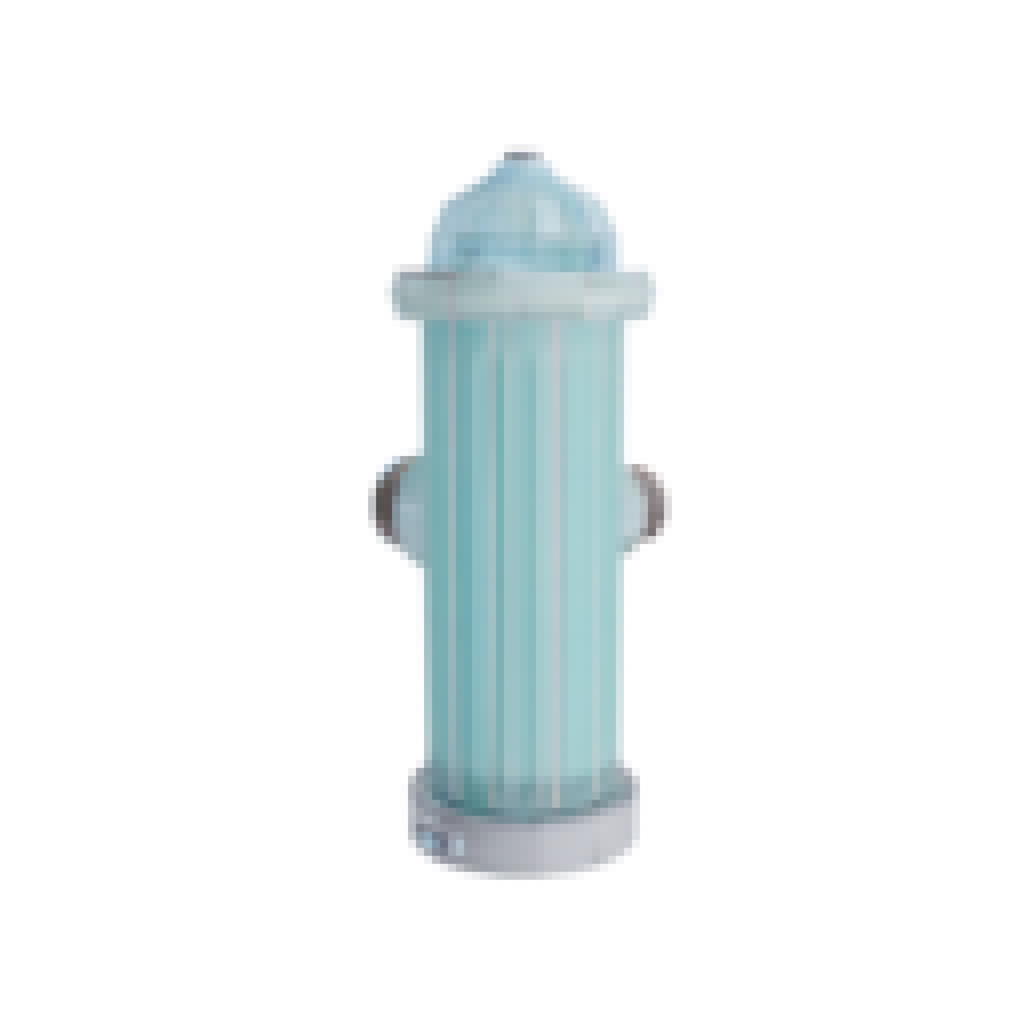} &
  \includegraphics[width=\shortsreswidth, valign=m]{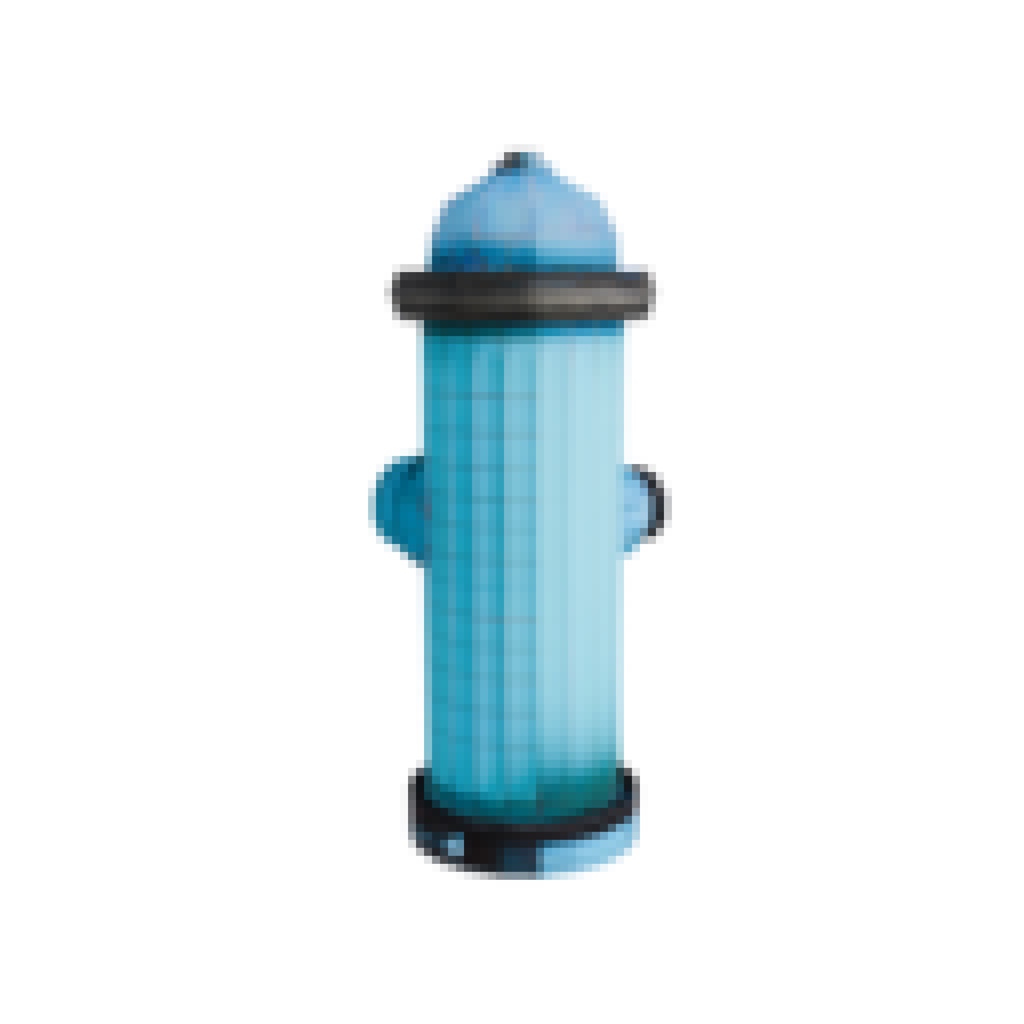} &
  \includegraphics[width=\shortsreswidth, valign=m]{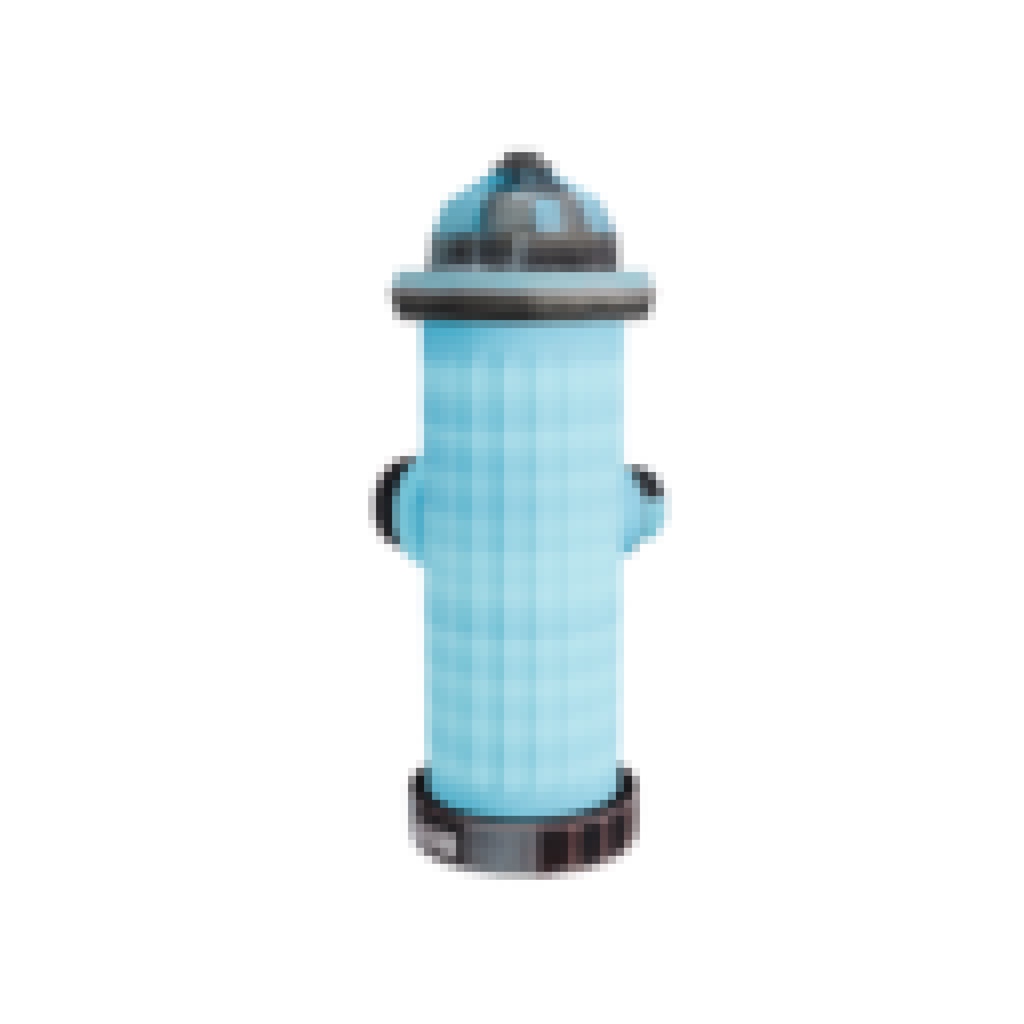} &
  \includegraphics[width=\shortsreswidth, valign=m]{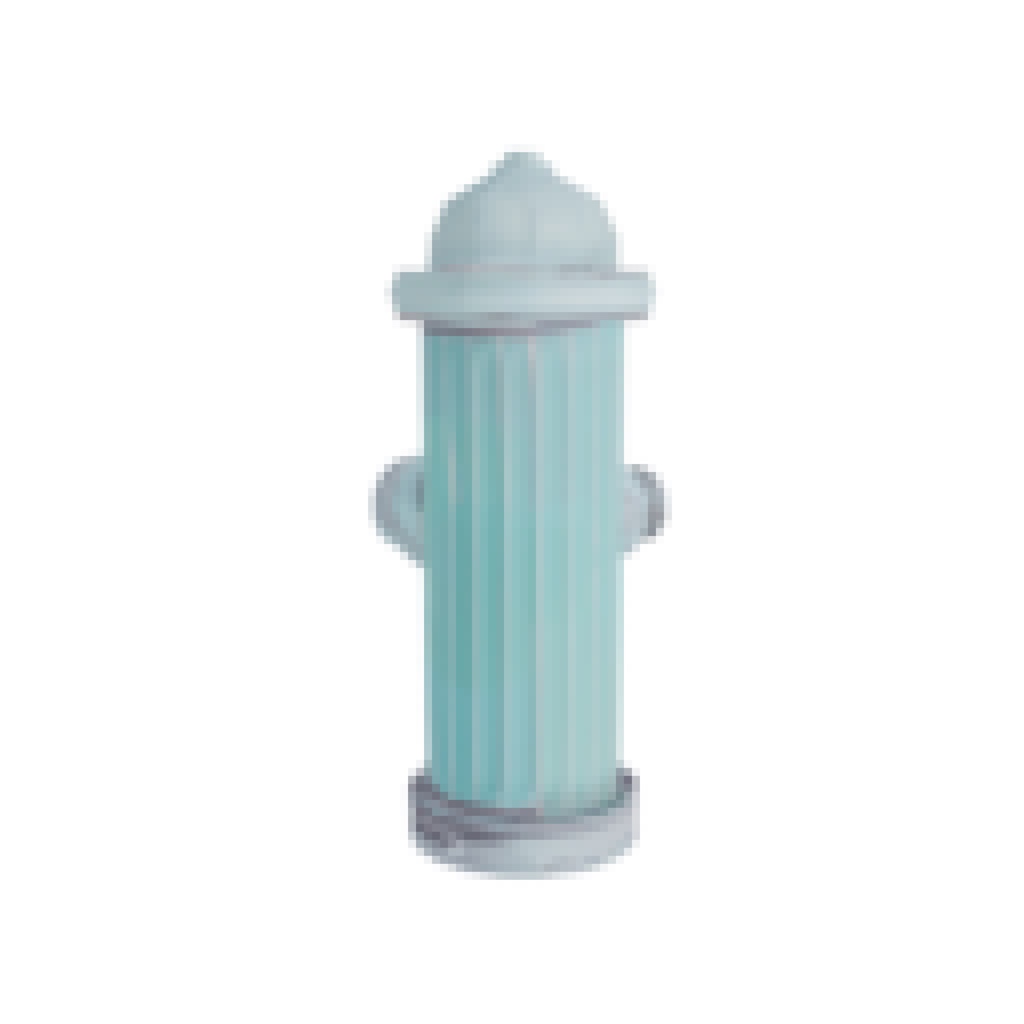}  \\
  \includegraphics[width=\shortsnormalswidth, valign=m]{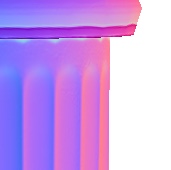} &
  \includegraphics[width=\shortsviewwidth, valign=m]{img/close-ups/_basecolor/white.jpg} &
  \includegraphics[width=\shortsreswidth, valign=m]{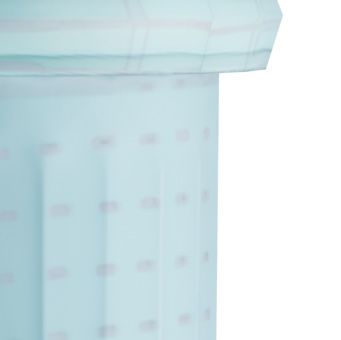} &
  \includegraphics[width=\shortsreswidth, valign=m]{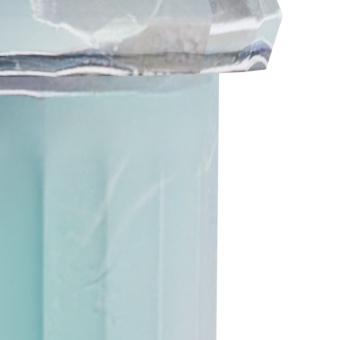} &
  \includegraphics[width=\shortsreswidth, valign=m]{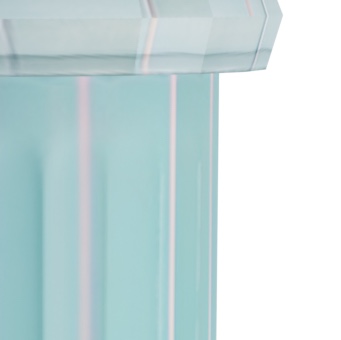} &
  \includegraphics[width=\shortsreswidth, valign=m]{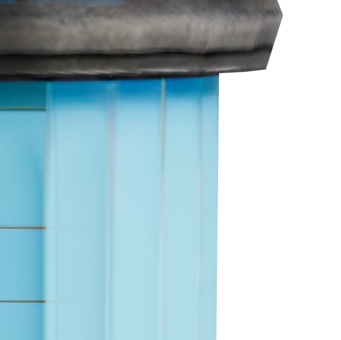} &
  \includegraphics[width=\shortsreswidth, valign=m]{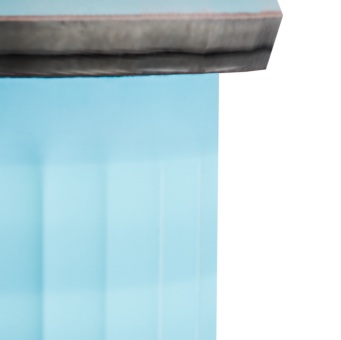} &
  \includegraphics[width=\shortsreswidth, valign=m]{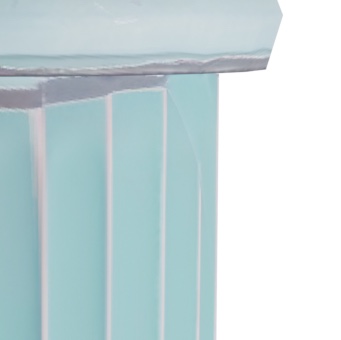}  \\

  \end{tabular}
  \caption{\footnotesize{Additional results on the baseline comparison, both global view and zoomed-in views.}}
  \label{fig:supp_comparison_gallery_1}
\end{figure*}

\endgroup
\providecommand{\shortsnormalswidth}{0.1\linewidth}
\providecommand{\shortsviewwidth}{0.1\linewidth}
\providecommand{\shortsreswidth}{0.1\linewidth}

\begingroup
\setlength{\tabcolsep}{2pt}
\renewcommand{\arraystretch}{0}

\begin{figure*}[h!tbp]
  \centering
  \begin{tabular}{cccccccc}
  \small Geometry & \small Single View & \small Paint 3D & \small TexGen & \small MV-Adapter & \small Hunyuan2.1 & \small Trellis & \small Ours  \\
  \includegraphics[width=\shortsnormalswidth, valign=m]{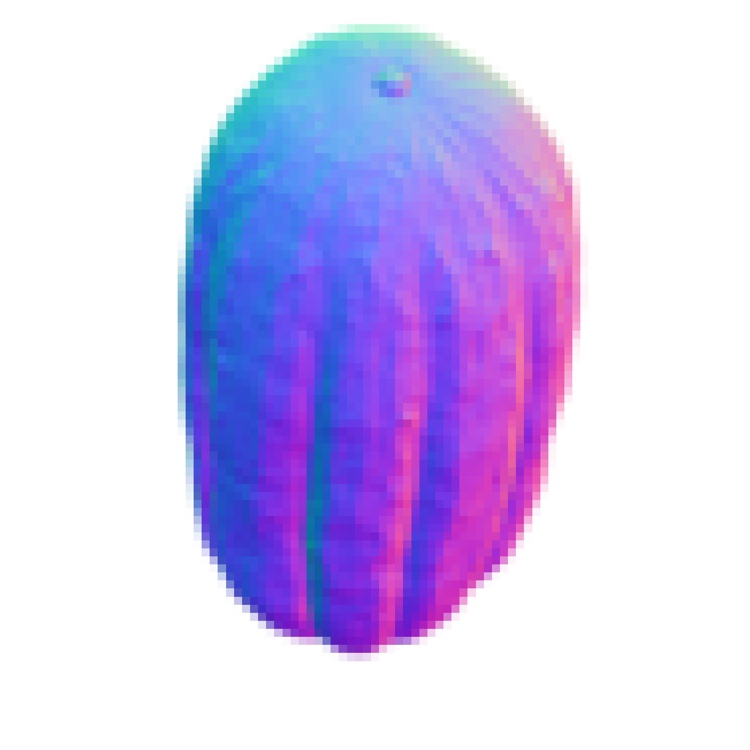} &
  \includegraphics[width=\shortsviewwidth, valign=m]{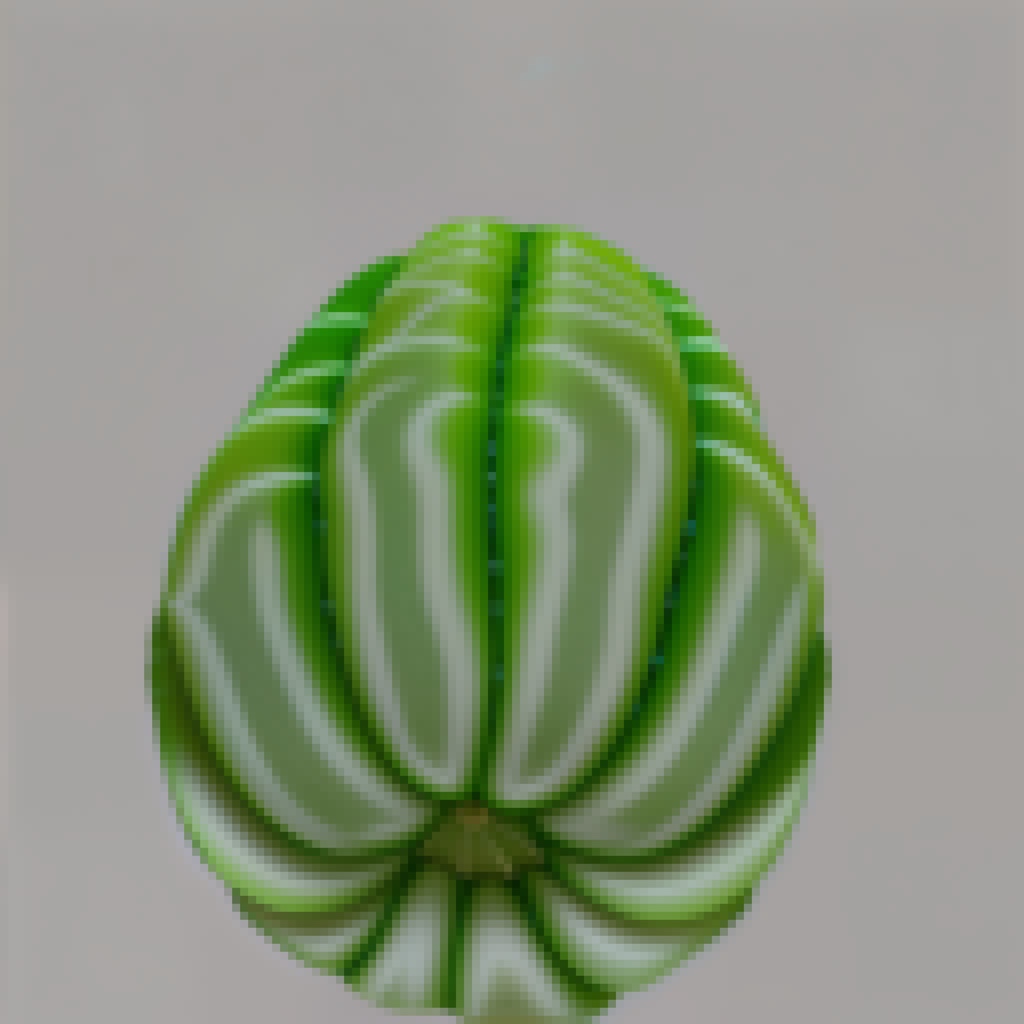} &
  \includegraphics[width=\shortsreswidth, valign=m]{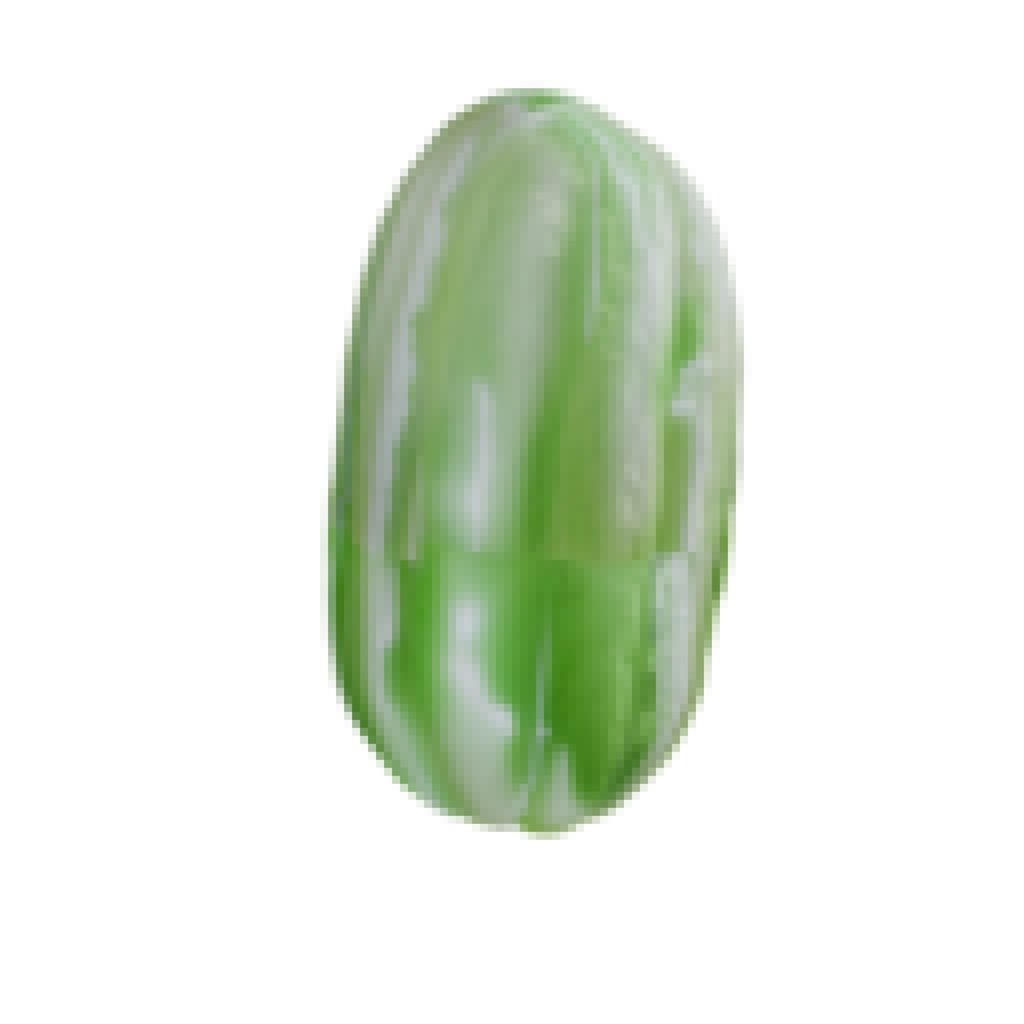} &
  \includegraphics[width=\shortsreswidth, valign=m]{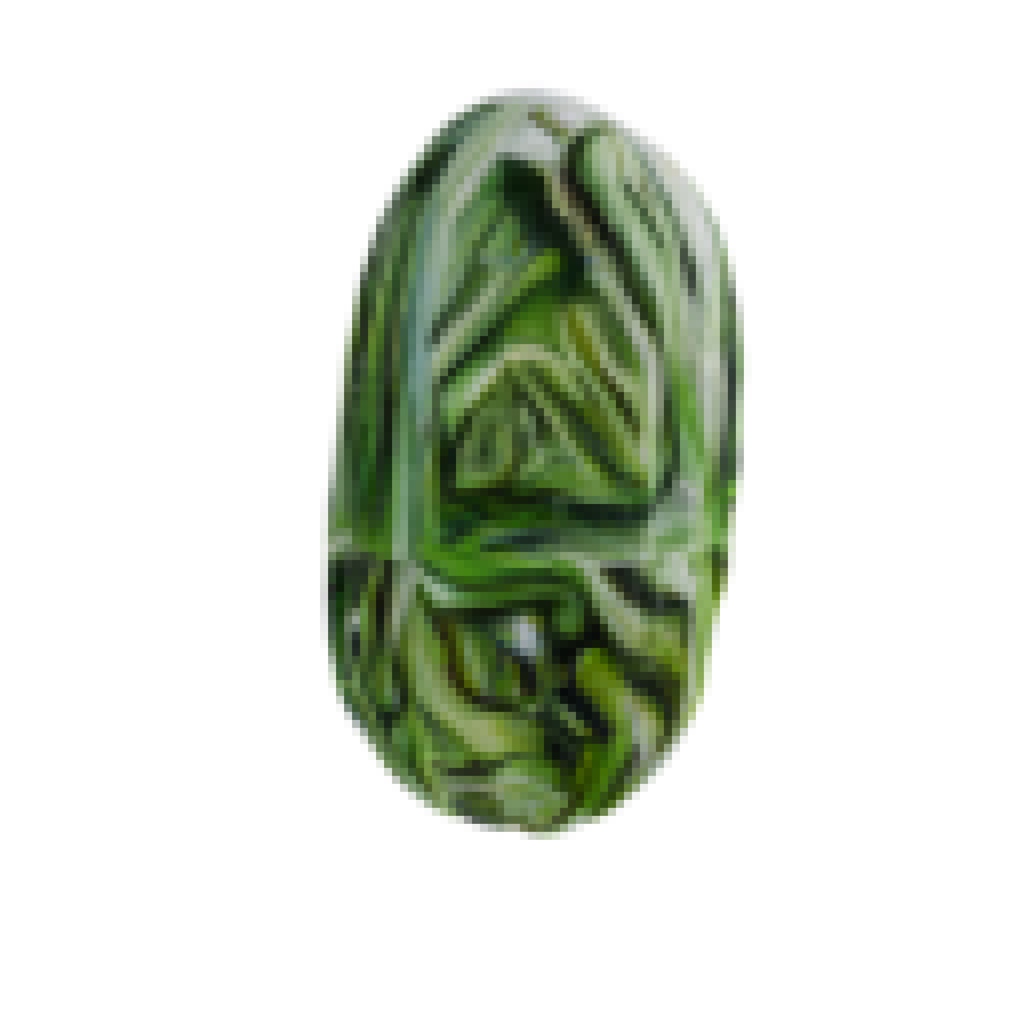} &
  \includegraphics[width=\shortsreswidth, valign=m]{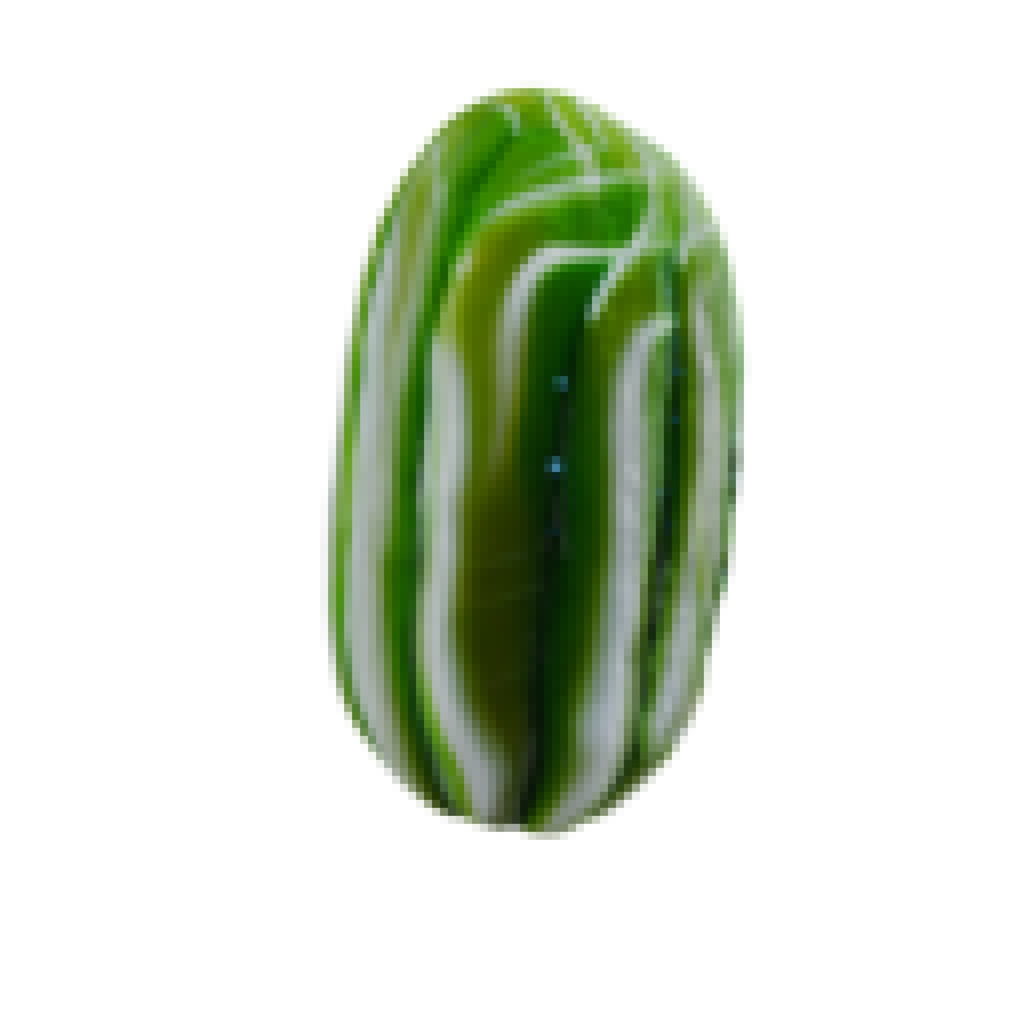} &
  \includegraphics[width=\shortsreswidth, valign=m]{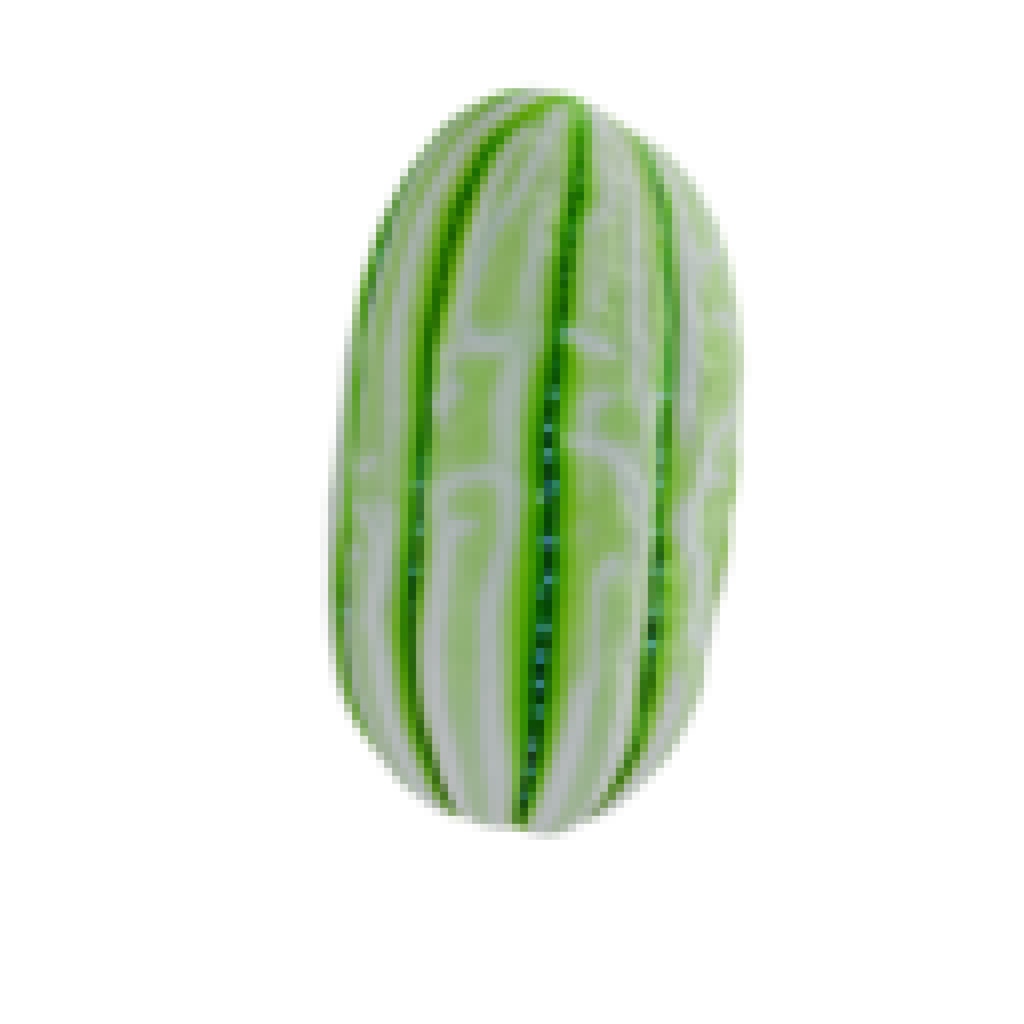} &
  \includegraphics[width=\shortsreswidth, valign=m]{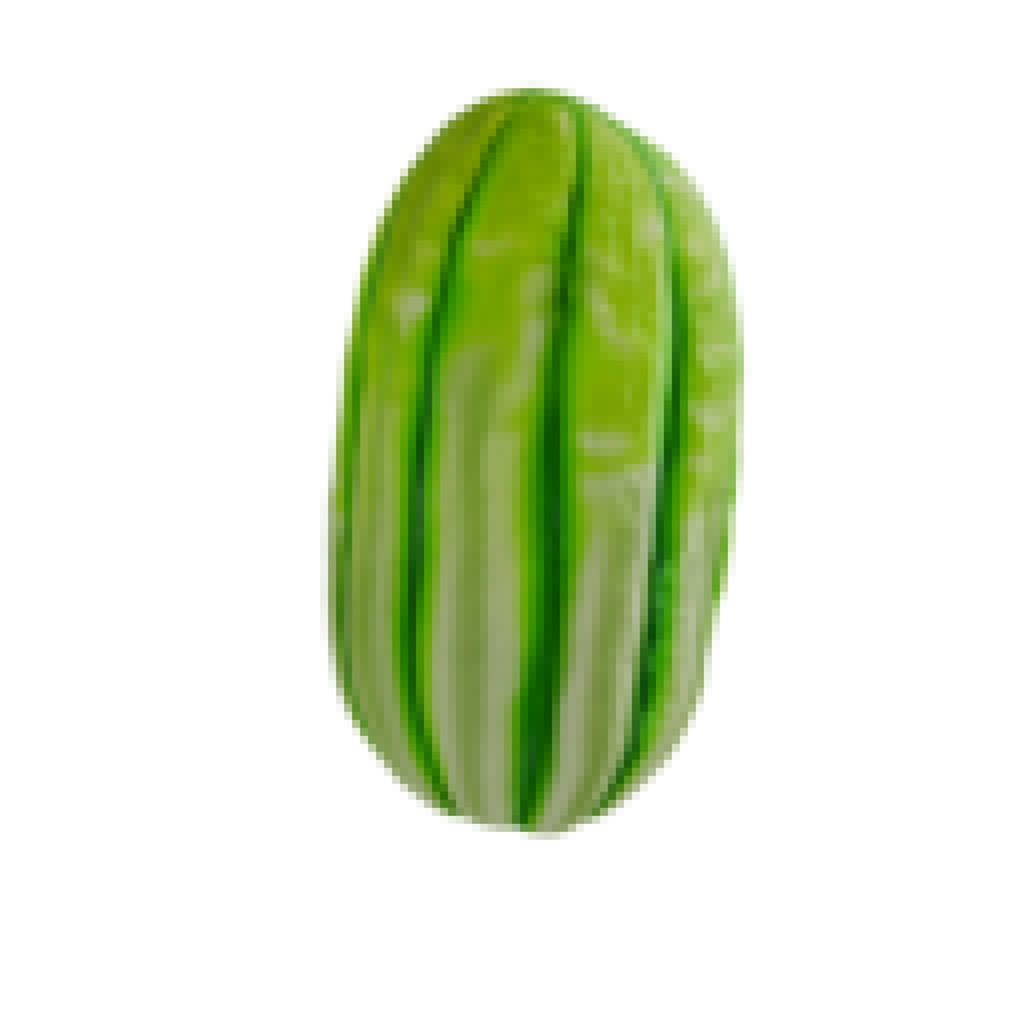} &
  \includegraphics[width=\shortsreswidth, valign=m]{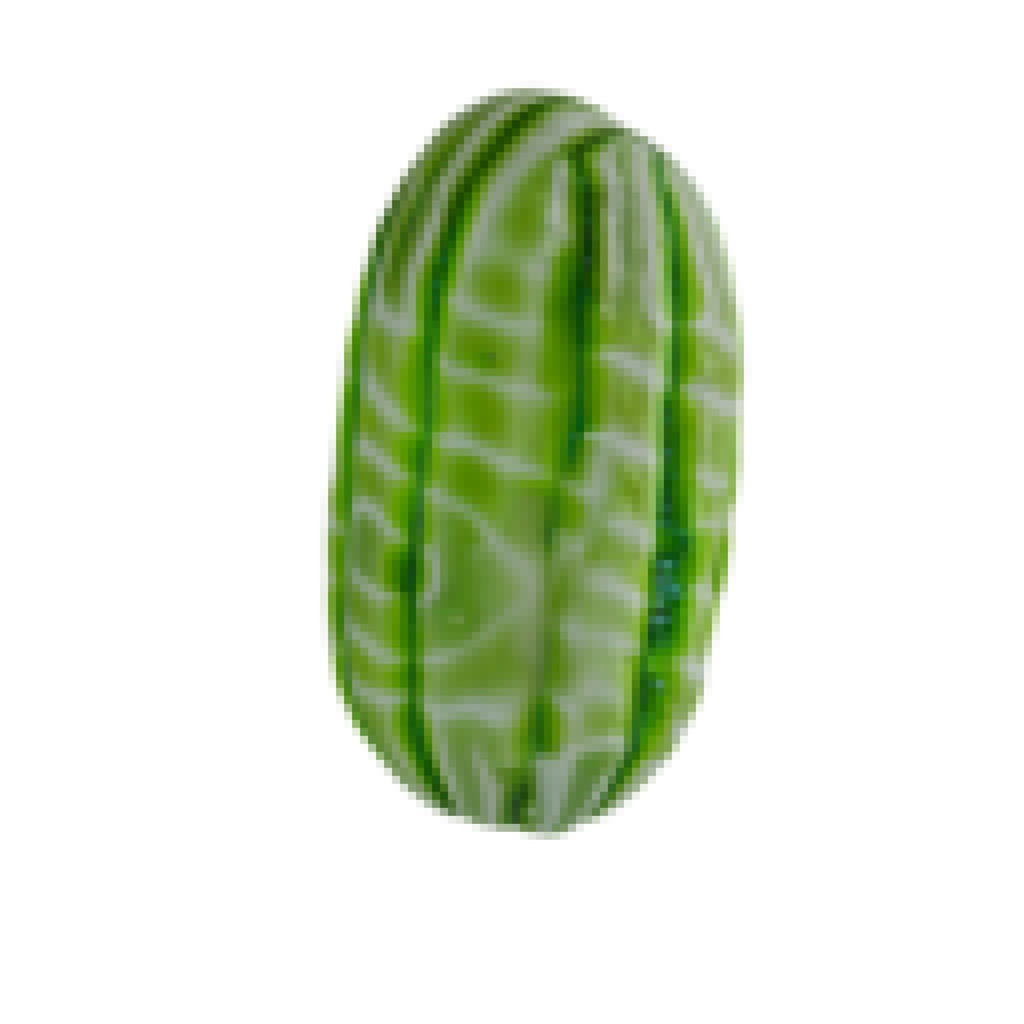}  \\
  \includegraphics[width=\shortsnormalswidth, valign=m]{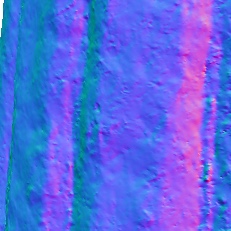} &
  \includegraphics[width=\shortsviewwidth, valign=m]{img/close-ups/_basecolor/white.jpg} &
  \includegraphics[width=\shortsreswidth, valign=m]{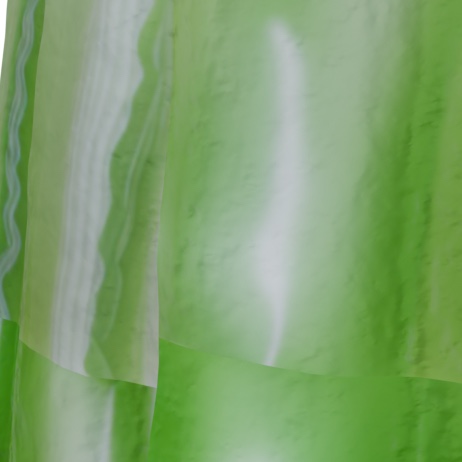} &
  \includegraphics[width=\shortsreswidth, valign=m]{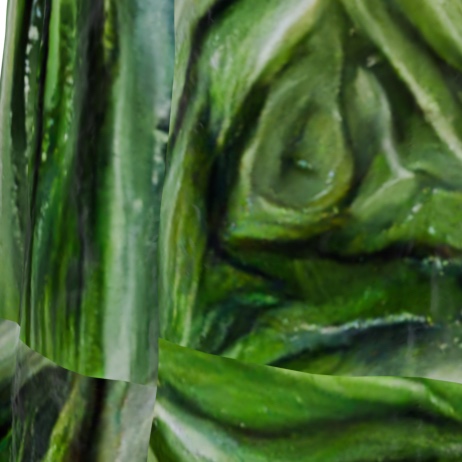} &
  \includegraphics[width=\shortsreswidth, valign=m]{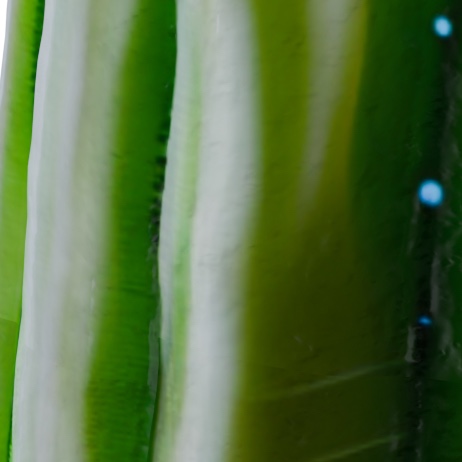} &
  \includegraphics[width=\shortsreswidth, valign=m]{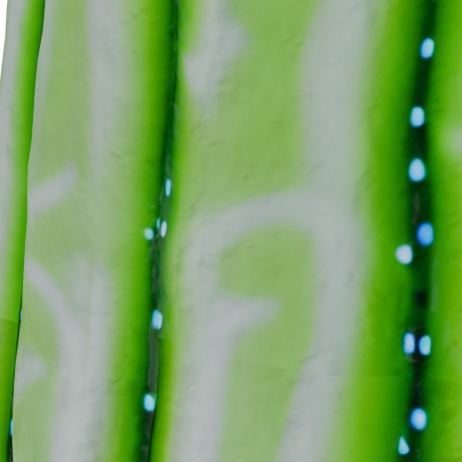} &
  \includegraphics[width=\shortsreswidth, valign=m]{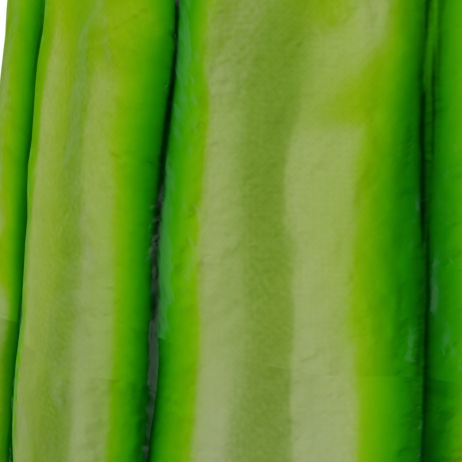} &
  \includegraphics[width=\shortsreswidth, valign=m]{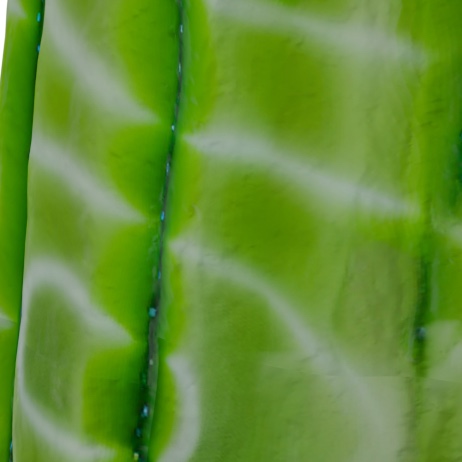}  \\
\includegraphics[width=\shortsnormalswidth, valign=m]{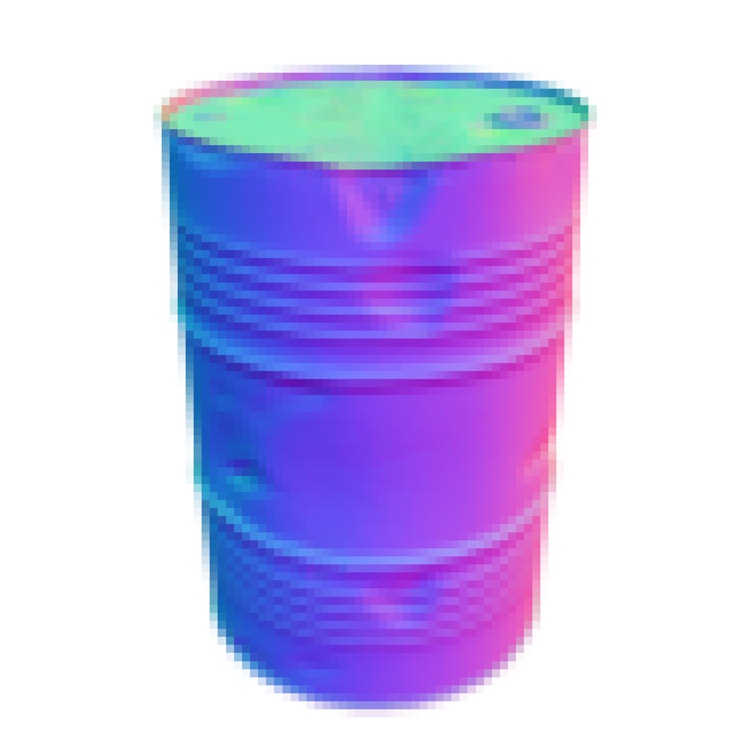} &
  \includegraphics[width=\shortsviewwidth, valign=m]{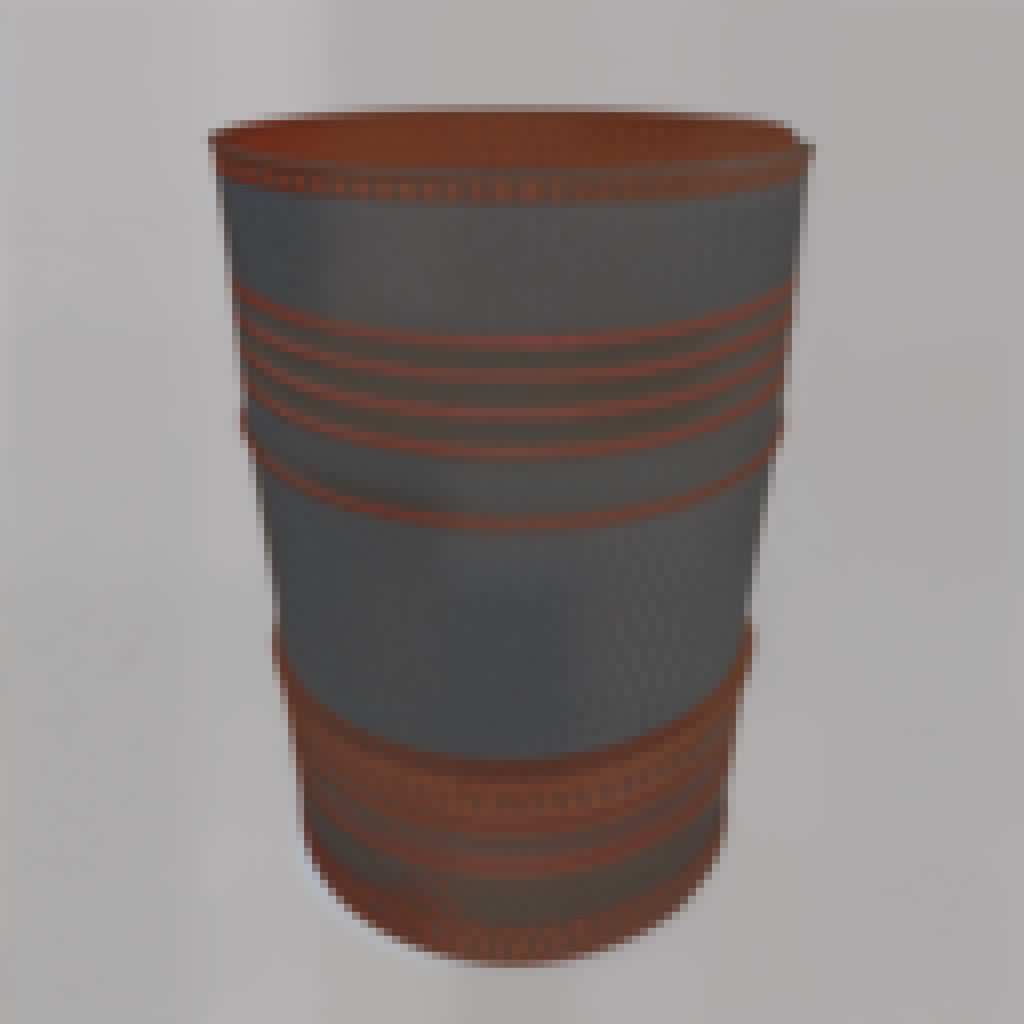} &
  \includegraphics[width=\shortsreswidth, valign=m]{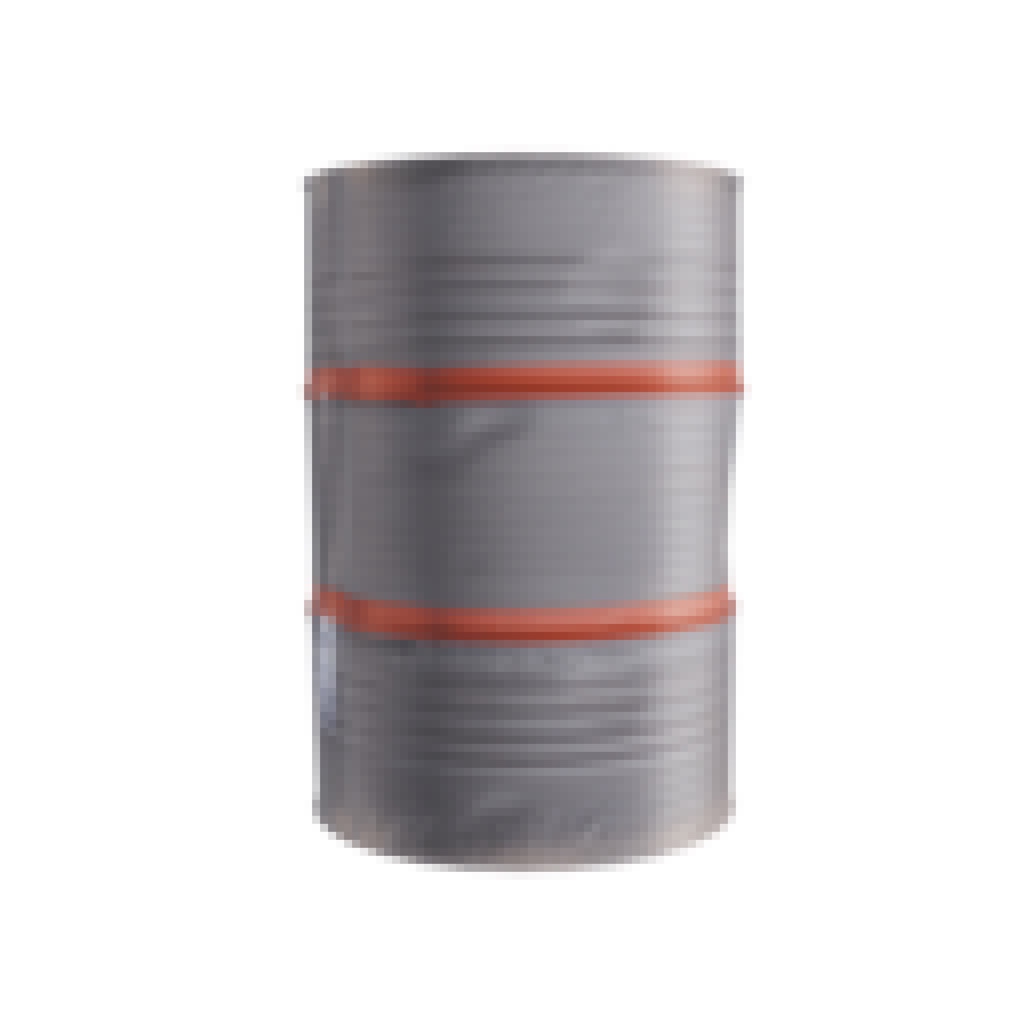} &
  \includegraphics[width=\shortsreswidth, valign=m]{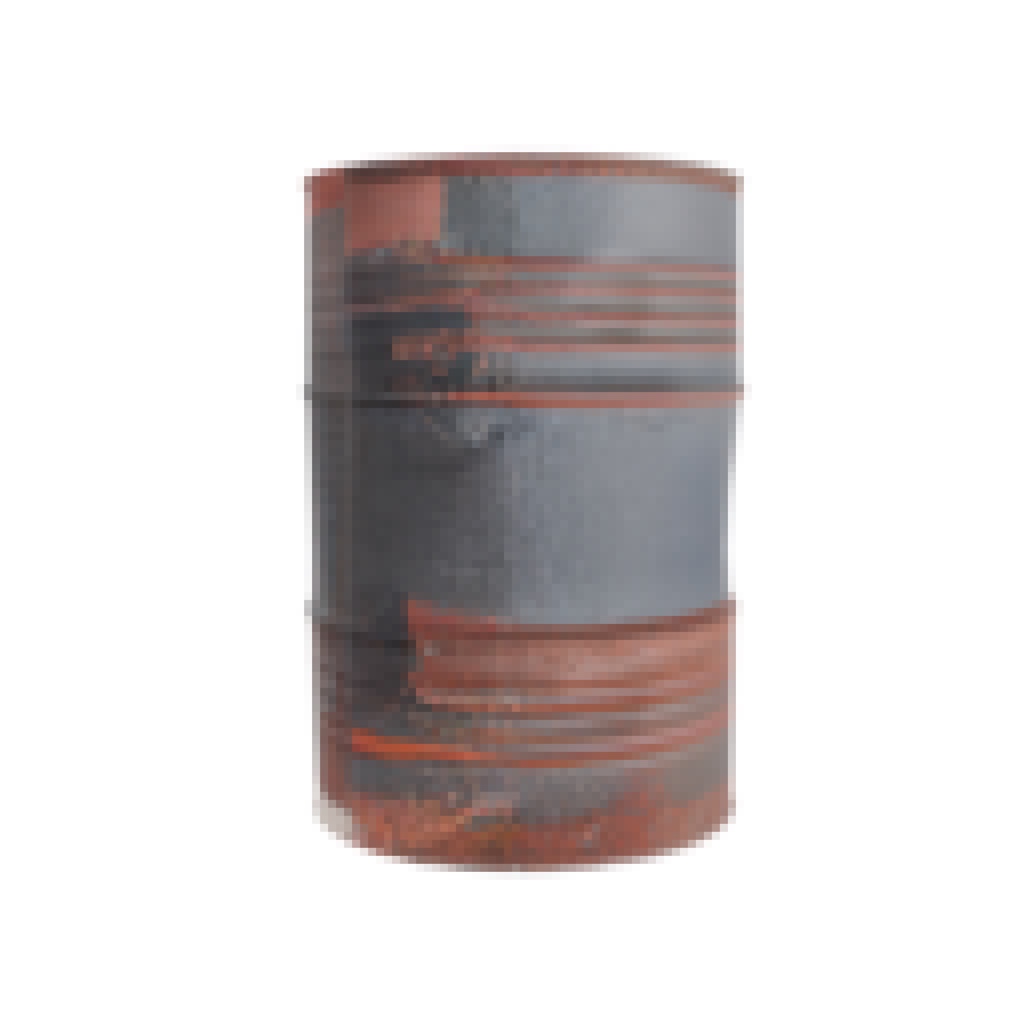} &
  \includegraphics[width=\shortsreswidth, valign=m]{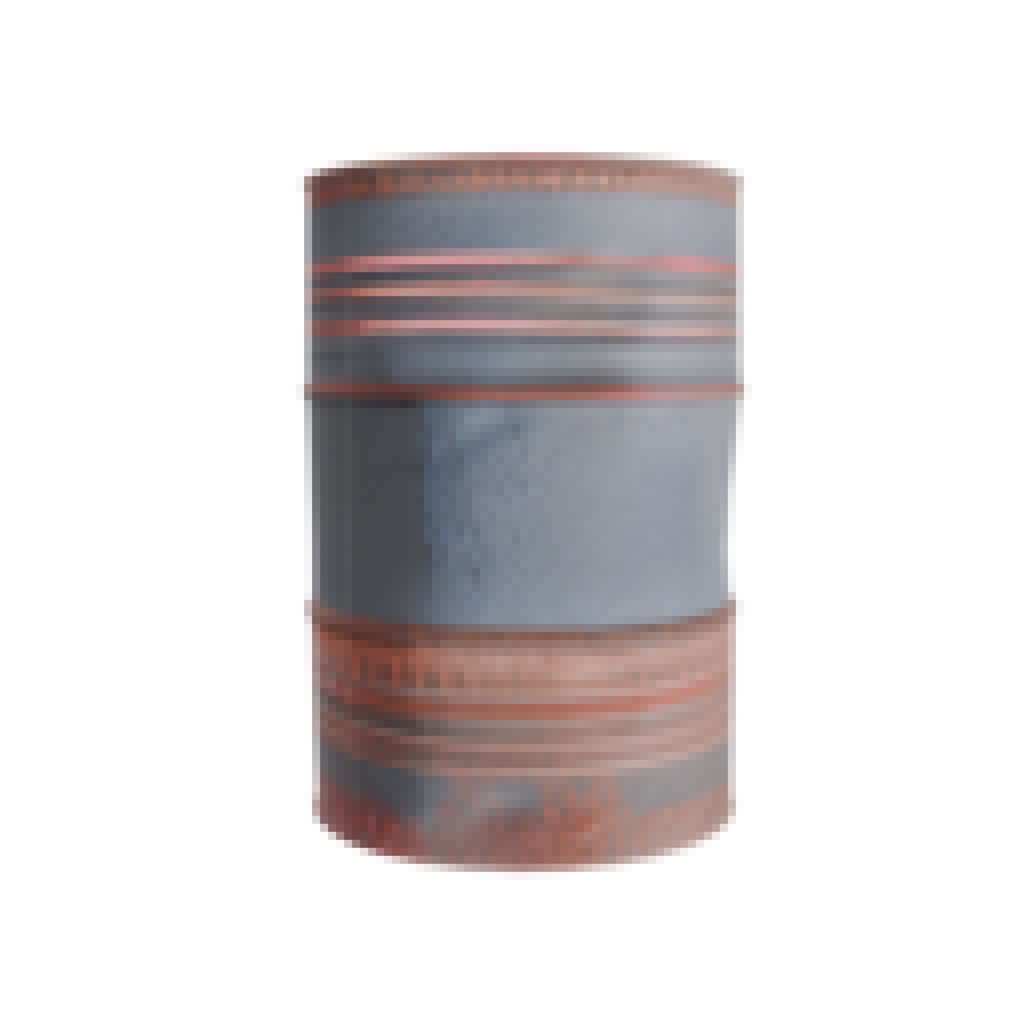} &
  \includegraphics[width=\shortsreswidth, valign=m]{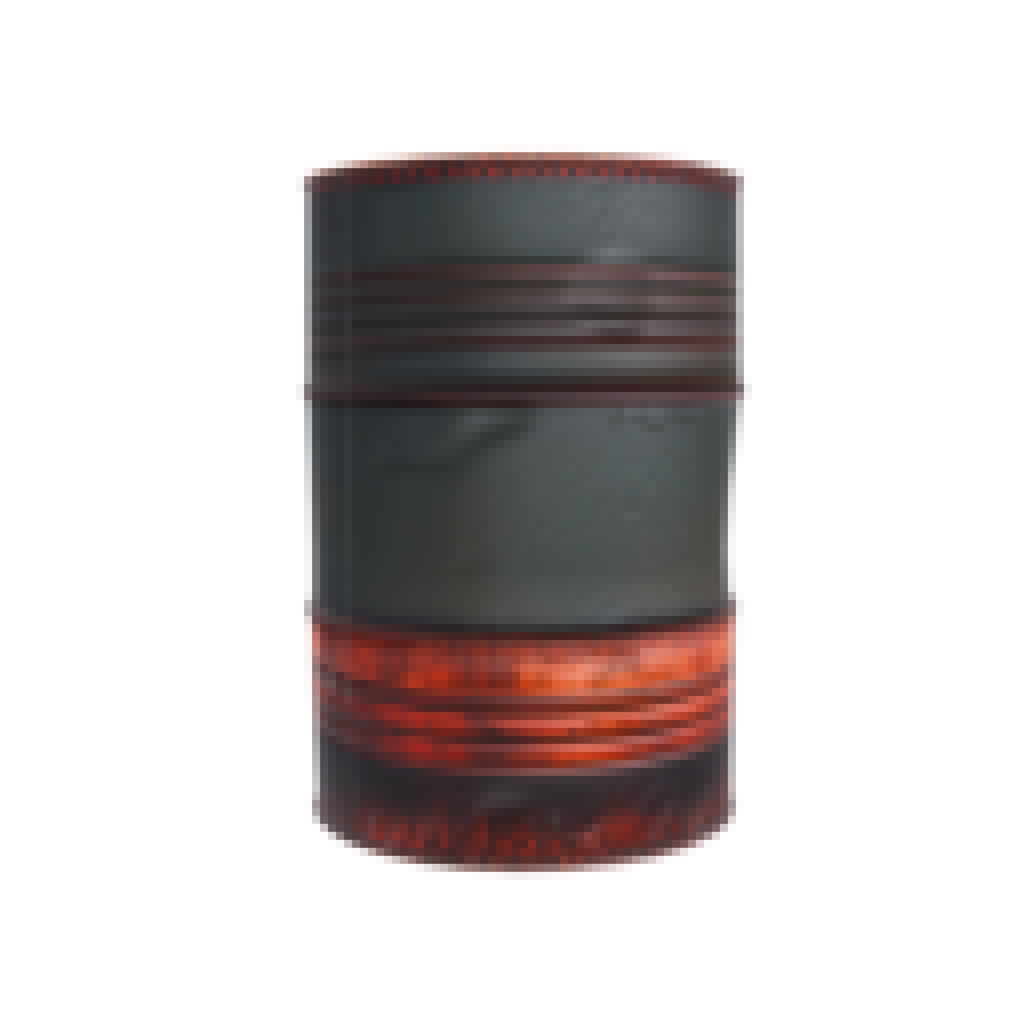} &
  \includegraphics[width=\shortsreswidth, valign=m]{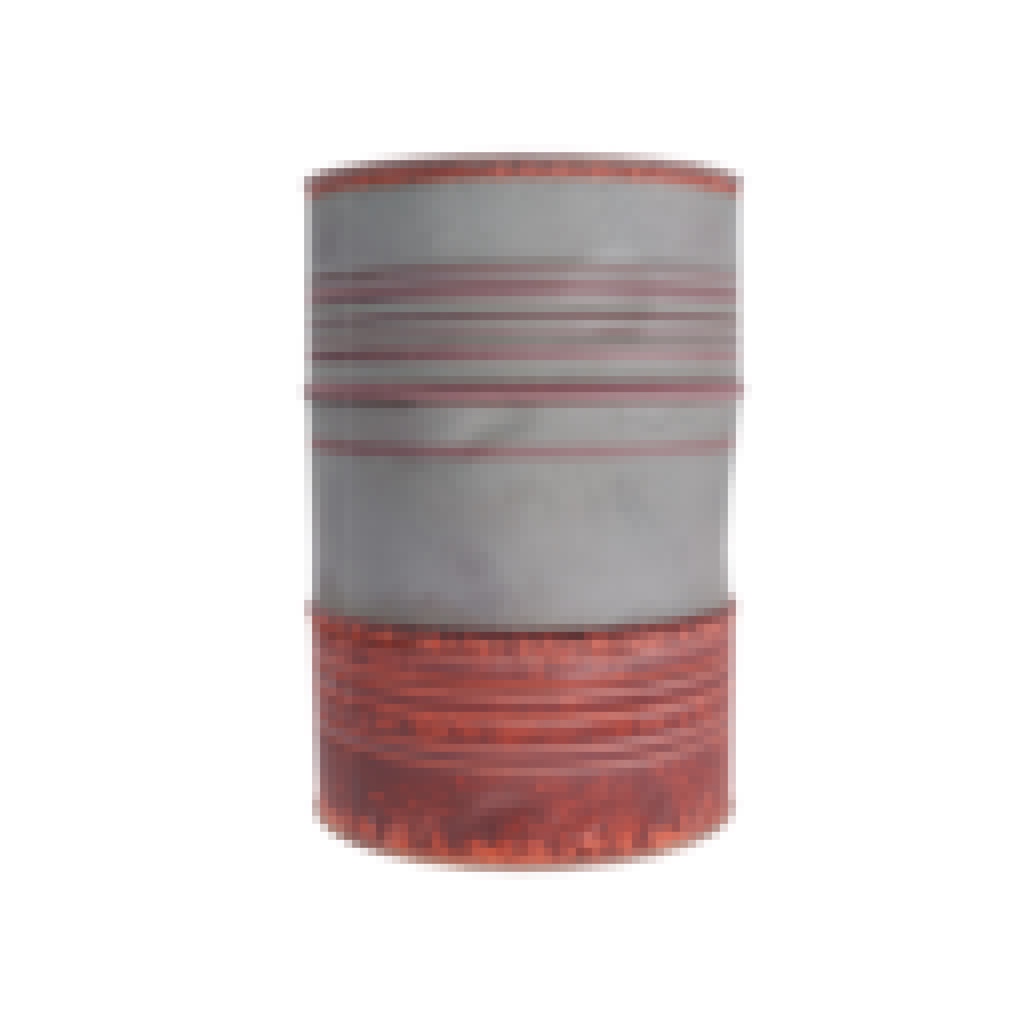} &
  \includegraphics[width=\shortsreswidth, valign=m]{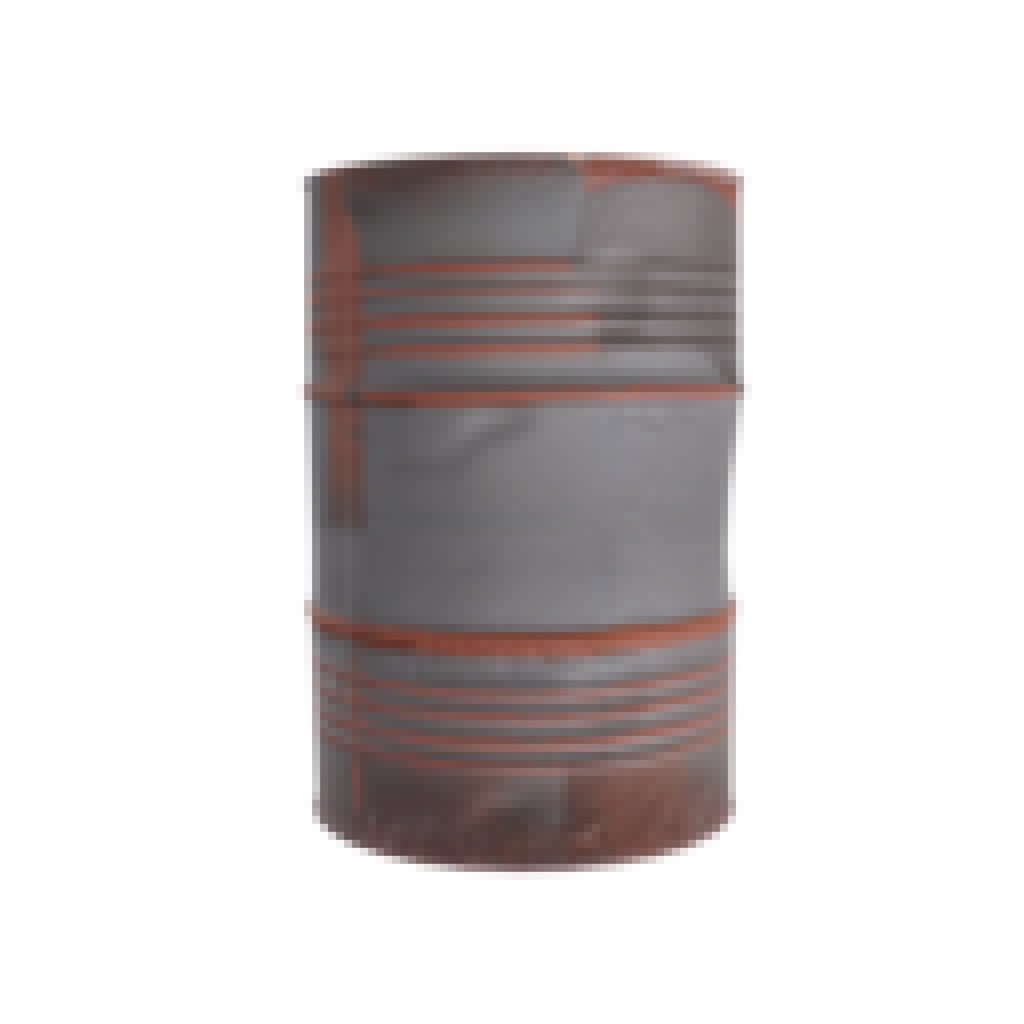}  \\
  \includegraphics[width=\shortsnormalswidth, valign=m]{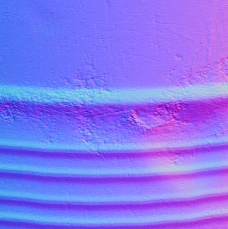} &
  \includegraphics[width=\shortsviewwidth, valign=m]{img/close-ups/_basecolor/white.jpg} &
  \includegraphics[width=\shortsreswidth, valign=m]{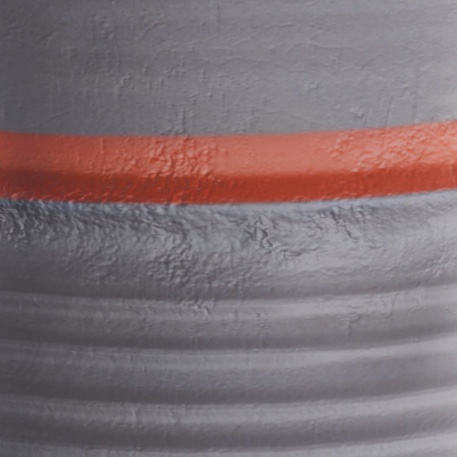} &
  \includegraphics[width=\shortsreswidth, valign=m]{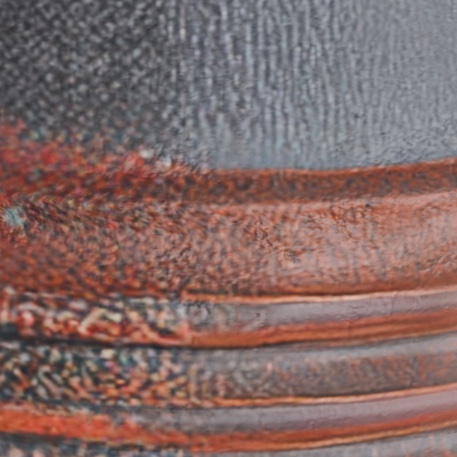} &
  \includegraphics[width=\shortsreswidth, valign=m]{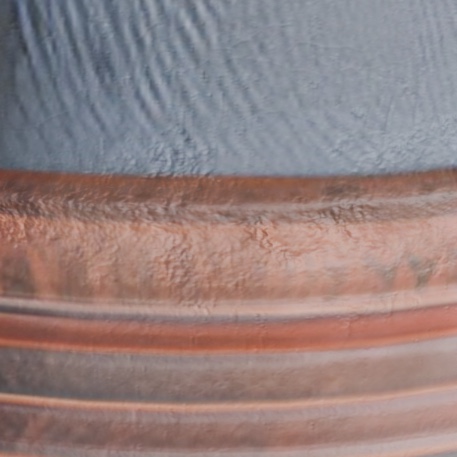} &
  \includegraphics[width=\shortsreswidth, valign=m]{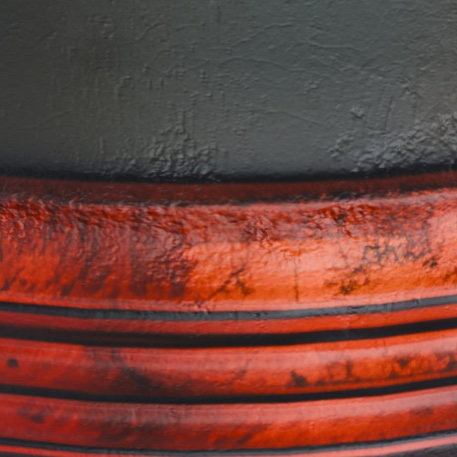} &
  \includegraphics[width=\shortsreswidth, valign=m]{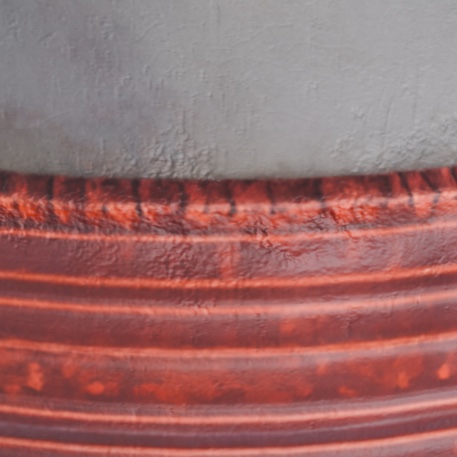} &
  \includegraphics[width=\shortsreswidth, valign=m]{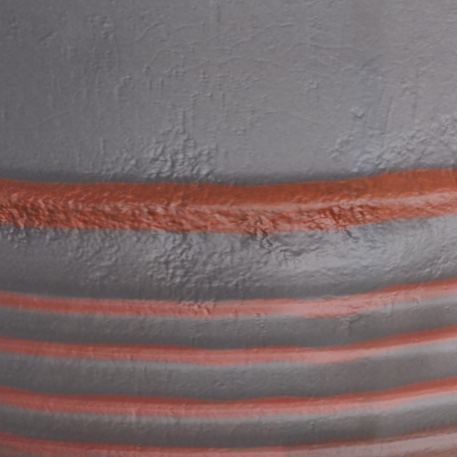}  \\
  \includegraphics[width=\shortsnormalswidth, valign=m]{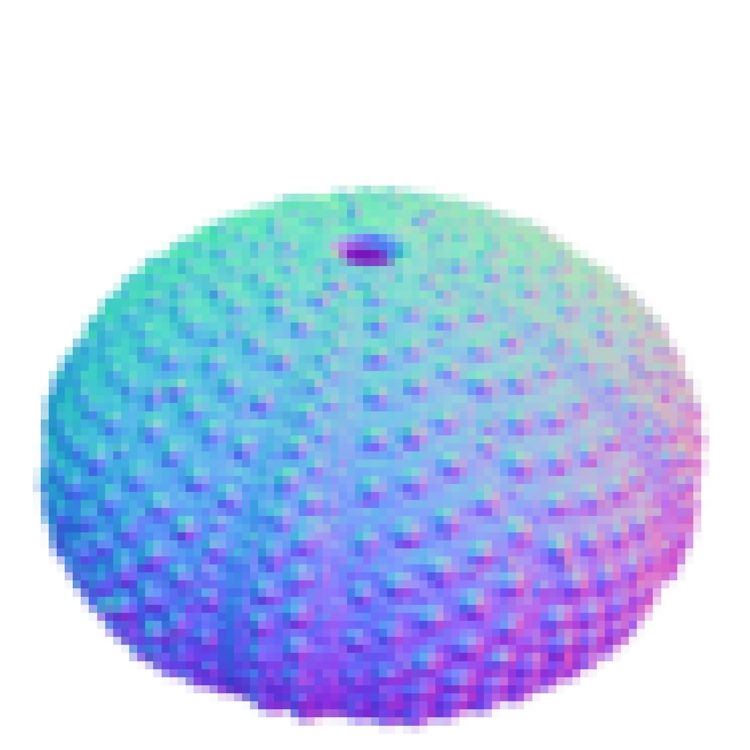} &
  \includegraphics[width=\shortsviewwidth, valign=m]{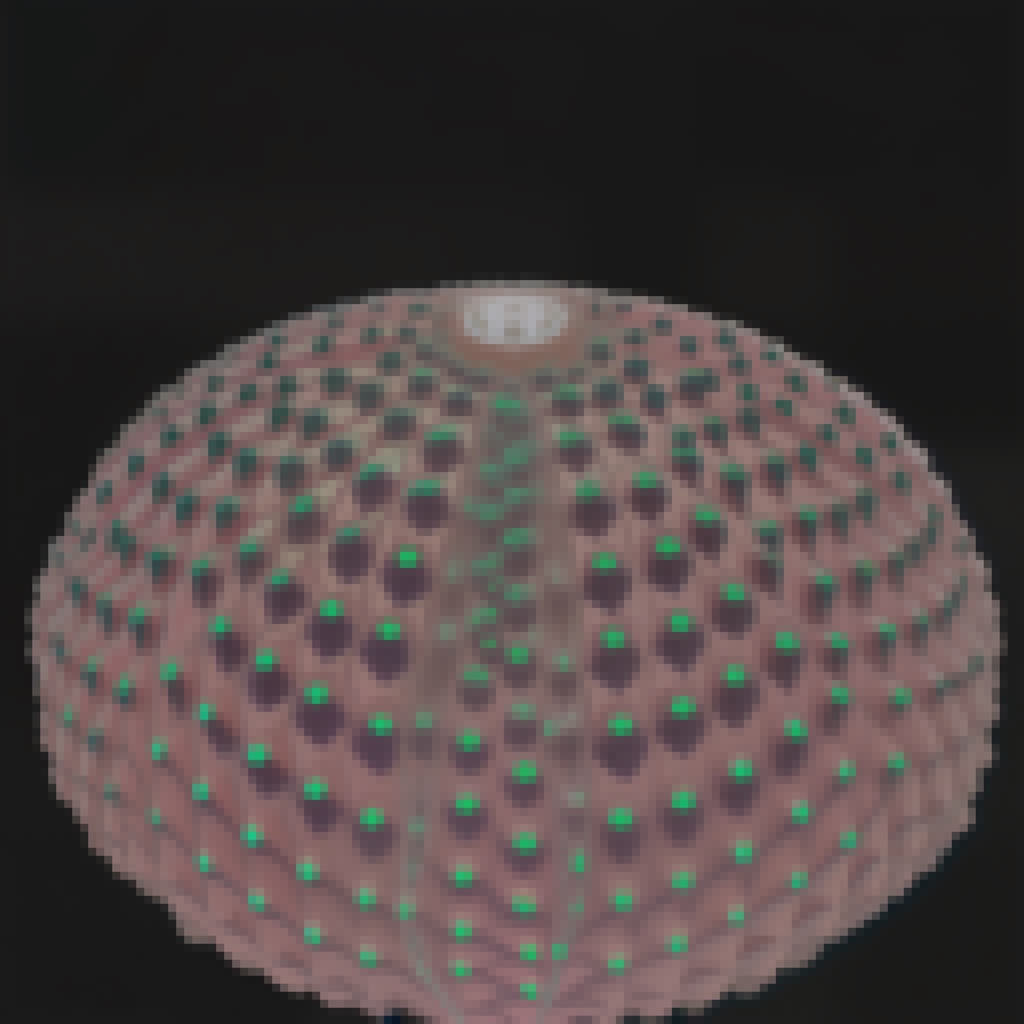} &
  \includegraphics[width=\shortsreswidth, valign=m]{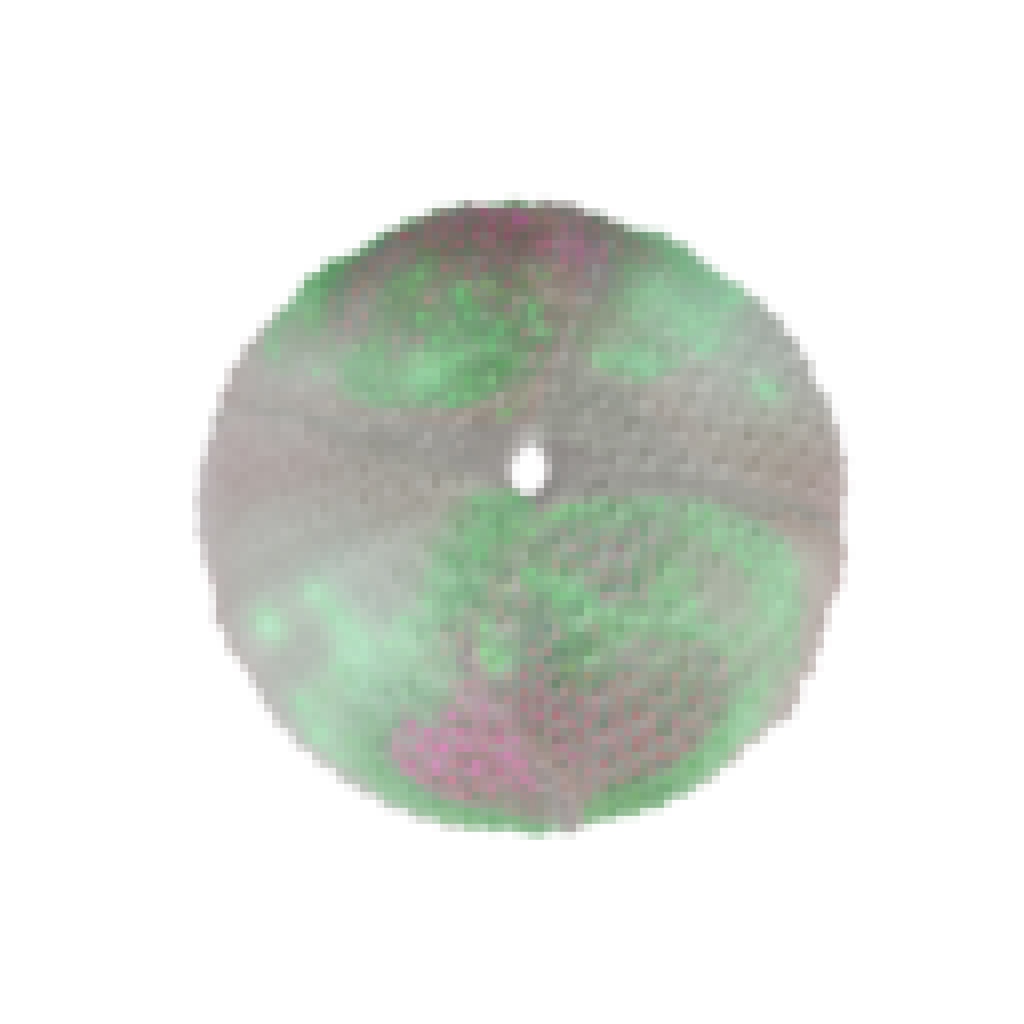} &
  \includegraphics[width=\shortsreswidth, valign=m]{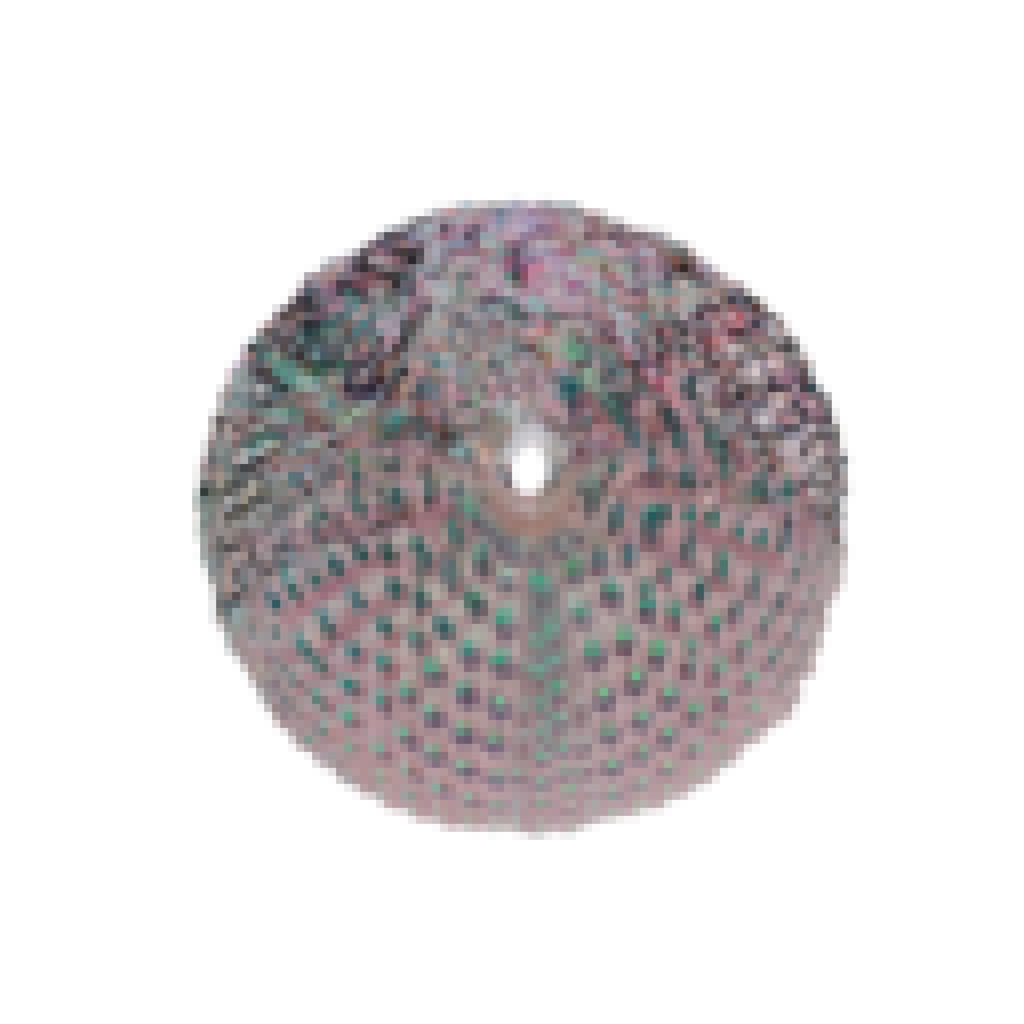} &
  \includegraphics[width=\shortsreswidth, valign=m]{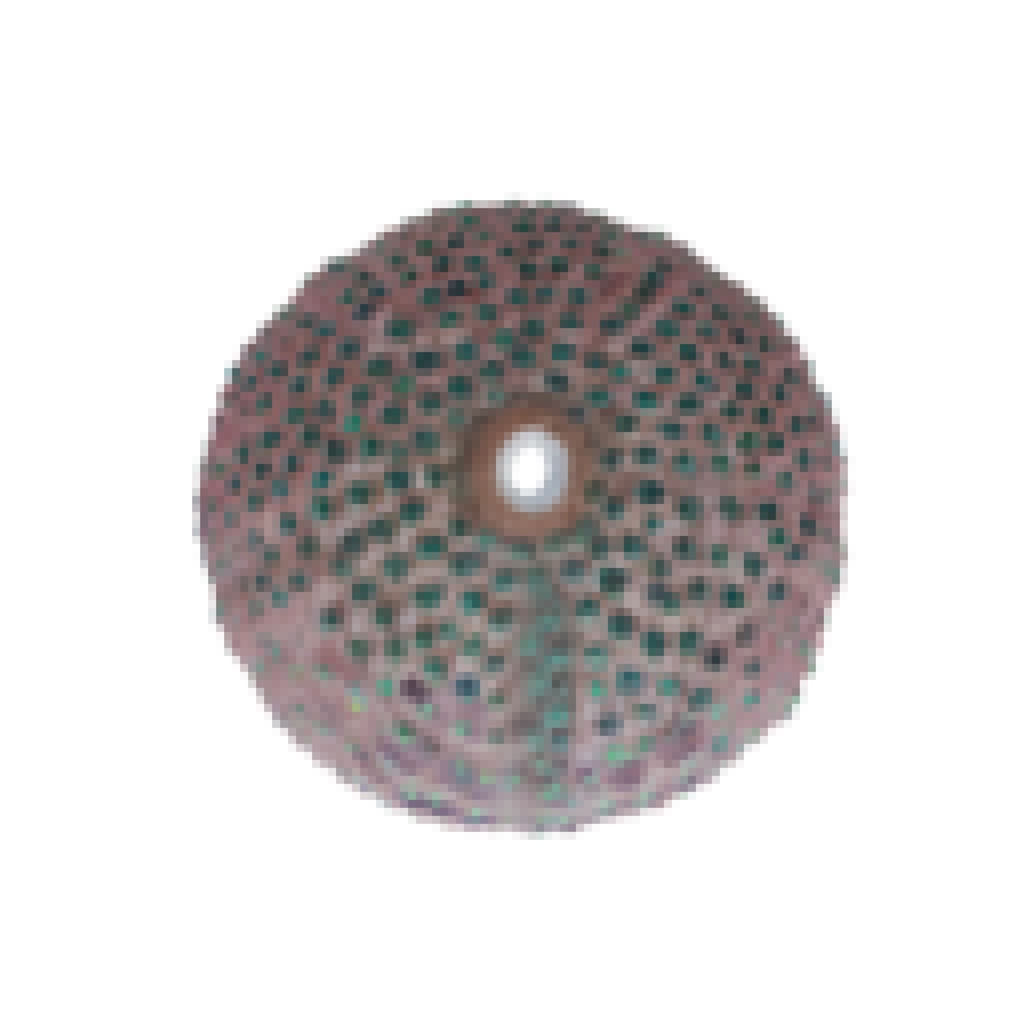} &
  \includegraphics[width=\shortsreswidth, valign=m]{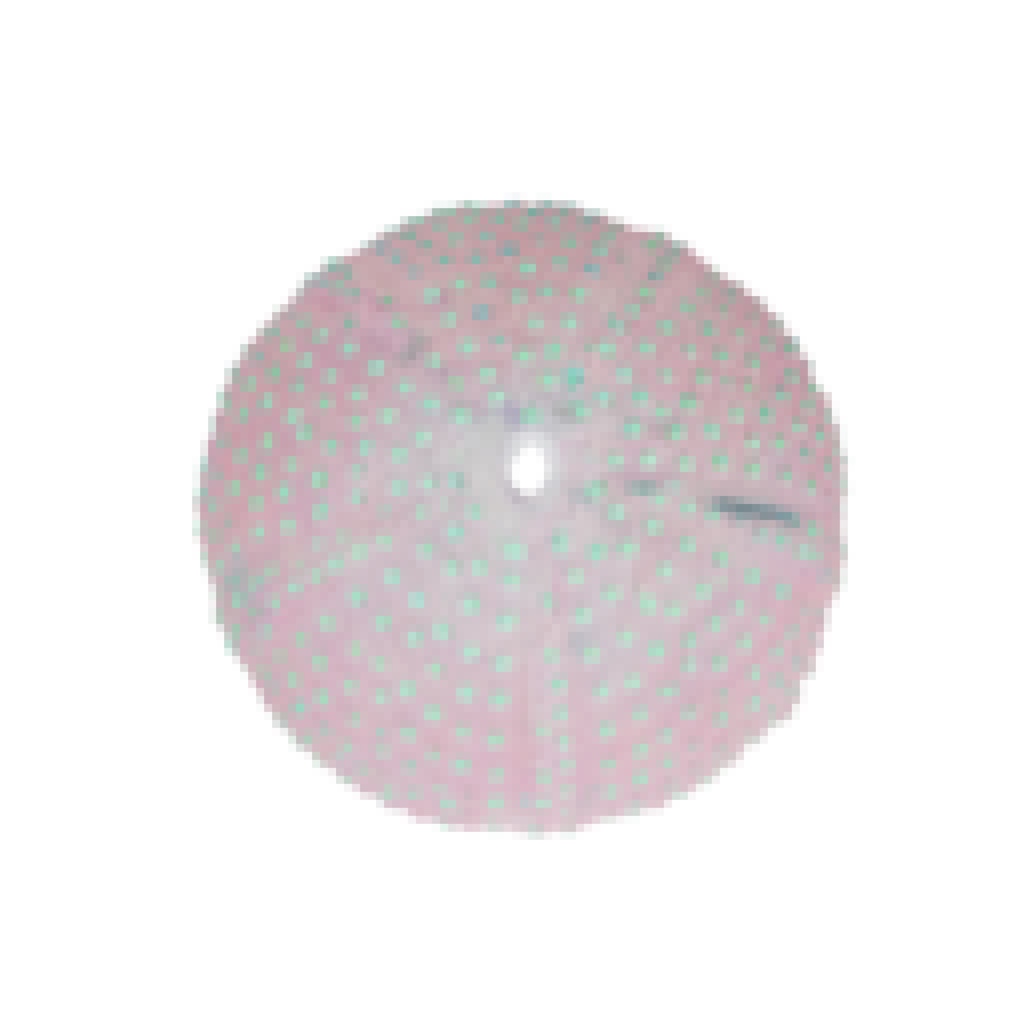} &
  \includegraphics[width=\shortsreswidth, valign=m]{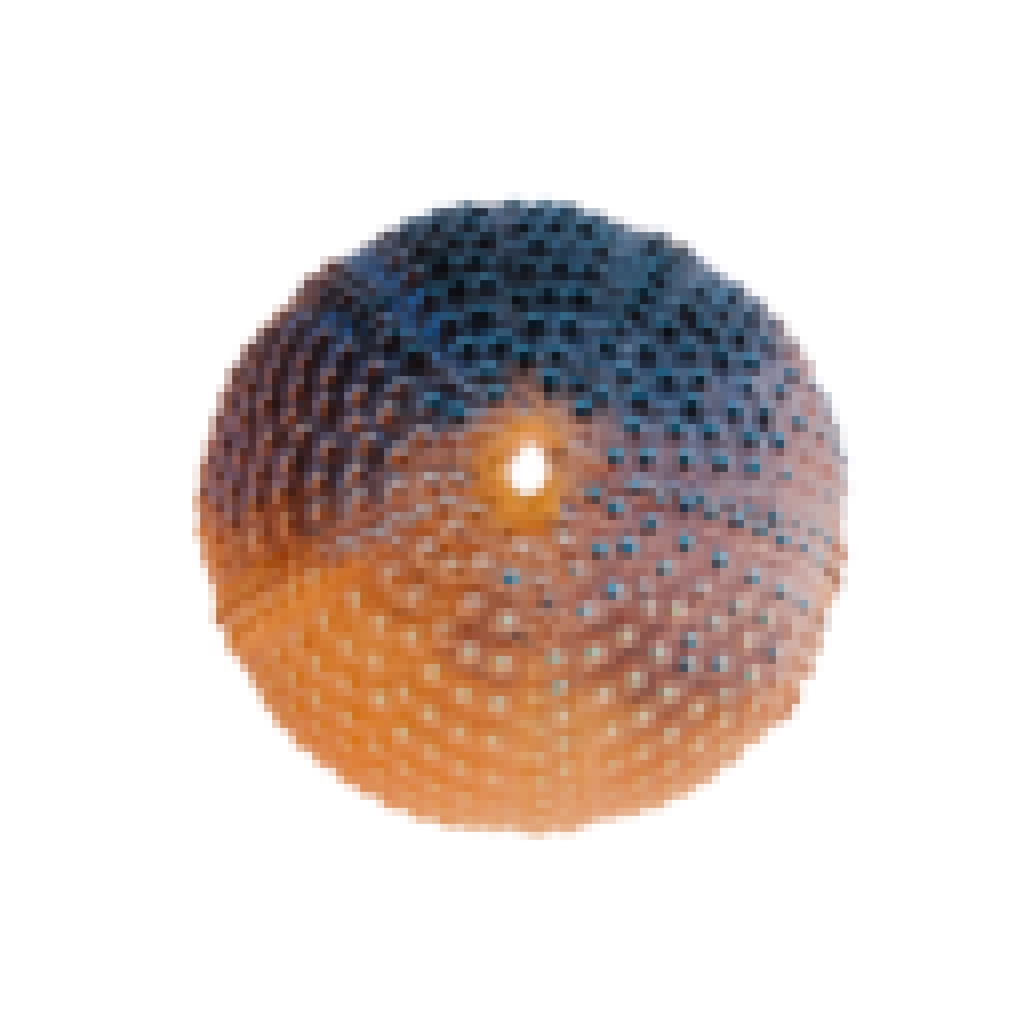} &
  \includegraphics[width=\shortsreswidth, valign=m]{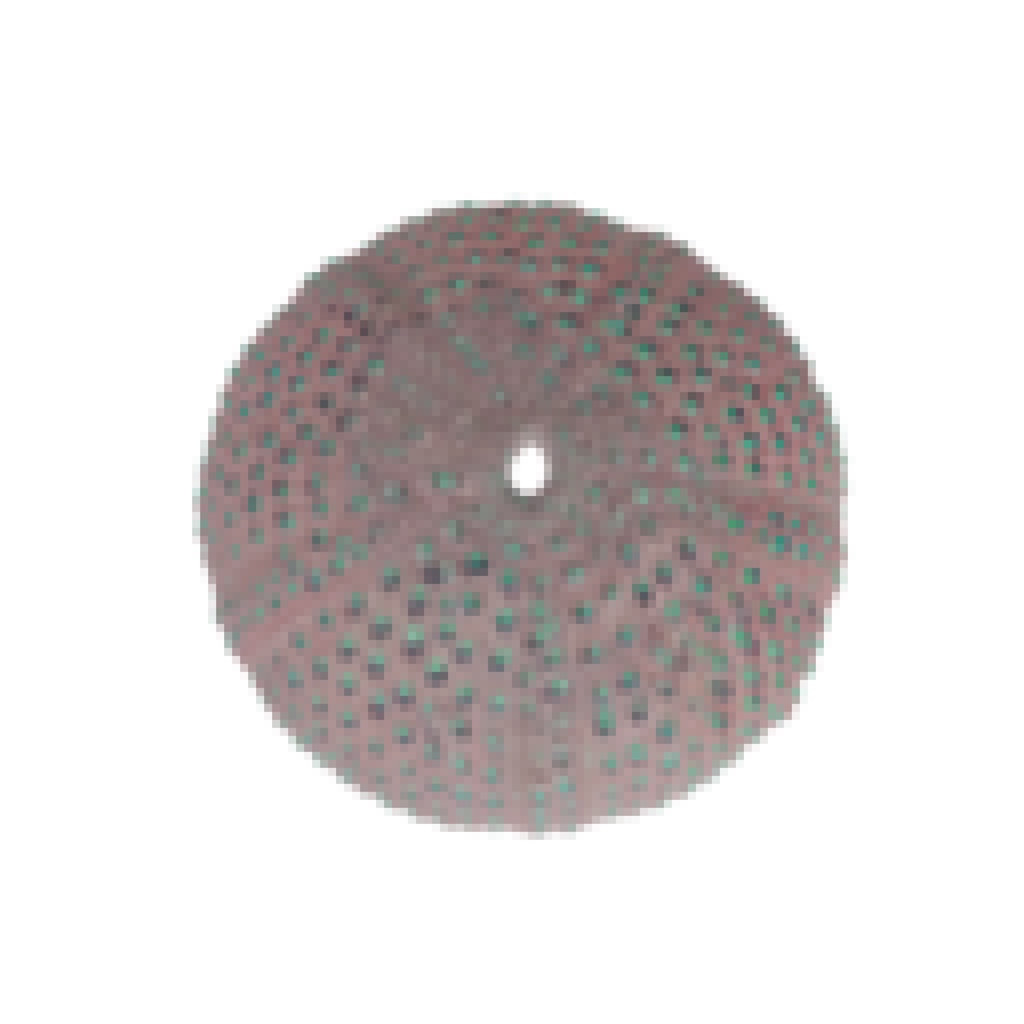}  \\
  \includegraphics[width=\shortsnormalswidth, valign=m]{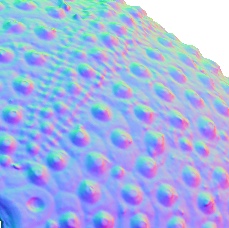} &
  \includegraphics[width=\shortsviewwidth, valign=m]{img/close-ups/_basecolor/white.jpg} &
  \includegraphics[width=\shortsreswidth, valign=m]{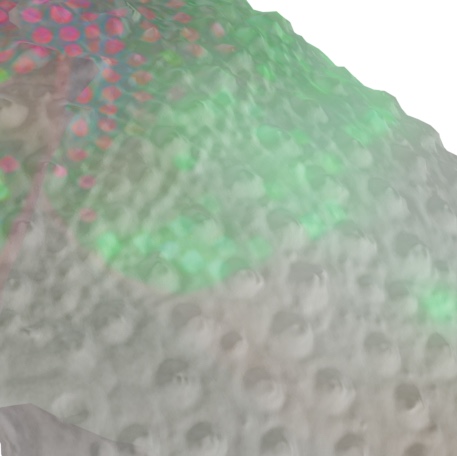} &
  \includegraphics[width=\shortsreswidth, valign=m]{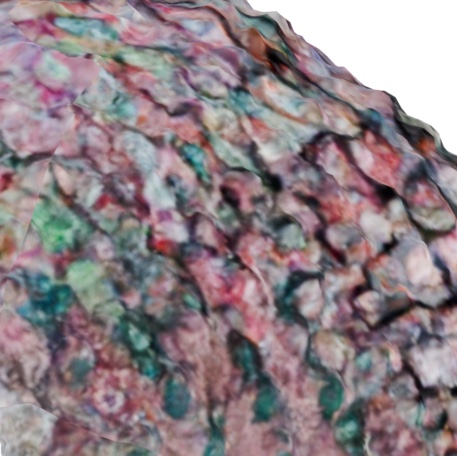} &
  \includegraphics[width=\shortsreswidth, valign=m]{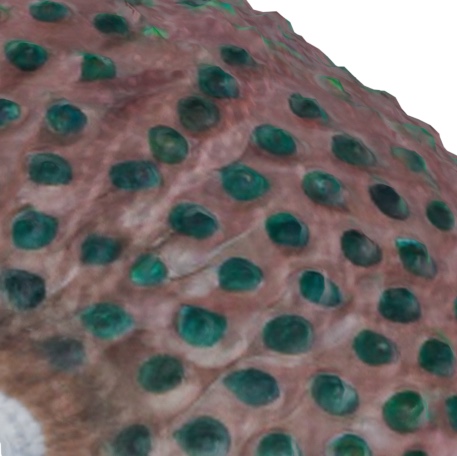} &
  \includegraphics[width=\shortsreswidth, valign=m]{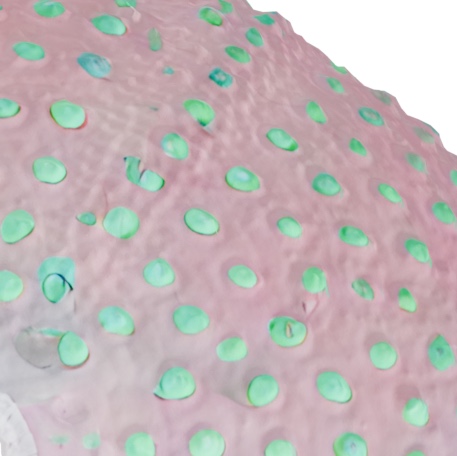} &
  \includegraphics[width=\shortsreswidth, valign=m]{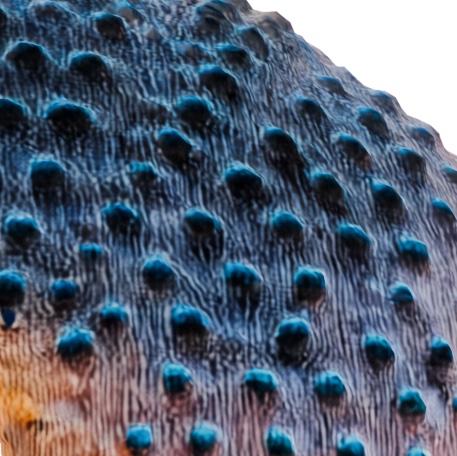} &
  \includegraphics[width=\shortsreswidth, valign=m]{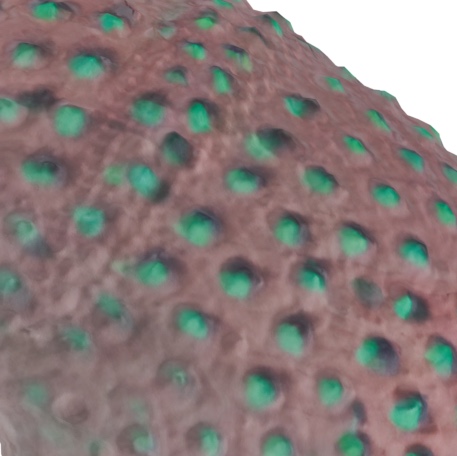}  \\
  \includegraphics[width=\shortsnormalswidth, valign=m]{img/normals/turtle.jpg} &
  \includegraphics[width=\shortsviewwidth, valign=m]{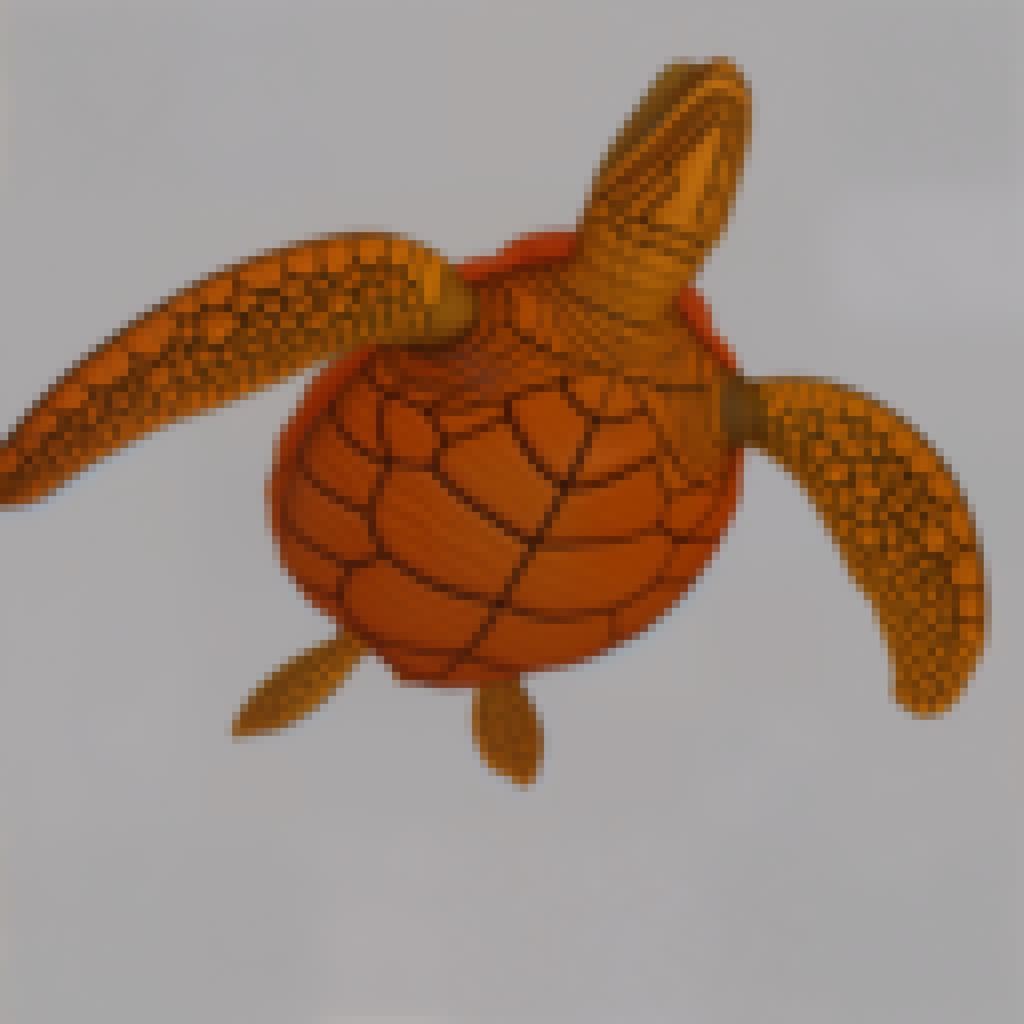} &
  \includegraphics[width=\shortsreswidth, valign=m]{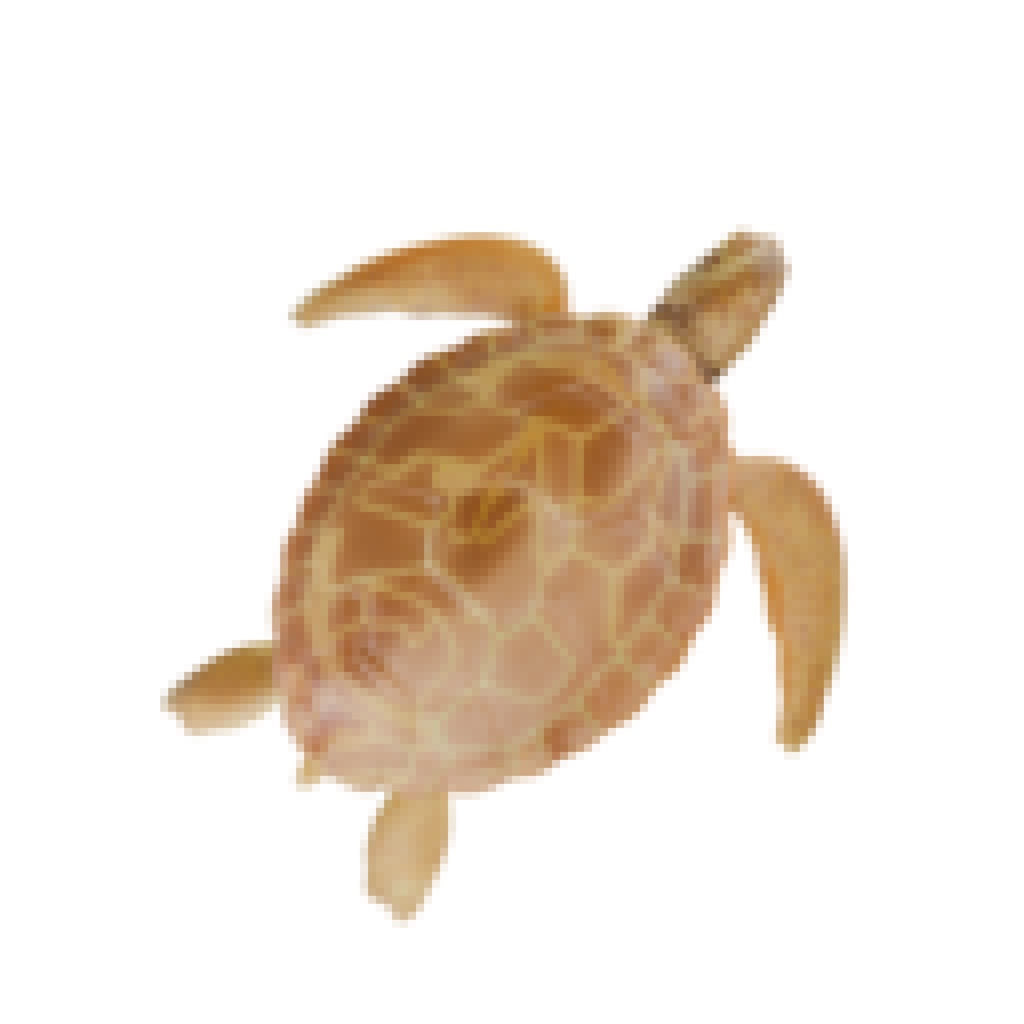} &
  \includegraphics[width=\shortsreswidth, valign=m]{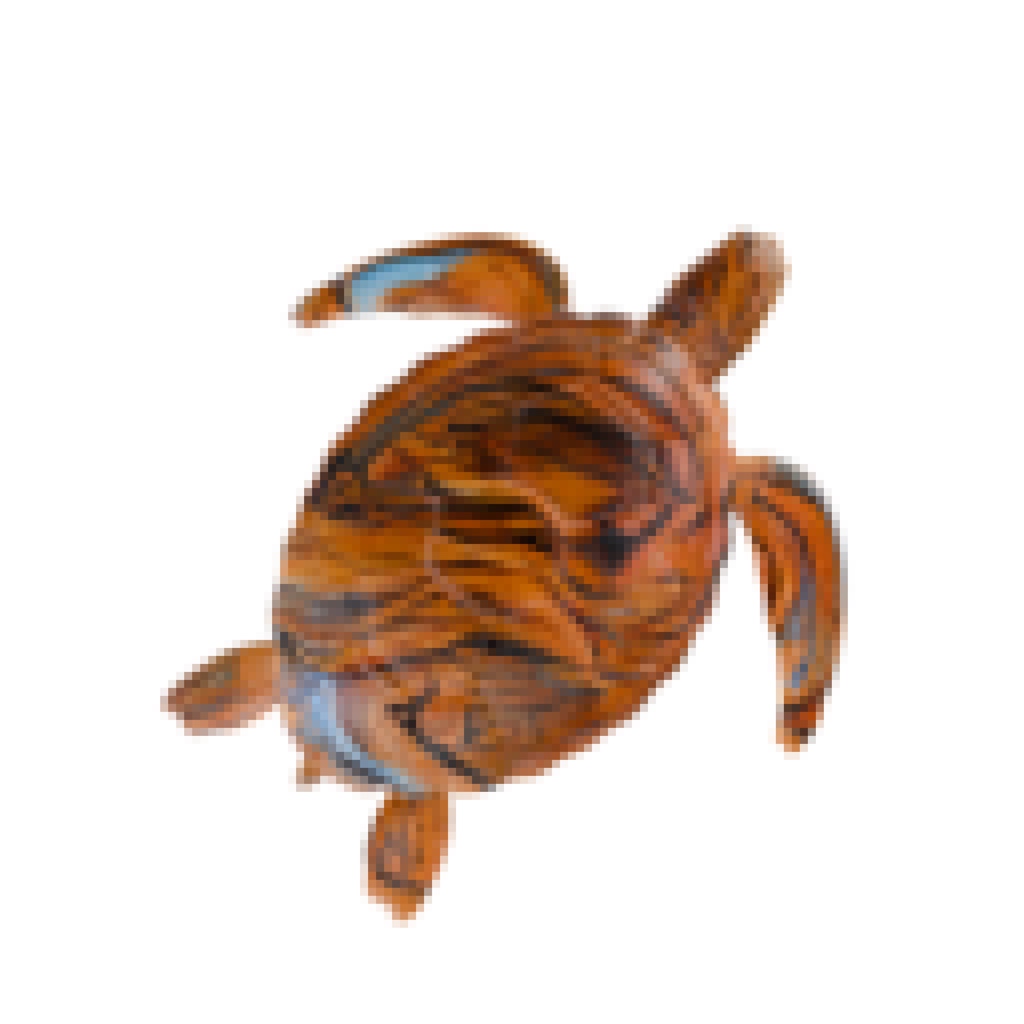} &
  \includegraphics[width=\shortsreswidth, valign=m]{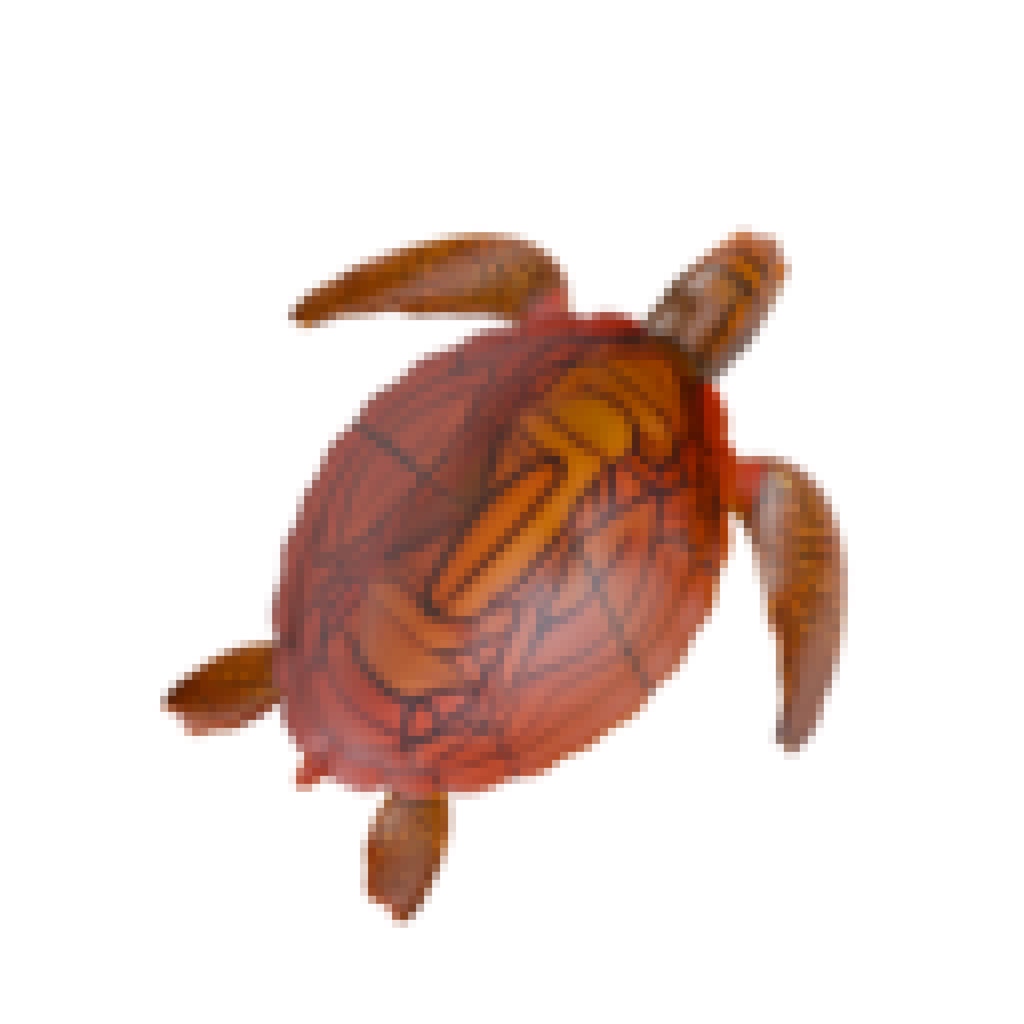} &
  \includegraphics[width=\shortsreswidth, valign=m]{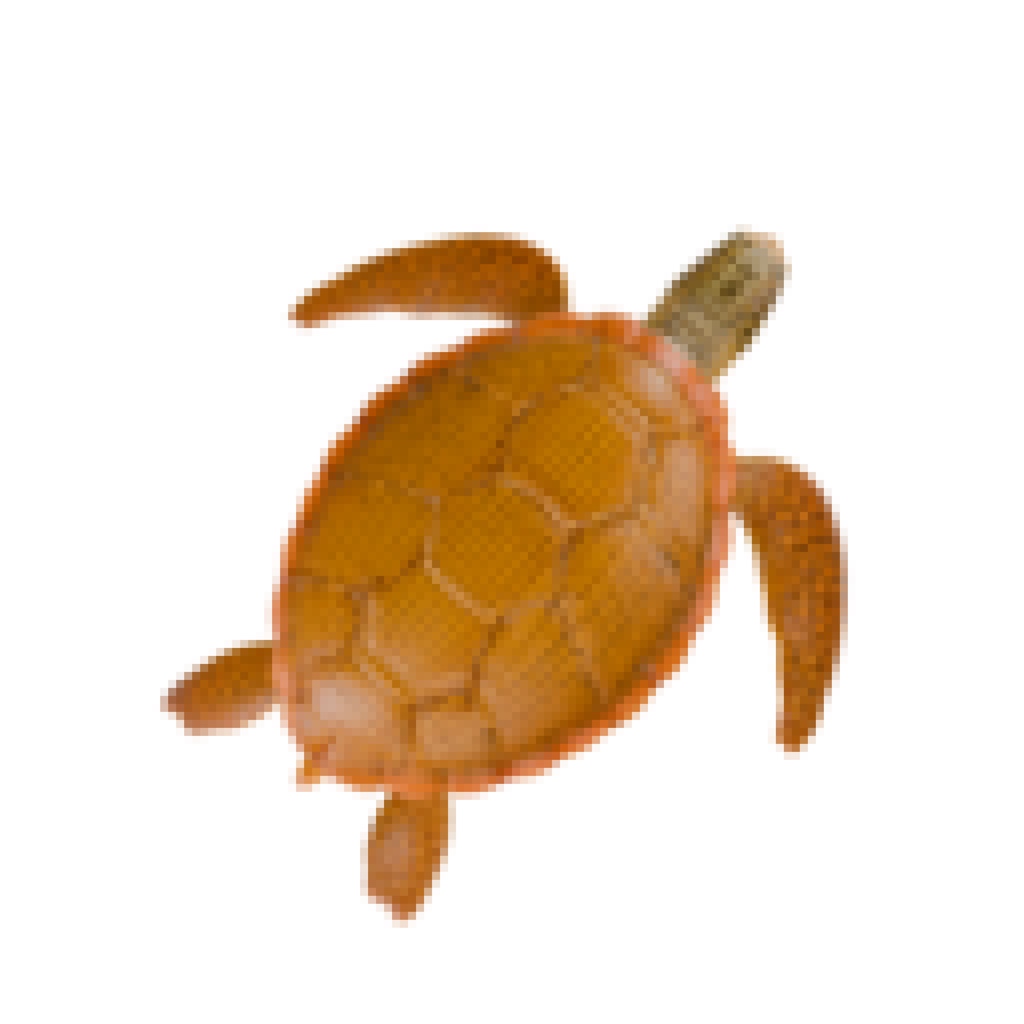} &
  \includegraphics[width=\shortsreswidth, valign=m]{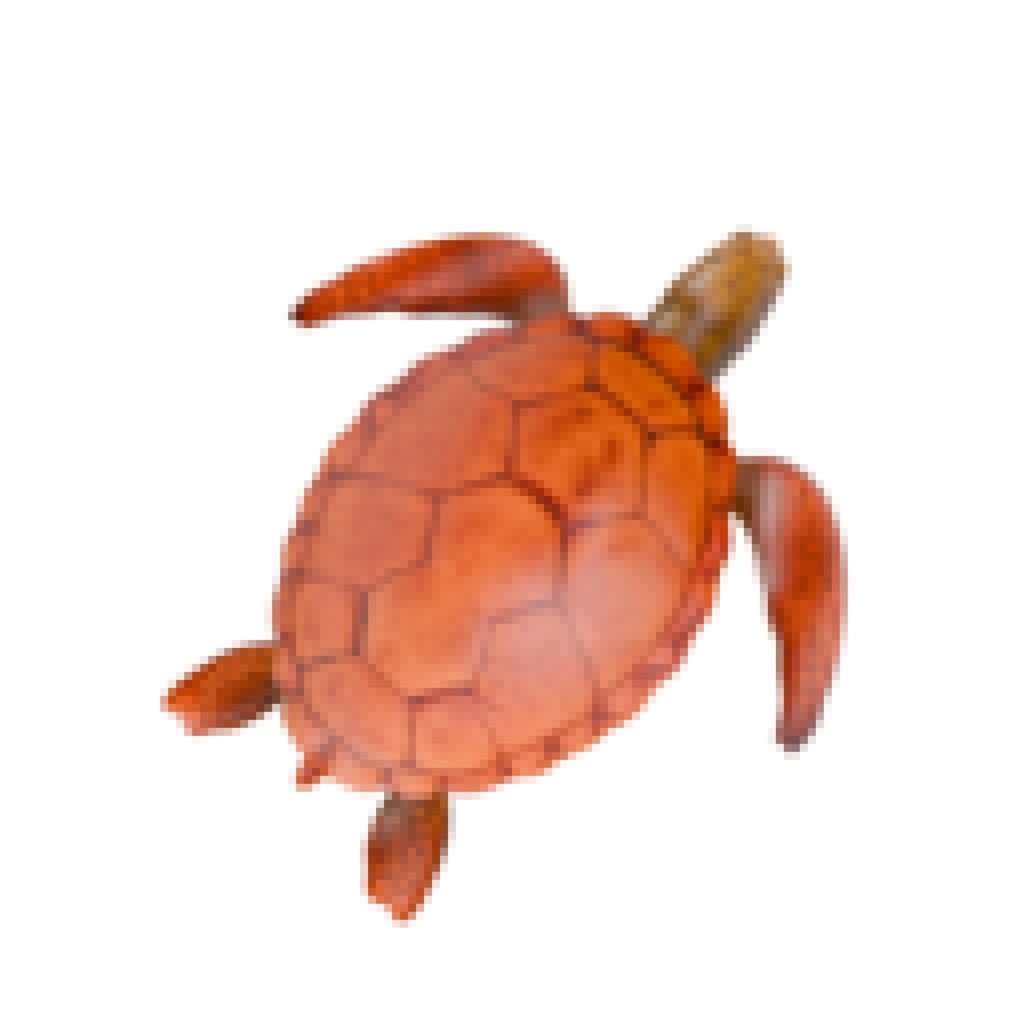} &
  \includegraphics[width=\shortsreswidth, valign=m]{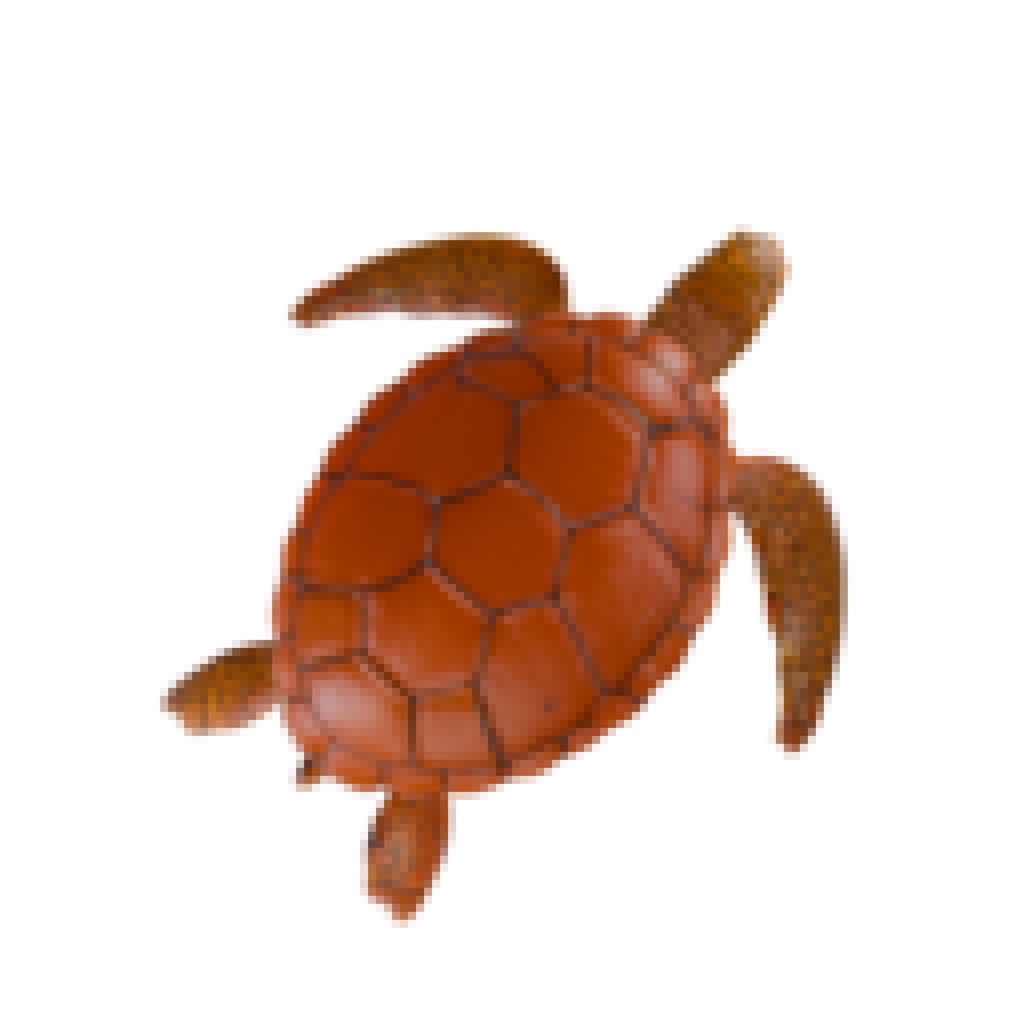}  \\
  \includegraphics[width=\shortsnormalswidth, valign=m]{img/close-ups/_normals/turtle/cam0003.jpg} &
  \includegraphics[width=\shortsviewwidth, valign=m]{img/close-ups/_basecolor/white.jpg} &
  \includegraphics[width=\shortsreswidth, valign=m]{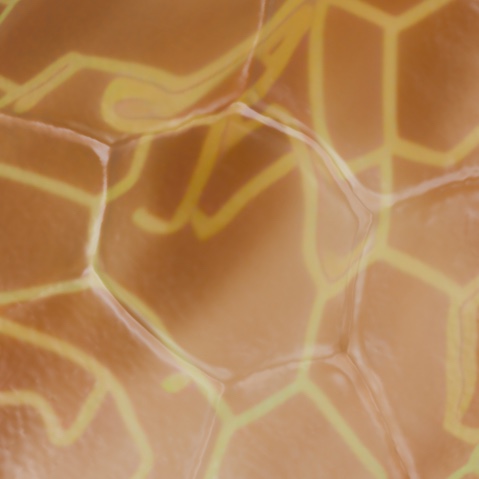} &
  \includegraphics[width=\shortsreswidth, valign=m]{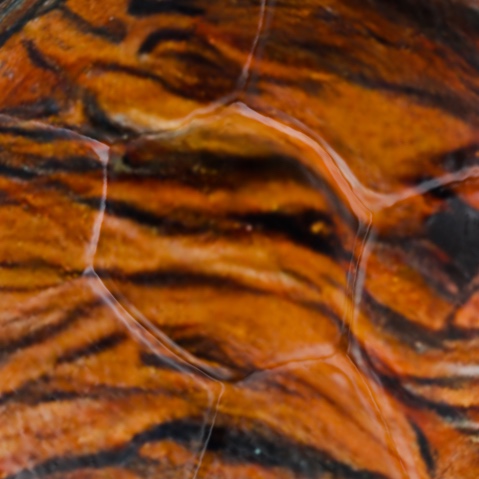} &
  \includegraphics[width=\shortsreswidth, valign=m]{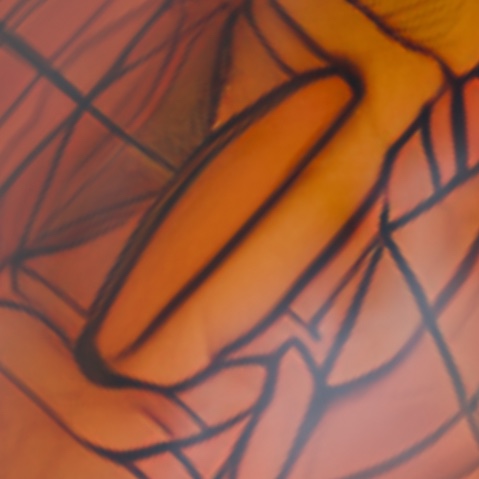} &
  \includegraphics[width=\shortsreswidth, valign=m]{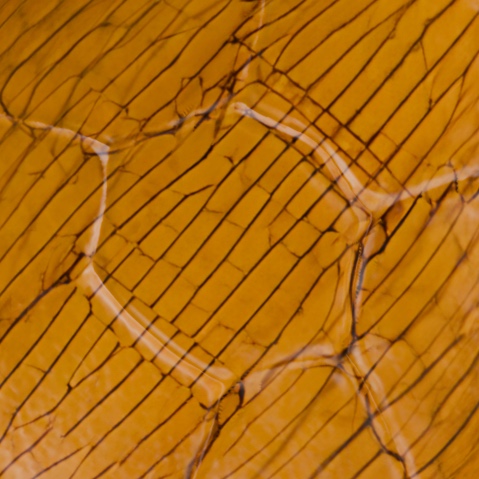} &
  \includegraphics[width=\shortsreswidth, valign=m]{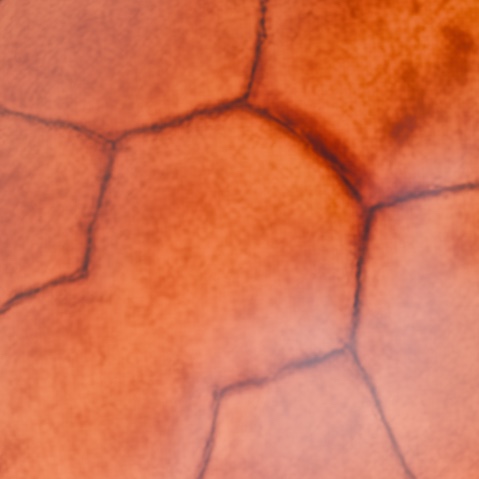} &
  \includegraphics[width=\shortsreswidth, valign=m]{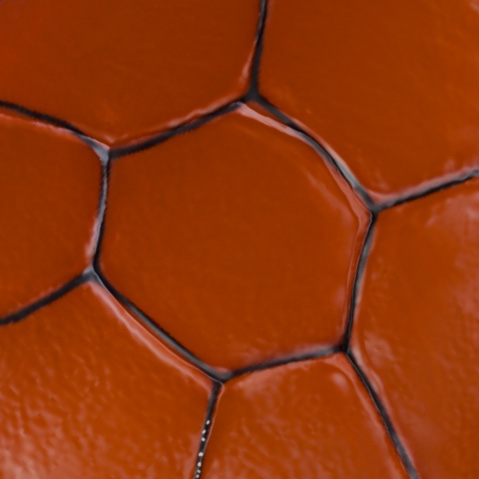}  \\

  \end{tabular}
  \caption{\footnotesize{Additional results on the baseline comparison, both global view and zoomed-in views.}}
  \label{fig:supp_comparison_gallery_2}
\end{figure*}

\endgroup

\newcommand{\normalswidth}{0.10\linewidth}
\newcommand{\viewwidth}{0.10\linewidth}

\begin{figure*}[h!tbp]
	\centering
    \subfloat[Texture Completion ]
    {
        \begin{tabular}{ccccccc}
        3D model & single view &  completion & completion & single view & completion & completion\\
        &  &  (opposite side) & (other view) &  & (opposite side) & (other view)\\
        \includegraphics[trim={0cm 0cm 0cm 0cm}, clip, width=\normalswidth, valign=m]{img/normals/cabbage.jpg} &
        \includegraphics[trim={0cm 0cm 0cm 0cm}, clip, width=\viewwidth, valign=m]{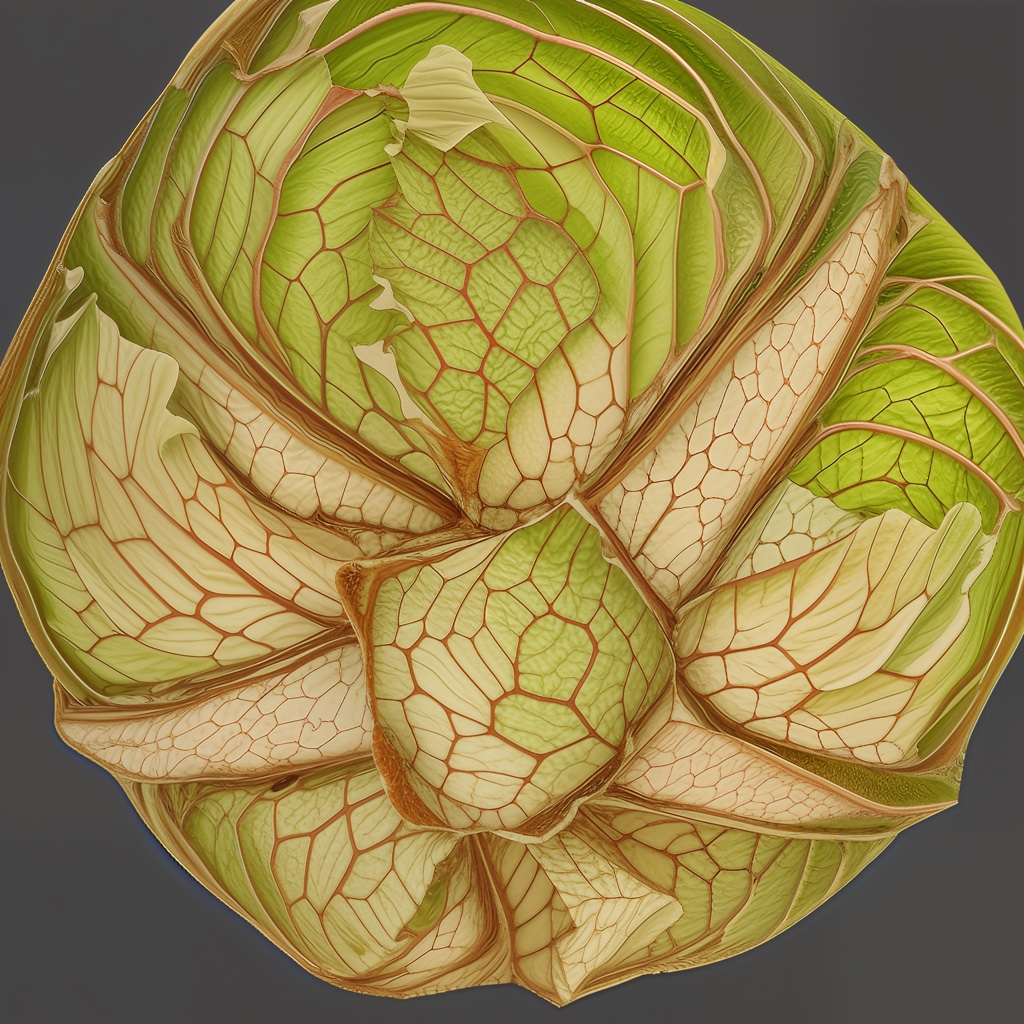} &
        \includegraphics[trim={0cm 0cm 0cm 0cm}, clip, trim={0cm 0cm 0cm 0cm}, clip, width=0.121\linewidth, valign=m, valign=m]{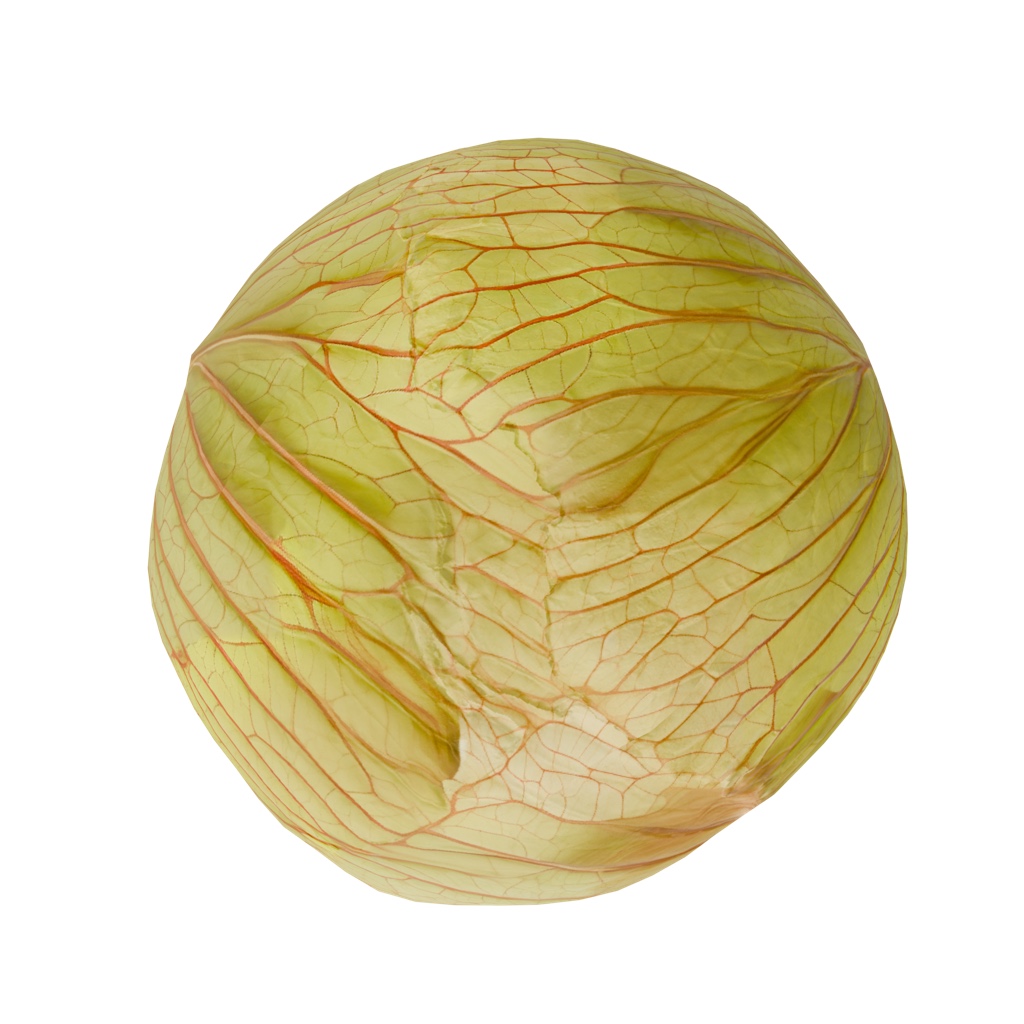} &
        \includegraphics[trim={0cm 0cm 0cm 0cm}, clip, trim={0cm 0cm 0cm 0cm}, clip, width=0.121\linewidth, valign=m, valign=m]{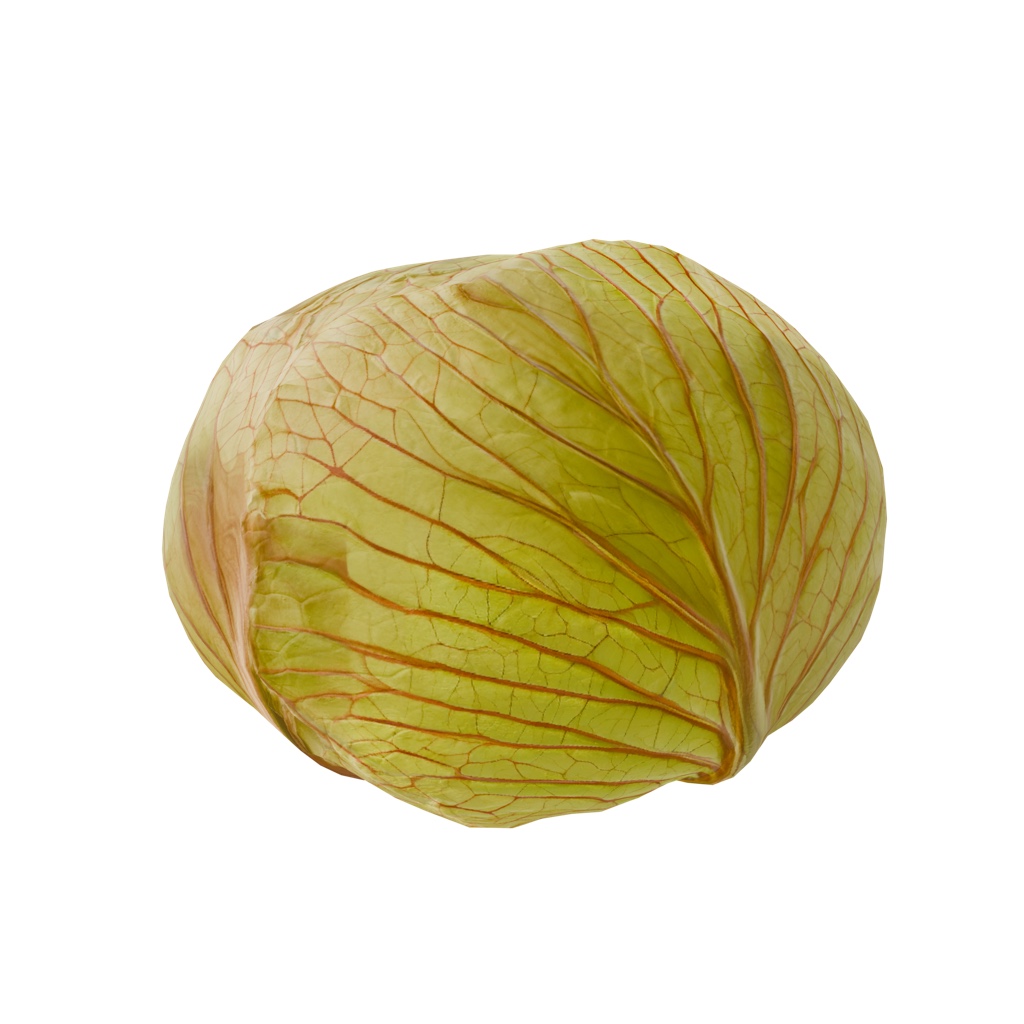} &
        \includegraphics[trim={0cm 0cm 0cm 0cm}, clip, width=\viewwidth, valign=m]{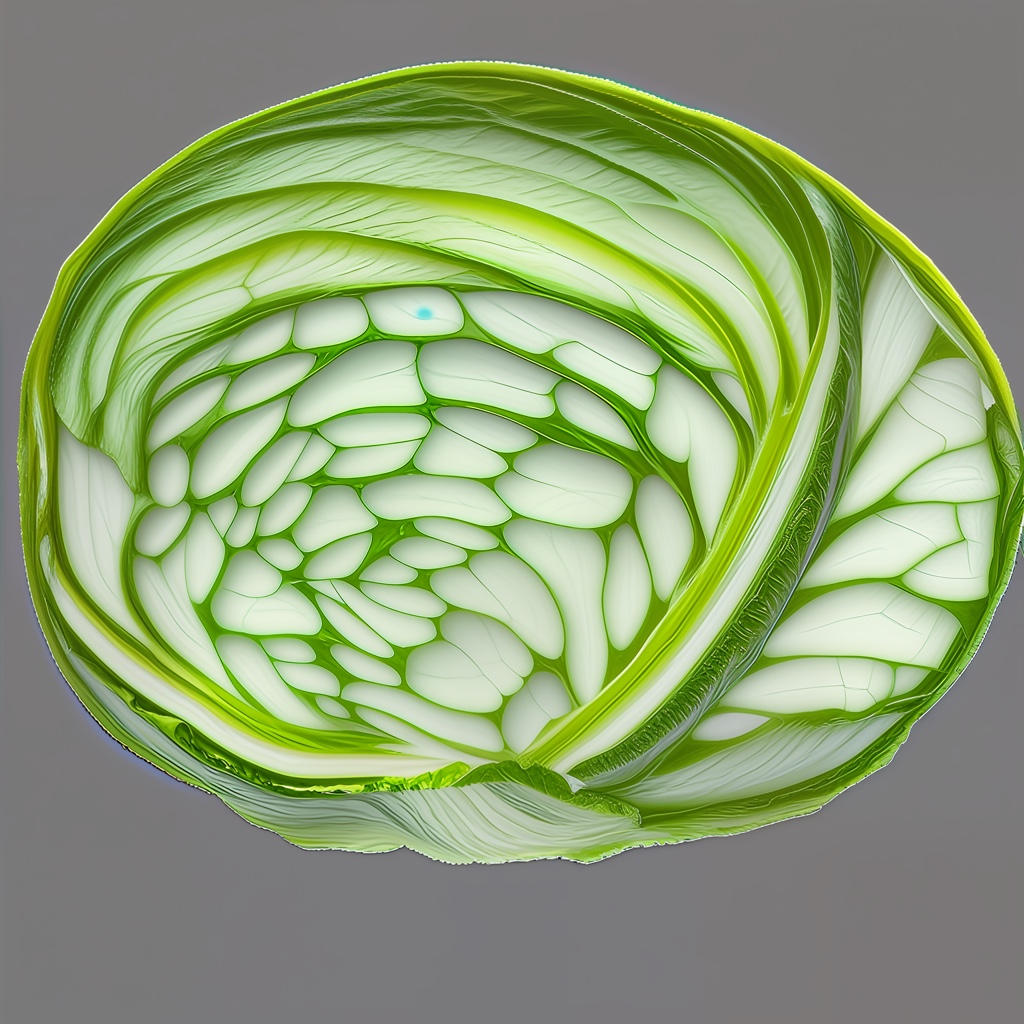} &
        \includegraphics[trim={0cm 0cm 0cm 0cm}, clip, trim={0cm 0cm 0cm 0cm}, clip, width=0.121\linewidth, valign=m, valign=m]{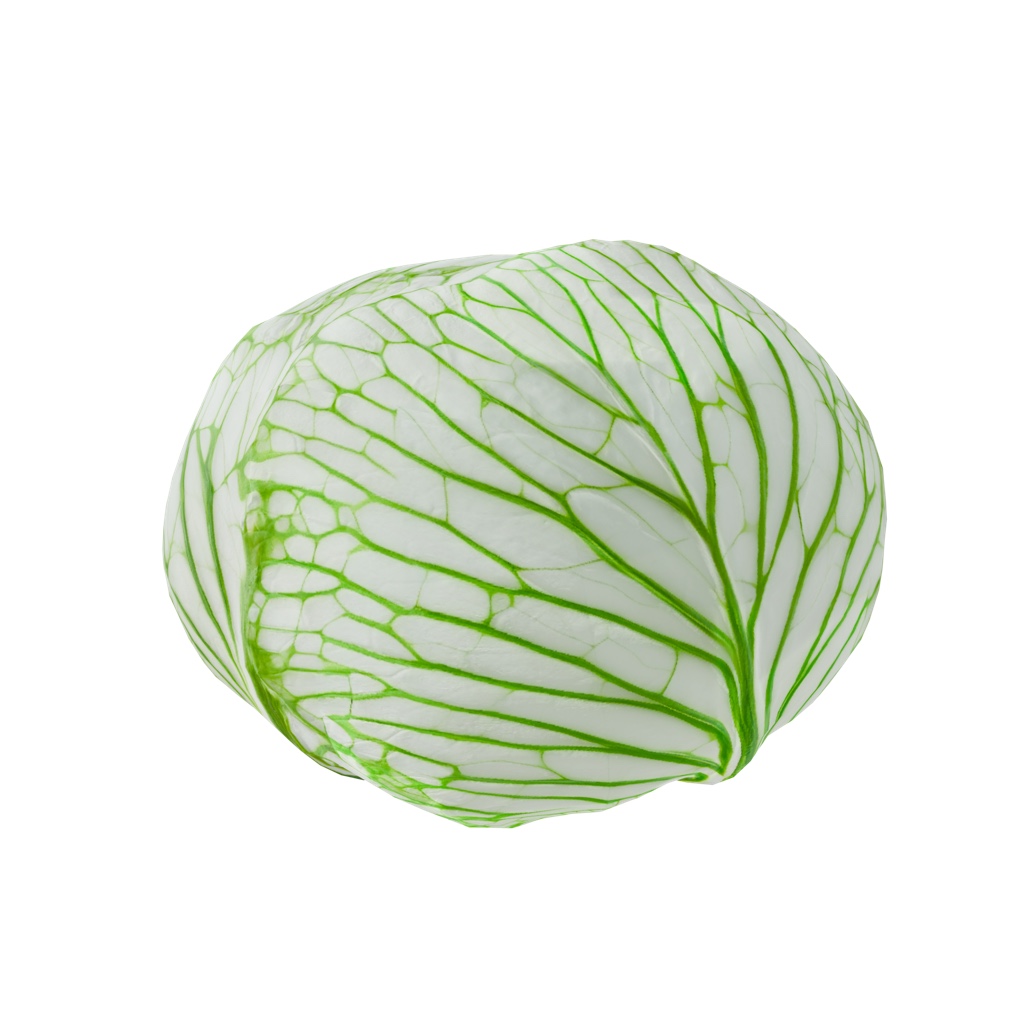} &
        \includegraphics[trim={0cm 0cm 0cm 0cm}, clip, trim={0cm 0cm 0cm 0cm}, clip, width=0.121\linewidth, valign=m, valign=m]{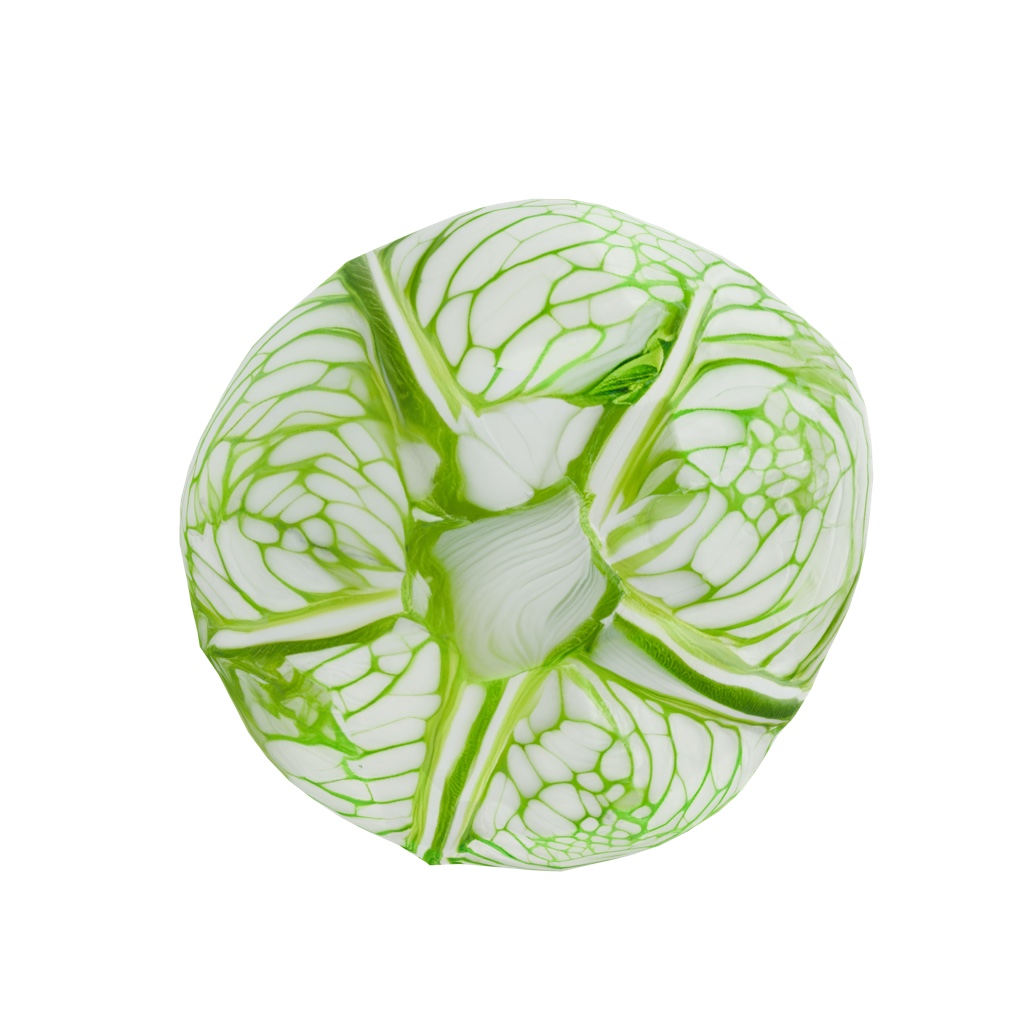} \\
    
        \includegraphics[trim={0cm 0cm 0cm 0cm}, clip, width=\normalswidth, valign=m]{img/normals/croissant.jpg} &
        \includegraphics[trim={0cm 0cm 0cm 0cm}, clip, width=\viewwidth, valign=m]{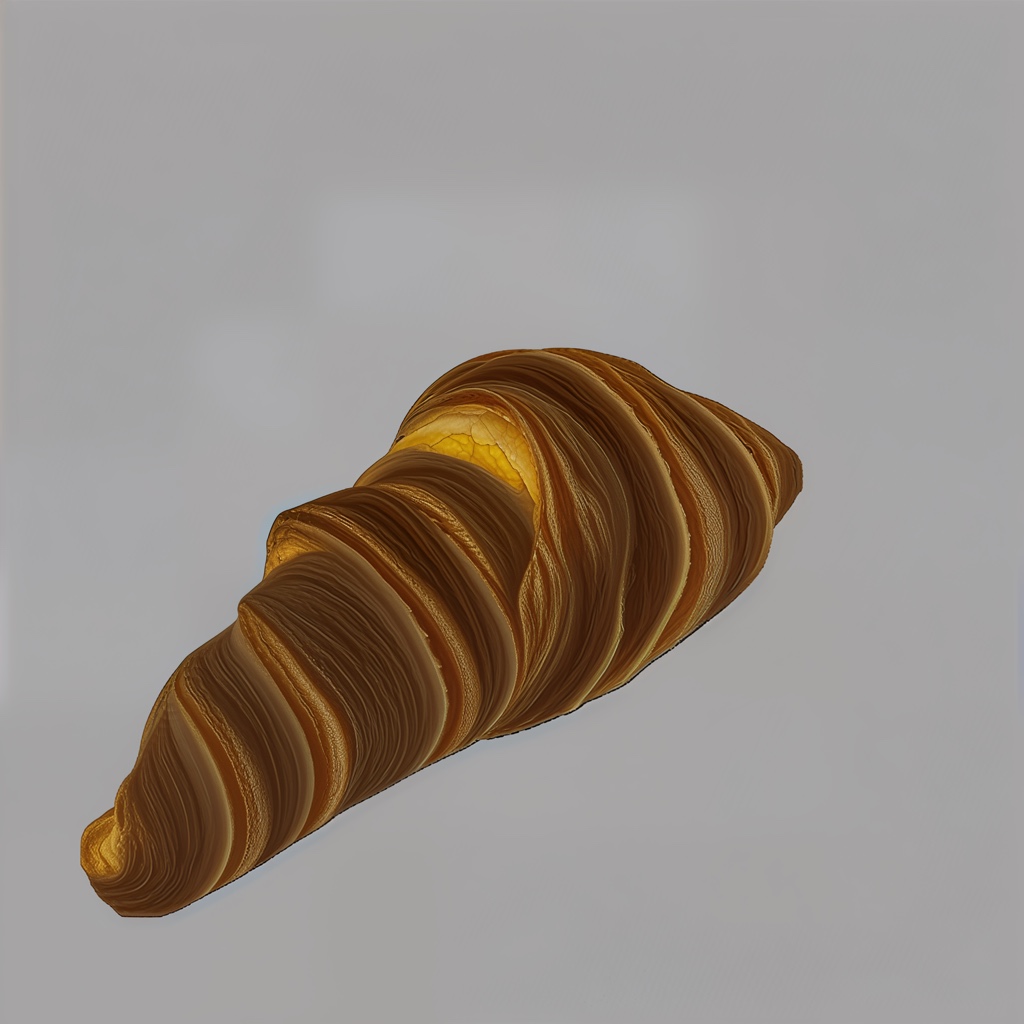} &
        \includegraphics[trim={0cm 0cm 0cm 0cm}, clip, trim={0cm 0cm 0cm 0cm}, clip, width=0.121\linewidth, valign=m, valign=m]{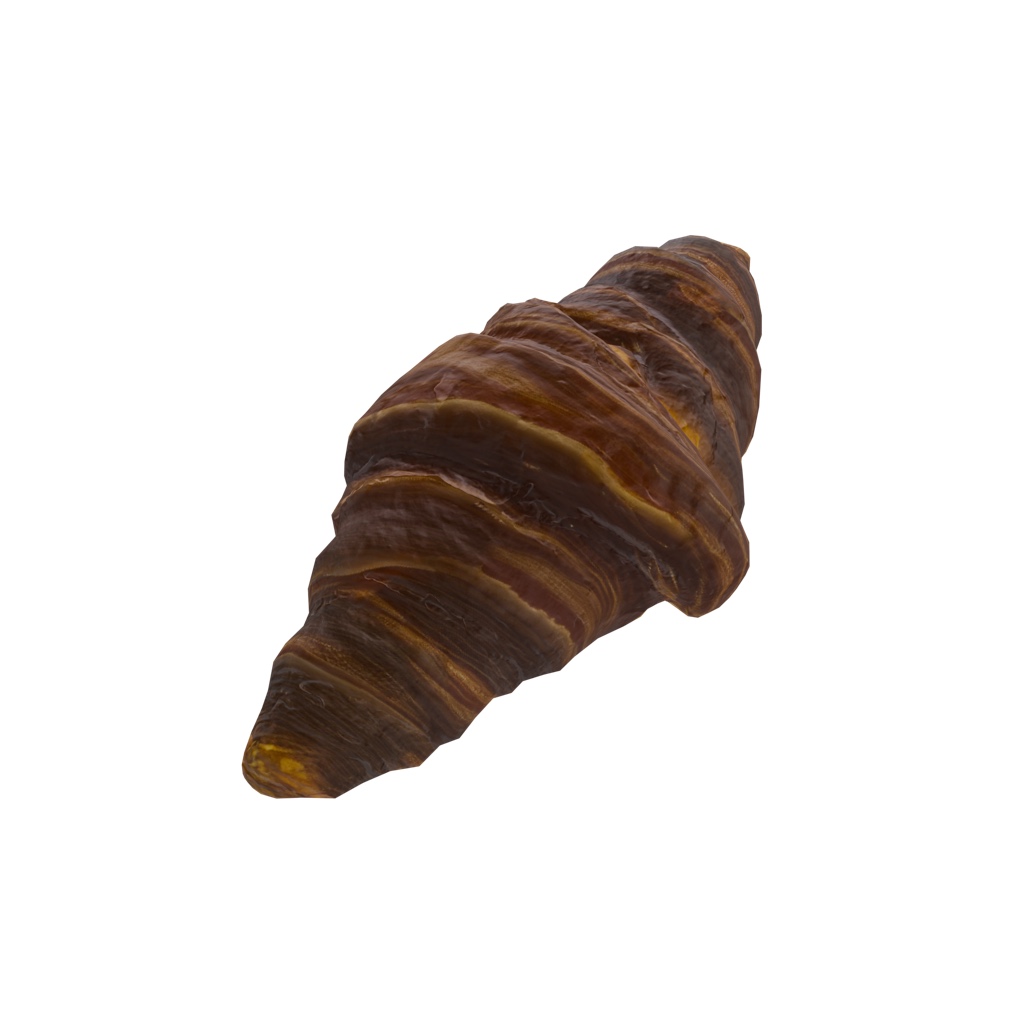} &
        \includegraphics[trim={0cm 0cm 0cm 0cm}, clip, trim={0cm 0cm 0cm 0cm}, clip, width=0.121\linewidth, valign=m, valign=m]{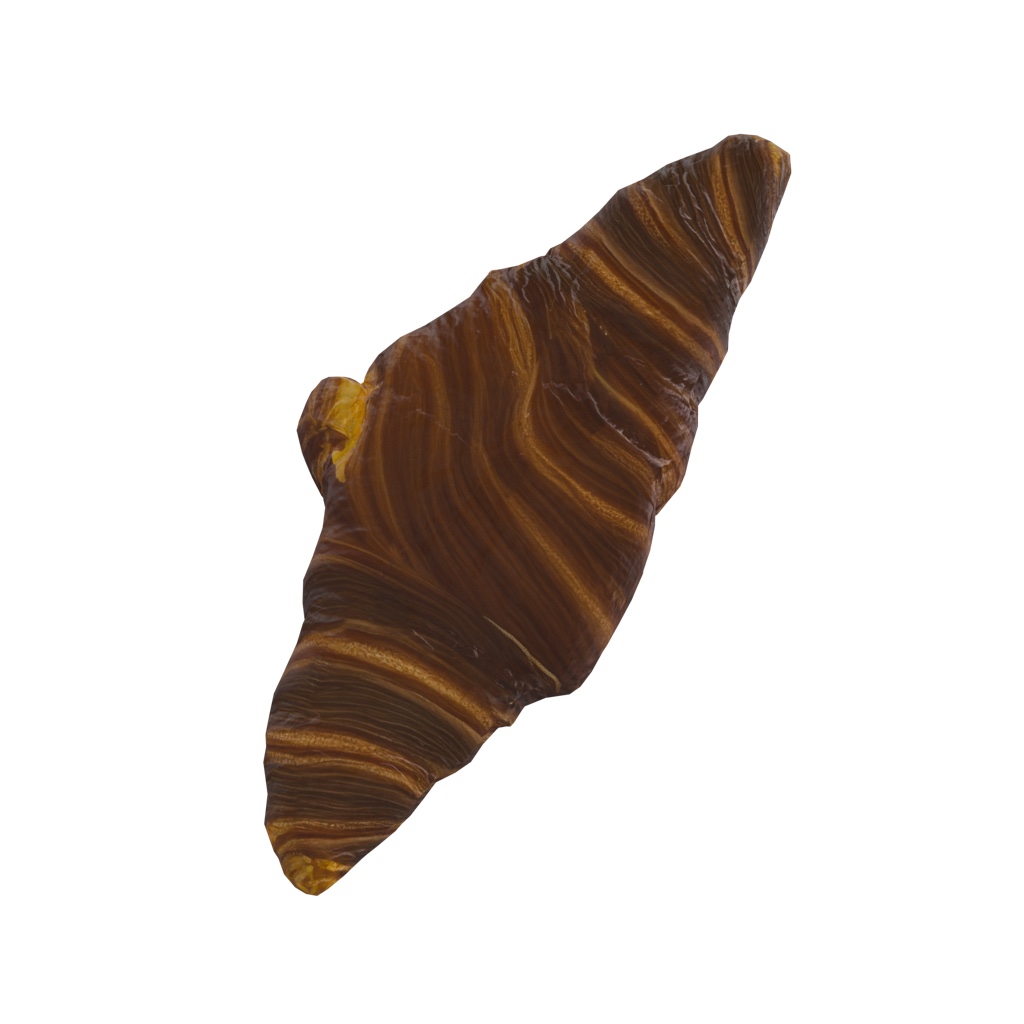} &
        \includegraphics[trim={0cm 0cm 0cm 0cm}, clip, width=\viewwidth, valign=m]{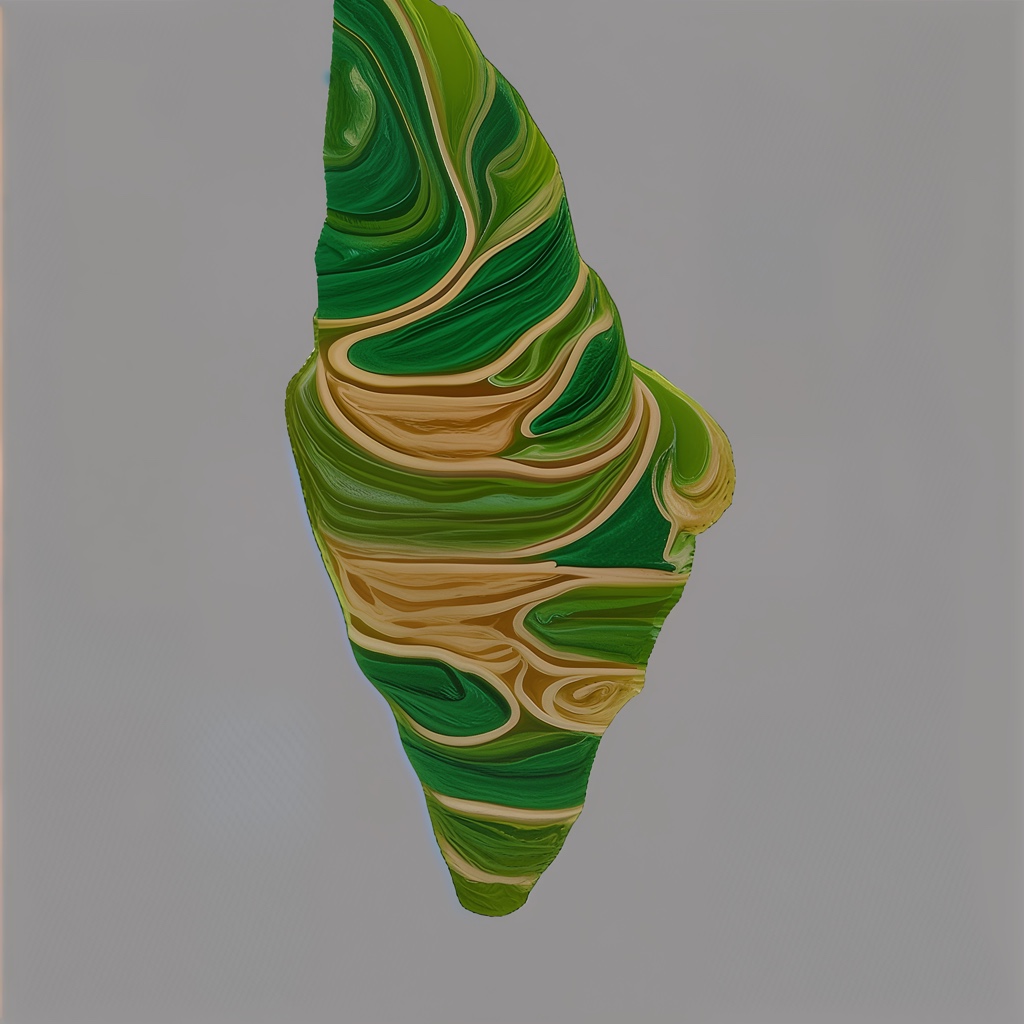} &
        \includegraphics[trim={0cm 0cm 0cm 0cm}, clip, trim={0cm 0cm 0cm 0cm}, clip, width=0.121\linewidth, valign=m, valign=m]{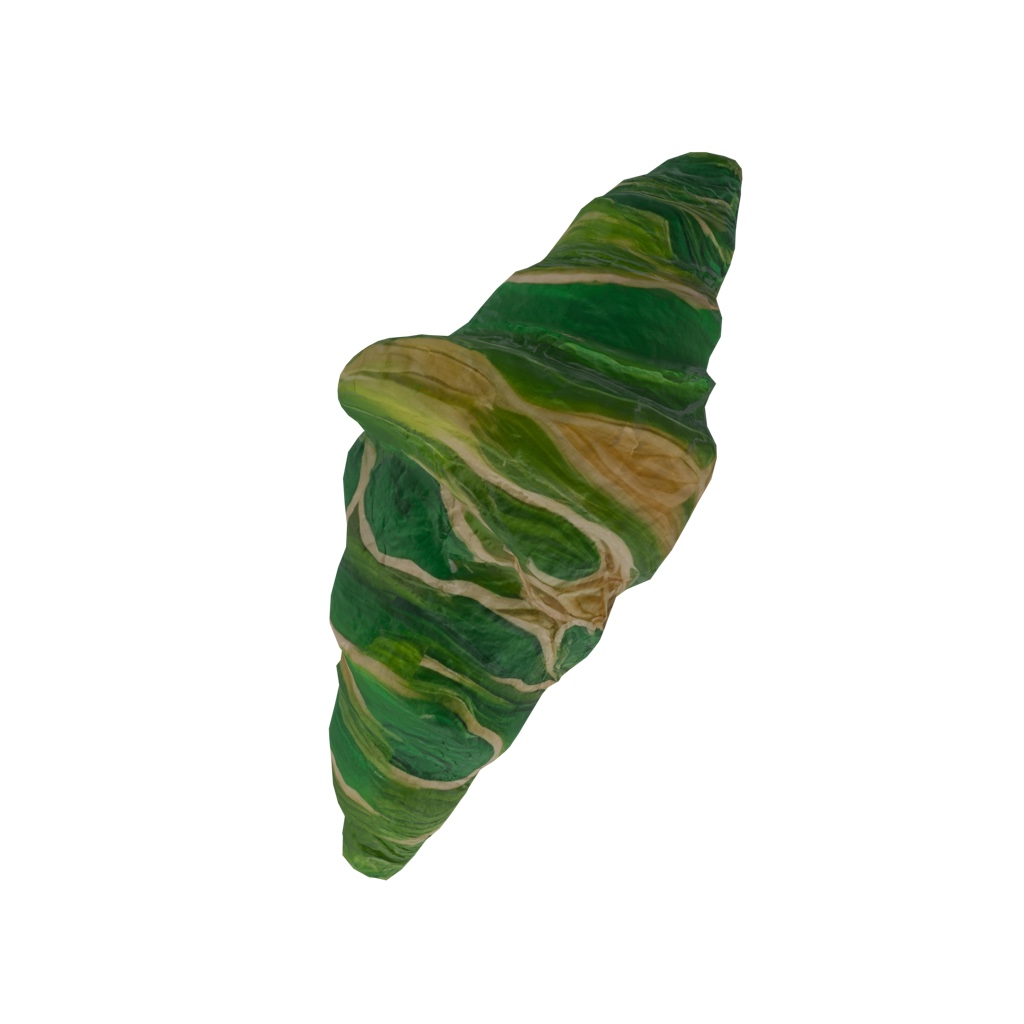} &
        \includegraphics[trim={0cm 0cm 0cm 0cm}, clip, trim={0cm 0cm 0cm 0cm}, clip, width=0.121\linewidth, valign=m, valign=m]{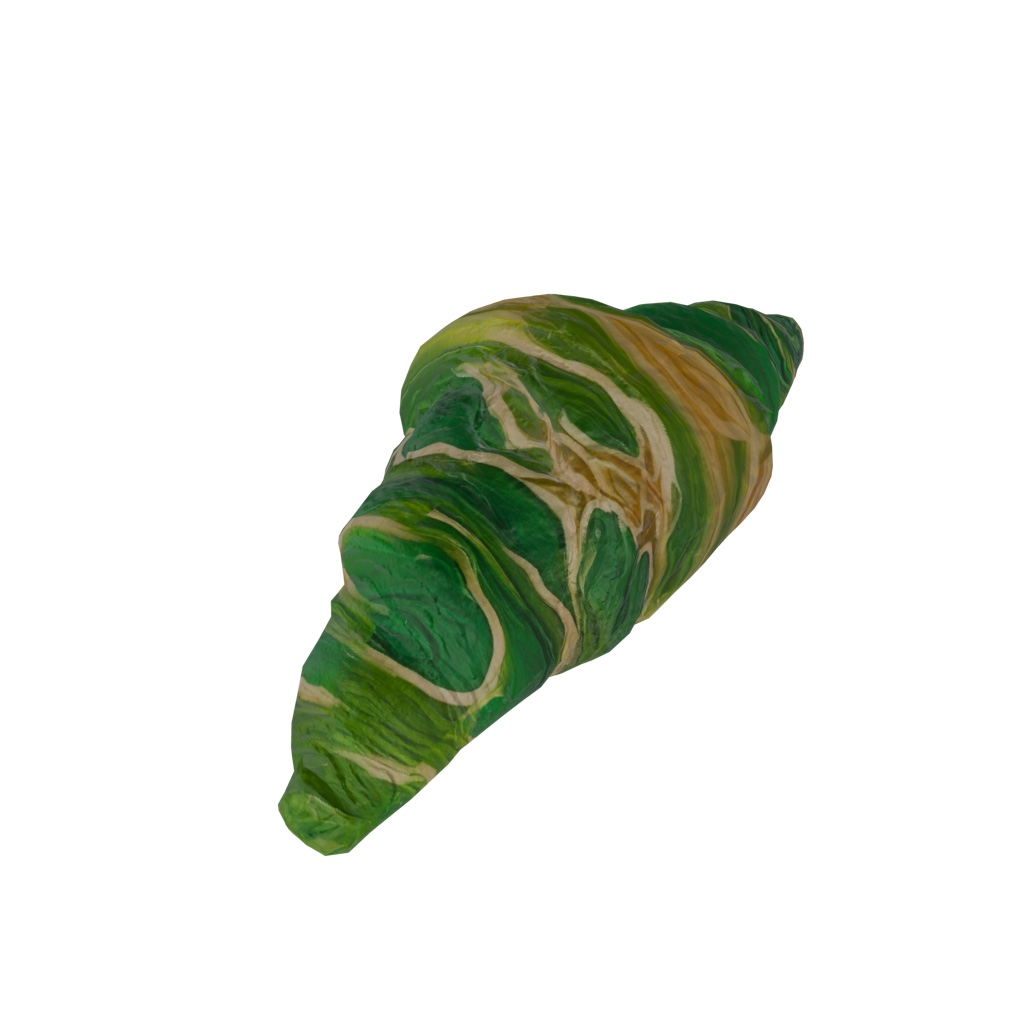} \\
    
        \includegraphics[trim={0cm 0cm 0cm 0cm}, clip, width=\normalswidth, valign=m]{img/normals/gourd.jpg} &
        \includegraphics[trim={0cm 0cm 0cm 0cm}, clip, width=\viewwidth, valign=m]{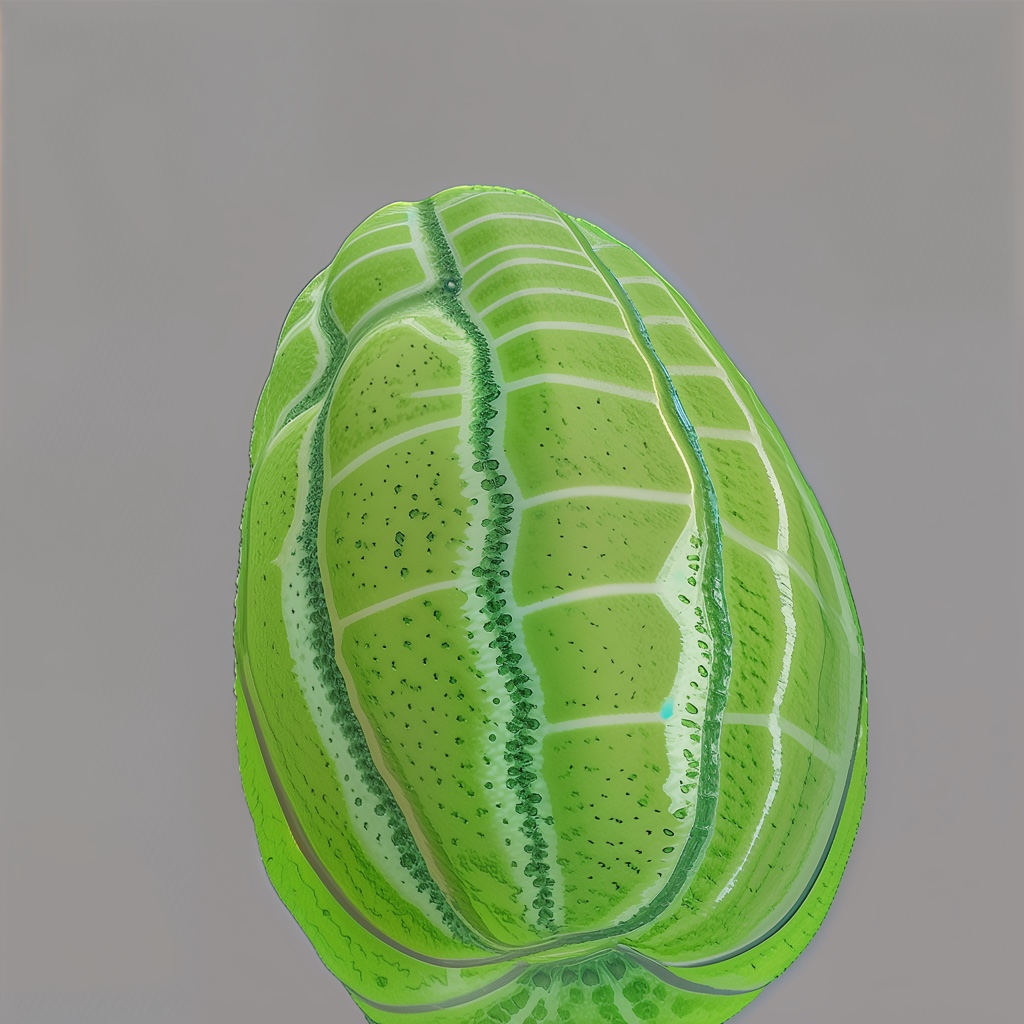} &
        \includegraphics[trim={0cm 0cm 0cm 0cm}, clip, width=0.121\linewidth, valign=m]{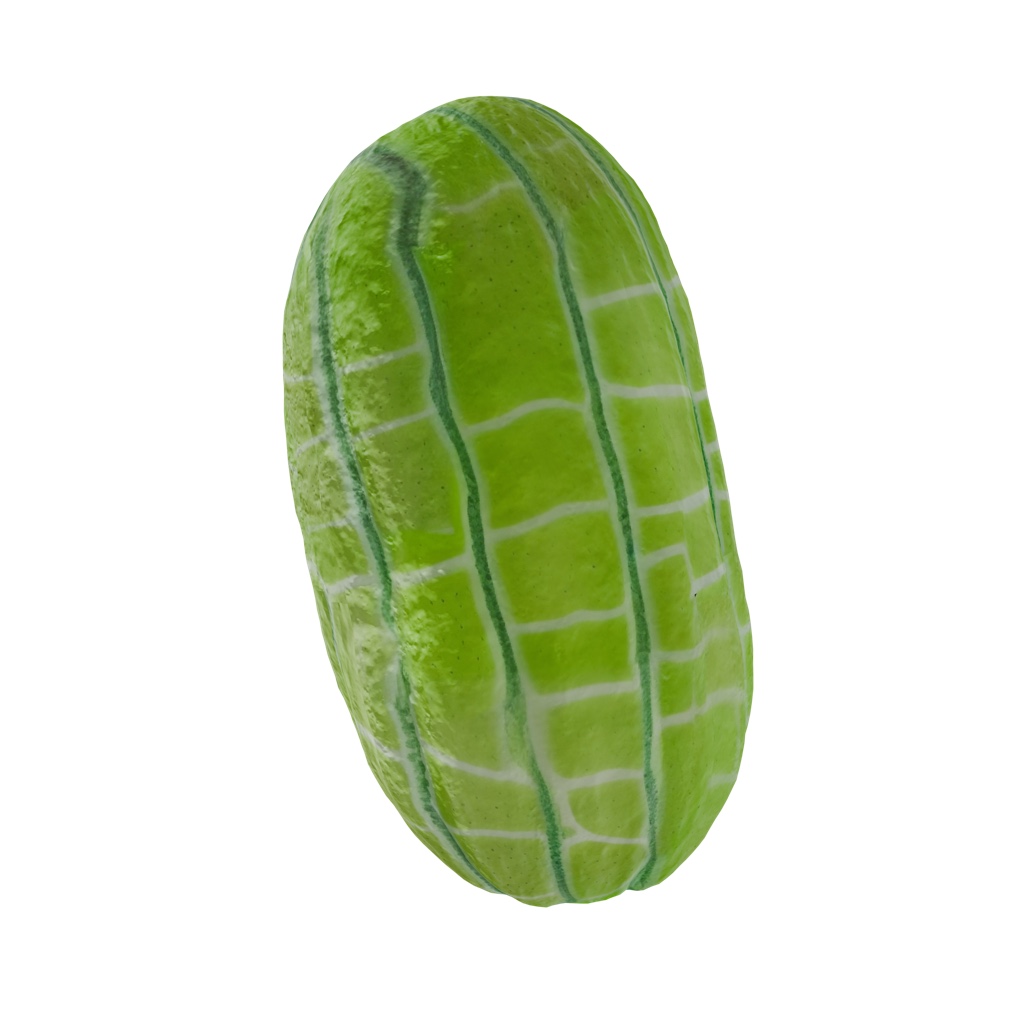} &
        \includegraphics[trim={0cm 0cm 0cm 0cm}, clip, width=0.121\linewidth, valign=m]{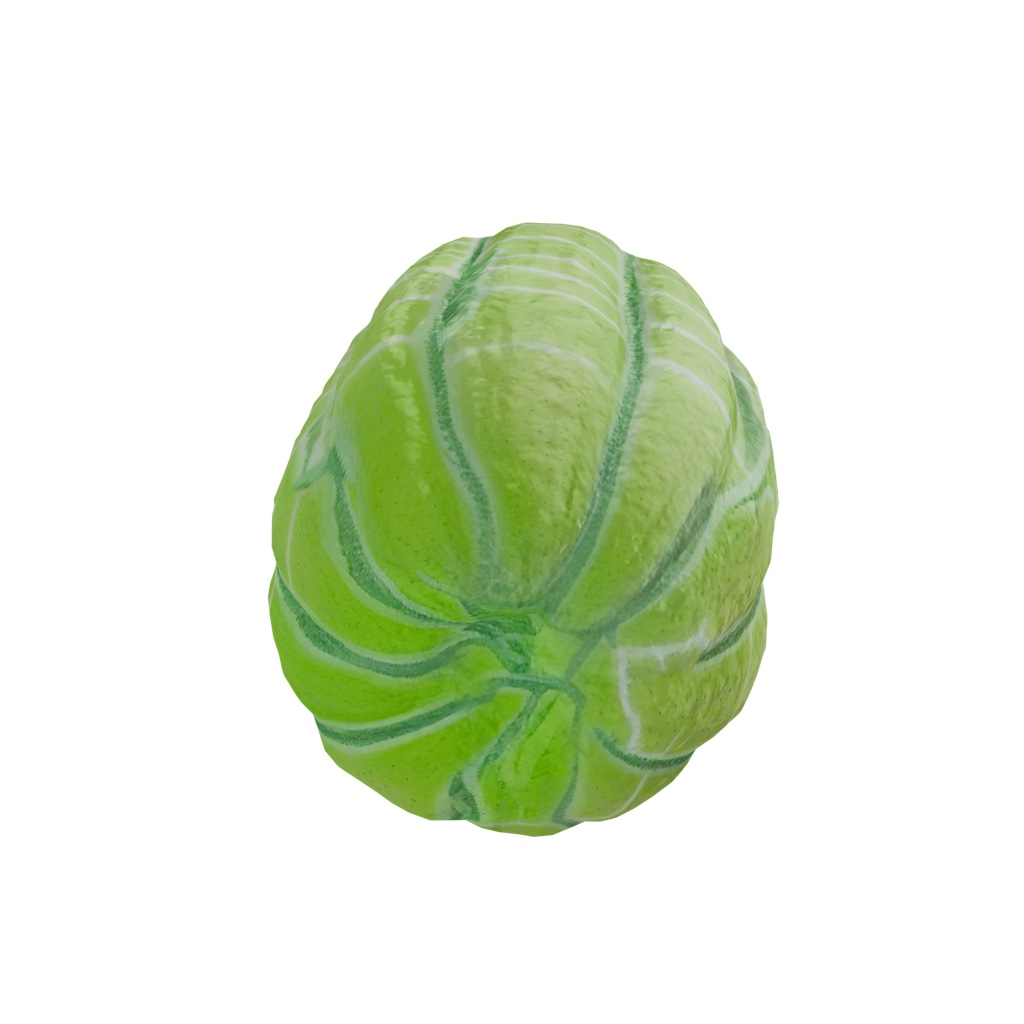} &
        \includegraphics[trim={0cm 0cm 0cm 0cm}, clip, width=\viewwidth, valign=m]{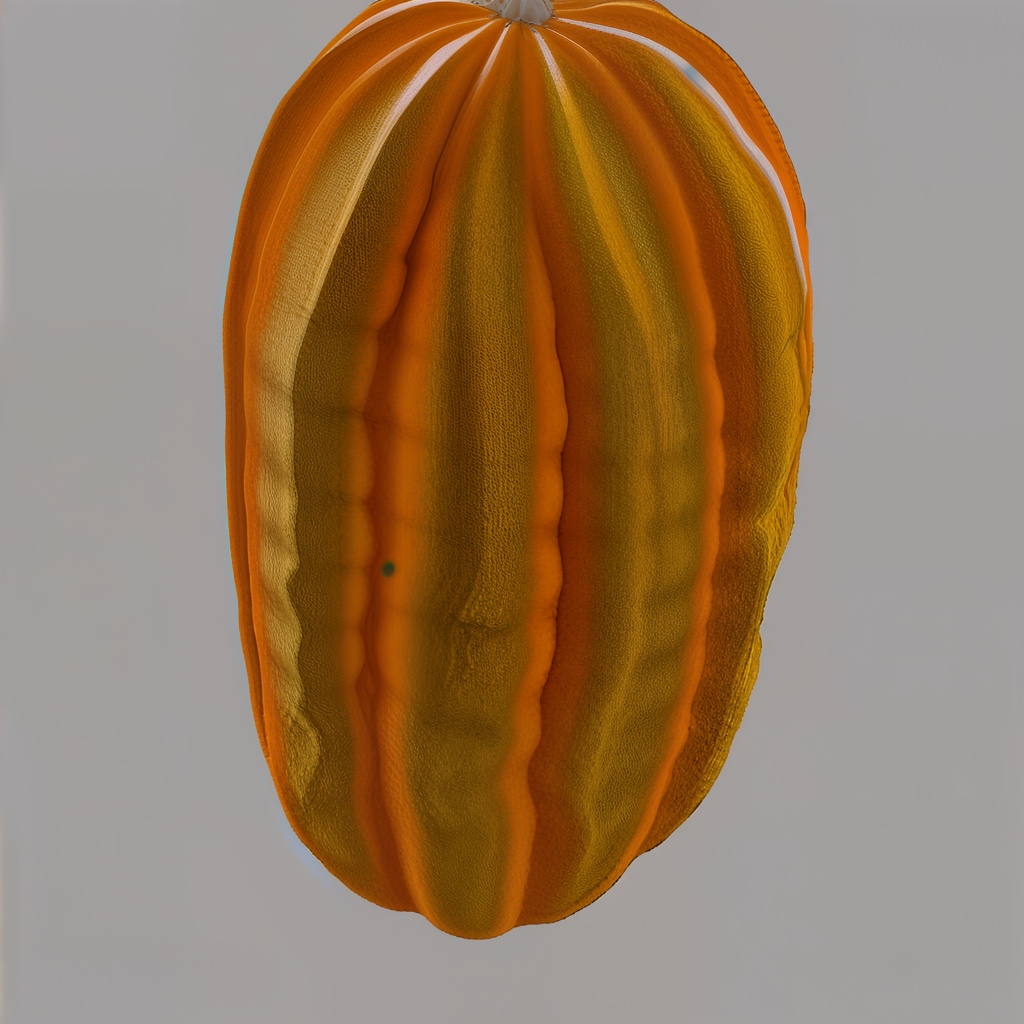} &
        \includegraphics[trim={0cm 0cm 0cm 0cm}, clip, width=0.121\linewidth, valign=m]{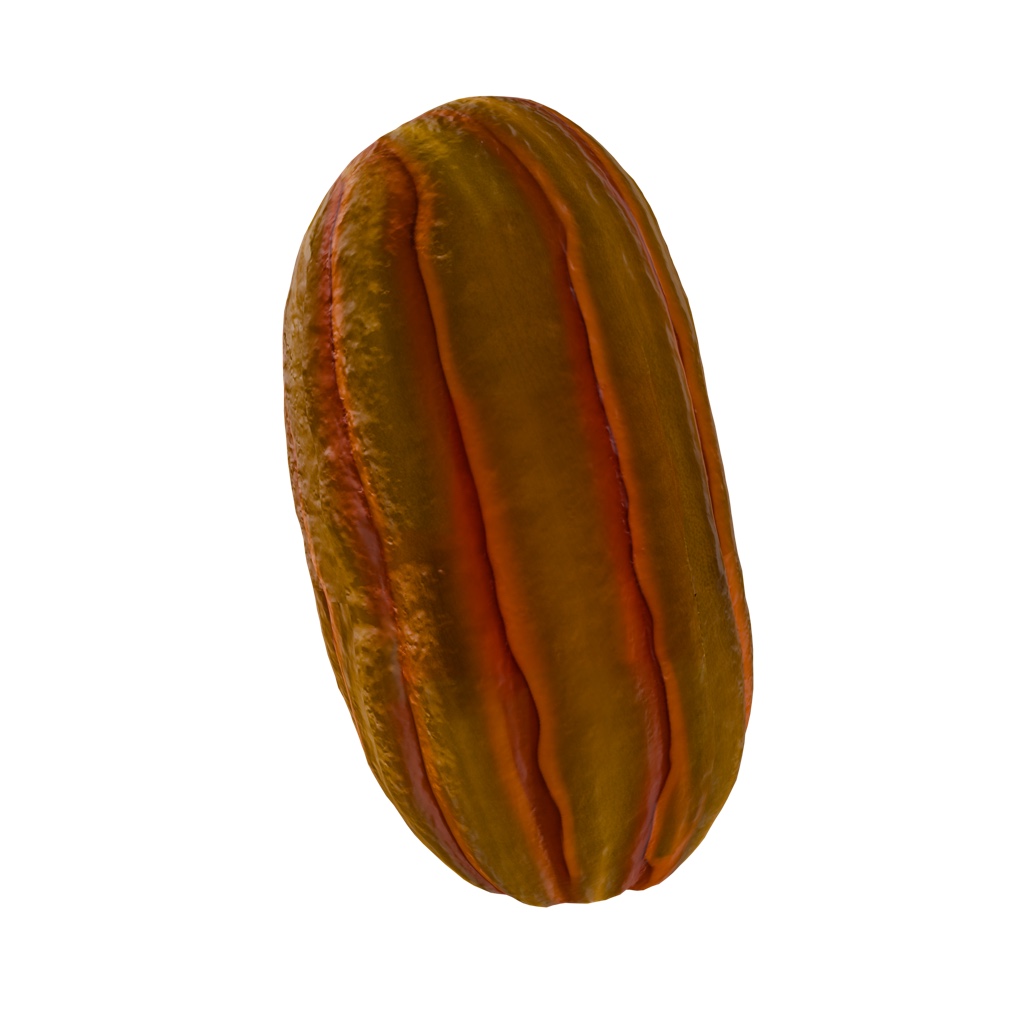} &
        \includegraphics[trim={0cm 0cm 0cm 0cm}, clip, width=0.121\linewidth, valign=m]{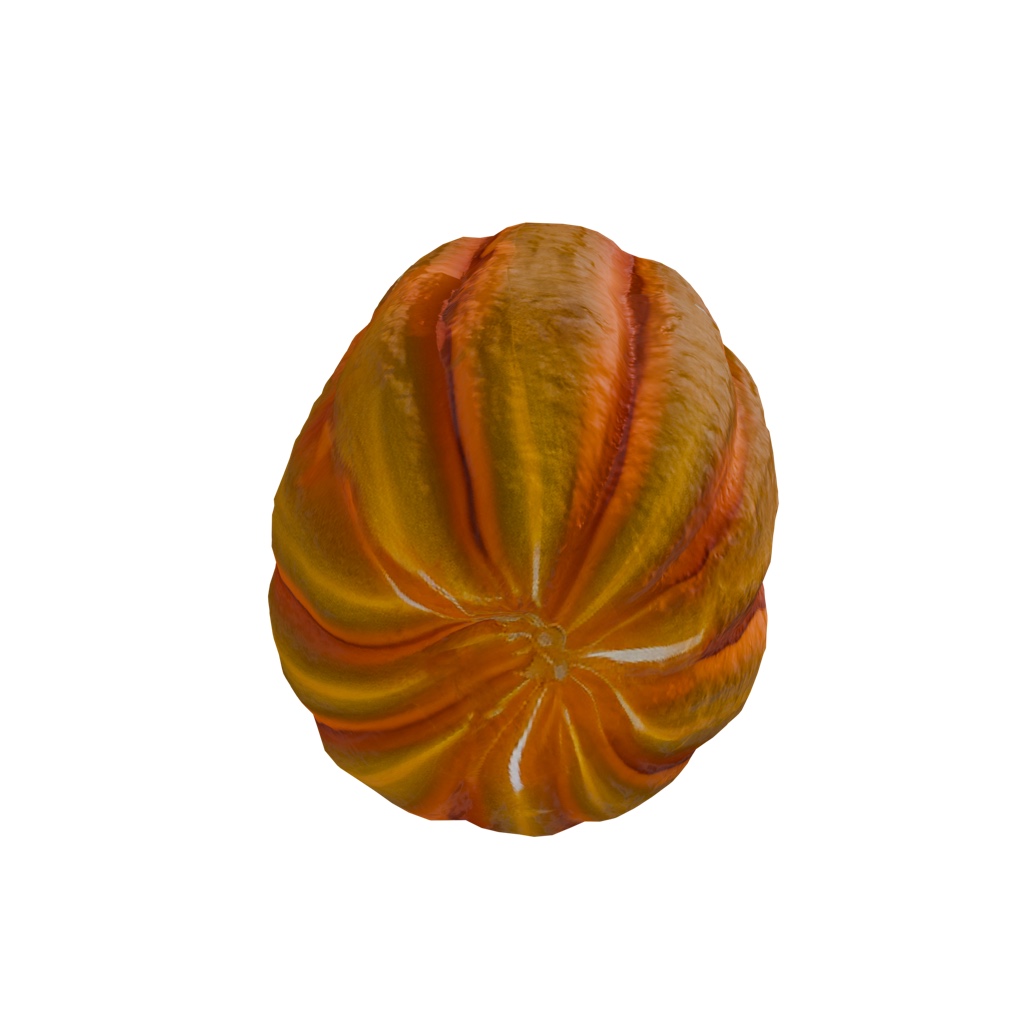} \\
    
      \includegraphics[trim={0cm 0cm 0cm 0cm}, clip, width=\normalswidth, valign=m]{img/normals/sea_urchin_shell.jpg} &
      \includegraphics[trim={0cm 0cm 0cm 0cm}, clip, width=\viewwidth, valign=m]{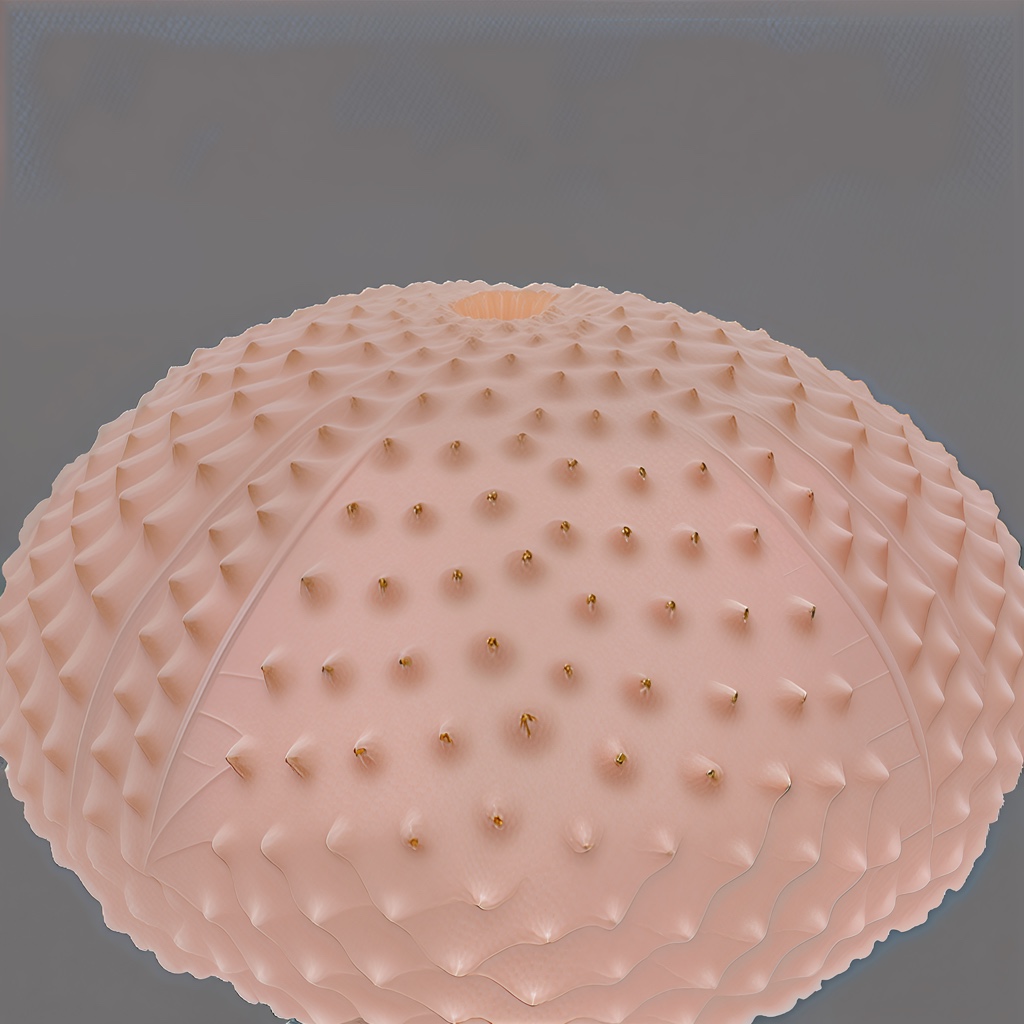} &
      \includegraphics[trim={6cm 6cm 6cm 6cm}, clip, width=0.121\linewidth, valign=m]{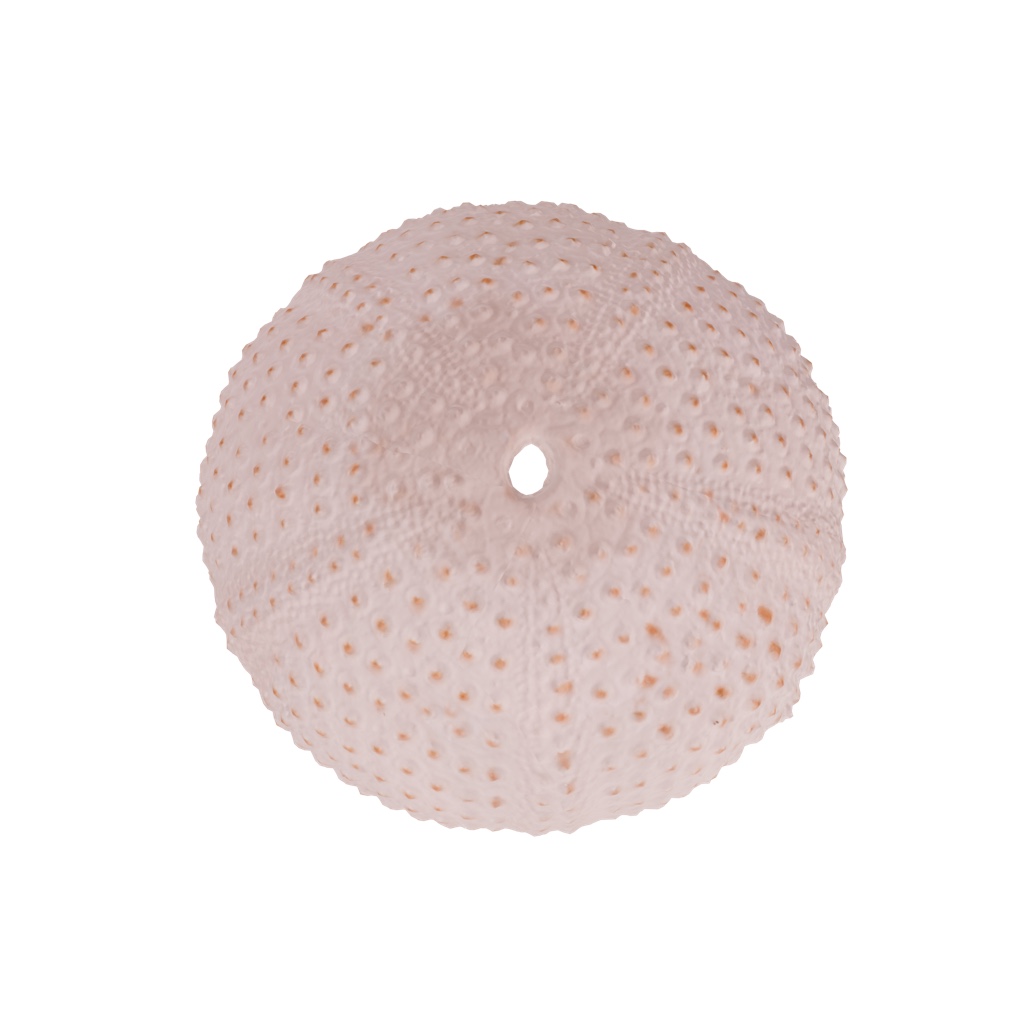} &
      \includegraphics[trim={6cm 6cm 6cm 6cm}, clip, width=0.121\linewidth, valign=m]{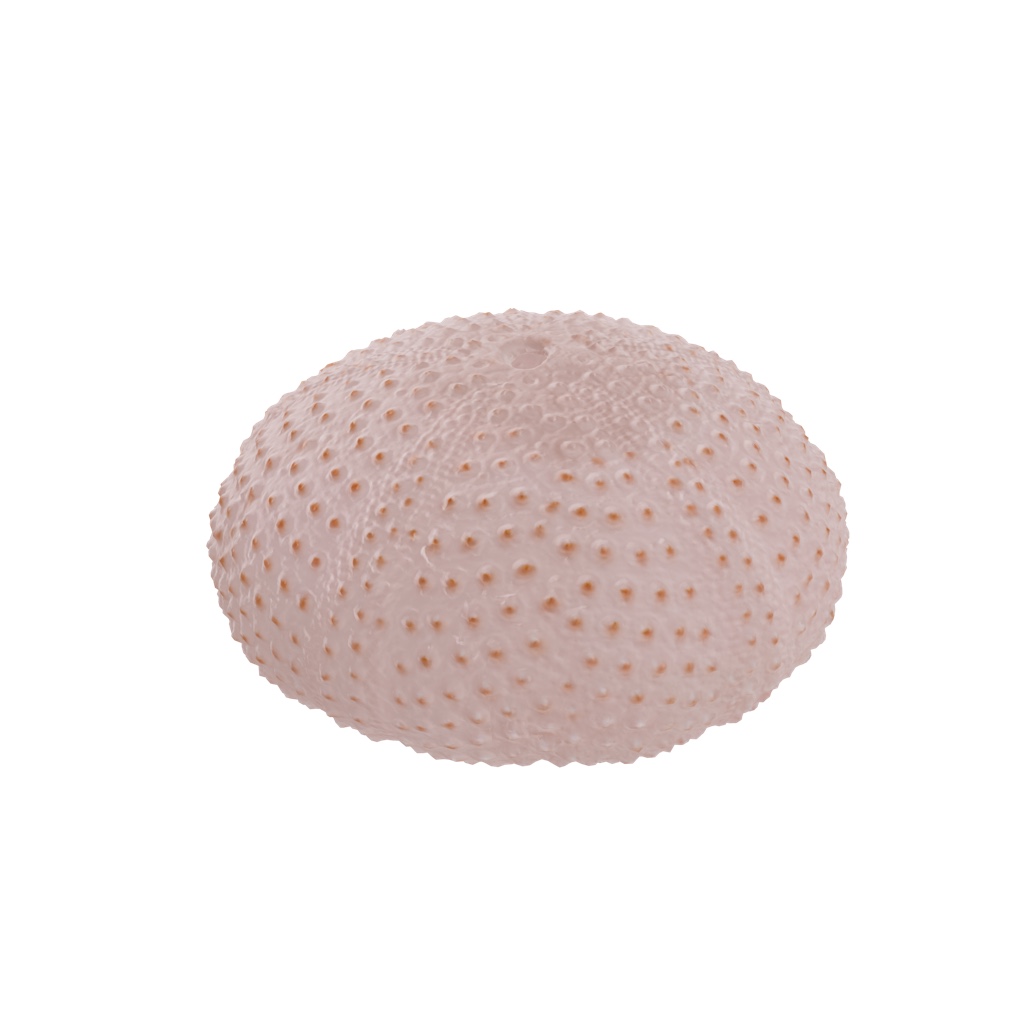} &
      \includegraphics[trim={0cm 0cm 0cm 0cm}, clip, width=\viewwidth, valign=m]{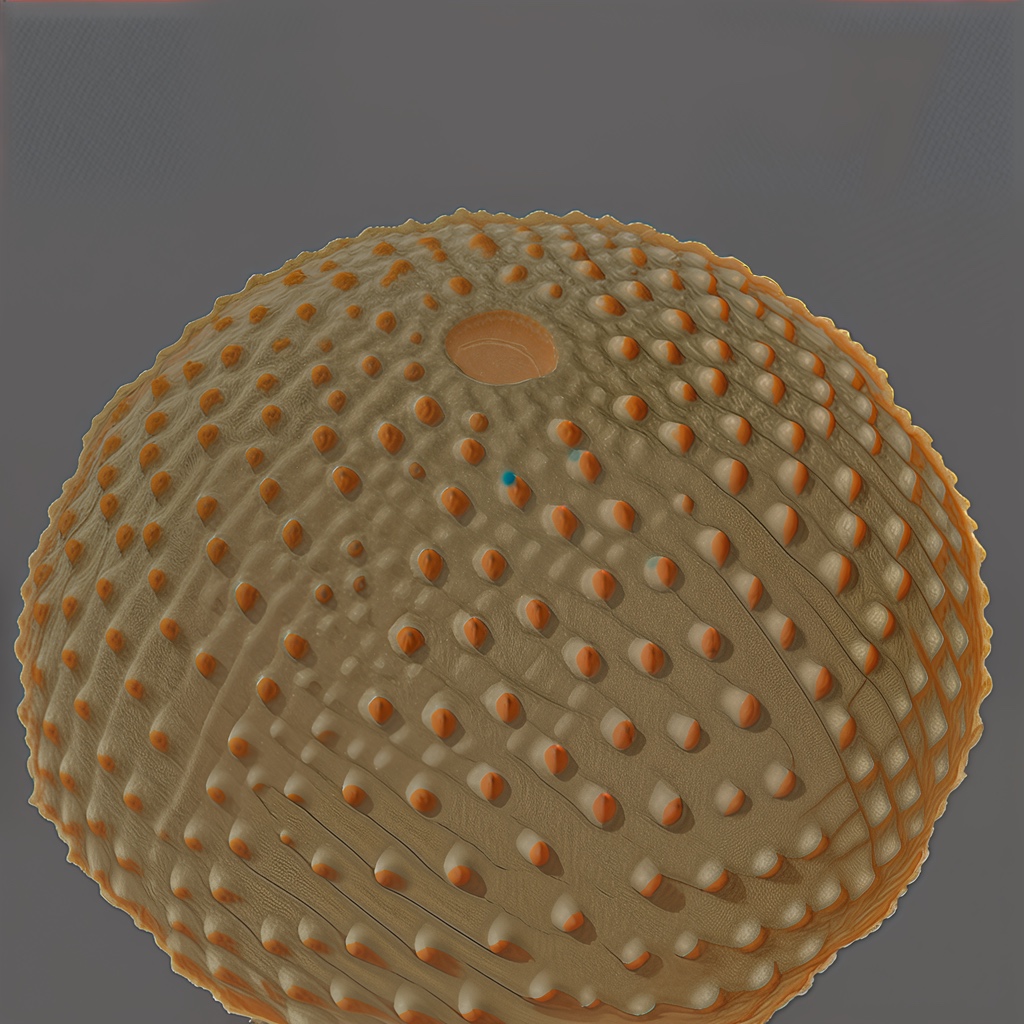} &
      \includegraphics[trim={6cm 6cm 6cm 6cm}, clip, width=0.121\linewidth, valign=m]{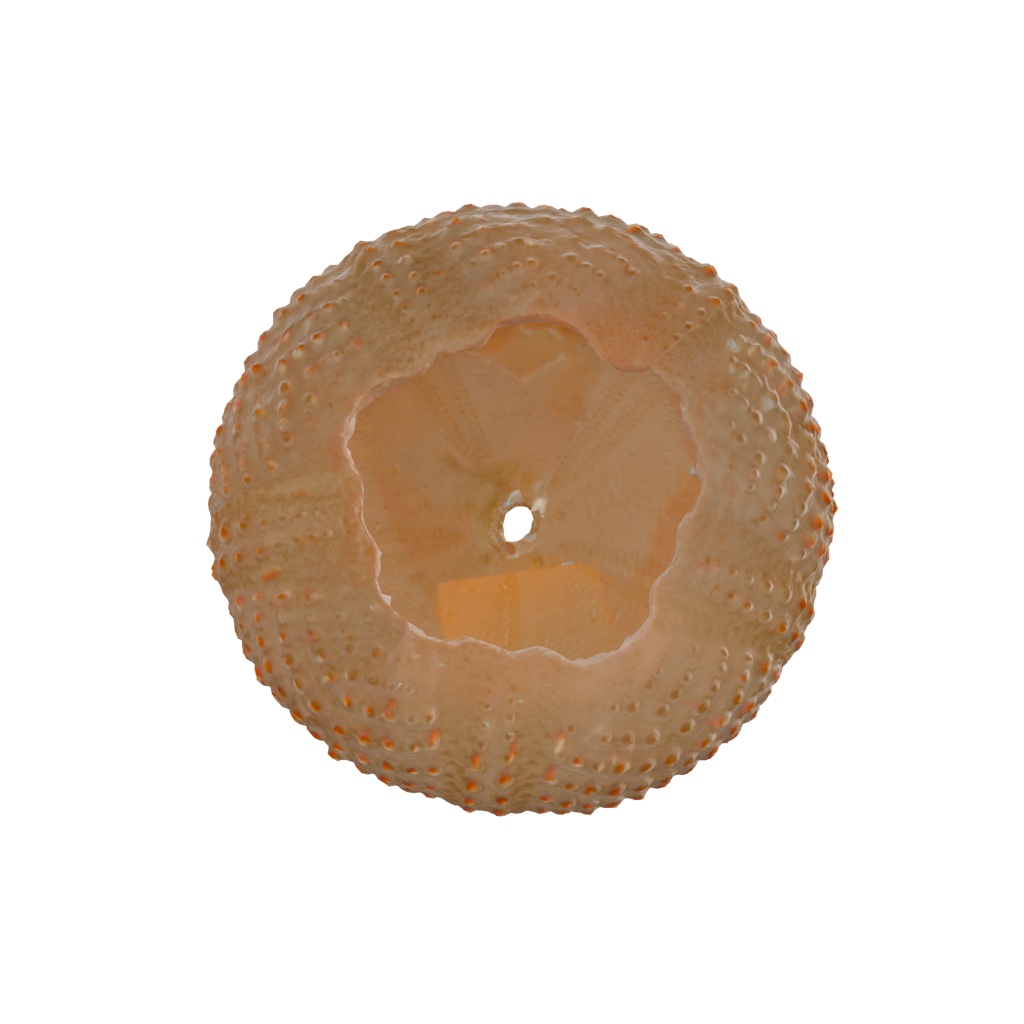} &
      \includegraphics[trim={6cm 6cm 6cm 6cm}, clip, width=0.121\linewidth, valign=m]{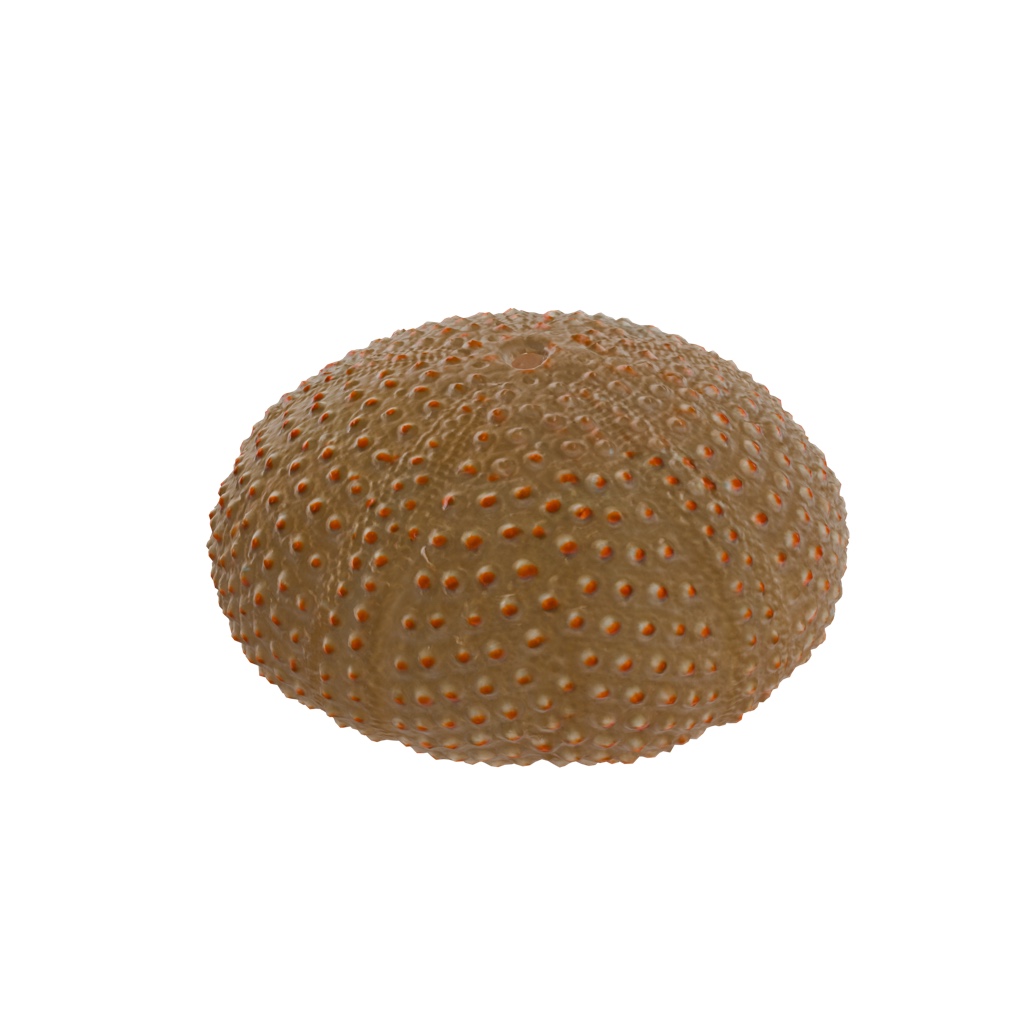} \\
    
      \includegraphics[trim={0cm 0cm 0cm 0cm}, clip, width=\normalswidth, valign=m]{img/normals/turtle.jpg} &
      \includegraphics[trim={0cm 0cm 0cm 0cm}, clip, width=\viewwidth, valign=m]{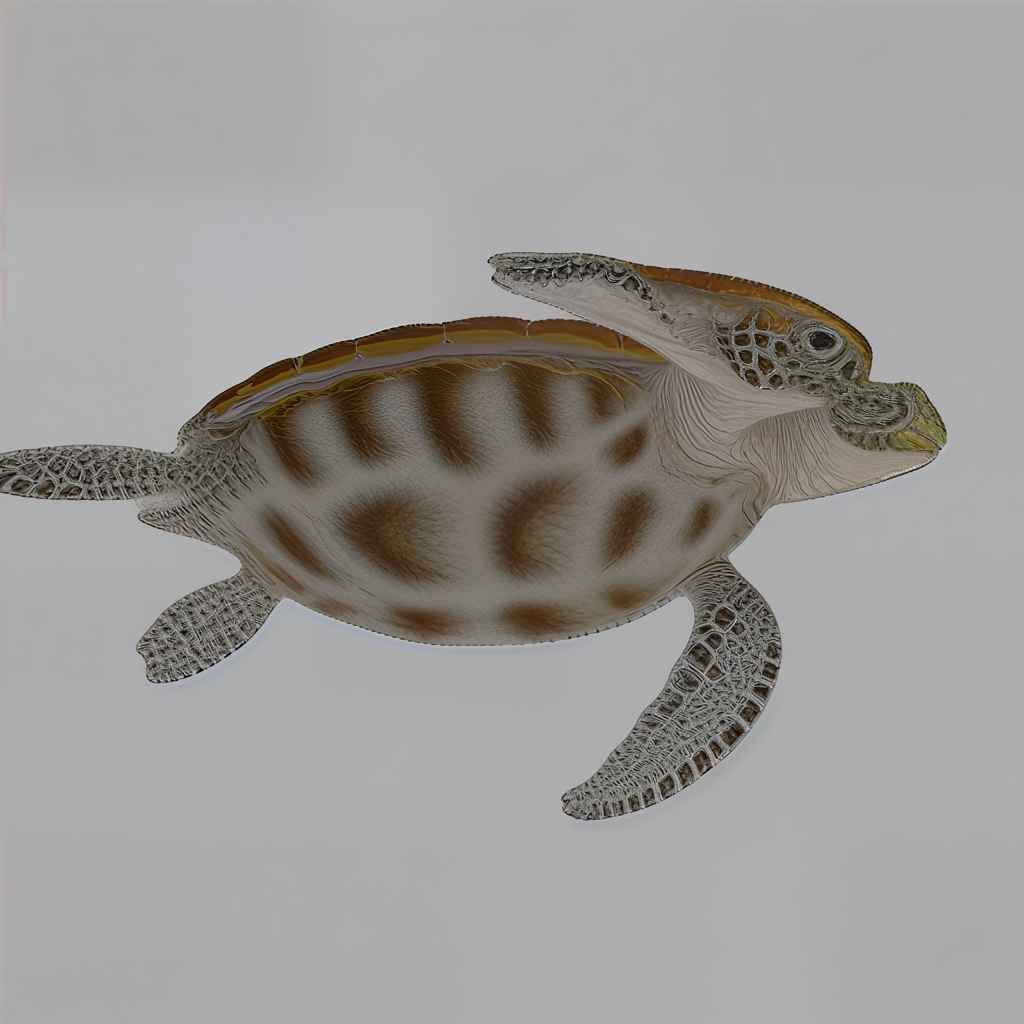} &
      \includegraphics[trim={0cm 12cm 3cm 0cm}, clip, width=0.121\linewidth, valign=m]{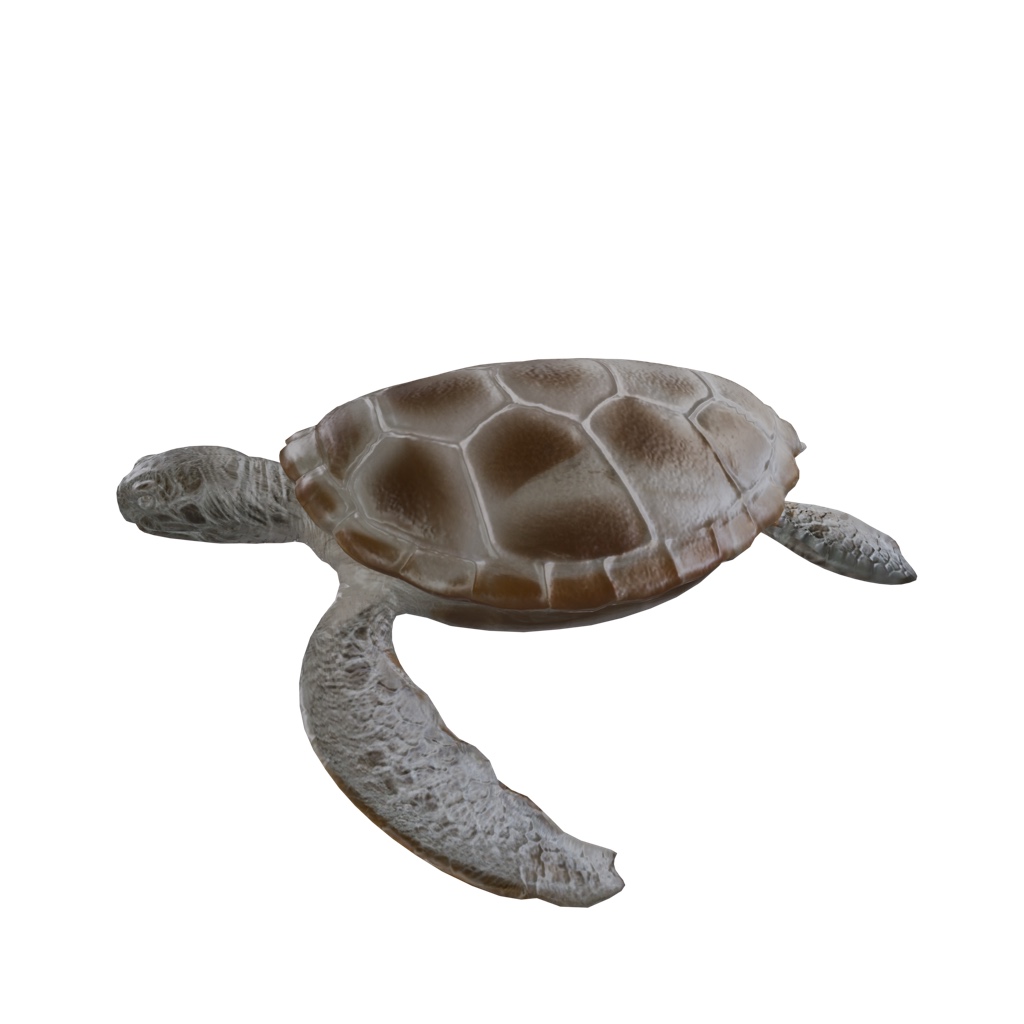} &
      \includegraphics[trim={0cm 0cm 0cm 0cm}, clip, width=0.121\linewidth, valign=m]{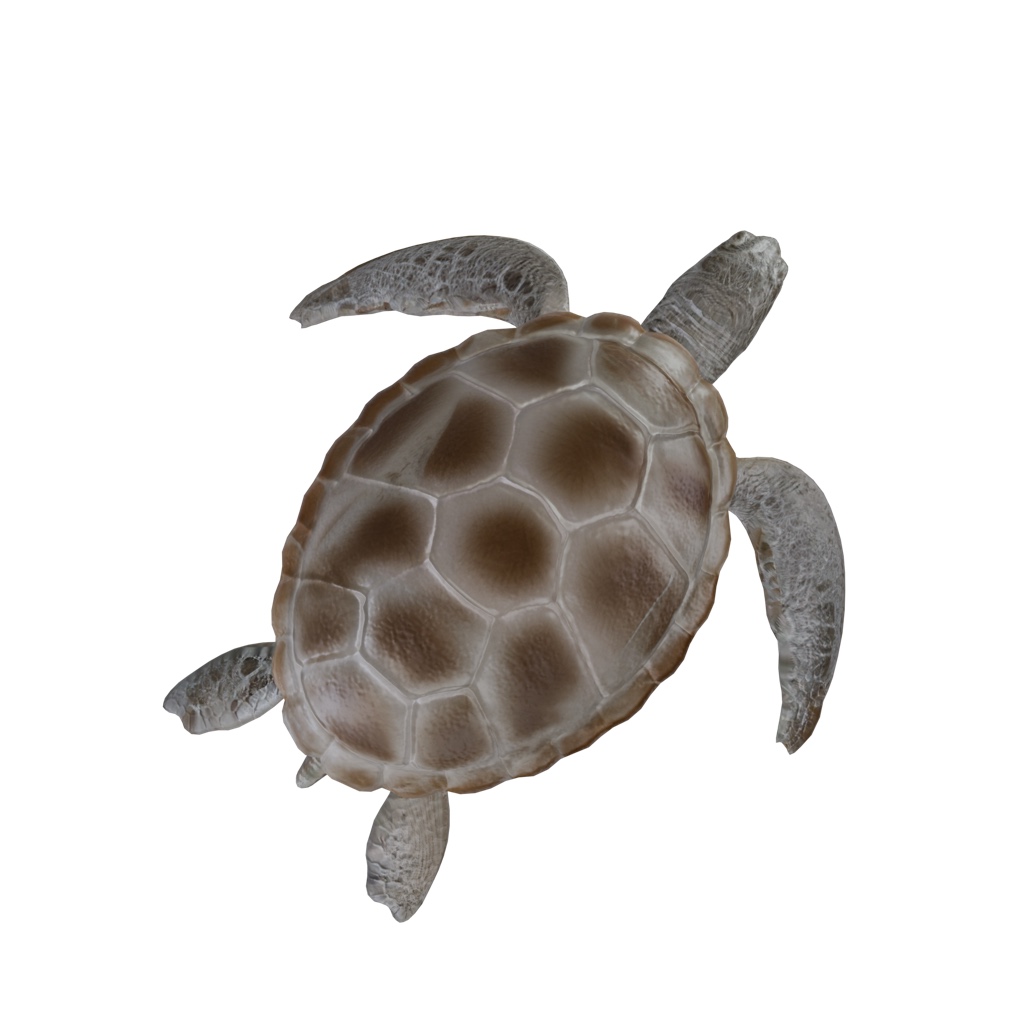} &
      \includegraphics[trim={0cm 0cm 0cm 0cm}, clip, width=\viewwidth, valign=m]{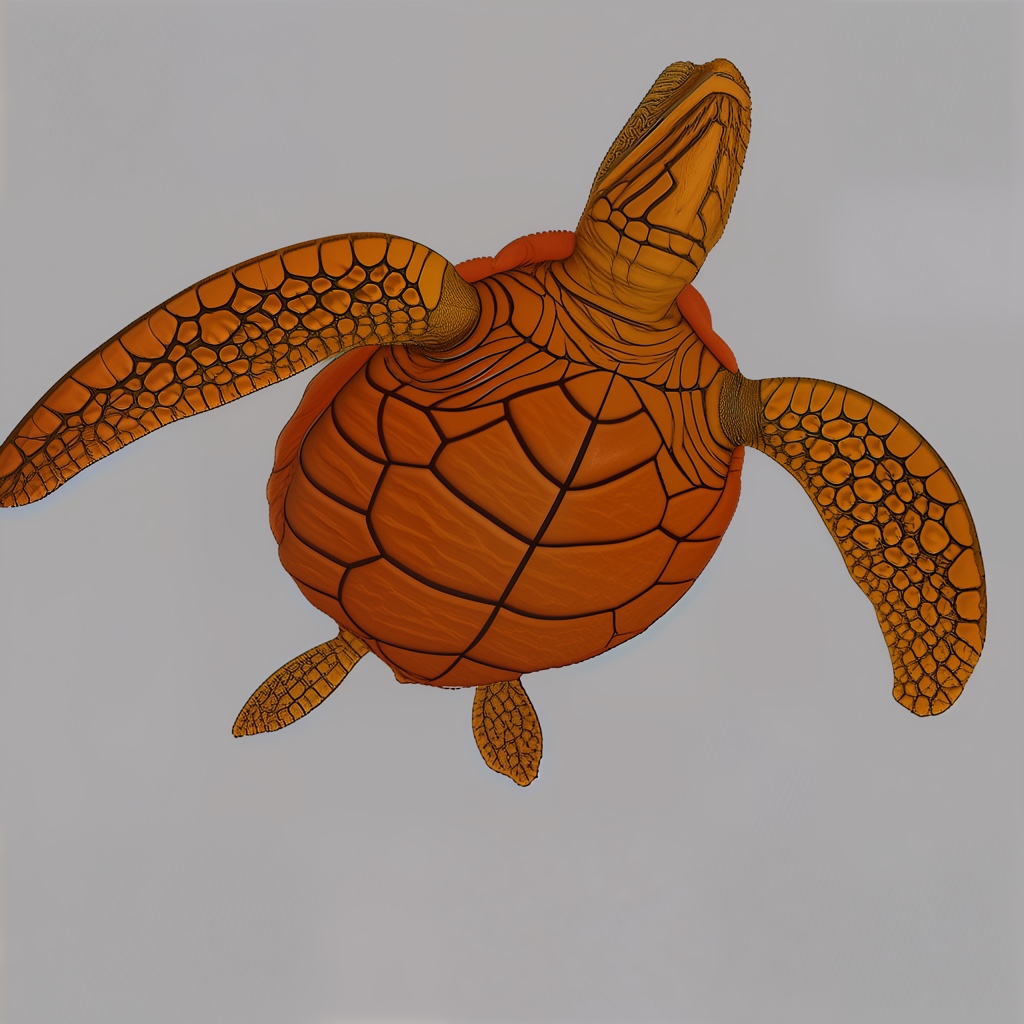} &
      \includegraphics[trim={0cm 12cm 3cm 0cm}, clip, width=0.121\linewidth, valign=m]{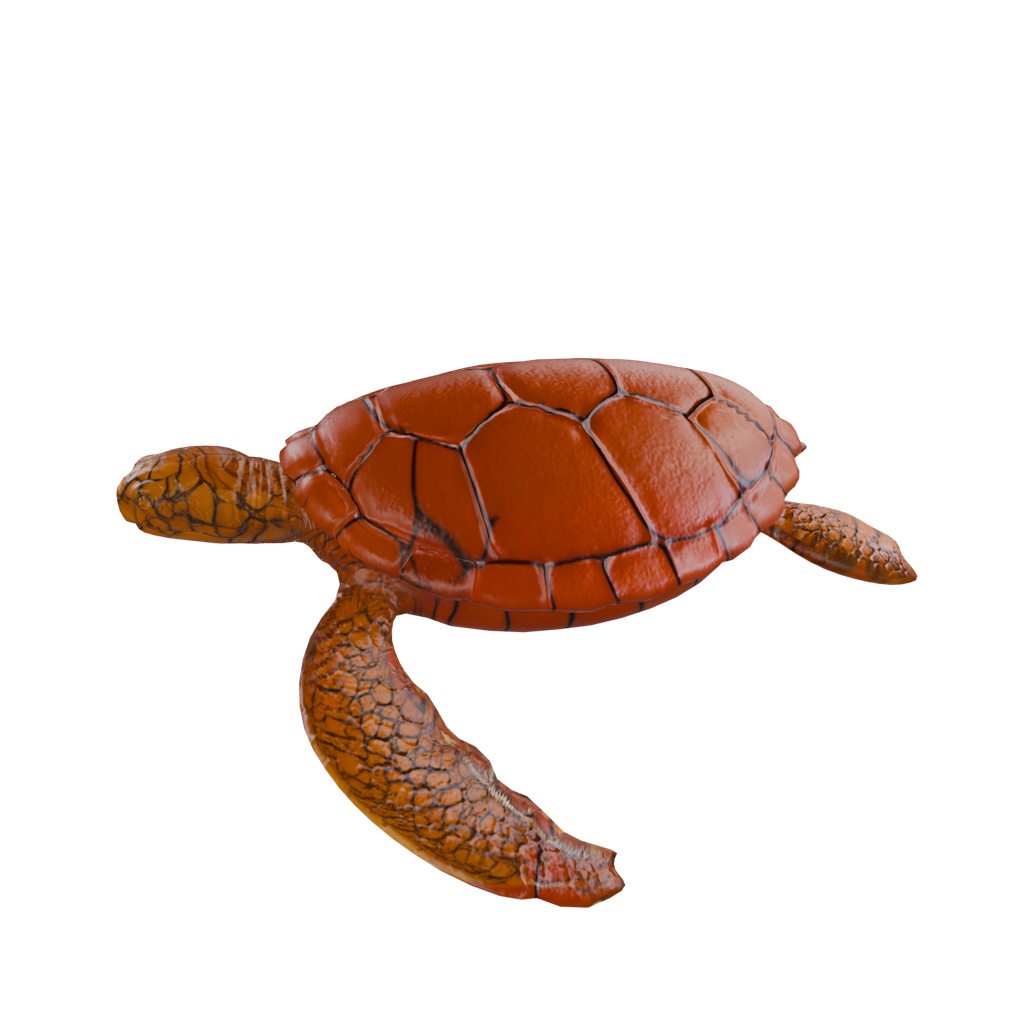} &
      \includegraphics[trim={0cm 0cm 0cm 0cm}, clip, width=0.121\linewidth, valign=m]{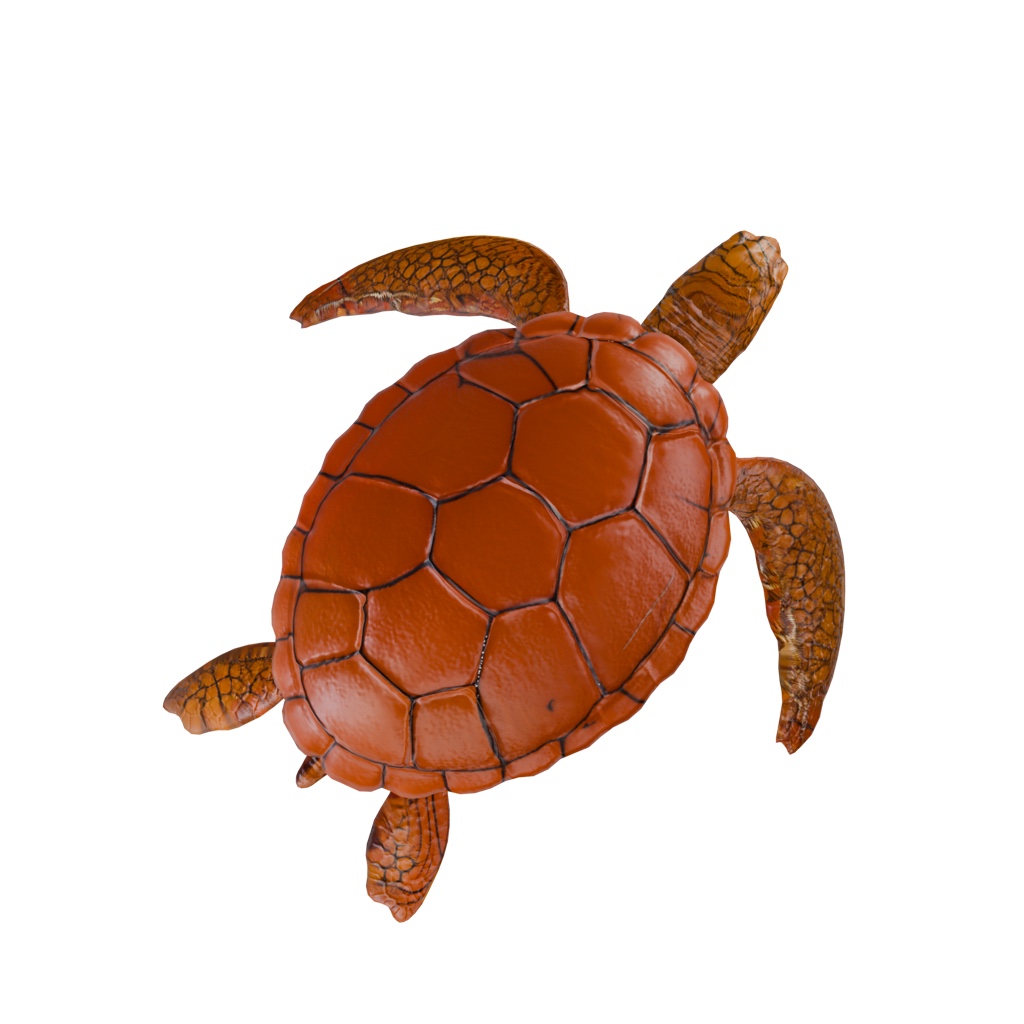} \\
    
      \includegraphics[trim={0cm 0cm 0cm 0cm}, clip, width=\normalswidth, valign=m]{img/normals/fire_hydrant.jpg} &
      \includegraphics[trim={0cm 0cm 0cm 0cm}, clip, width=\viewwidth, valign=m]{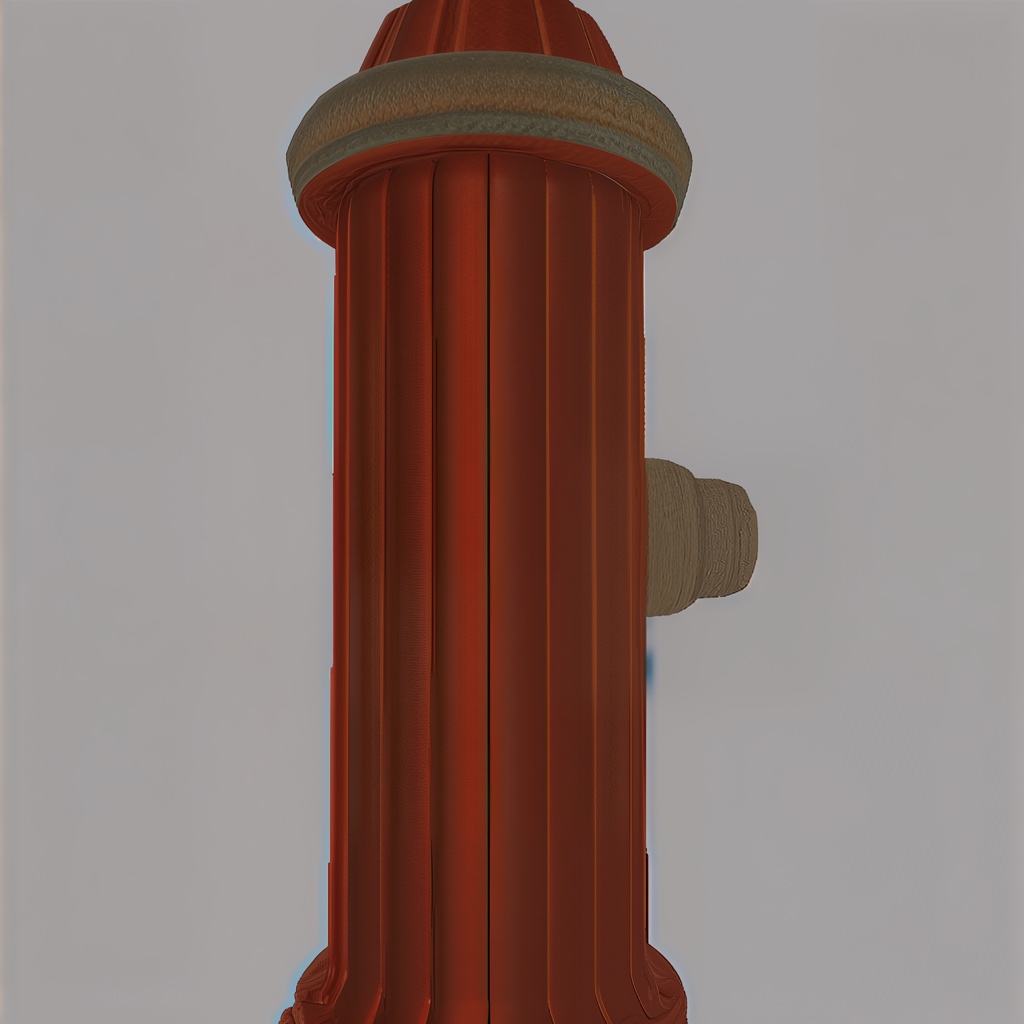} &
      \includegraphics[trim={0cm 0cm 0cm 0cm}, clip, width=0.121\linewidth, valign=m]{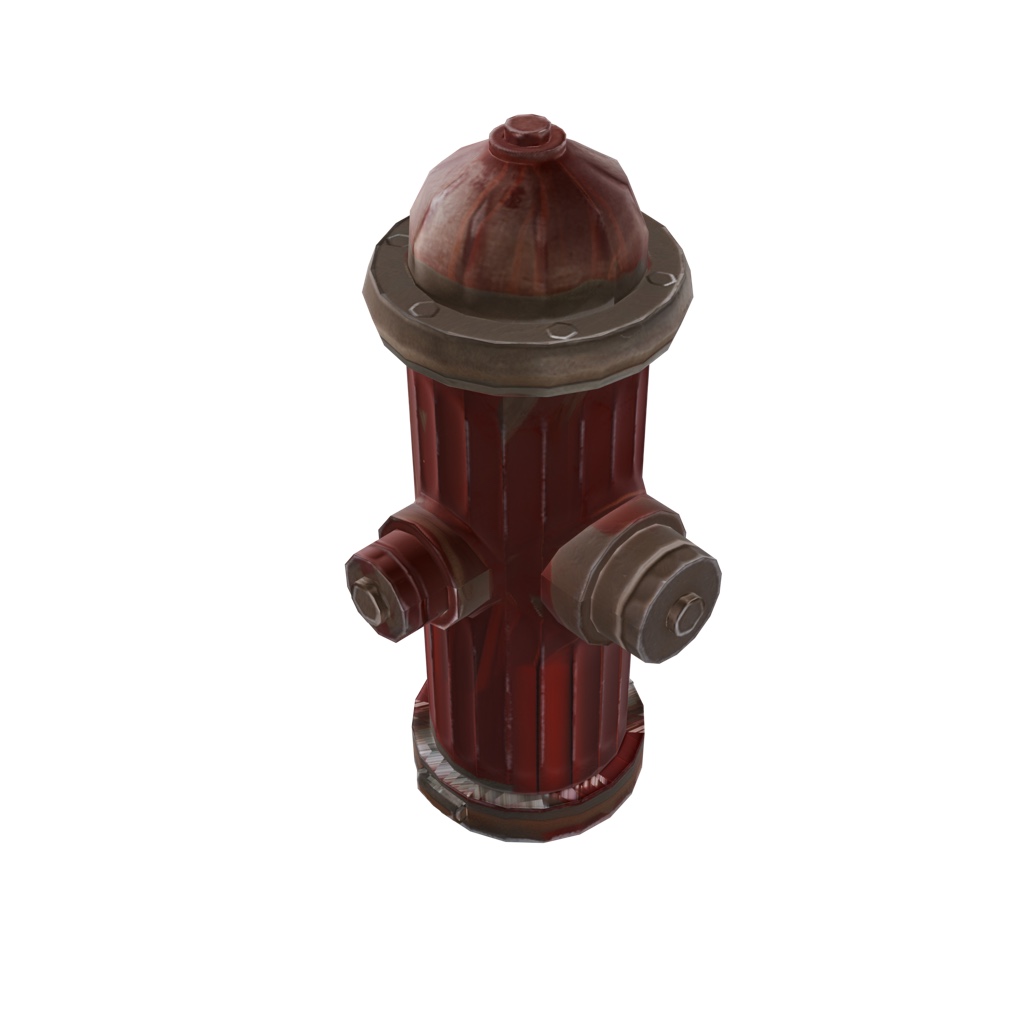} &
      \includegraphics[trim={0cm 0cm 0cm 0cm}, clip, width=0.121\linewidth, valign=m]{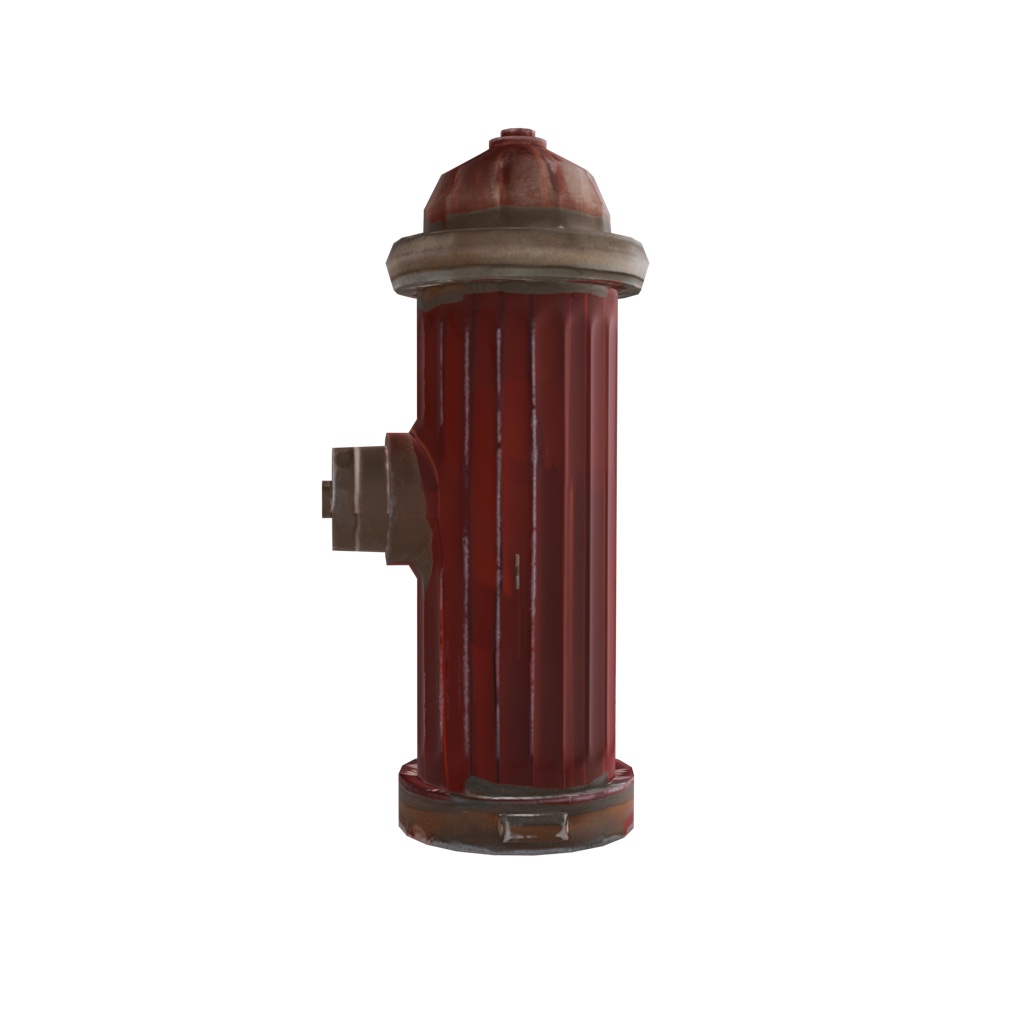} &
      \includegraphics[trim={0cm 0cm 0cm 0cm}, clip, width=\viewwidth, valign=m]{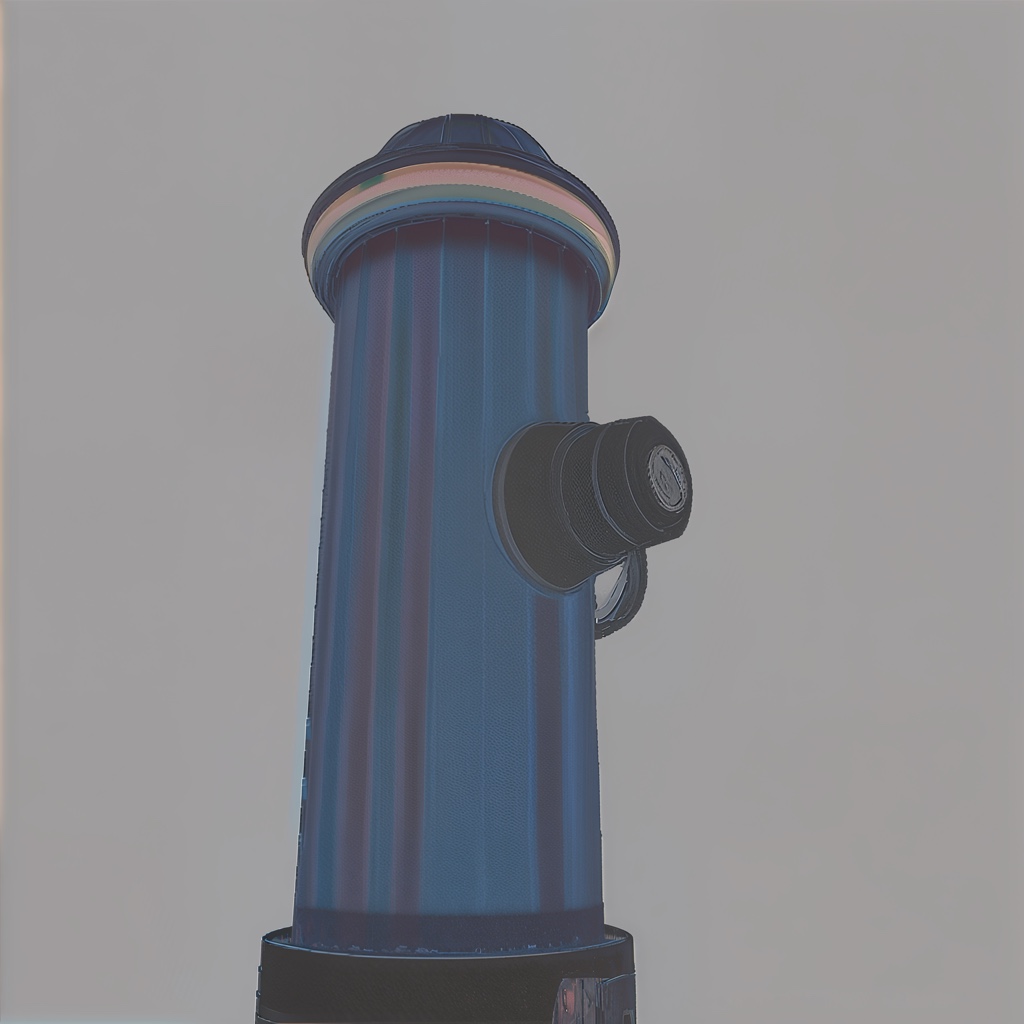} &
      \includegraphics[trim={0cm 0cm 0cm 0cm}, clip, width=0.121\linewidth, valign=m]{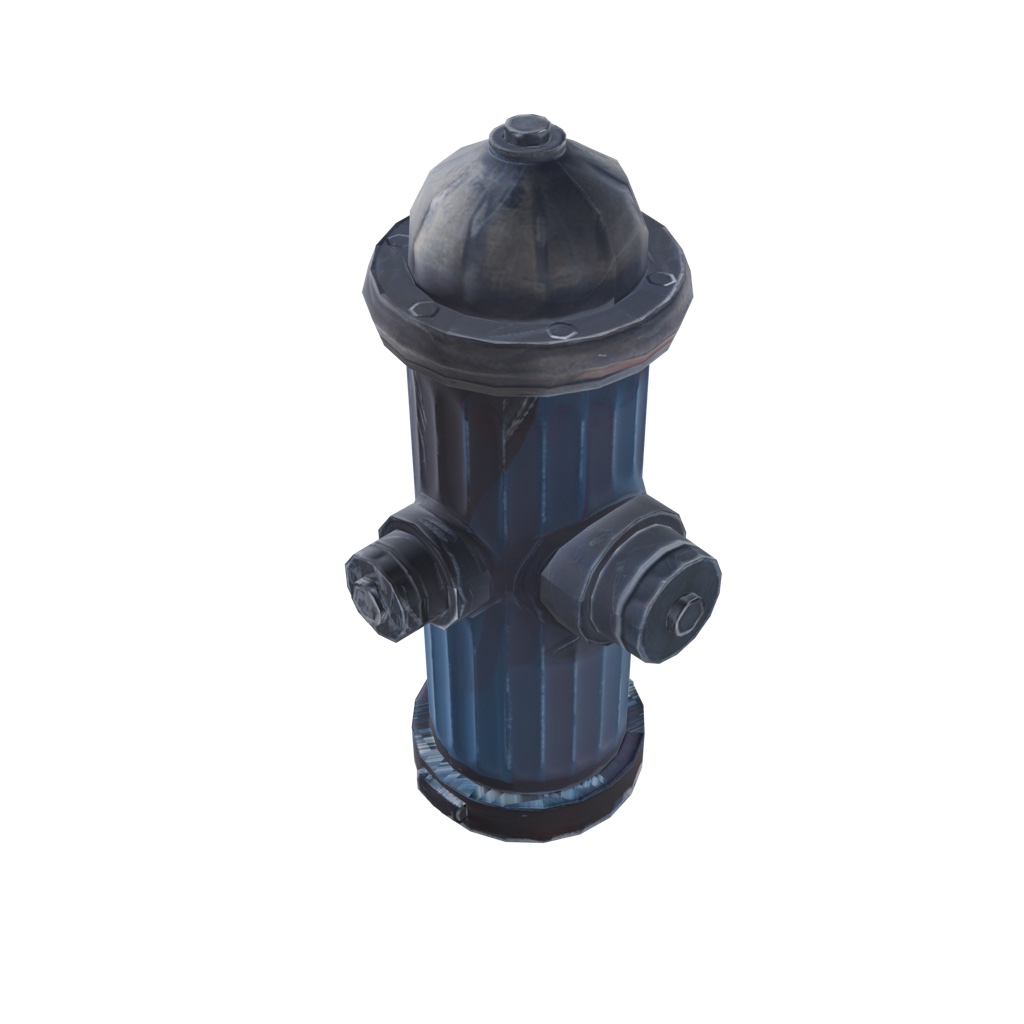} &
      \includegraphics[trim={0cm 0cm 0cm 0cm}, clip, width=0.121\linewidth, valign=m]{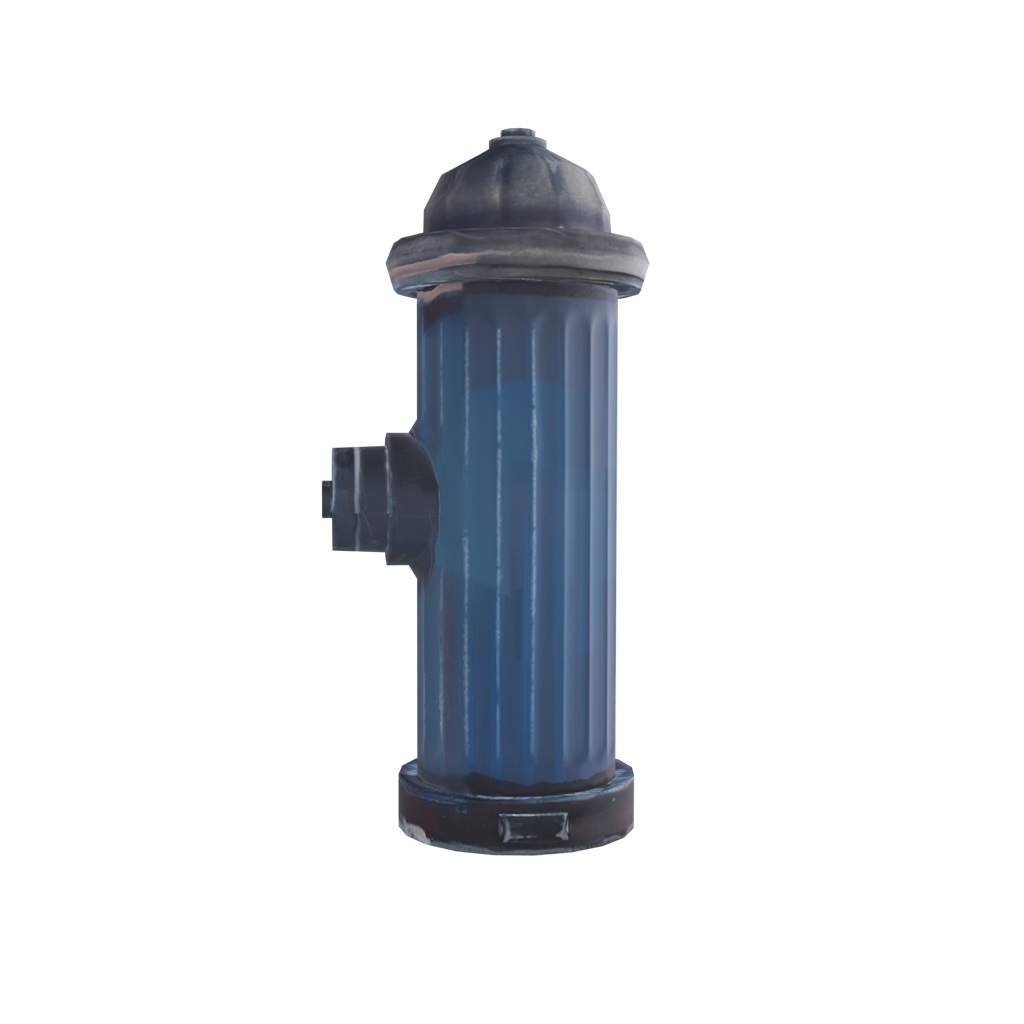} \\

      \includegraphics[trim={0cm 0cm 0cm 0cm}, clip, width=\normalswidth, valign=m]{img/normals/dirty_tire.jpg} &
      \includegraphics[trim={0cm 0cm 0cm 0cm}, clip, width=\viewwidth, valign=m]{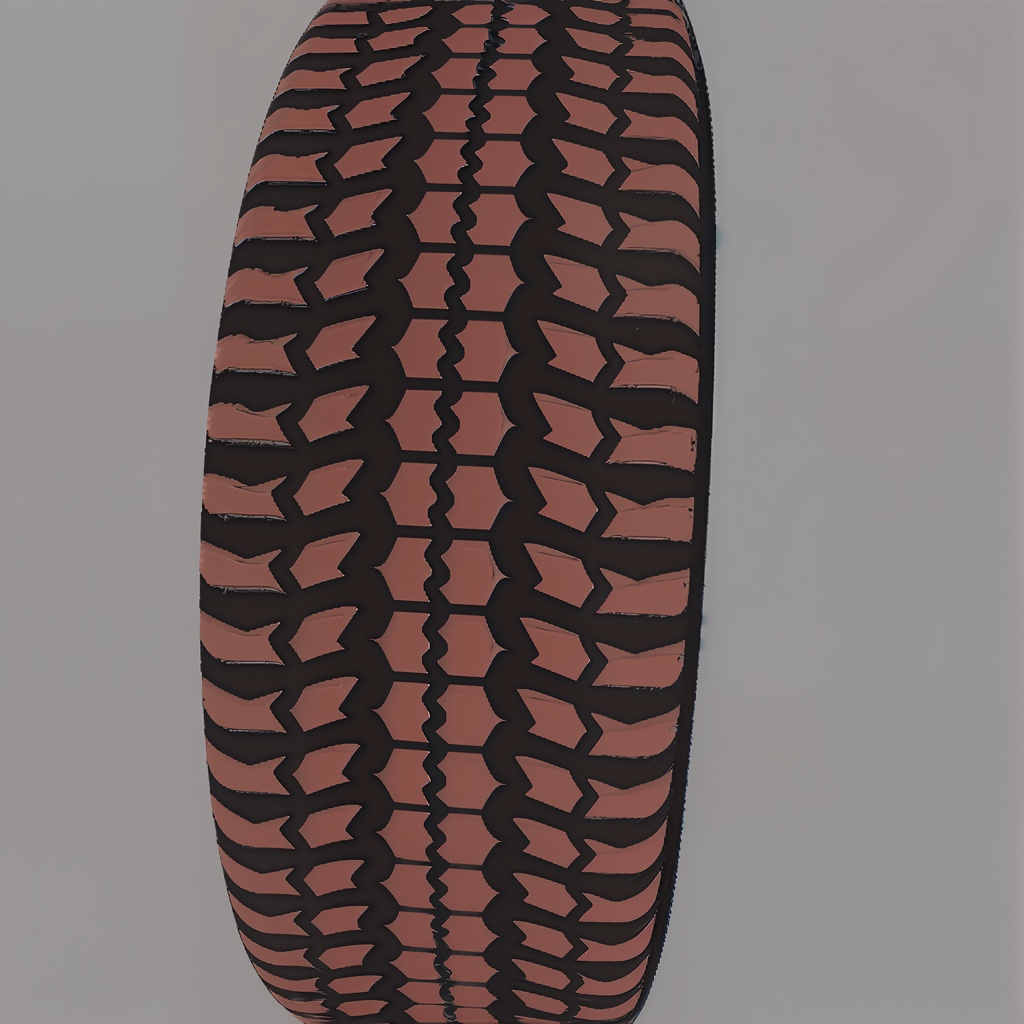} &
      \includegraphics[trim={0cm 0cm 0cm 0cm}, clip, width=0.121\linewidth, valign=m]{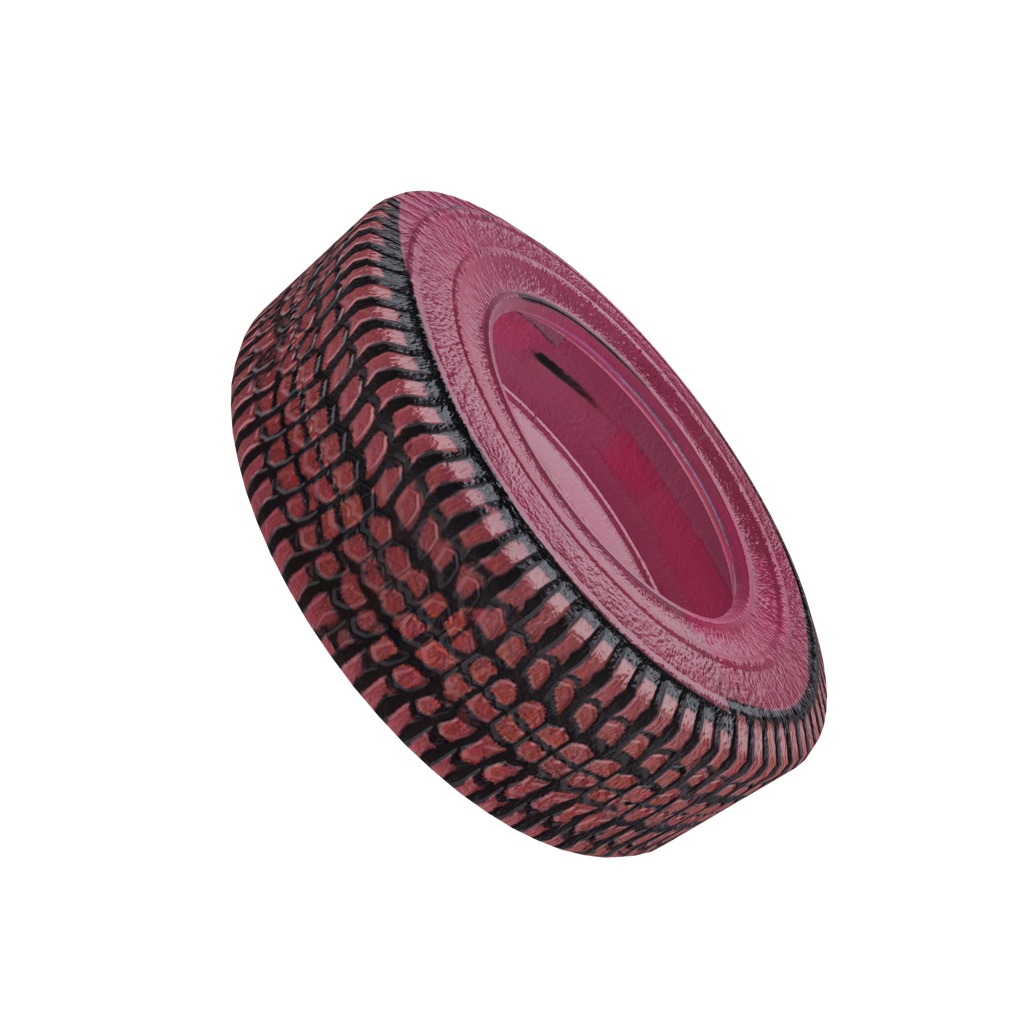} &
      \includegraphics[trim={0cm 0cm 0cm 0cm}, clip, width=0.121\linewidth, valign=m]{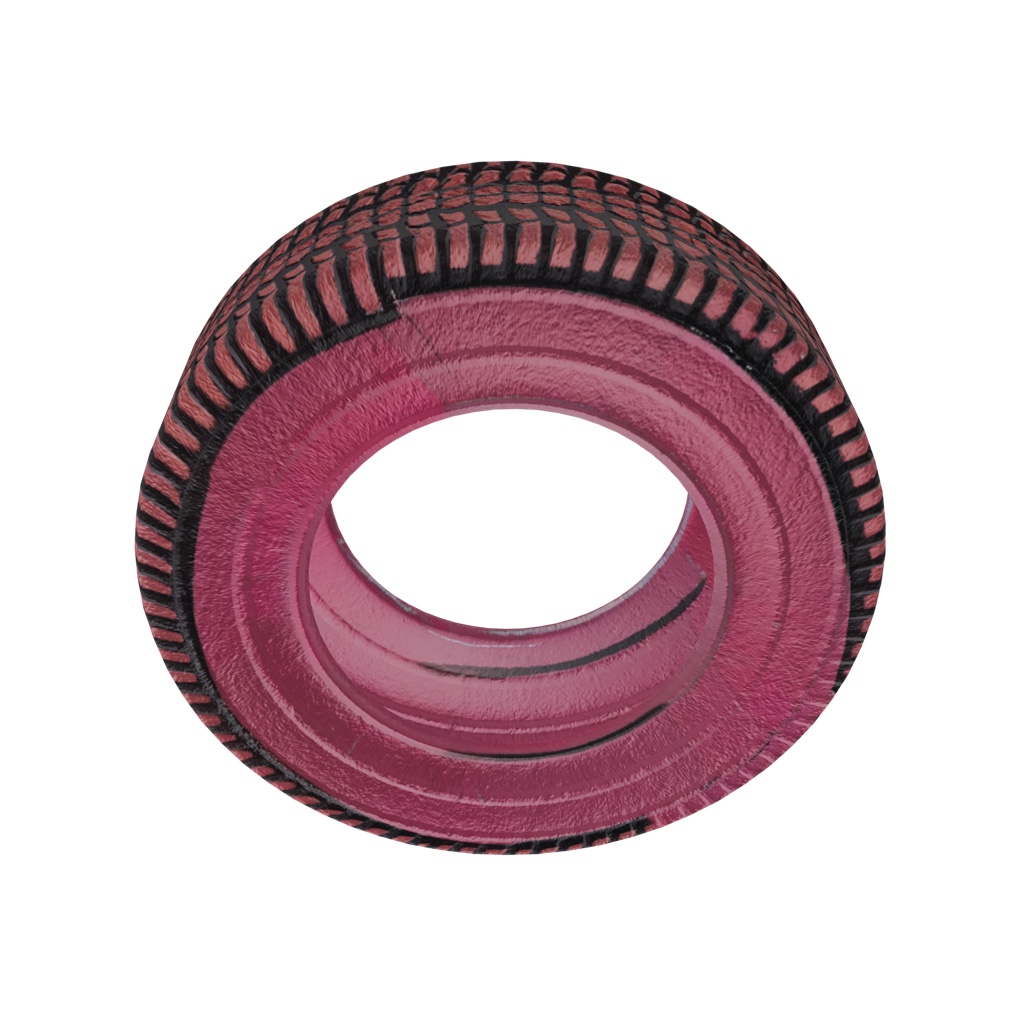} &
      \includegraphics[trim={0cm 0cm 0cm 0cm}, clip, width=\viewwidth, valign=m]{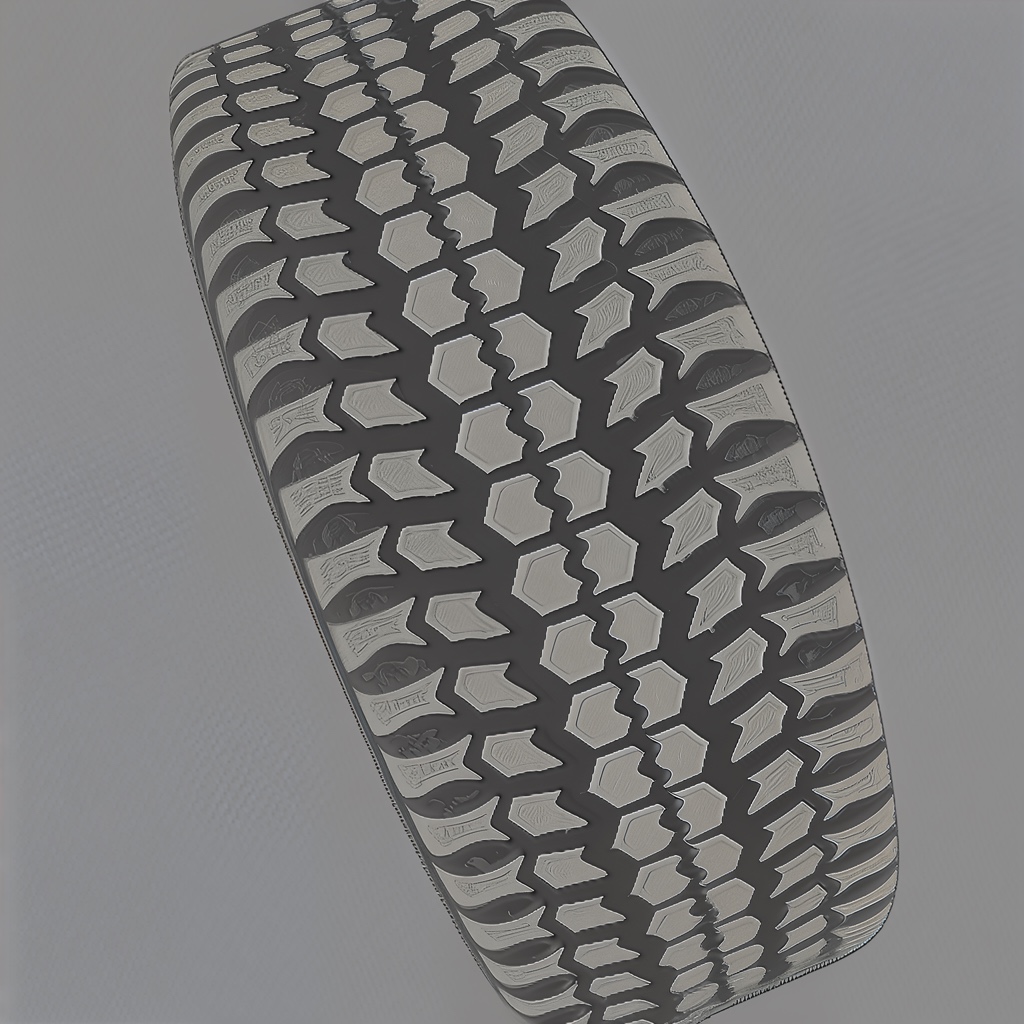} &
      \includegraphics[trim={0cm 0cm 0cm 0cm}, clip, width=0.121\linewidth, valign=m]{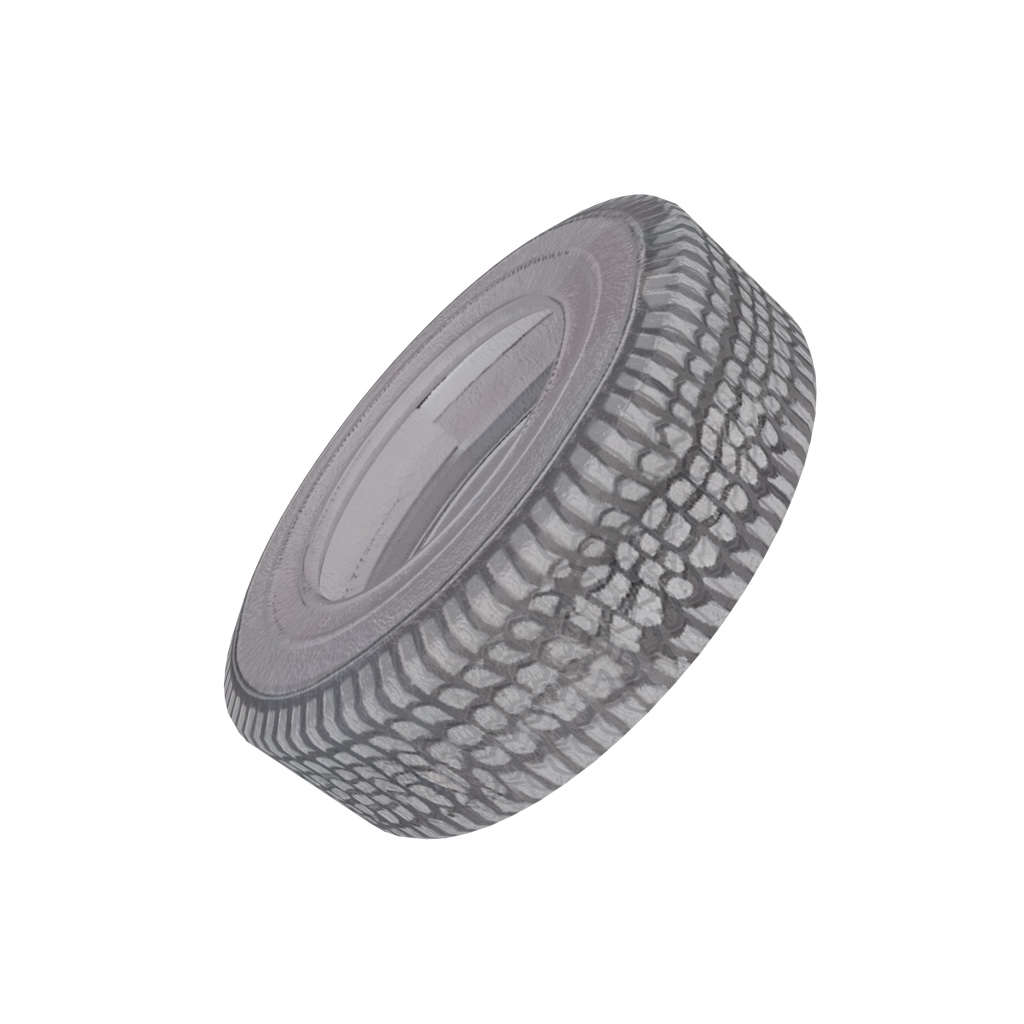} &
      \includegraphics[trim={0cm 0cm 0cm 0cm}, clip, width=0.121\linewidth, valign=m]{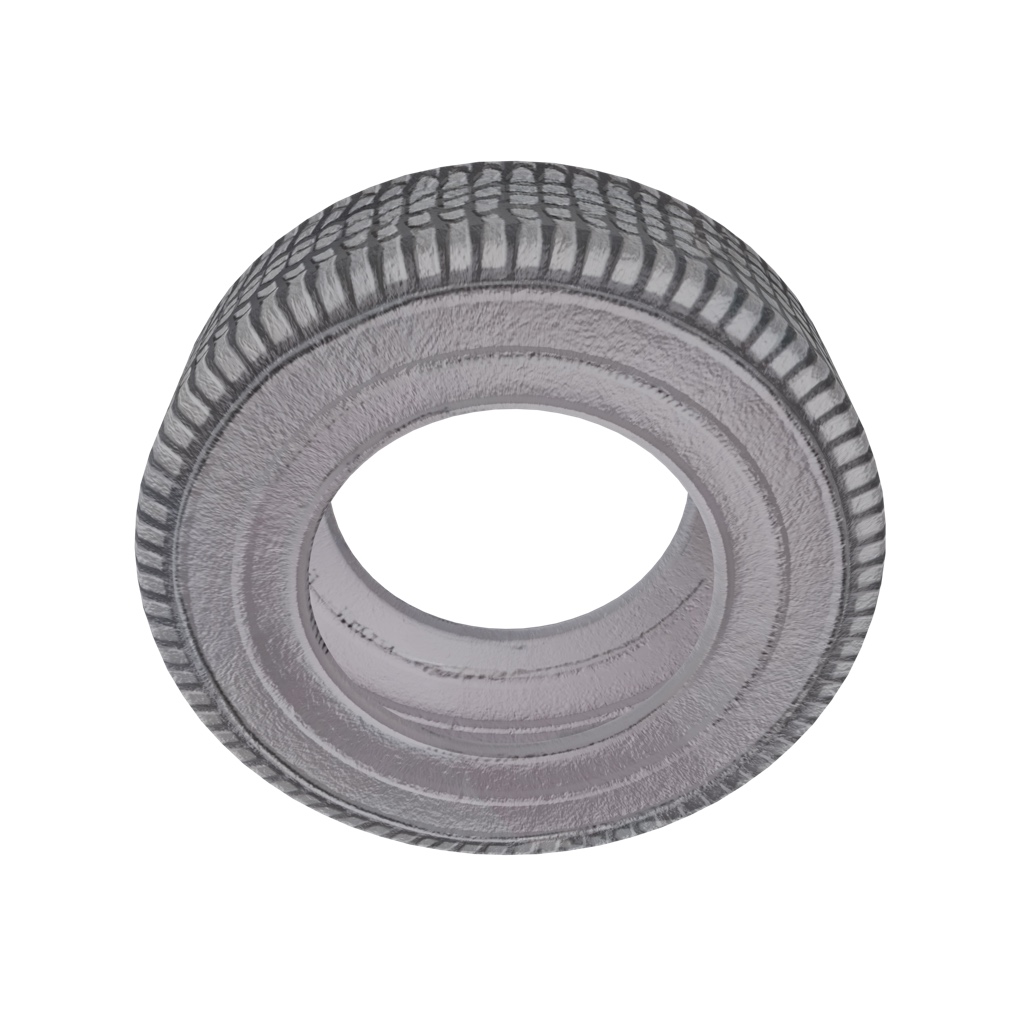} \\

      \includegraphics[trim={0cm 0cm 0cm 0cm}, clip, width=\normalswidth, valign=m]{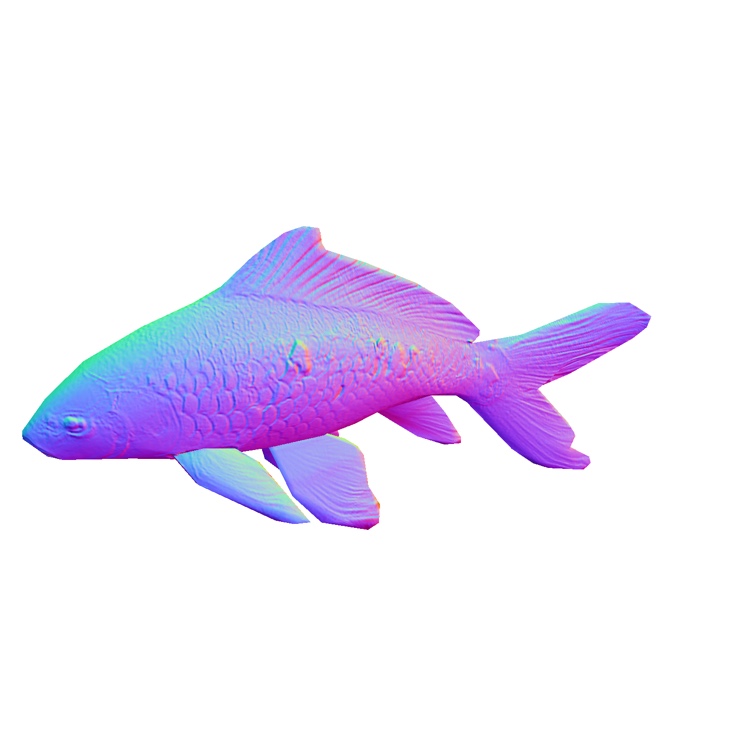} &
      \includegraphics[trim={0cm 0cm 0cm 0cm}, clip, width=\viewwidth, valign=m]{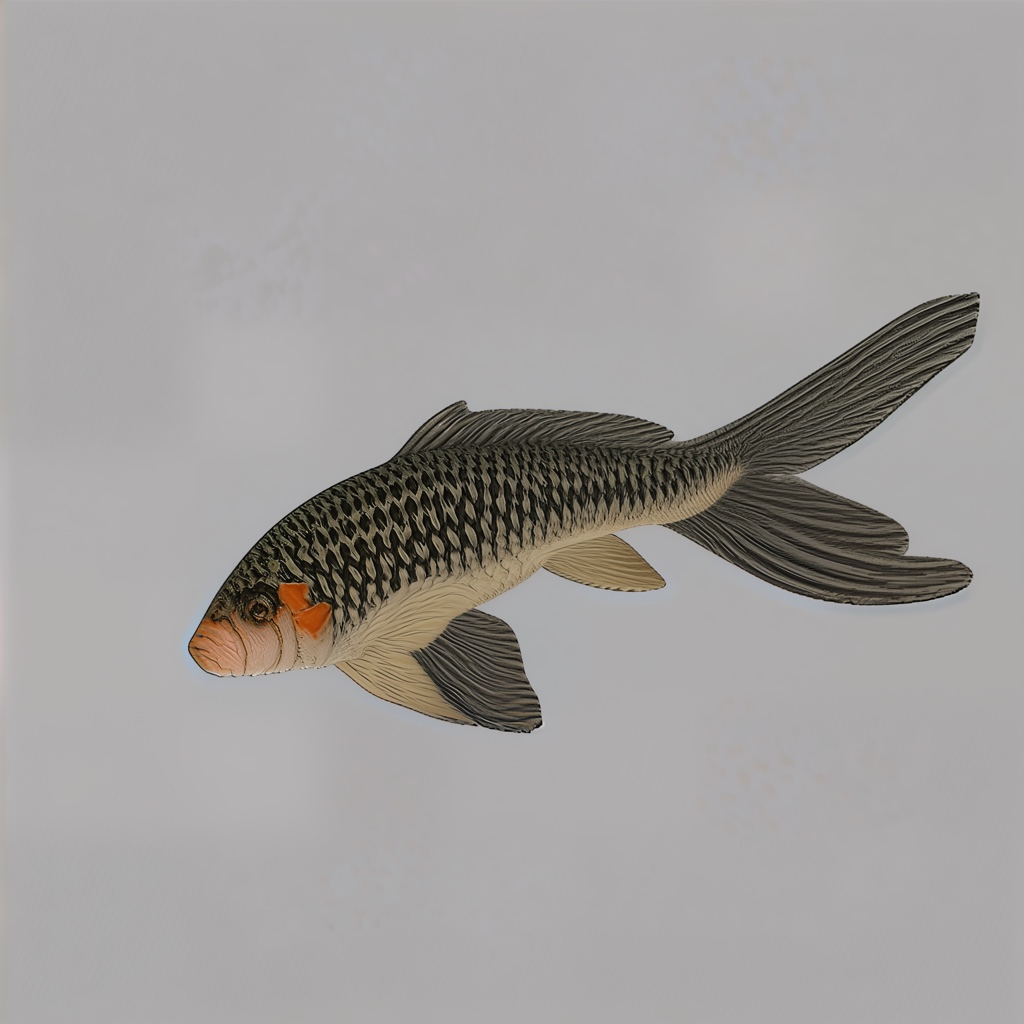} &
      \includegraphics[trim={0cm 8cm 0cm 8cm}, clip, width=0.121\linewidth, valign=m]{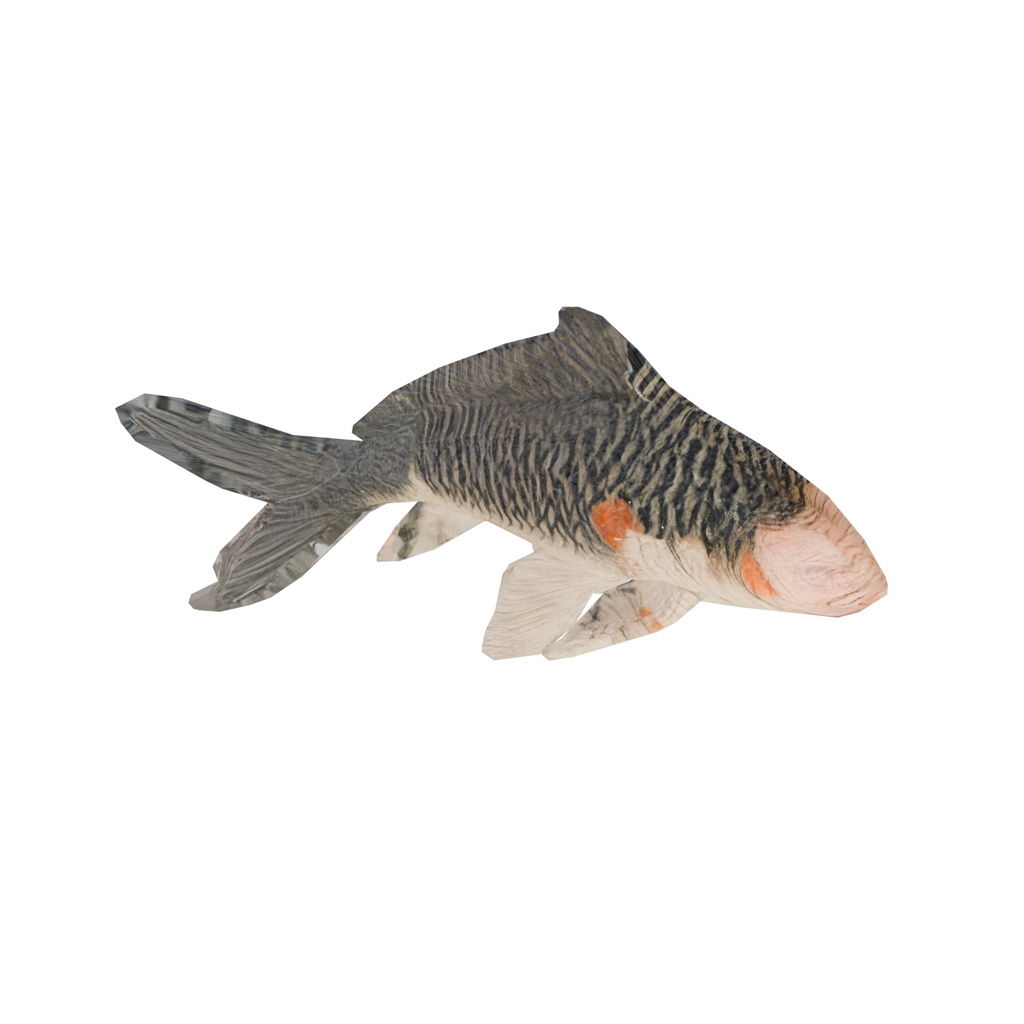} &
      \includegraphics[trim={0cm 8cm 0cm 8cm}, clip, width=0.121\linewidth, valign=m]{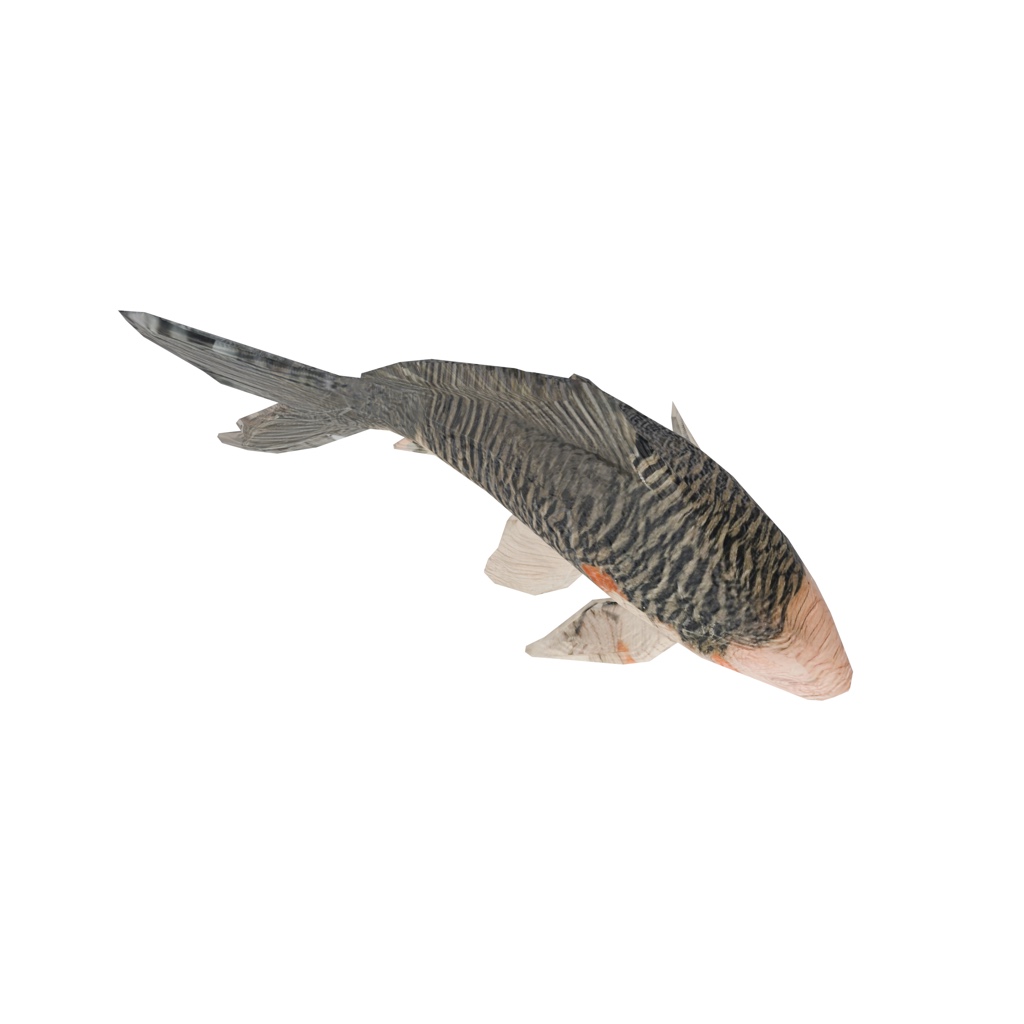} &
      \includegraphics[trim={0cm 0cm 0cm 0cm}, clip, width=\viewwidth, valign=m]{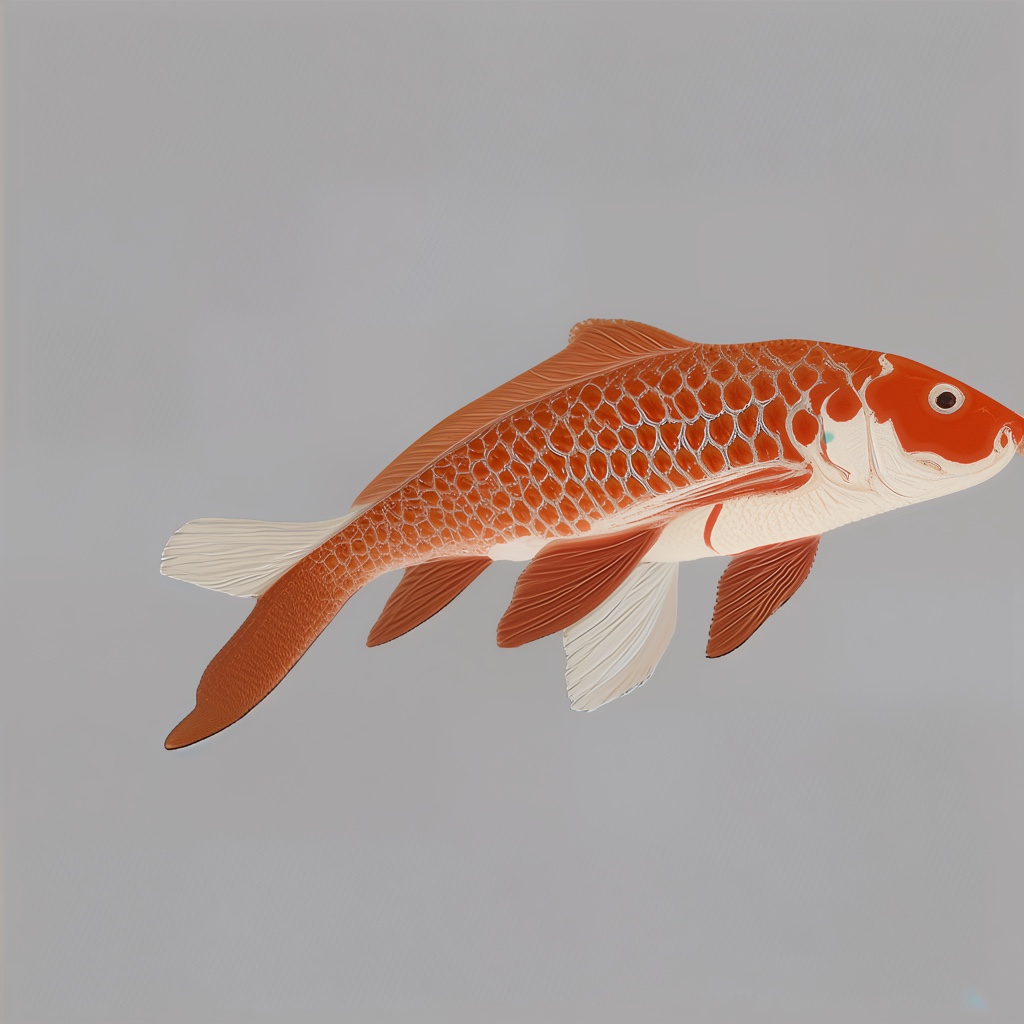} &
      \includegraphics[trim={0cm 8cm 0cm 8cm}, clip, width=0.121\linewidth, valign=m]{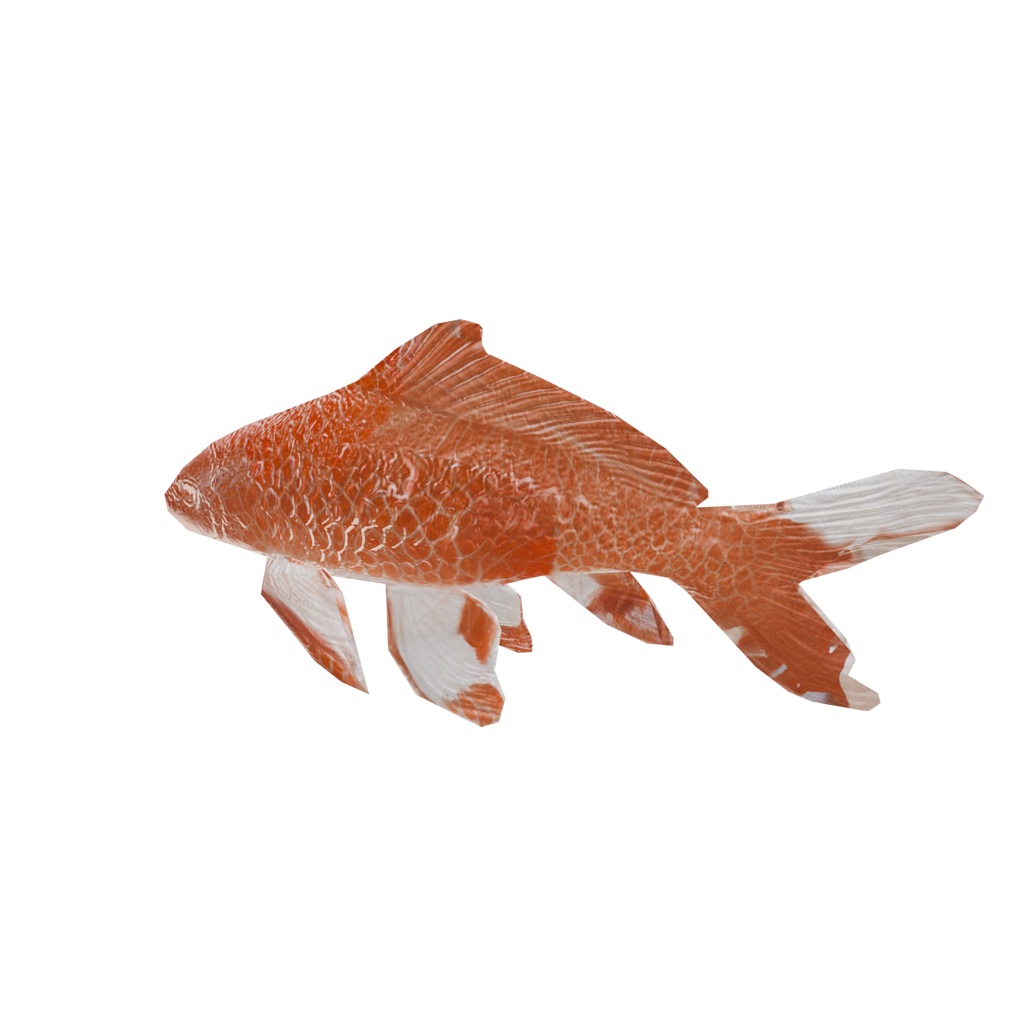} &
      \includegraphics[trim={0cm 8cm 0cm 8cm}, clip, width=0.121\linewidth, valign=m]{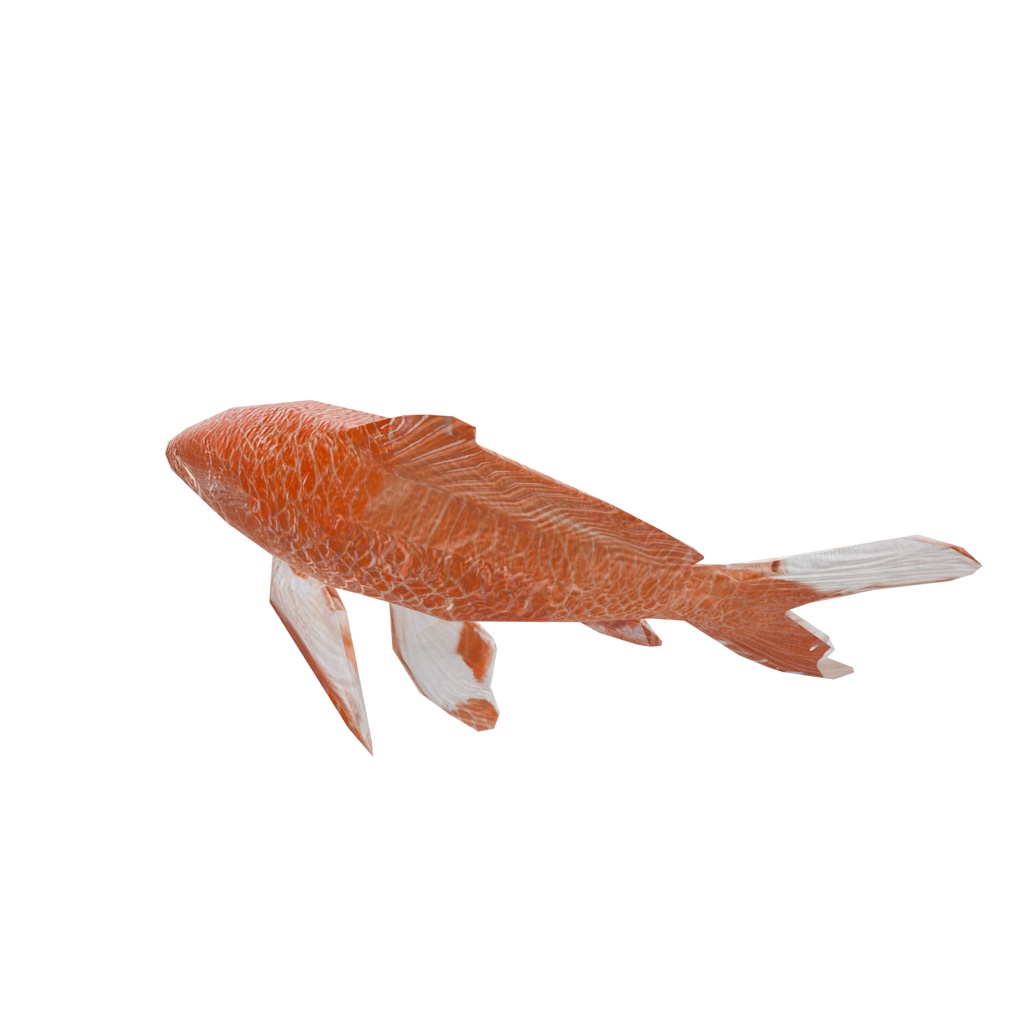} \\

      \includegraphics[trim={0cm 0cm 0cm 0cm}, clip, width=\normalswidth, valign=m]{img/normals/rusty_barrel_metal.jpg} &
      \includegraphics[trim={0cm 0cm 0cm 0cm}, clip, width=\viewwidth, valign=m]{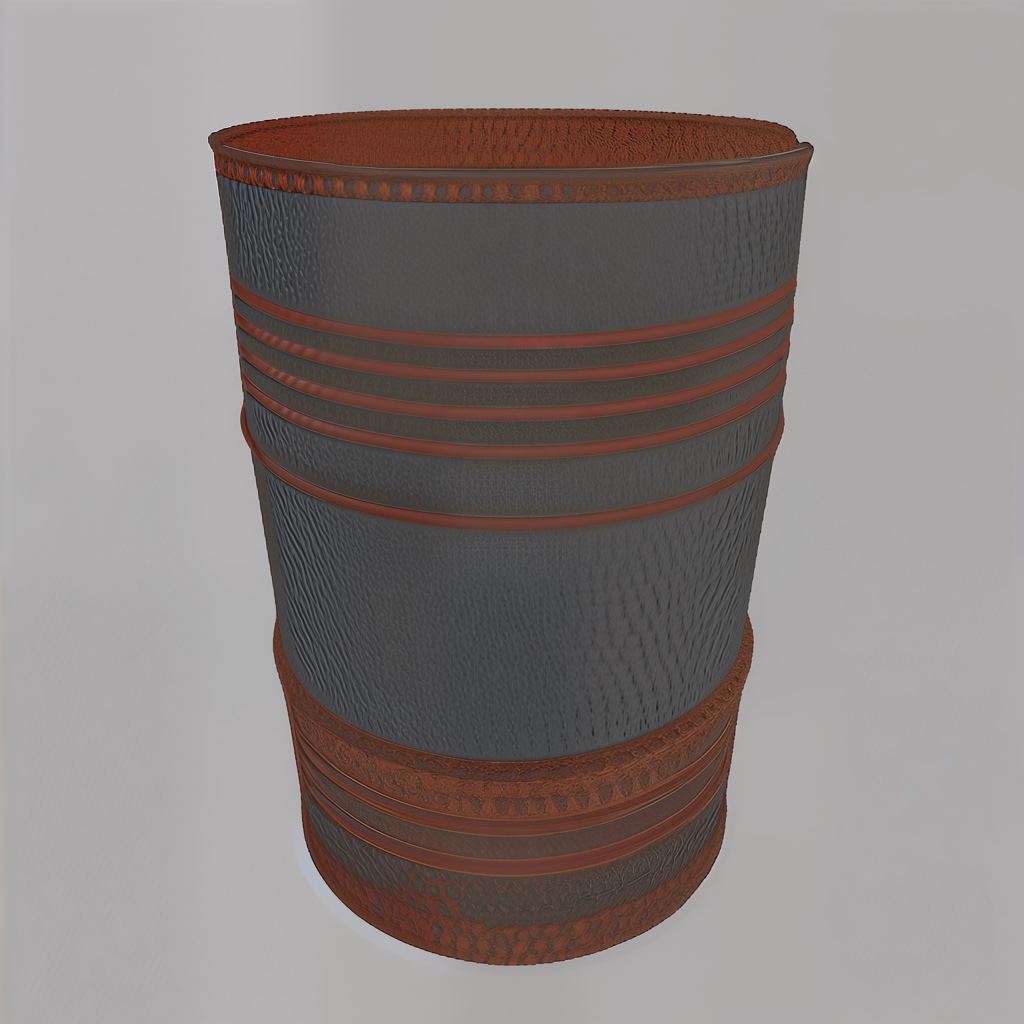} &
      \includegraphics[trim={0cm 0cm 0cm 0cm}, clip, width=0.121\linewidth, valign=m]{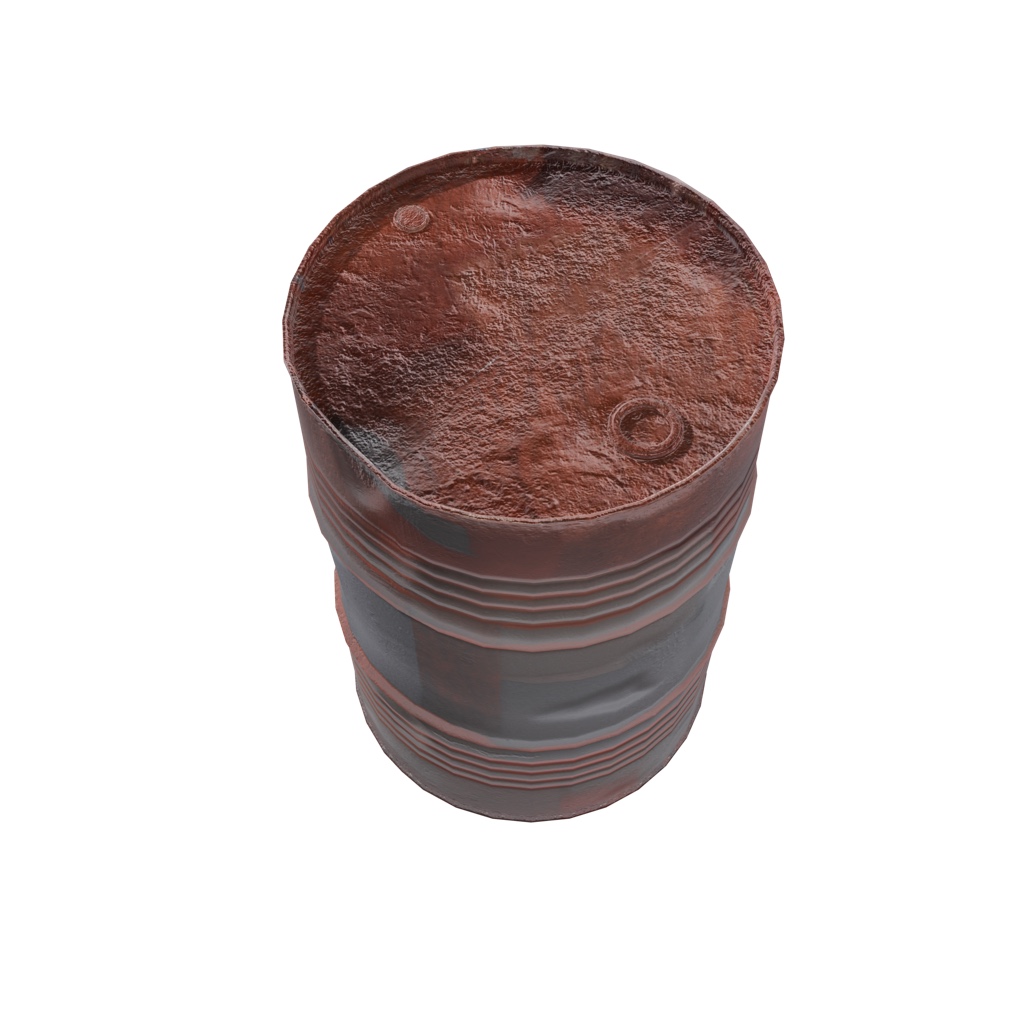} &
      \includegraphics[trim={0cm 0cm 0cm 0cm}, clip, width=0.121\linewidth, valign=m]{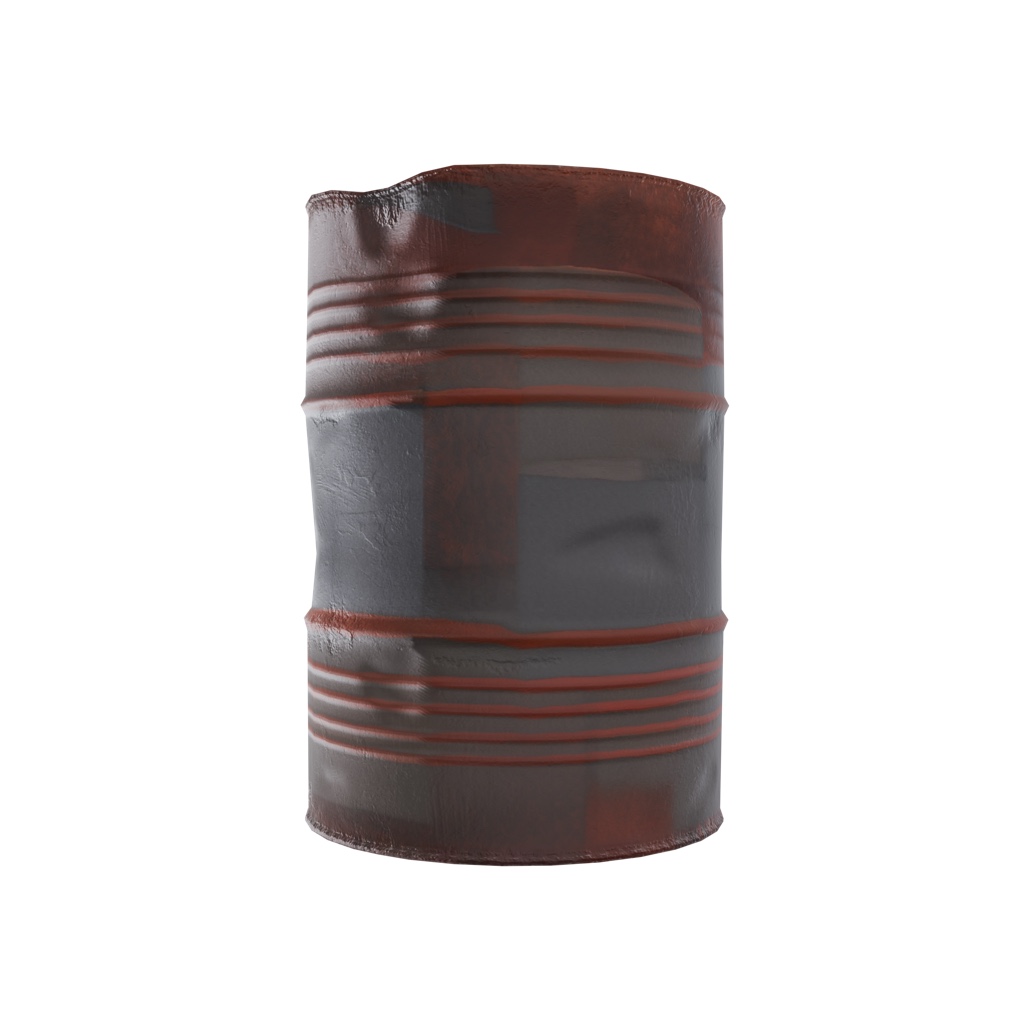} &
      \includegraphics[trim={0cm 0cm 0cm 0cm}, clip, width=\viewwidth, valign=m]{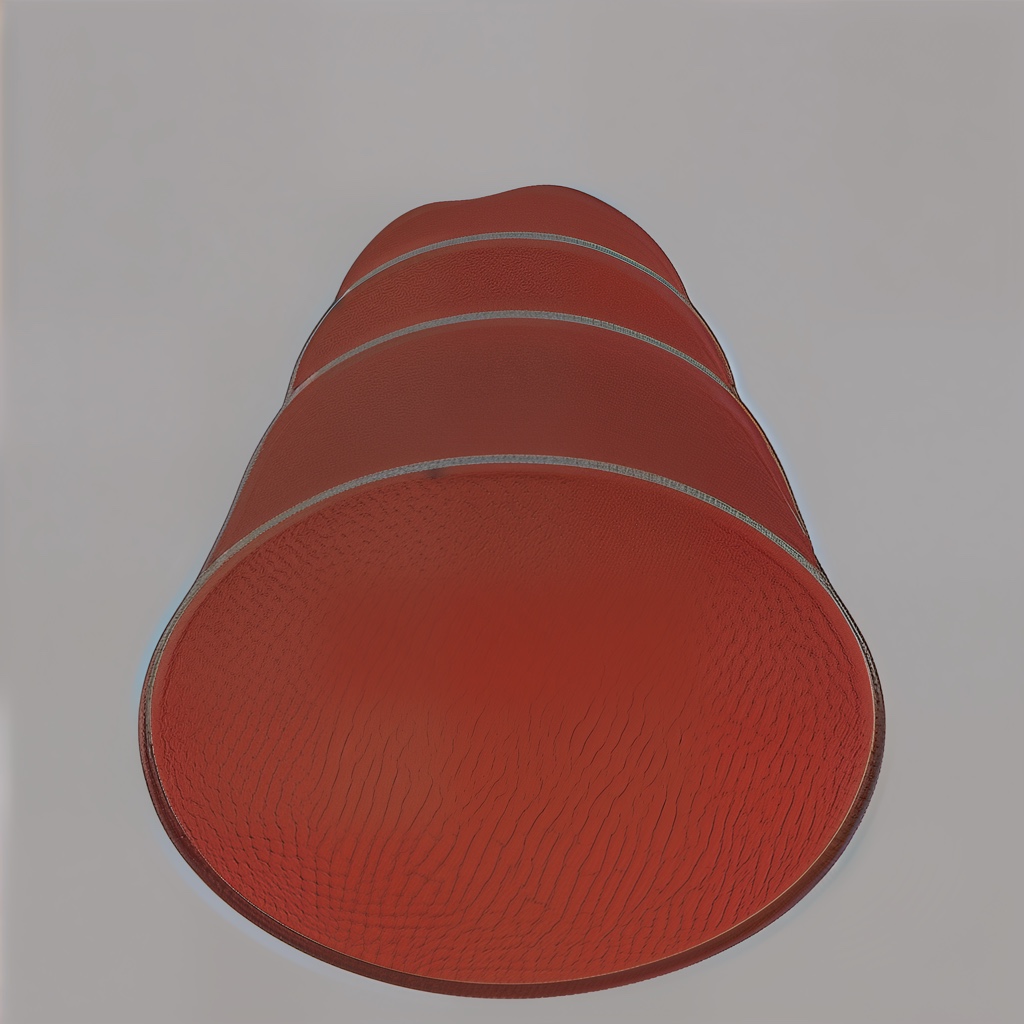} &
      \includegraphics[trim={0cm 0cm 0cm 0cm}, clip, width=0.121\linewidth, valign=m]{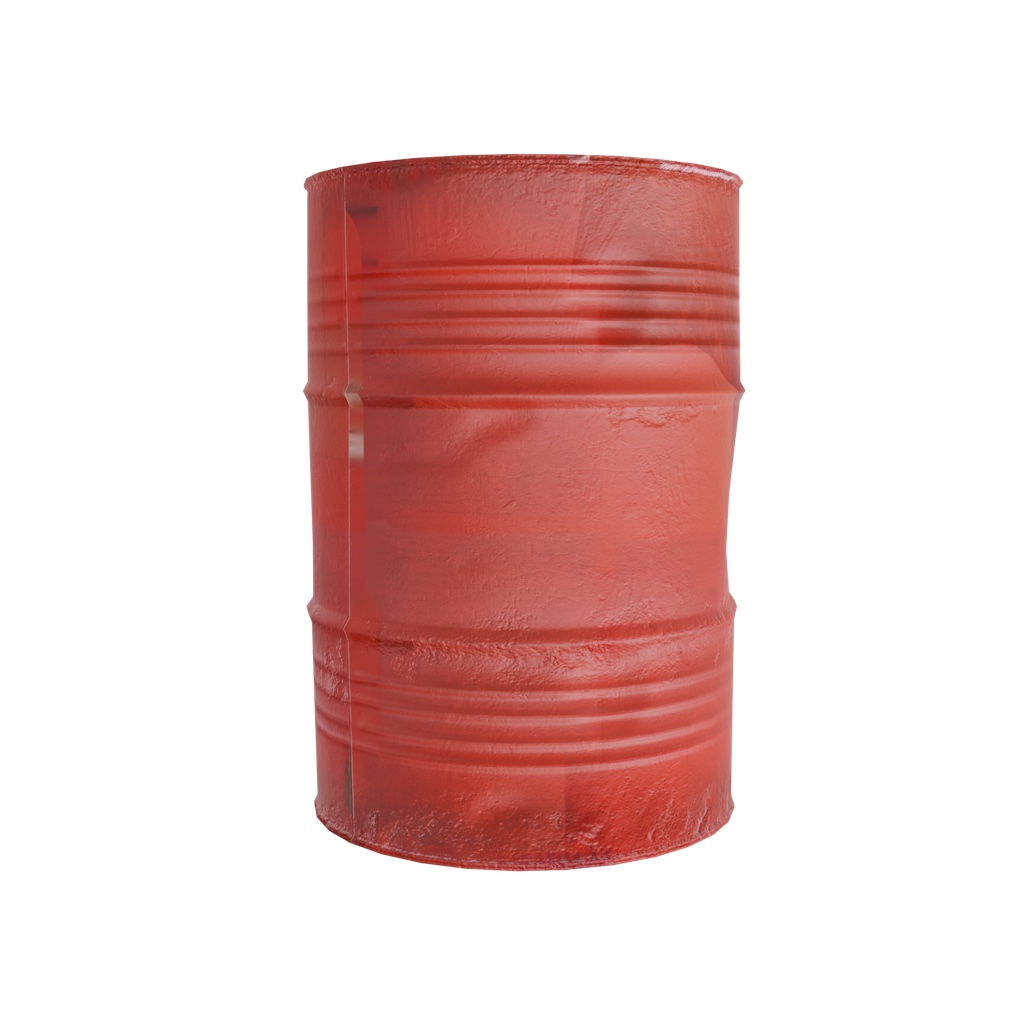} &
      \includegraphics[trim={0cm 0cm 0cm 0cm}, clip, width=0.121\linewidth, valign=m]{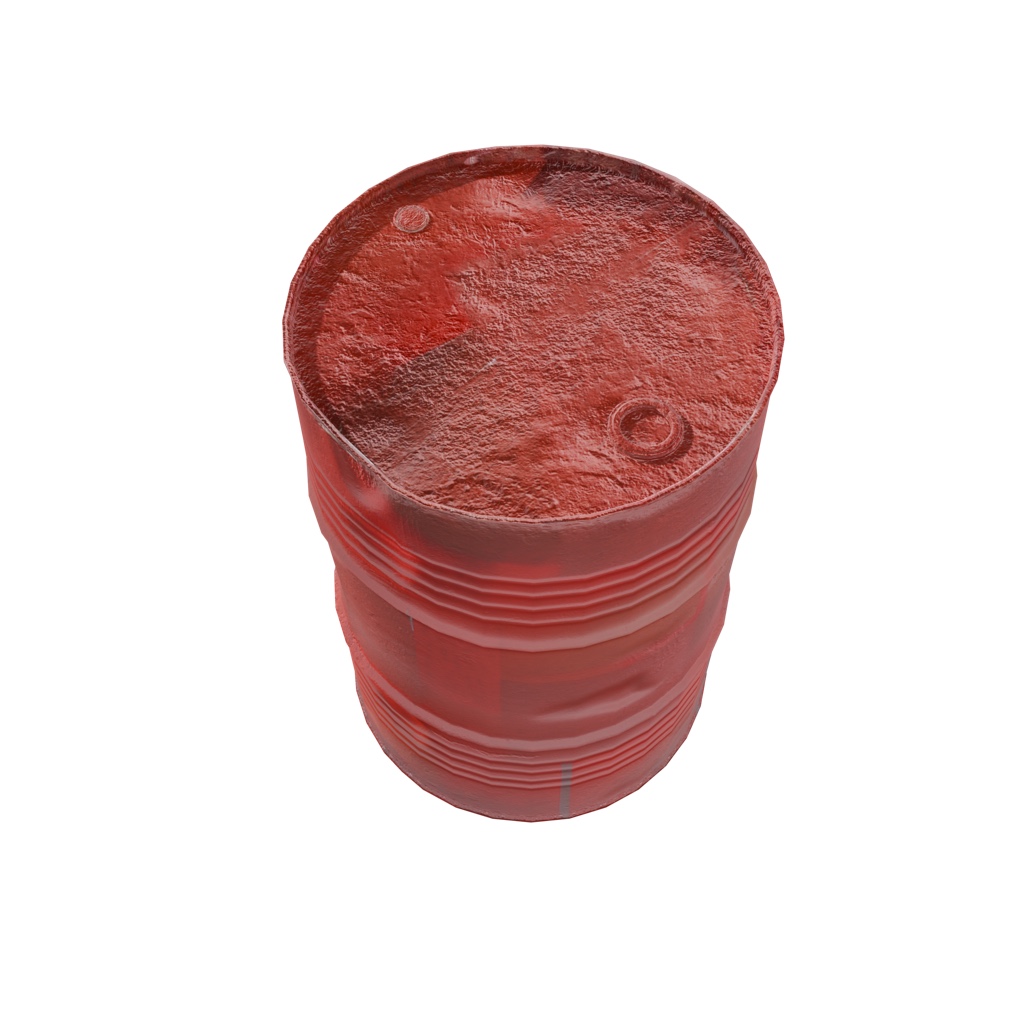} \\
  
        \end{tabular}
    } \\
	\caption{We show more automatic texture completion results with GLOSS. We show views of the 3D model that are opposite to the conditional view. We also show the textured 3D model from a view other than these two.}
	\label{fig:supp_completion_gallery}
\end{figure*}

\end{document}